\documentclass[11pt]{book}
\usepackage[utf8]{inputenc}

\title{Nonlinear Schr\"{o}dinger equations with trapping potentials in higher dimensions}
\author{Filip Ficek}
\date{September 2020}

\usepackage[b5paper, top = 0.9 in, bottom = 0.9 in, left = 0.9 in, right=0.9 in]{geometry}

\usepackage{graphicx} 
\usepackage{amsmath} 
\usepackage{amssymb}
\usepackage{amsthm}
\usepackage{enumitem}
\usepackage[overload]{empheq}
\usepackage{mathrsfs}  
\usepackage{subcaption}
\usepackage[T1]{fontenc}
\usepackage{pdfpages}

\usepackage{appendix}

\usepackage{changepage} 
\newcommand{\marwidFF}{30pt} 
\newcommand{\breakFF}{5pt} 

\newtheorem{thm}{Theorem}[chapter]

\newtheorem{lem}{Lemma}[chapter]

\theoremstyle{remark}

\newcommand{\arXiv}[1]{\href{http://www.arXiv.org/abs/#1}{arXiv:#1}}
\usepackage[colorlinks=true, linkcolor=blue, bookmarks=true]{hyperref}

\begin{document}
\pagenumbering{gobble}

\includepdf{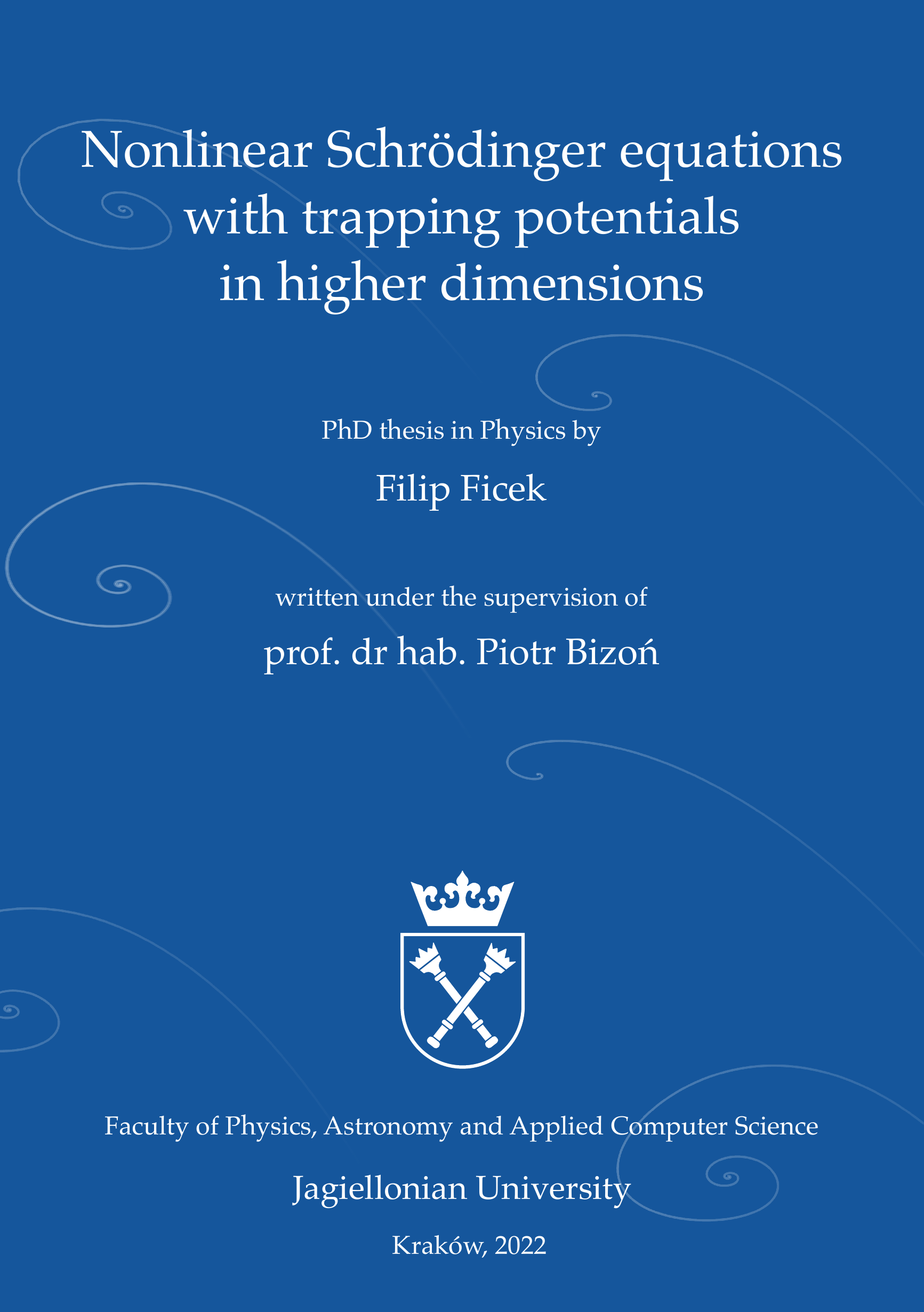}

\newpage\null\thispagestyle{empty}\newpage

\begin{center}
    {\huge \textbf{Oświadczenie}}
\end{center}
\vspace{1.5cm}

Ja niżej podpisany Filip Ficek (nr indeksu:\ 1087631) doktorant Wydziału Fizyki, Astronomii i Informatyki Stosowanej Uniwersytetu Jagiellońskiego oświadczam, że przedłożona przeze mnie rozprawa doktorska pt. ,,Nonlinear Schr\"{o}dinger equations with trapping potentials in higher dimensions’’ jest oryginalna i przedstawia wyniki badań wykonanych przeze mnie osobiście, pod kierunkiem prof.\ dr.\ hab.\ Piotra Bizonia. Pracę napisałem samodzielnie.
\vspace{0.5cm}

Oświadczam, że moja rozprawa doktorska została opracowana zgodnie z Ustawą o prawie autorskim i prawach pokrewnych z dnia 4 lutego 1994 r.\ (Dziennik Ustaw 1994 nr 24 poz.\ 83 wraz z późniejszymi zmianami).

\vspace{0.5cm}	
Jestem świadom, że niezgodność niniejszego oświadczenia z prawdą ujawniona w dowolnym czasie, niezależnie od skutków prawnych wynikających z ww.\ ustawy, może spowodować unieważnienie stopnia nabytego na podstawie tej rozprawy.
\vspace{1.5cm}

\noindent Kraków, dnia ……………………\qquad\qquad\qquad ……………………\\
\begin{flushright}
podpis doktoranta\quad\, \textcolor{white}{.}
\end{flushright}

\newpage\null\thispagestyle{empty}\newpage

\begin{center}
    {\large \textbf{Streszczenie}}
\end{center}

Nieliniowe równania Schr\"{o}dingera od strony matematycznej zwykle badane są za pomocą metod wariacyjnych, które zdają się zawodzić w wyższych wymiarach. Niniejsza rozprawa próbuje obejść ten problem poprzez skupienie się na rozwiązaniach sferycznie symetrycznych, co pozwala na zastosowanie klasycznych metod teorii równań różniczkowych zwyczajnych i układów dynamicznych. Zaprezentowane wyniki dotyczą między innymi istnienia i jednoznaczności stanów stacjonarnych, ich częstotliwości oraz stabilności. Opisana jest także dynamika w przybliżeniu układu rezonansowego. Główny nacisk został położony na równanie Schr\"{o}dingera-Newtona-Hooke'a, które przedstawione jest jako nierelatywistyczna granica niewielkich zaburzeń czasoprzestrzeni anty-de Sittera.
\vspace{2.5cm}

\begin{center}
    {\large \textbf{Abstract}}
\end{center}
From the mathematical side, nonlinear Schr\"{o}dinger equations are usually investigated via variational methods, that cease to work in higher dimensions. This thesis tries to overcome this problem by focusing on spherically symmetric solutions. Then, one can use classical methods coming from the fields of ordinary differential equations and dynamical systems. The results presented here include existence and uniqueness of the stationary solutions, their frequency, and stability. The dynamical properties of the resonant approximation are also explored. The main focus is given to the Schr\"{o}dinger-Newton-Hooke equations that is shown to be a nonrelativistic limit of perturbations of the anti-de Sitter spacetime.

\newpage\null\thispagestyle{empty}\newpage
\vspace*{50mm}
{\LARGE 
\textit{\noindent Everything in the world is exactly the same.}}\\
{\LARGE \null\hfill Ye}

\newpage\null\thispagestyle{empty}\newpage

\pagenumbering{arabic}
\setcounter{page}{7}
\tableofcontents

\newpage
This thesis is based on the following publications (although it also contains a number of results that are yet to be published):
\begin{enumerate}
    \item[{[F1]}] P.~Bizo\'n, O.~Evnin, F.~Ficek, {\em A nonrelativistic limit for AdS perturbations}, Journal of High Energy Physics \textbf{12}, 112 (2018), \arXiv{1810.10574}.
    \item[{[F2]}] P.~Bizo\'n, F.~Ficek, D.~E.~Pelinovsky, S.~Sobieszek, {\em Ground state in the energy super-critical Gross-Pitaevskii equation with a harmonic potential}, Nonlinear Analysis \textbf{210}, 112358 (2021), \arXiv{2009.04929}.
    \item[{[F3]}] F.~Ficek, {\em Schr\"odinger-Newton-Hooke system in higher dimensions: Stationary states}, Physical Review D \textbf{103}, 104062 (2021), \arXiv{2104.00149}.
\end{enumerate}

The following publications were written and published during my PhD studies, however, they are not covered here:
\begin{enumerate}
    \item F.~Ficek, P.~Fadeev, V.~V.~Flambaum, D.~F.~Jackson Kimball, M.~G.~Kozlov, Y.~V.~Stadnik,  D.~Budker, {\em Constraints on exotic spin-dependent interactions between matter and antimatter from antiprotonic helium spectroscopy}, Physical Review Letters \textbf{120}, 183002 (2018), \arXiv{1801.00491}.
    \item F.~Ficek, D.~Budker, {\em Constraining exotic interactions}, Annalen der Physik \textbf{531}, 1800273 (2019), \arXiv{1808.01233}.
    \item P.~Fadeev, Y.~V.~Stadnik, F.~Ficek, M.~G.~Kozlov, V.~V.~Flambaum, D.~Budker, {\em Revisiting spin-dependent forces mediated by new bosons: Potentials in the coordinate-space representation for macroscopic and atomic-scale experiments}, Physical Review A \textbf{99}, 022113 (2019),\\ \arXiv{1810.10364}.
    \item P.~Fadeev, F.~Ficek, M.~G.~Kozlov, V.~V.~Flambaum, D.~Budker, {\em Pseudovector and pseudoscalar spin-dependent interactions in atoms}, accepted for publication in Physical Review A.

\end{enumerate}
\chapter{Introduction}
The main scope of this dissertation is the equation
\begin{align}\label{eqn:introSNH}
i\partial_t \psi &= -\Delta\psi+|x|^2\psi- \left(\int_{\mathbb{R}^d} \frac{|\psi(t,y)|^2}{|x-y|^{d-2}}\, dy\right) \psi.
\end{align}
It belongs to the wide class of nonlinear Schr\"{o}dinger equations (in short NLS) and its main features are nonlocality of the nonlinearity (due to the integral, the value of the last term depends not only on the value of $\psi$ in the given point, but also on its values in other points) and the presence of the trapping potential, specifically harmonic potential. Equations with such nonlinearity can be found in the literature under different names, such as Schr\"{o}dinger-Newton \cite{Bah14, Cho08, Guz04, Har03, Mor98}, Schr\"{o}dinger-Poisson \cite{Iva08, Lub08, Par20}, Hartree \cite{Cao11, Cao12, Car05, Fen16, Fro03, Hua13, Wan08}, or Choquard \cite{Che17, Mor17, Van17}. These names are usually used regardless of the fact whether the external potential term is present or not. 
In the following, I will call Eq.\ (\ref{eqn:introSNH}) the Schr\"{o}dinger-Newton-Hooke equation (or SNH), where "Hooke" emphasizes the presence of the harmonic term (a similar joke name can be already found in the literature, e.g.\ \cite{Gib03}). The same name will also refer to the time-independent version of this equation (mostly in Chapter \ref{sec:stationary}), but I will do my best to keep it clear from the context. On the rare occasions of using the term Schr\"{o}dinger-Newton (SN), I will mean the equation
\begin{align}\label{eqn:introSN}
i\partial_t \psi &= -\Delta\psi - \left(\int_{\mathbb{R}^d} \frac{|\psi(t,y)|^2}{|x-y|^{d-2}}\, dy\right) \psi
\end{align}
or its time-independent version.

\vspace{\breakFF}
\begin{adjustwidth}{\marwidFF}{\marwidFF}
\small\qquad
Even though the main focus of this thesis is SNH equation, on many occasions I would like to show how the reasonings presented here may be applied to other NLS equations (justifying the rather general title of the thesis). In such cases I will usually be considering the Gross-Pitaevskii equation with harmonic trapping (which will be denoted by GP from now on):
\begin{align}\label{eqn:introGP}
i\partial_t \psi &= -\Delta\psi+|x|^2\psi-|\psi|^2 \psi.
\end{align}
Even though at first sight this system looks simpler than SNH equation (it is local, for starters), interestingly enough some of the presented results are actually harder to obtain in the case of GP equation. 

Some minor focus will be also given to other NLS equations so it is convenient to introduce them in general as
\begin{align}\label{eqn:introNLS}
i\partial_t \psi &= -\Delta\psi+V(|x|)\,\psi-F(\psi).
\end{align}
Here $V$ denotes the external potential, while $F$ is the nonlinearity. I will consider almost exclusively cases where $V$ is a trapping potential (i.e. $\lim_{|x|\to\infty}V(x)=\infty$) and $F$ is either the nonlinearity characteristic of Schr\"{o}dinger-Newton systems or is given by $F(\psi)=|\psi|^{p-1}\psi$ for some $p>1$ (GP equation is an example of such). To differentiate between parts of the thesis regarding SNH and other systems, for the latter I decided to use such narrower paragraphs as this one.
\end{adjustwidth}\vspace{\breakFF}

The Schr\"{o}dinger-Newton equations in general describe bosonic systems with the attractive interaction between the constituting particles, usually of an electric or gravitational nature. As such, they represent various quantum mechanical systems, but also appear at the interface of quantum theory and gravity. Probably the earliest occurrence of such systems can be dated to the year 1937 and the works of Fr\"{o}hlich on the interplay between deformations in a crystal structure caused by a movement of a charged particle and the behaviour of this particle \cite{Fro37, Fro54}. This idea and the connected notion of a polaron were later developed by many other researchers, c.f.\ \cite{Kup63} (especially pages 1--32). The Schr\"{o}dinger-Newton equations also have found use in the description of such systems as one-component plasma \cite{Lie77} or light beams propagating in nonlinear media \cite{Alb15, Ass19, Par20}.

Systems modeled by the Schr\"{o}dinger-Newton equations, but with the nonlinear term coming from the gravitational interaction seem to have appeared slightly later. In 1969 Ruffini and Bonazzola were studying a system of self-gravitating scalar bosons arriving at SN equation \cite{Ruf69}. Such configuration, now called a boson star, was later investigated by many others (see \cite{Jet92} for an overview), also recently, in connection with such questions as Dark Matter \cite{Bra16, Eby16} and the nature of super-massive objects in galactic nuclei \cite{Tor00}. In this context it is worth to mention a recent rise of interest in ultralight axion models described by SN with an additional nonlinear term \cite{Par20, Mar16, Sch14}. Finally, Schr\"{o}dinger-Newton equations naturally emerge in many attempts to build a theory of quantum gravity. Either in studies of the quantum collapse nature \cite{Pen96, Mor98}, as a semi-classical limit of full quantum gravity \cite{Giu12, Bah14}, or in other contexts \cite{Dio84, Jon95}.

All these different applications share one common feature: as they come out from the quantum-mechanical models, they limit the considerations to at most three-dimensional cases. However, in this work the main focus is on SNH system in higher dimensions (in the majority of this thesis I assume the spatial dimension $d$ to be greater than six, but in Section \ref{sec:resonant} I also tackle the $d=4$ case). From the physical point of view, such considerations may be motivated by their connection with some open questions such as the soliton resolution problem and the stability of the anti-de Sitter spacetime. The discussion of this connection constitutes the content of Chapter \ref{sec:motivation}.

\vspace{\breakFF}
\begin{adjustwidth}{\marwidFF}{\marwidFF}
\small\qquad
The additional, personally maybe the most important motivation staying behind this thesis comes from the fact that there are almost no results regarding any NLS with trapping potentials in supercritical dimensions (more precisely, energy-supercritical dimensions, the proper definitions are introduced in Section \ref{sec:symmetries}). The only work in this field that I am aware of is the series of papers by Selem and his coauthors \cite{Sel11, Sel12, Sel13} where they consider Eq.\ (\ref{eqn:introGP}), for which four is the energy-critical dimension. I believe that this situation comes from both the lack of an apparent physical motivation to deal with NLS in dimensions higher than three and the breakdown of typically used mathematical tools in supercritical dimensions. The latter is discussed in greater detail in Section \ref{sec:subcritical}.
\end{adjustwidth}\vspace{\breakFF}

When working on this thesis, I had to make some decisions regarding the presentation of the topic, some of them were not easy. First of all, I decided it to be a standalone work rather then a compilation of publications. Thanks to this, I was able to give more elaborate reasonings than in the articles (including some more technical parts covered by appendices). It also let me incorporate some additional materials including results that were not yet published. Initially, I wanted this thesis to cover as wide range of NLS equations as possible. However, keeping a high level of generality without losing clarity due to the additional technical assumptions needed in such an approach proved to be difficult. This is why I eventually decided to focus on SNH equation and discuss other NLS equations on the side. Finally, even though this work is rather mathematical, this is still a PhD thesis in theoretical physics. Hence, the presented reasonings are not always backed by strict proofs, but sometimes they rely on heuristic or numerical arguments. 

Including this short introduction, this work consists of six chapters. As already mentioned, Chapter \ref{sec:motivation} aims at giving some physical motivation to consider SNH in higher dimensions. The main result presented there is a derivation of Eq.\ (\ref{eqn:introSNH}) as a nonrelativistic limit of perturbations of the anti-de Sitter spacetime coming from \cite{Biz18}. Chapter \ref{sec:background} introduces some preliminary concepts (including the notion of criticality) used in the following parts and gives a short overview of the existing results. In Chapter \ref{sec:stationary} I focus on stationary solutions of NLS in critical and supercritical dimensions. The attention is turned mainly to their existence, uniqueness, and frequencies. This chapter is based on \cite{Biz21} and \cite{Fic21} but in spirit is much closer to the latter. Stability of these stationary states is investigated in Chapter \ref{sec:dynamics}. It also follows the second part of \cite{Biz18} and describes the resonant approximation of SNH showing its interesting properties in four dimensions. The thesis concludes in Chapter \ref{sec:conclusion} where I sum up the most important points and try to outline some future prospects.

Before I finish this chapter and move on to using the first-person plural, I would like to directly acknowledge some of the people who supported me during my work on this thesis. Primarily, I am thankful to my supervisor, Piotr Bizo\'{n}, who introduced me to this subject. Most of the problems I encountered in my work were completely new to me, giving me an opportunity to familiarize myself with many interesting topics. He was aiding me during this adventure with both physical insights and mathematical tools. Secondly, I would like to thank Konrad Szyma\'{n}ski for his constant support, on both scientific and personal levels. I'm also grateful to Oleg Evnin, Dmitry Pelinovsky, and Szymon Sobieszek with whom I had the joy to collaborate. I greatly appreciate all hints and help I got from Patryk Mach in all these years. In times when I got temporarily stuck, I could always redirect some of the productivity to other projects done with the Budker Group at Johannes Gutenberg-Universit\"{a}t Mainz, I am thankful in particular to Pavel Fadeev and Dmitry Budker for this opportunity. In the end, I wanted to thank people who were guiding me onto the academic path from my very childhood -- my parents S\l{}awomir and Agnieszka, my grandfather Kazimierz, and my aunt Jadwiga. 

This work was funded by the Polish National Science Centre within Grants No.\ 2020/36/T /ST2/00323 and No.\ 2017/26/A/ST2/00530. I also acknowledge the financial support coming from the project Kartezjusz. Finally, I am very grateful for the hospitality and support shown by the Mittag-Leffler Institute within the {\em General Relativity, Geometry and Analysis: beyond the first 100 years after Einstein} program.

\chapter{Motivation}\label{sec:motivation}
In the previous section we touched upon some of the problems motivating studies of SN and SNH systems from the physical viewpoint. However, as already mentioned, they focus almost exclusively on three-dimensional cases, while in this thesis we are interested in higher dimensions. Here we would like to provide some physical motivation to tackle such topic. We begin with a soliton resolution conjecture. Its short description, together with a connected issue of a weak turbulence, can be found in Section \ref{sec:turbulence}. These topics are currently of considerable interest and SNH system is only a very small part of a whole panorama of dispersive equations that can be studied in this context. To fix our attention on SNH system, in Section \ref{sec:SNHderivation} we briefly discuss the problem of stability of the anti-de Sitter (AdS) spacetime and connect it with Eq.\ (\ref{eqn:introSNH}) by the virtue of the nonrelativistic limit of AdS spacetime perturbations. The derivation described there comes from \cite{Biz18}.

\section{Soliton resolution problem and weak turbulence}\label{sec:turbulence}
One of the most remarkable features of nonlinear dispersive models is the presence of solitons -- the nonlinear bound states (i.e.\ spatially localized stable solutions). They emerge as an interplay between the dispersive nature of the equation and the focusing behaviour of the nonlinearity. Analysis of equations exhibiting soliton solutions is currently a very broad field laying mostly within the theory of partial differential equations but using tools from many other disciplines such as dynamical systems and algebraic geometry. Despite the rapid development since the 80s, there are still many open problems \cite{Mar18}, some of them of the fundamental nature. Among them, there is a question of long time behaviour of solutions. One can, for example, ask whether the solitons are asymptotically stable, i.e.\ that small perturbations of their profile get radiated to infinity restoring the initial shape of the soliton. 

A more complicated question is the soliton resolution conjecture. It states that for generic initial data, after a sufficiently long time the solution separates into a collection of decoupled solitons and radiation escaping to infinity. It suggests a dichotomy: in general, the only possible behaviours of the system are wave packets or radiation. An interesting interpretation of this effect, alluding to the common interplay between structured and random components (here played by the bound states and radiative solutions, respectively), can be found in the Simons Lecture given by Terrence Tao in 2007 \cite{TaoBlog}. In the case of linear Schr\"{o}dinger equation a similar result is known as the RAGE theorem \cite{Amr73, Ens78,Rue69}, but for nonlinear dispersive equations there is no general theory. The existing proofs mostly apply to completely integrable models such as one-dimensional Nonlinear Schr\"{o}dinger or Korteweg-de Vries equations, where one can use methods like inverse scattering \cite{Kow17}.

The key role in the just described phenomena is played by the fact that radiation can escape to infinity. Hence, it is clear that the long time behaviour of solutions may drastically differ for systems bounded in some way that does not allow for this getaway. Such confinement may be achieved by the compactness of the domain, existence of the reflecting boundary, or presence of the trapping potential. Then, the excess energy that would normally escape from the soliton, can interact with it through the nonlinear term almost indefinitely. This process can be observed in the momentum space where it is realised as mode mixing: the energy initially focused in the lower modes is gradually moving to the higher ones. Such energy cascade is something characteristic for a weak turbulence \cite{Eis06,Jos17,Pom92} -- the transfer of energy into increasingly lower spatial scales. For systems with two conserved quantities such as mass and energy (we define and discuss these notions in Section \ref{sec:symmetries}) one can then observe that the transfer of energy to higher modes is accompanied by the concentration of mass in lower modes. As a result, we observe the resolution once again, this time in the momentum space, where the energy migrates to higher and higher frequencies instead of spatial infinity. The full comprehension of these processes is still out of grasp (these mechanisms are probably best understood in the case of 2-dimensional Euler equations \cite{Way11}). One of the effects still waiting for a satisfying explanation is the problem of energy returns (apparent return of the energy to the initial state, concentrated in the lower modes, instead of the expected thermalisation), connected with the seventy-year-old Fermi–Pasta–Ulam–Tsingou problem \cite{Fer55}. The Schr\"{o}dinger-Newton-Hooke model may serve as another model that can be investigated in this context, possibly giving some insight to this phenomena.

\section{Non-relativistic limit of AdS perturbations}\label{sec:SNHderivation}
An important open problem in mathematical general relativity is the question of stability of the Anti-de Sitter spacetime (AdS). Its significance has at least two sources. First, it is a rather natural question within the classical theory of gravity. The AdS spacetime is a maximally symmetric solution of the vacuum Einstein equations with a negative cosmological constant. The other maximally symmetric vacuum solutions are Minkowski (no cosmological constant) and de Sitter (positive cosmological constant) spacetimes. However, while the latter two have been known to be stable under small perturbations for over three decades \cite{Chr, Fri86}, a comparable result is still unavailable for the former. The second, probably more popular motivation came to life in 1998 with the influential article by Maldacena \cite{Mal98}. The AdS/CFT correspondence conjectured there, postulating an equivalence between string theory on an asymptotically AdS spacetime and conformal field theory on its boundary, became one of the most popular topics in modern physics with countless publications and many uses in nuclear physics and condensed matter physics. The better understanding of the dynamics at the gravity side of this duality may have profound consequences for the future applications of this correspondence.

Having these in mind, it is rather baffling that the problem of AdS stability came under more thorough investigations relatively late. One of the foundational works in this direction was an article published by Bizo\'{n} and Rostworowski in 2011 \cite{Biz11}, where the authors conjured the instability of AdS spacetime in $d\geq 3$ spatial dimensions based on the numerical results. The instability manifested in the energy cascade pumping energy to increasingly lower scales, as described in the previous section, and supposedly resulting in the formation of a black hole. This result inspired some new research in this area in the following years, including many papers on the weak turbulence and collapse in asymptotically AdS spacetimes and other related models \cite{Bia17, Bia19, BiaIP, Fre16, Jal11, Lie13}. However, regarding the question of stability, for a long time all investigations were based on either numerical work or heuristic reasonings. The first rigorous proof of AdS instability was given by Moschidis in 2018 \cite{Mos20} and it dealt with a special case of a spherically symmetric Einstein--massless Vlasov system with an additional presence of an inner mirror near the origin. Even though the assumption of the inner mirror was dropped in the following work \cite{MosIP}, these results still regard a very specific case in which the considered matter is a massless dust satisfying the Vlasov equation.

As the described problem seems to be very difficult and it is unclear whether the existent tools are sufficient to tackle it in full generality, it may be constructive to try to understand its simplified versions. One of such would be a nonrelativistic limit of weak field perturbations of AdS. Here we would like to show that the equation describing such model takes the form of SNH system. This derivation, coming from \cite{Biz18}, is similar in spirit to the one presented in \cite{Giu12}. However, they slightly differ by the facts that we do not assume the spherical symmetry and we take into account a negative cosmological constant.

Let us consider a scalar field $\phi$ on some background $d+1$--dimensional spacetime with metric $g_{\mu\nu}$. Then, our system is described by two sets of equations, the first one being the Einstein field equations
\begin{align} 
G_{\mu\nu}+\Lambda g_{\mu\nu}=\frac{8\pi G}{c^4} T_{\mu\nu},
\label{eqn:1einstein}
\end{align}
where $G_{\mu\nu}$ and $T_{\mu\nu}$ are Einstein and energy-momentum tensors, respectively, $\Lambda$ is a negative cosmological constant, while $G$ is the gravitational constant and $c$ is the speed of light. We use here the spacelike convention, i.e.\ our metric tensor has a signature ($-,+,...,+$). The Greek indices in tensors stand for general coordinates, while $t$ will denote time and the spatial cartesian coordinates will be denoted by Latin indices. The second equation is the Klein-Gordon equation
\begin{align} 
g^{\mu\nu}\nabla_\mu\nabla_\nu\phi-\frac{m^2 c^2}{\hbar^2}\phi=0,
\label{eqn:1kleingordon}
\end{align}
where $m$ is the mass associated with the scalar field $\phi$, $\nabla_\mu$ is a covariant derivative, and $\hbar$ denotes the reduced Planck constant. The typical modus operandi in theoretical physics would now consist of assuming some natural system of units and getting rid of the physical constants (e.g.\ Planck units, where $c=\hbar=G=1$). However, ultimately we will be interested in taking the nonrelativistic limit, i.e.\ $c\to\infty$, so it is more convenient to keep all physical constants up to that point. Equations (\ref{eqn:1einstein}) and (\ref{eqn:1kleingordon}) are coupled to each other through the presence of the energy-momentum tensor in the case of Eq.\ (\ref{eqn:1einstein}) and the metric tensor together with covariant derivatives in Eq.\ (\ref{eqn:1kleingordon}). As the only matter present in the spacetime is the scalar field described by Eq.\ (\ref{eqn:1kleingordon}), the energy-momentum tensor takes the form
\begin{align*} 
T_{\mu\nu}=\frac{\hbar^2}{2m}\left[\partial_\mu \phi\, \partial_\nu\bar{\phi}+\partial_\mu\bar{\phi}\, \partial_\nu\phi-g_{\mu\nu}\left(\partial_\rho\phi\, \partial^\rho\bar{\phi}+\frac{m^2c^2}{\hbar^2}|\phi|^2\right)\right].
\label{eqn:energymomentum}
\end{align*}

Let us now assume that the scalar field $\phi$ is small, so we can approximate the metric by the first post-Newtonian corrections which are isotropic at this order \cite{Mis}:
\begin{align}\label{eqn:ansatz1}
ds^2=-c^2\left(1+\frac{2A(t,x)}{c^2}\right)dt^2+\left(1+\frac{2B(t,x)}{c^2}\right)\sum_{j=1}^d (dx^j)^2.
\end{align}
The functions $A$ and $B$ are unknown variables that we are solving for. We also introduce a new function $\psi$ defined by
\begin{align}\label{eqn:ansatz2}
\phi(t,x)=e^{-i\frac{mc^2}{\hbar}t}\psi(t,x).
\end{align}
These simple ansatze are sufficient for our purpose, as when performing the limit $c\to\infty$ we will be interested only in the lowest order in $c^{-1}$ (higher orders would require expanding the functions $\psi$, $A$, and $B$ in series with respect to $c^{-1}$). 

Next, we insert the ansatze (\ref{eqn:ansatz1}) and (\ref{eqn:ansatz2}) into Eqs.\ (\ref{eqn:1einstein}) and (\ref{eqn:1kleingordon}) and expand the results into the series with respect to $c^{-1}$. In the following we will be writing down explicitly only the lowest nontrivial orders of such expansions. We use the big-$\mathcal{O}$ notation, where $\mathcal{O}(c^k)$ means terms of the order higher than or equal to $c^k$ in $c^{-1}$. The Klein-Gordon equation (\ref{eqn:1kleingordon}) gives
\begin{align}\label{eqn:2kleingordon}
    2i\frac{m}{\hbar}\partial_t \psi+\Delta\psi-2\frac{m^2}{\hbar^2}A\psi+\mathcal{O}(c^{-2})=0,
\end{align}
where the laplacian $\Delta$ is just $\sum_{j=1}^d \partial^2_j$, since we are working in cartesian coordinates. Here the mass term in Eq.\ (\ref{eqn:1kleingordon}) of order $c^2$, cancels out with a term coming from the double differentiation of the exponent in our ansatz for $\phi$. As a result, we arrive at Eq.\ (\ref{eqn:2kleingordon}), in which one can recognize the Schr\"{o}dinger equation with external potential $m A$. 

Similar analysis can be done for the Einstein equations, although keep in mind that they are in fact a system of $(d+1)(d+2)/2$ equations. As our ansatze have only three unknown functions, taking all these equations is superfluous. Indeed, as we will see, it is enough to take just the diagonal components: $tt$- and $jj$- (for $j$ going through each spatial dimension). Then the $tt$-components of the tensors present in Eq.\ (\ref{eqn:1einstein}) are
\begin{align*}
G_{tt}&=-(d-1)\Delta B+\mathcal{O}(c^{-2}),\\
g_{tt}&=-c^2+\mathcal{O}(1),\\
T_{tt}&=m|\psi|^2c^4+\mathcal{O}(c^2).
\end{align*}
Keeping in mind that the energy-momentum tensor $T_{\mu\nu}$ in Eq.\ (\ref{eqn:1einstein}) is divided by $c^4$, we see that the right hand side of this equation is of order $c^0$. At the same time, it seems that the left hand side is dominated by $\Lambda g_{\mu\nu}$ having the order $c^2$. This problematic situation can be resolved by a more careful treatment of $\Lambda$ in the nonrelativistic limit. It turns out that as $c\to\infty$, $\Lambda$ should rather behave like $c^{-2}$ instead of being constant, as it is a necessary condition for the preservation of the boost symmetries in this limit \cite{Gib03}. Hence, we assume
\begin{align*}\label{eqn:lambda}
\lim_{c\to\infty} \Lambda c^2 = -\frac{d(d-1)}{2}\Omega^2,
\end{align*}
where $\Omega$ is some nonzero constant. The minus sign comes from the fact that we are interested in a negative cosmological constant, while the additional coefficient is for the sake of simplicity in further calculations. Let us also point out that any other type of behaviour of $\Lambda$ as $c\to\infty$ would lead to either blow up or vanishing of the cosmological constant term in the lowest order of Eq. (\ref{eqn:1einstein}). Now we can safely go to the limit $c\to\infty$ in $tt-$component of Eq.\ (\ref{eqn:1einstein}) getting
\begin{align}\label{eqn:EttEnd}
-(d-1)\Delta B+\frac{d(d-1)}{2}\Omega^2=8\pi G m |\psi|^2.
\end{align}

The third equation will be obtained from the $jj-$components of Eq.\ (\ref{eqn:1einstein}). One can show that as $c\to\infty$, it holds
\begin{align*}
G_{jj}&=\left[\Delta A -\partial_j^2 A+(d-2)(\Delta B-\partial_j^2 B)\right] c^{-2} +\mathcal{O}(c^{-4}),\\
g_{jj}&=1+\mathcal{O}(c^{-2}),\\
T_{jj}&=\frac{\hbar^2}{m}|\partial_j \psi|^2+i\frac{\hbar}{2m}\left(\bar{\psi}\partial_t\psi-\psi\partial_t\bar{\psi}\right)-\sum_{k=1}^d|\partial_k \psi|^2+\mathcal{O}(c^{-2}).
\end{align*}
Now taking into account the behaviour of $\Lambda$ in this limit, it turns out that the left hand side of Eq.\ (\ref{eqn:1einstein}) is of order $c^{-2}$, while the right hand side of $c^{-4}$. We can multiply both sides by $c^2$ obtaining the orders of $c^0$ and $c^{-2}$, respectively. It means that in the lowest order the matter term becomes irrelevant and in the limit we obtain
\begin{align*}
\Delta A - \partial_j^2 A+(d-2)(\Delta B-\partial_j^2 B)-\frac{d(d-1)}{2}\Omega^2=0.
\end{align*}
This way we got $d$ equations, one for each choice of $j$, but all of them include derivatives with respect to the particular coordinates: $\partial_j^2 A$ and $\partial_j^2 B$. We can change these terms into laplacians by summing these equations over all spatial coordinates $j$. Such procedure yields
\begin{align}\label{eqn:EjjEnd}
\Delta A +(d-2)\Delta B -\frac{d^2}{2}\Omega^2=0.
\end{align}
This equation, together with Eqs.\ (\ref{eqn:2kleingordon}) and (\ref{eqn:EttEnd}) constitute the full system of equations for $\psi$, $A$, and $B$. Let us now point out that when writing Eq.\ (\ref{eqn:1kleingordon}) we silently assumed the minimal coupling, while probably the most reasonable choice would be the conformal one \cite{Wal}. We did it deliberately, as it turns out that the additional term present in the  Klein-Gordon equation with the conformal coupling is of lower order in $c^{-1}$ then the remaining terms. Hence, it gives no contribution to our equations.

Moving on with this system, right away we can get rid of $B$, present only as $\Delta B$ in Eqs.\ (\ref{eqn:EttEnd}) and (\ref{eqn:EjjEnd}), and arrive at a system consisting of Eq.\ (\ref{eqn:2kleingordon}) and
\begin{align*}
\Delta A = \frac{8\pi G m(d-2)}{d-1} |\psi|^2 + d\, \Omega^2.
\end{align*}
This relation can be partially integrated by introducing a new function $v=A-\frac{1}{2}\Omega^2 |x|^2$. Replacing $A$ with $v$ removes the second term on the right hand side, but introduces to Eq.\ (\ref{eqn:2kleingordon}) a new expression. After some cleaning up, we finally get
\begin{subequations}\label{eqn:derivationSNH1}
\begin{align*}[left ={ \empheqlbrace}]
i\hbar\partial_t\psi&=-\frac{\hbar^2}{2m}\Delta\psi+\frac{1}{2}m\Omega^2 |x|^2 \psi +mv\psi,\\
\Delta v &= \frac{8\pi G m(d-2)}{d-1} |\psi|^2.
\end{align*}
\end{subequations}
This system is a Schr\"{o}dinger equation with a harmonic potential and an additional term that can be interpreted as the gravitational potential coming from the Newtonian self-gravitation of the field $\psi$. At this point, there is no need to keep the physical constants present, so we remove them by appropriate rescalings. As a result, we can write this system in a simple form that will be used in the rest of this thesis
\begin{subequations}\label{eqn:2SNHt}
\begin{align}[left ={ \empheqlbrace}]
i\partial_t\psi&=-\Delta\psi+|x|^2 \psi +v\psi,\label{eqn:2SNHta}\\
\Delta v &= |\psi|^2.\label{eqn:2SNHtb}
\end{align}
\end{subequations}

The last thing to do is to show that in fact this system is equivalent to Eq.\ (\ref{eqn:introSNH}). We do it using the Green function of the laplacian (we assume $d\geq 3$):
\begin{align}\label{eqn:green}
G(x,y)=-\frac{\Gamma(d/2)}{2\pi^{d/2}(d-2)} \frac{1}{|x-y|^{d-2}}.
\end{align}
The numerical constant present in this expression, equal to the inverse of the product of $(d-2)$ and the area of the $(d-1)$-dimensional unit sphere $S_{d-1}$, will be denoted by $A_d$:
\begin{align*}
A_d=\frac{1}{(d-2)S_{d-1}}=\frac{\Gamma(d/2)}{2\pi^{d/2}(d-2)}.
\end{align*}
Now we may solve for $v$ in Eq.\ (\ref{eqn:2SNHtb}), as it is equal to the integral of $|\psi|^2$ with $G$ playing the role of the kernel. Plugging it into Eq.\ (\ref{eqn:2SNHta}) finally gives
\begin{align}\label{eqn:SNHt}
i\partial_t \psi &= -\Delta\psi+|x|^2\psi-A_d \left(\int_{\mathbb{R}^d} \frac{|\psi(t,y)|^2}{|x-y|^{d-2}}\, dy\right) \psi.
\end{align}
This formula differs from Eq.\ (\ref{eqn:introSNH}) by the presence of the coefficient $A_d$ in the nonlinear term. This difference can be easily removed with a proper rescaling of $\psi$, however, we would like to retain the complete equality between the solutions of Eqs.\ (\ref{eqn:SNHt}) and Eq.\ (\ref{eqn:2SNHt}), so we shall keep $A_d$. We end up this section with the observation that the nonlinear term in Eq. (\ref{eqn:SNHt}) is just a convolution, so it is sometimes more convenient to write it as $A_d\, (|\psi|^2\ast |x|^{-(d-2)})\psi$.

\chapter{Background}\label{sec:background}
Before we proceed to the main content of this thesis, we want to establish some preliminary results, mainly regarding the notions of subcritical, critical, and supercritical dimensions. We do this in Section \ref{sec:symmetries}. In the following chapters we focus almost exclusively on NLS equations in supercritical dimensions, so as a reference we present some results concerning subcritical dimensions in Section \ref{sec:subcritical}.

\section{Symmetries and conserved quantities}\label{sec:symmetries}
When encountering a new problem, it is usually a good idea to identify its symmetries. The most basic symmetry of our problem, present also in other NLS equations, is the global phase freedom: if $\psi$ is a solution to Eq.\ (\ref{eqn:SNHt}), then for any $\alpha\in\mathbb{R}$ also $e^{i\alpha}\psi$ is the solution. By Noether's theorem this symmetry gives us the conservation of mass (sometimes also known as a number of particles or charge)
\begin{equation}\label{eqn:mass}
    \mathcal{M}(\psi)=\int_{\mathbb{R}^d} |\psi|^2\, dx.
\end{equation}
Another conserved quantity is energy
\begin{equation}\label{eqn:energy}
    \mathcal{E}(\psi)=\frac{1}{2}\int_{\mathbb{R}^d}|\nabla\psi|^2\, dx+ \frac{1}{2}\int_{\mathbb{R}^d} |x|^2 |\psi|^2\, dx - \frac{A_d}{4}\int_{\mathbb{R}^d} \left(|\psi|^2\ast|x|^{-(d-2)}\right)|\psi|^2\, dx.
\end{equation}
It comes from the time translation symmetry: if $\psi(t,x)$ is a solution, then so is $\psi(t-t_0,x)$. Equation (\ref{eqn:energy}) can be also regarded as a functional of two independent functions $\psi$ and $\bar{\psi}$. Then it plays the role of the hamiltonian of SNH system, since the equation of motion (\ref{eqn:SNHt}) can be obtained with the functional derivative as $i\partial_t \psi=\delta\mathcal{E}[\psi,\bar{\psi}]/\delta \bar{\psi}$. However, we will not be using this approach. Other symmetries typical for NLS equations, such as space translation or Galilean invariance are absent in our system because of the presence of the potential term. 

Of course, for mass $\mathcal{M}$ and energy $\mathcal{E}$ to be constant during the evolution, firstly they need to exist. Let $\psi$ be the solution to Eq.\ (\ref{eqn:SNHt}) with initial condition $\psi(0,x)=\psi_0(x)$. Then the aforementioned conserved quantities are well defined when $\psi_0$ belongs to appropriate function spaces. Since the mass is just the $L^2(\mathbb{R}^d)$ norm, it suffices to assume $\psi_0\in L^2(\mathbb{R}^d)$ to have it defined. In the case of energy this problem is a little bit more complicated. Due to the presence of a derivative term, one needs to ensure that $\nabla\psi$ exists (at least in a weak sense) and is square-integrable. For the second term in $\mathcal{E}$ to converge, we need $|x|\psi_0 \in L^2(\mathbb{R}^d)$. Of course the last term seems to be the most problematic, as it contains a convolution under the integral. We may deal with it by using the Hardy-Littlewood-Sobolev inequality \cite{Lie}, then the convergence of the last term follows if, for example, $\psi_0\in L^p$ where $p=4d/(d-2)$. The suitable choice of a functional space is crucial when applying the functional-analytic methods, as mentioned in the next section. However, in this thesis we employ another approach and consider only classical solutions (smooth enough that all appropriate derivatives exist) going to zero as $|x|\to\infty$. Then the harmonic term ensures the decay to be fast enough that all integrals in Eqs.\ (\ref{eqn:mass}) and (\ref{eqn:energy}) are convergent.

\vspace{\breakFF}
\begin{adjustwidth}{\marwidFF}{\marwidFF}
\small\qquad
The GP equation and other NLS systems also enjoy conservation of mass and energy. The mass is given by Eq.\ (\ref{eqn:mass}), while the specific form of the energy depends on the choice of the potential and nonlinearity. For example, if the evolution is driven by Eq.\ (\ref{eqn:introNLS}) with a power nonlinearity given by $F=|\psi|^{p-1}\psi$, then the energy is given by
\begin{equation*}
    \mathcal{E}(\psi)=\frac{1}{2}\int_{\mathbb{R}^d}|\nabla\psi|^2\, dx+ \frac{1}{2}\int_{\mathbb{R}^d} V(|x|) |\psi|^2 \, dx- \frac{2}{p+1}\int_{\mathbb{R}^d} |\psi|^{p+1}\, dx.
\end{equation*}
The term coming from the nonlinearity in the energy expression is here negative. Such systems (including also SNH system) are called \textit{focusing}. Taking $F=-|\psi|^{p-1}\psi$ gives the opposite sign in this term. In such case the nonlinearity is \textit{defocusing}. Focusing and defocusing systems differ in many ways, for instance, the former can have soliton solutions even without the trapping potential \cite{Fib}, while for the latter it is impossible.
\end{adjustwidth}\vspace{\breakFF}

Now we would like to consider a symmetry of a slightly different kind. 
Let $\psi$ be a solution of SN equation defined for some time interval $I$ including zero. Let us also impose some initial conditions at $t=0$, so in conclusion we have a function $\psi:I\times\mathbb{R}^d \to \mathbb{C}$ satisfying Eq.\ (\ref{eqn:introSN}) and $\psi(0,x)=\psi_0(x)$. Having $\psi$, we define a new function $\mbox{\L}\psi:\tan^{-1}(I)\times\mathbb{R}^d \to \mathbb{C}$ given by
\begin{align*}\label{eqn:lens}
    \mbox{\L}\psi(t,x) := \frac{1}{\cos^{d/2} t} \psi\left(\tan t, \frac{x}{\cos t}\right)e^{-\frac{i}{4}|x|^2 \tan t}.
\end{align*}
We shall call $\mbox{\L}$ the lens transform \cite{Car02, Fib, Tal70, Tao09}. An important feature of this transformation is that $\tan^{-1}(I)\subset(-\frac{\pi}{2},\frac{\pi}{2})$, regardless of the initial choice of $I$. Hence, for $I=\mathbb{R}$ or $I=[0,\infty)$ it compactifies the domain. Due to this property, the lens transform is sometimes compared to the pseudoconformal transform or even the Penrose compactification known from general relativity \cite{Tao09}.

By a simple but rather tedious calculation one can show that if $\psi$ is a solution to SN equation (\ref{eqn:introSN}), then $\mbox{\L}\psi$ is a solution to
\begin{align*}
    i\partial_t\mbox{\L}\psi = -\Delta\mbox{\L}\psi- \cos^{d-4}t\, \left(\int_{\mathbb{R}^d}\frac{|\mbox{\L}\psi(t,y)|^2}{|x-y|^{d-2}}\,dy\right) \mbox{\L}\psi +\frac{1}{4}|x|^2 \mbox{\L} \psi
\end{align*}
with the same initial condition as $\mbox{\L}\psi(0,x)=\psi(0,x)$. This is the NLS equation with a harmonic potential and a nonlinearity of the Schr\"{o}dinger-Newton type, but multiplied by $\cos^{d-4}t$. It is especially interesting when $d=4$, then the lens transform sends a solution of SN system into a solution of SNH system (the fact that the harmonic term differs by the factor $1/4$ from what we considered earlier is irrelevant). Let us point out that the reverse procedure is also possible by the means of the inverse lens transform
\begin{align*}
    \mbox{\L}^{-1}\psi(t,x) := \frac{1}{(1+t^2)^{d/4}} \psi\left(\tan^{-1} t, \frac{x}{\sqrt{1+t^2}}\right)e^{i\frac{|x|^2 t}{4(1+t^2)}}.
\end{align*}
This gives a bijection between solutions of two different systems defined on appropriate time domains. Note however, that while $\psi$ can be in principle a global-in-time solution, $\mbox{\L}\psi$ is defined only for $|t|<\pi/2$ and its behaviour after this time is a delicate matter.

\vspace{\breakFF}
\begin{adjustwidth}{\marwidFF}{\marwidFF}
\small\qquad
The lens transform also lets to switch between solutions to GP with and without the harmonic term, and then study a more complicated problem by the means of the simpler one with no potential, although in this case $d=2$ is the dimension where this procedure is possible \cite{Car02, Tao09}. The key feature of this transform, even in the slightly more general form \cite{Fib}, is that it lets to switch on and off the presence of the harmonic trapping. 
\end{adjustwidth}\vspace{\breakFF}

We mentioned the lens transform mainly to illustrate the connections between a nonlinear problem with and without the harmonic potential and the special role of four dimensions for SNH equation. We will not use this further in this thesis.

The last symmetry we want to discuss is the scaling symmetry present in the SN system. Let $\psi$ be a solution of SN equation (\ref{eqn:introSN}), then for any $\lambda>0$
\begin{align*}
    \psi_\lambda (t,x)=\lambda^{-2}\,\psi\left(\frac{t}{\lambda^2},\frac{x}{\lambda}\right)
\end{align*}
is also a solution to SN. The mass and energy of $\psi_\lambda$ can be then expressed using the mass and energy of the original solution $\psi$:
\begin{align*}
    \mathcal{M}(\psi_\lambda)=\lambda^{d-4}\,\mathcal{M}(\psi),\qquad \mathcal{E}(\psi_\lambda)=\lambda^{d-6}\,\mathcal{E}(\psi).
\end{align*}
For $d=4$ the mass is invariant with respect to this scaling, hence $d=4$ is called the mass-critical dimension. Analogously, $d=6$ is the energy-critical dimension. We also define subcritical and supercritical dimensions as dimensions lower or higher than the critical ones, respectively. Introduction of the potential term in SNH equation (\ref{eqn:introSNH}) breaks the scaling symmetry, however, the notion of subcritical/critical/supercritical dimensions is still useful, as we will see in the following.

The mass-invariance of SN is, next to the lens symmetry, another property characteristic to $d=4$. We study this case further in Section \ref{sec:resonant}. Apart from that section, we will be considering almost exclusively SNH equation in dimensions $d\geq 6$ (the reasons for this will be presented in the next section). Hence, from now on when writing about the critical/subcritical/supercritical dimensions, we mean the notion of the energy-criticality.

\vspace{\breakFF}
\begin{adjustwidth}{\marwidFF}{\marwidFF}
\small\qquad
The same calculations can be repeated for GP and other NLS equations with power nonlinearities. Then it turns out that in the absence of the potential term for any solution $\psi$ we can obtain a new solution by the rescaling $\psi_\lambda(t,x)=\lambda^{-2/(p-1)} \psi(t/\lambda^2,x/\lambda)$. This also leads to the analogous notion of mass-critical and energy-critical dimensions, in this case equal to $4/(p-1)$ and $2(p+1)/(p-1)$, respectively (notice that the difference between the energy- and mass-critical dimensions is always equal to two). As we will see, there are lots of qualitative similarities between the behaviour of various NLS equations in their respective subcritical, critical and supercritical dimensions. In particular, as our main scope are supercritical dimensions, we will observe many parallels between SNH system in $d\geq 6$ and GP equation in $d\geq 4$. We end this section with the observation that the critical dimensions agree for SN equation and NLS equation with $F=|\psi|\psi$. Interestingly enough, in the following we will encounter more similarities between these cases, but also some similarities between SNH and GP equations.
\end{adjustwidth}\vspace{\breakFF}

\section{Subcritical dimensions}\label{sec:subcritical}
In this section we briefly present some known results regarding SNH in subcritical dimensions (the comprehensive review of SNH, SN, and similar nonlocal NLS equations can be found in \cite{Mor17}). But first we want to try to explain an  almost complete lack of results regarding critical and supercritical NLS equations. 

The most popular approach that can be encountered in practically every mathematical investigation of SNH system, is based on the calculus of variations. Instead of looking for solutions "explicitly", one constructs an appropriate functional (on a carefully chosen function space) and proves that it possesses an extremum and that this extremum is a solution of the initial problem. This strategy, seemingly simple at first sight, has some crucial points limiting its applicability. We will discuss them using SNH equation as an example.

For NLS equations with harmonic potential the natural choice for the space to work in is
\begin{align*}
\Sigma := \left\{ u\in H^1(\mathbb{R}^d)\, \left|\, \int_{\mathbb{R}^d} |x|^2|u|^2 \, dx <\infty\right. \right\},
\end{align*}
where $H^1(\mathbb{R}^d)$ is a Sobolev space. Now, the construction of the appropriate functional $\Sigma\to\mathbb{R}$ depends on the particular result one wants to get. For example, the existence of stationary solutions with frequency $\omega$ (more details on these notions are given in Chapter \ref{sec:stationary}), provided that $\omega$ is in a right range, can be obtained by minimizing the functional (see \cite{Cao12} for details)
\begin{align*}
    \mathcal{S}(u)& :=\mathcal{E}(u)-\frac{\omega}{2}\mathcal{N}(u),
\end{align*}
under the condition $\mathcal{I}[u]=0$, where ($\Vert\cdot\Vert_2$ denotes here $L^2(\mathbb{R}^d)$ norm)
\begin{align*}
    \mathcal{I}(u)&:=\Vert \nabla u\Vert_2^2+ \Vert x u\Vert_2^2 -\omega \Vert u\Vert_2^2 - A_d\int_{\mathbb{R}^d} \left(|u|^2\ast|x|^{-(d-2)}\right)|u|^2\, dx.
\end{align*}
It is rather simple to show that such minimum would be a desired stationary state, the only problem is to prove its existence. It is usually done by constructing a minimizing sequence and proving its convergence. Such approach needs some tools from functional analysis. In particular, one has to assure some notion of compactness. Here it can be provided by the fact that the embedding $\Sigma\hookrightarrow L^q(\mathbb{R}^d)$ is compact when $2\leq q\leq 2d/(d-2)$ for $d\geq 3$ \cite{Ree, Zha00}. One also must have a way to appropriately estimate the nonlinear term, it can be done with Hardy-Littlewood-Sobolev and Gagliardo–Nirenberg inequalities. All these auxiliary results have some technical assumptions. In the case of SNH it turns out that they are applicable only for $d\leq 5$, which coincides with the subcritical dimensions. In the end we get the following result for subcritical dimensions: for every $\omega<d$ there exists a positive, radially-symmetric, monotonically decreasing stationary state with frequency $\omega$ \cite{Cao12, Fro03, Van17}.

These methods can be also used to give other existence results, such as the existence of positive stationary states with any prescribed mass $\mathcal{M}>0$ if $d=3$ \cite{Cao12, Luo19}. Similar tools from functional analysis give also results regarding the stability of the stationary solutions. Then it turns out that they are stable if $d=3$ or $d=4$, but for $d=5$ there exists a boundary frequency $\omega_0$ dividing stable and unstable positive stationary states \cite{Cao12, Fen16, Hua13}. Going beyond the stationary solutions, in $d\leq 5$ any function from $\Sigma$ poses good initial data to the Cauchy problem for SNH equation \cite{Caz, Wan08}. However, the existence of such solutions is assured only locally, in fact for some choices of the initial data the solution blows up in finite time \cite{Fen16, Hua13, Luo19, Wan08}. Let us also just mention that another possible approach that has also been used to investigate SNH system is the semiclassical approximation \cite{Cao11, Car05}, however, this method lays outside of the scope of this thesis.

\vspace{\breakFF}
\begin{adjustwidth}{\marwidFF}{\marwidFF}
\small\qquad
The literature describing other NLS systems is very vast, so let us focus here just on GP equation (for more general results and treatment of other systems we refer to monographs such as \cite{Caz, Fib, Pel, Tao}). Then $d=4$ is a critical dimension and for $d<4$ one observes similar behaviour as for the subcritical dimensions of SNH system. In particular, then there exist positive stationary solutions with any frequency $\omega<d$ \cite{Fuk01, Hir02}. One can even show their uniqueness, in the sense that such solutions are uniquely determined by their frequency \cite{Hir02, Hir07}. The results regarding the stability of these stationary solutions are also analogous to the ones seen for SNH system: in $d=3$, i.e.\ a dimension one less than the critical dimension, they are stable only at some range $(\omega_0,d)$, while for smaller dimensions ($d=1$ and $d=2$) there is no instability \cite{Fuk01, Fuk03A, Fuk03B}. Finally, information about local and global existence of time-dependent solutions to GP equation can be found in \cite{Car02, Caz, Fuk01, Oh89}.

The references provided here represent just a tip of the iceberg of results, not only for general NLS equations, but also for SNH and GP systems in subcritical dimensions. However, we wanted to focus here just on the results most relevant to the content of this thesis, so we can later compare behaviour of the discussed systems in subcritical and supercritical dimensions. For more information on other aspects of the mentioned systems we refer to the cited monographs and references therein.
\end{adjustwidth}\vspace{\breakFF}.

\chapter{Stationary solutions}\label{sec:stationary}
Now we proceed to the study of stationary solutions to SNH equation (\ref{eqn:introSNH}), that is solutions of the form
\begin{align*} 
\psi(t,x)=e^{-i\omega t}u(x),
\label{eqn:stationary}
\end{align*}
where $\omega$ is a real number called the frequency. This ansatz plugged into Eq.\ (\ref{eqn:2SNHt}) changes it into an elliptic equation with the nonlocal nonlinearity:
\begin{align}\label{eqn:SNHstatInt}
 -\Delta u+|x|^2 u-A_d \left(\int_{\mathbb{R}^d} \frac{u(y)^2}{|x-y|^{d-2}}\, dy\right) u= \omega\, u.
\end{align}
We are interested in bound states which are solutions to this equation that decay to zero $\lim_{|x|\to\infty}u(x)=0$. Among them, the most important are positive solutions (satisfying $u>0$) that will be called the {\em ground states}, in opposition to the remaining solutions called the {\em excited states}. Good understanding of the stationary states is not only a feat in itself, but is also the first step in the study of the dynamics of the system. Stationary states not only are the simplest solutions to the dynamical equation, but also are possible attractors in the evolution of the system.

\vspace{\breakFF}
\begin{adjustwidth}{\marwidFF}{\marwidFF}
\small\qquad
The same stationarity ansatz gives the respective nonlinear elliptic equations for the stationary solutions of other NLS equations. For example, in the case of GP equation (\ref{eqn:introGP}) we obtain 
\begin{align} 
-\Delta u + |x|^2 u-|u|^2 u =\omega\, u.
\label{eqn:4GP}
\end{align}
In this chapter, this equation will be covered in greater detail, along the main discussion regarding SNH. In some places we also consider bound states of other NLS equations, in general having the form coming from Eq.\ (\ref{eqn:introNLS})
\begin{align} 
-\Delta u + V u- F(u)=\omega\, u .
\label{eqn:4NLS}
\end{align}
\end{adjustwidth}\vspace{\breakFF}

As we have seen in Section \ref{sec:subcritical}, the variational methods are not applicable in energy-supercritical dimensions so we need to use other tools. To this end, we consider only spherically symmetric solutions. In the case of ground states there is no loss of generality here, as the ground states must be spherically symmetric. We discuss this property, together with some connected secondary results, in Section \ref{sec:spherical_symmetry}. When studying stationary solutions, assuming spherical symmetry reduces the relevant partial differential equations to the ordinary ones. This will allow us to use the theory of ODEs and dynamical systems to get results such as existence and uniqueness, as we show in Sections \ref{sec:existence} and \ref{sec:uniqueness}. The main result presented there is that for any fixed central value $u(0)>0$ there exists a unique ground state $u$ of Eq.\ (\ref{eqn:SNHstatInt}), characterised by a unique frequency $\omega$. In Section \ref{sec:omegab} we investigate the possible values of this frequency and its dependence on $u(0)$ in various dimensions, both analytically and numerically.

\section{Spherical symmetry}\label{sec:spherical_symmetry}
As we have just mentioned, the assumption of spherical symmetry may be partially excused by the fact that the most important solutions, the ground states, are bound to be spherically symmetric. Results of this type may be proved using many different approaches, such as the ones based on rearrangements and polarizations \cite{Bro07, Lie, Mor17}. Here, we choose to employ the moving planes method, as it seems to be working perfectly well also in supercritical dimensions.

The moving planes method originated from the works of Alexandroff and Serrin \cite{Ser71} (we refer to \cite{Ber91, Bre99, Fra} for more information on its history and various applications) and was used by Gidas, Ni, and Nirenberg \cite{Gid79} to show that for a wide class of nonlinear elliptic equations with Dirichlet boundary conditions in a ball, positive solutions must be spherically symmetric. Since then, many similar results concerning both bounded and unbounded domains were obtained. These proofs use a simple geometric fact that a function is spherically symmetric around zero if and only if it has a reflection symmetry with respect to every hyperplane crossing the zero. 

To sketch the reasoning behind this method, let us choose any unit vector $n\in\mathbb{R}^d$ and construct a family of hyperplanes $T_s=\{x\in\mathbb{R}^d\, | \, n\cdot x = s\}$ and half-spaces $H_s=\{x\in\mathbb{R}^d\, | \, n\cdot x \geq s\}$. Let now $u$ be the ground state solution of some nonlinear elliptic equation $\Delta u+f(|x|,u)=0$, with $f$ satisfying appropriate technical conditions. For any $s>0$ one may define in $H_s$ the function $w_s(x)=u(x)-u(r_s(x))$, where $r_s(x)=x-2(n\cdot x-s)n$ is a reflection of $x$ through $T_s$ (if the domain of $u$ is bounded, this step needs some additional attention). Now, depending on the specific case, one uses boundary conditions, asymptotic behaviour, or some technical assumptions on $f$ to show that for sufficiently large $s$ it holds $w_s<0$ in $H_s$. The maximum principle lets us to extend this result to some smaller values of $s$. With bootstrapping, we may repeat this procedure down to $s=0$, where by continuity it gives $w_0\leq0$ in $H_0$. Exactly the same line of action applied to $-n$ gives the opposite inequality, hence $w_0=0$ and $u$ is symmetric with respect to a hyperplane crossing the zero and orthogonal to $n$. Repeating this for every unit $n\in\mathbb{R}^d$ gives the spherical symmetry. For a more detailed description of this method we refer, for instance, to the book by Fraenkel \cite{Fra}. Let us just point out that in its simplest version the moving plane method is purely classical -- it uses only the maximum principle and needs no functional-analytic tools.

\vspace{\breakFF}
\begin{adjustwidth}{\marwidFF}{\marwidFF}
\small\qquad
Before we discuss the application of the moving plane method to the more complicated case of SNH equation, we would like to explore GP equation and other systems with local nonlinearities. Many of them can be covered by the rather general result obtained by Li and Ni in 1993 \cite{Li93}. In the slightly simplified version it considers the equation $\Delta u+f(|x|,u)=0$ on $\mathbb{R}^d$, with $f$ continuous, non-increasing in $|x|$, and locally Lipschitz in $u$. If there exists $M>0$ such that $\partial_u f(|x|,u)\leq 0$ for $|x|>M$ and $u<1/M$, then the positive decaying solutions must be spherically symmetric about some point, and monotonically decreasing with respect to it. One can easily check that for Eq.\ (\ref{eqn:4GP}) there is $f(|x|,u)=\omega u-|x|^2 u+u^3$ and $\partial_u f = \omega-|x|^2+3u^2$. Hence, for any fixed $\omega$ the assumptions of the cited theorem are satisfied and for the ground states there is no loss of generality in introducing $r=|x|$ and reducing Eq.\ (\ref{eqn:4GP}) to 
\begin{align} 
- u''-\frac{d-1}{r} u' + r^2 u-u^3=\omega\, u.
\label{eqn:4GPrad}
\end{align}
This theorem applies to a much wider class of nonlinear Schr\"{o}dinger equations with different radial potentials and different nonlinearities.
\end{adjustwidth}\vspace{\breakFF}

In case of SNH equation it may seem like one needs to use some special argumentation allowing for the presence of the integral in the nonlinearity. Indeed, there exist such proofs \cite{Che06, Mor17, Ma10}, but to deal with this integral they use tools like Hardy-Littlewood-Sobolev inequality, hence they require some additional assumptions on the dimension $d$. In particular, for Eq.\ (\ref{eqn:SNHstatInt}) this approach works only in dimensions $d<6$. Because of this, we take a different approach and use the local form we already encountered in Chapter \ref{sec:motivation}. Let us recall that Eq.\ (\ref{eqn:SNHt}) is equivalent to Eqs.\ (\ref{eqn:2SNHt}). Hence 
\begin{subequations}\label{eqn:SNHstatSys}
\begin{align}[left ={ \empheqlbrace}]
-\Delta u+|x|^2 u +v u,\label{eqn:SNHstatSysa} = \omega\, u\\
\Delta v = u^2.\label{eqn:SNHstatSysb}
\end{align}
\end{subequations}
Again, $v$ plays here the role of the potential and we are interested in solutions where both $u$ and $v$ tend to zero in infinity. Such systems of nonlinear elliptic equations has been thoroughly investigated (we refer to Chapter 7 of \cite{Dam} for a nice summary). Here we will focus on the result by Busca and Sirakov \cite{Bus00}, as it leads to the spherical symmetry of the ground states of Eqs.\ (\ref{eqn:SNHstatSys}).

In \cite{Bus00} the authors consider systems of arbitrarily many equations, however, for the sake of simplicity, we focus here on the systems of two nonlinear elliptic equations: $\Delta u_i +f_i(r,u_1,u_2)=0$ with $i=1,2$. Before stating the main result and applying it to SNH, let us introduce some definitions. We will say that our system is {\em cooperative} when $\partial f_1 / \partial u_2 \geq 0$ and $\partial f_2 / \partial u_1 \geq 0$. One can in some sense strengthen this condition and assume that there exists $R>0$ such that for $|x|>R$ and $|u_i|<1/R$ these two inequalities are strict. Such system is called {\em strongly-coupled} and cannot be reduced to two separated systems. Finally, we may imagine that also for $|x|>R$ and $|u_i|<1/R$ the matrix $\partial f_i/\partial u_j$ has no positive principal minors (i.e., $\partial f_1 / \partial u_1 \leq 0$, $\partial f_2 / \partial u_2 \leq 0$, and $(\partial f_1 / \partial u_1)(\partial f_2 / \partial u_2) -(\partial f_1 / \partial u_2) (\partial f_2 / \partial u_1)  \leq 0$). This condition in a way generalises the assumption $\partial_u f(|x|,u)\leq 0$ present in the theorem by Li and Ni \cite{Li93}. Now, Theorem 1 of \cite{Bus00} says that if the system of two nonlinear elliptic equations satisfies this condition, is cooperative, and strongly-coupled, then its positive decaying solutions $u_1$ and $u_2$ are spherically symmetric with respect to some point. If $f$ depends on $r=|x|$, then this point is $x=0$.

In order to apply this result, we need to make a small tweak in Eqs.\ (\ref{eqn:SNHstatSys}). As the function $v$ present there plays the role of a gravitational potential vanishing in infinity, one can expect it to be negative. However, the theorem covers positive solutions. Hence, for a brief moment let us introduce $u_2:=-v$ and also use $u_1$ instead of $u$. Then the functions $f_i$ have the form $f_1=-|x|^2 u_1+ \omega u_1+ u_1 u_2$ and $f_2=u_2^2$. It is straightforward to check that they satisfy the assumptions of the theorem by Busca and Sirakov, hence positive decaying solutions $u_1$ and $u_2$ must be spherically symmetric. Going back to the original variables, it means that every positive solution $u$ with the respective negative potential $v$ satisfying Eqs.\ (\ref{eqn:SNHstatSys}) are bound to be spherically symmetric. Hence, when studying the ground states it is enough to introduce $r=|x|$ and consider
\begin{subequations}\label{eqn:SNHrad}
\begin{align}[left ={ \empheqlbrace}]
-u''-\frac{d-1}{r} u'+r^2 u +v u &=\omega u,\label{eqn:SNHrada}\\
v''+\frac{d-1}{r}v' &= u^2.\label{eqn:SNHradb}
\end{align}
\end{subequations}
For the solutions of this system to be smooth near zero, we impose the condition $u'(0)=v'(0)=0$.

Reducing the problem to the system of ODEs greatly simplifies the situation and lets us use rather elementary tools. For instance, in the next section we will prove the existence of stationary solutions using the shooting method. This technique relies on two classical results: the local existence of solutions to the ordinary differential equations and their continuous dependence on initial conditions and parameters. Their full statements and proofs can be found in any textbook on ordinary differential equations (e.g.\ \cite{Cod, Har}). However, in the following we will be interested in posing the initial conditions at $r=0$, where Eq.\ (\ref{eqn:SNHrad}) is singular. Such case is not covered by these standard results and needs a separate treatment. For the convenience of the reader, we present the relevant theorems with proofs in Appendix \ref{sec:appODE}.

\begin{figure}
\centering
\includegraphics[width=0.4\textwidth]{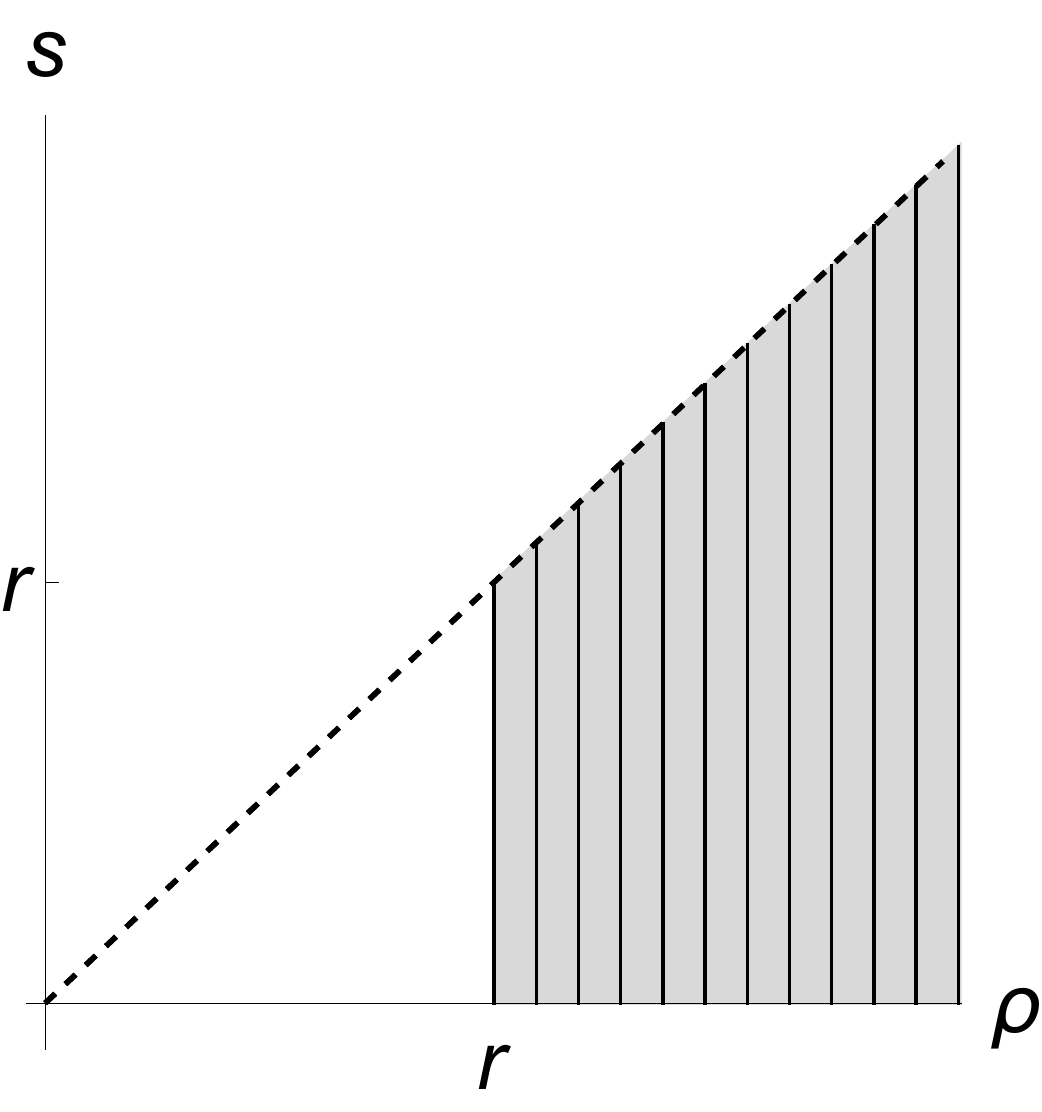}\qquad\qquad
\includegraphics[width=0.4\textwidth]{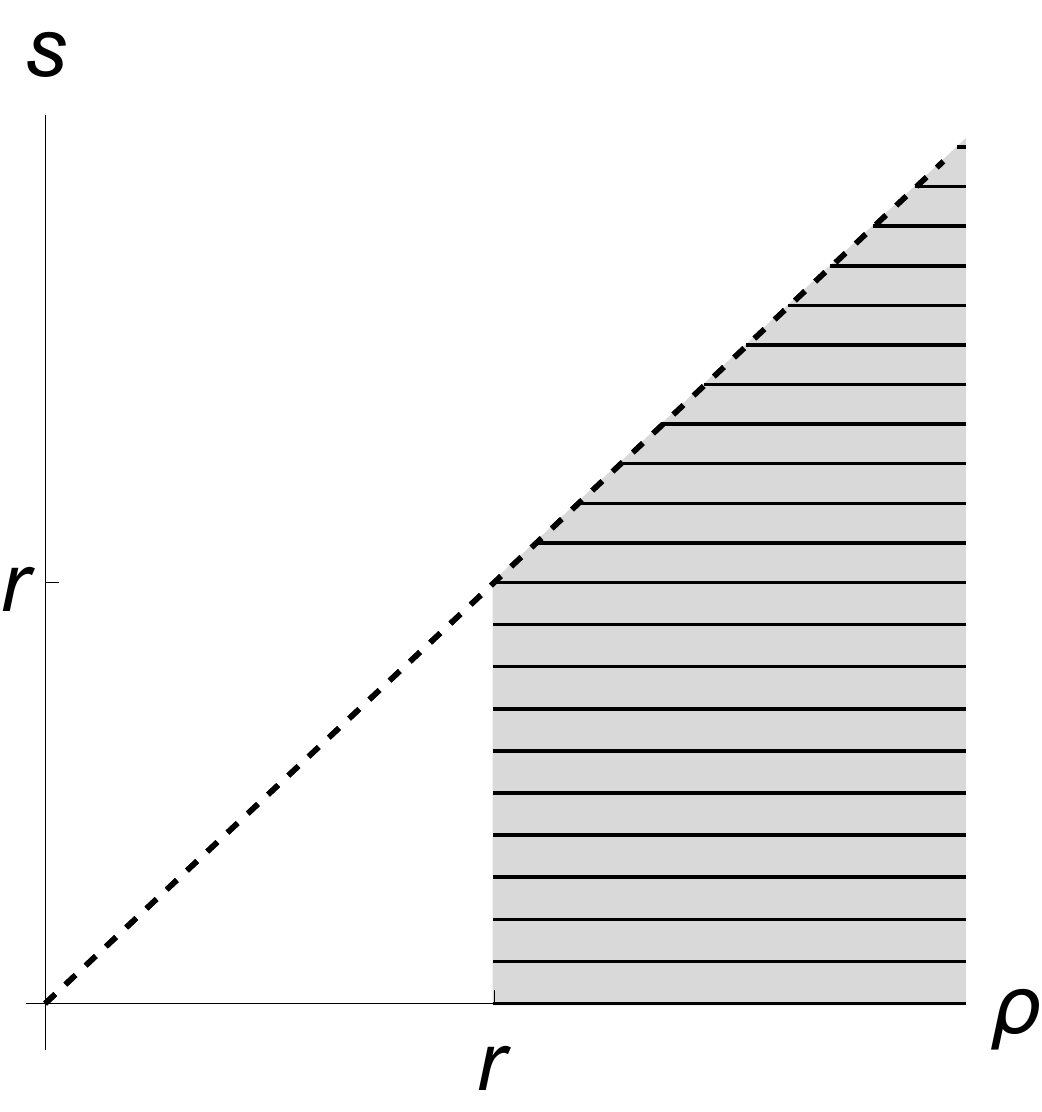}
\caption{Visualisation of the rearrangement of the order of integration in the derivation of the Newton formula.}
\label{fig:newton}
\end{figure}

We finish this chapter with a useful formula simplifying the integral nonlinearity in SNH equation (\ref{eqn:SNHstatInt}) under spherical symmetry. It will prove especially useful when considering the properties of small solutions and the dynamics of SNH system. It is sometimes called the Newton formula and it states that for a radial function $f$ decaying sufficiently fast, it holds for $d\geq 3$:
\begin{align}\label{eqn:newton}
 \int_{\mathbb{R}^d} \frac{f(|y|)}{|x-y|^{d-2}}\, dy
 =S_{d-1} \int_0^{\infty} \frac{f(s) s^{d-1}}{\max\{r^{d-2},s^{d-2}\}}\, ds.
\end{align}
This equality can be shown using the fact that in $d$ dimensions the Green function of the laplacian is given by Eq.\ (\ref{eqn:green}). Let us denote the left hand side of Eq.\ (\ref{eqn:newton}) by $\mathcal{I}$, so it holds $\Delta\mathcal{I}=-(d-2)S_{d-1}f$.
Hence, in spherical symmetry
\begin{align*}
\frac{1}{r^{d-1}}\frac{d}{dr}\left( r^{d-1} \frac{d \mathcal{I}}{dr} \right) = -(d-2)S_{d-1}f(r)
\end{align*}
The function $\mathcal{I}$ can be extracted from this expression by two integrations:
\begin{align*}
\mathcal{I}(r) = &(d-2)S_{d-1}\int_r^\infty \frac{1}{\rho^{d-1}} \left(\int_0^\rho f(s)\, s^{d-1}\, ds\right)\, d\rho\\
=&(d-2)S_{d-1}\int_0^r \left(\int_r^\infty \frac{f(s)\,s^{d-1}}{\rho^{d-1}}\,d\rho\right)\, ds \\
&+ (d-2)S_{d-1}\int_r^\infty \left(\int_s^\infty \frac{f(s)\,s^{d-1}}{\rho^{d-1}}\,d\rho\right)\, ds\\
=&S_{d-1}\int_0^r \frac{f(s)\, s^{d-1}}{r^{d-2}}\,ds + S_{d-1}\int_r^\infty \frac{f(s)\, s^{d-1}}{s^{d-2}}\,ds\\
=& S_{d-1}\int_0^\infty \frac{f(s)\, s^{d-1}}{\mbox{max}\{r,s\}^{d-2}}\,ds.
\end{align*}
The second equality comes from the rearrangement of the order of integration, as shown in the Fig.\ \ref{fig:newton}. This formula lets us to rewrite Eq.\ (\ref{eqn:SNHstatInt}) under the spherical symmetry as
\begin{align}\label{eqn:SNHradInt}
-u''-\frac{d-1}{r}u'+r^2 u-\frac{1}{d-2} \left(\int_0^\infty \frac{u(s)^2\, s^{d-1}}{\max\{r^{d-2},s^{d-2}\} }\, ds\right) u =\omega u.
\end{align}

\section{Existence}\label{sec:existence}
We are now ready to tackle one of the main results of this thesis: the existence of spherically-symmetric stationary states of SNH system. In particular, we prove that for every central value $u(0)=b>0$ there exists a bound state with an arbitrary number of zeroes. We also extend these results to singular solutions, where $u(0)=\infty$. Our approach is based on the shooting method (similar techniques were used in the case of ground states of SN equation in \cite{Cho08}).
\subsection{Preliminaries}\label{sec:4pre}
By the shooting method we understand the following idea: instead of viewing Eq.\ (\ref{eqn:SNHradInt}) as a boundary problem satisfying $u(0)=b$ and $\lim_{r\to\infty}u(r)=0$, we want to treat it as an initial value problem with initial conditions $u(0)=b$, $u'(0)=0$. Then we look for the value of $\omega$ giving us a desired solution decaying to zero in infinity. This approach is not well suited to the equation (\ref{eqn:SNHradInt}), as it is in fact an integro-differential equation. Hence, we focus on solutions of the equivalent system (\ref{eqn:SNHrad}) with the potential $v$ decaying to zero in infinity. Unfortunately, now we encounter another problem when posing initial values at $r=0$: we do not know a priori a value of $v(0)$. As a result, we have two unknown parameters, $\omega$ and $v(0)$, that need to be chosen in such a way that both $u$ and $v$ tend to zero in infinity. This complicates the situation, since two-dimensional shooting needs more sophisticated topological tools (see \cite{Ams, Has}). However, one can get rid of $\omega$ by formally replacing the function $v$ with $h=\omega-v$ (for the future convenience, we have also flipped the sign of the potential function). This way we get a new system:
\begin{subequations}\label{eqn:SNHradh}
\begin{align}[left ={ \empheqlbrace}]
u''+\frac{d-1}{r} u'-r^2 u +h u &=0,\label{eqn:SNHradha}\\
h''+\frac{d-1}{r}h' + u^2 &= 0.\label{eqn:SNHradhb}
\end{align}
\end{subequations}
with initial conditions
\begin{align}\label{eqn:SNHradhinitial}
    u(0)=b,\qquad h(0)=c,\qquad u'(0)=h'(0)=0.
\end{align}
Thus, we have obtained a new initial value problem with a single unknown parameter (for given $b$) denoted by $c$, that we will now investigate.

The possibility of returning from solutions of Eqs.\ (\ref{eqn:SNHradh}) to solutions of Eqs.\ (\ref{eqn:SNHrad}) depends on the existence of $\lim_{r\to\infty} h(r)$, as it is equal to the frequency $\omega$ in system (\ref{eqn:SNHrad}). A quick look at Eq.\ (\ref{eqn:SNHradhb}) suggests that when $u$ is vanishing, for sufficiently large $r$ it holds $|h(r)|< A r^2$ with $A$ being some constant, so for large $r$ the harmonic term dominates in Eq.\ (\ref{eqn:SNHradha}) and the nonlinear term can be neglected. It means that $u$ decays exponentially and $\lim_{r\to\infty} h(r)$ is finite, as can be seen by integrating Eq.\ (\ref{eqn:SNHradhb}) twice. It also lets us to approximate Eq.\ (\ref{eqn:SNHrada}) for large $r$ simply by the linear part (being just the quantum linear oscillator). Hence, there the solution shall behave like
\begin{figure}
\centering
\includegraphics[width=0.65\textwidth]{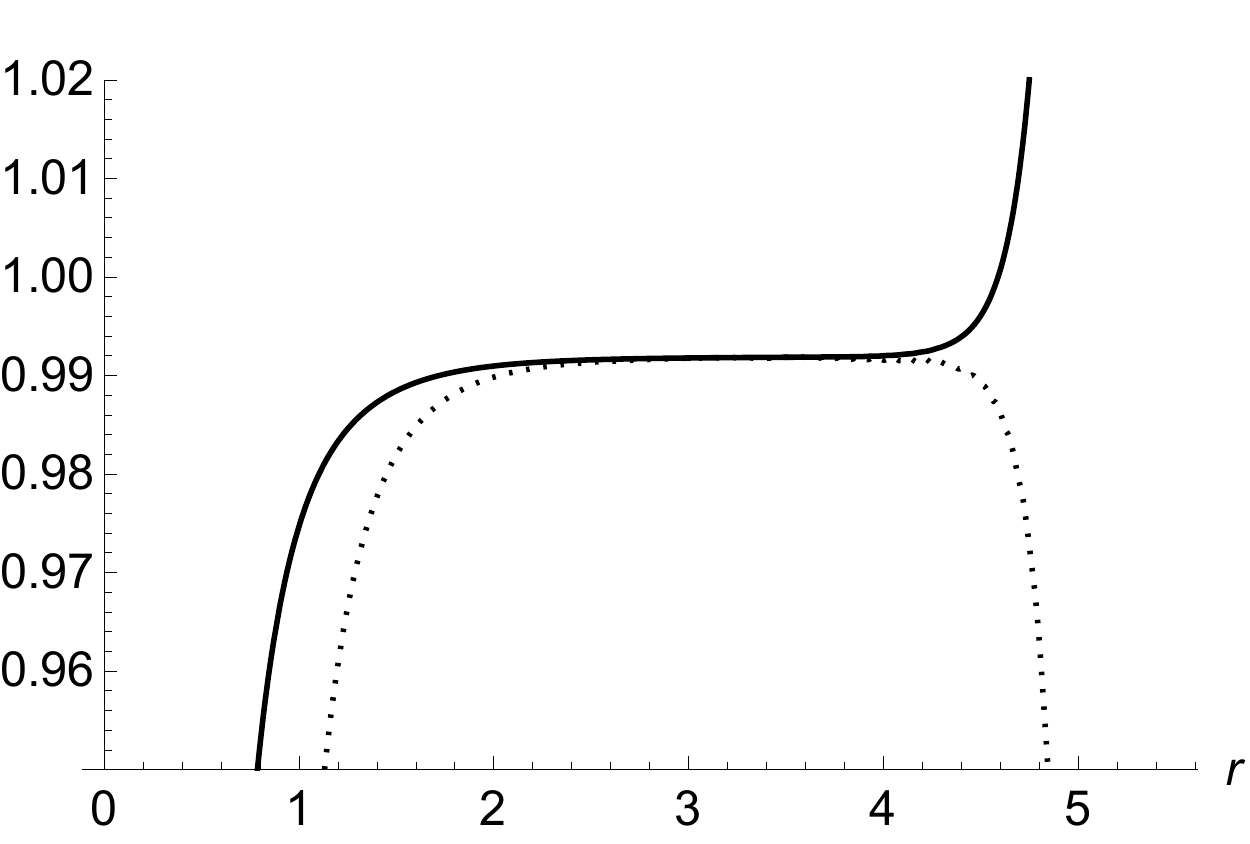}
\caption{Plots of $u(r)/u^{(1)}_{as}(r)$ (the solid line) and $u'(r)/u^{(1)\,\prime}_{as}(r)$ (the dotted line) for $d=7$, $b=1$, and $\omega=6.982625$.}
\label{fig:shootuas}
\end{figure}
\begin{equation}\label{eqn:solas}
u^{(C)}_{as}(r) = C\, e^{-r^2/2}\, U\left(\frac{d-\omega}{4},\frac{d}{2},r^2\right),
\end{equation}
where $U$ denotes the confluent hypergeometric function of the second kind and $C$ is some constant. This observation can be used to expand the numerically obtained solution $u$ into the whole half-line by gluing it with $u^{(C)}_{as}$ with appropriate $C$. To find the point of gluing and the value of $C$ we plot the functions $u(r)/u^{(1)}_{as}(r)$ and $u'(r)/u^{(1)\,\prime}_{as}(r)$, as seen in Fig.\ \ref{fig:shootuas}. If our numerical solution is a sufficiently good approximation to the bound state, there exists an interval of intermediate values of $r$ where both functions are more or less constant and equal to each other (for smaller values of $r$ functions $u^{(C)}_{as}$ are not a good approximation of the solution, while for larger $r$ the plots departure from zero due to inaccuracies in the shooting parameter value). Taking some point $R$ from this region to be the gluing point and the value of $u(R)/u^{(1)}_{as}(R)$ to be $C$ gives us the approximation of $u$ on the whole half-line.

\vspace{\breakFF}
\begin{adjustwidth}{\marwidFF}{\marwidFF}
\small\qquad
The same asymptotic behaviour can be observed also for GP, as stated and proved in Lemma 3.3 of \cite{Biz21}. This result regards both the solution and its derivative, meaning that here also $u\sim u^{(C)}_as$ and $u'\sim u^{(C)\,\prime}_{as}$ with the same constant $C$. The rigorous proof of this fact is based on the appropriate redefinition of dependent and independent variables and a careful analysis of the resulting dynamical system using the fixed-point arguments.
\end{adjustwidth}\vspace{\breakFF}

\begin{figure}
\centering
\begin{subfigure}{0.47\textwidth}
\includegraphics[width=\textwidth]{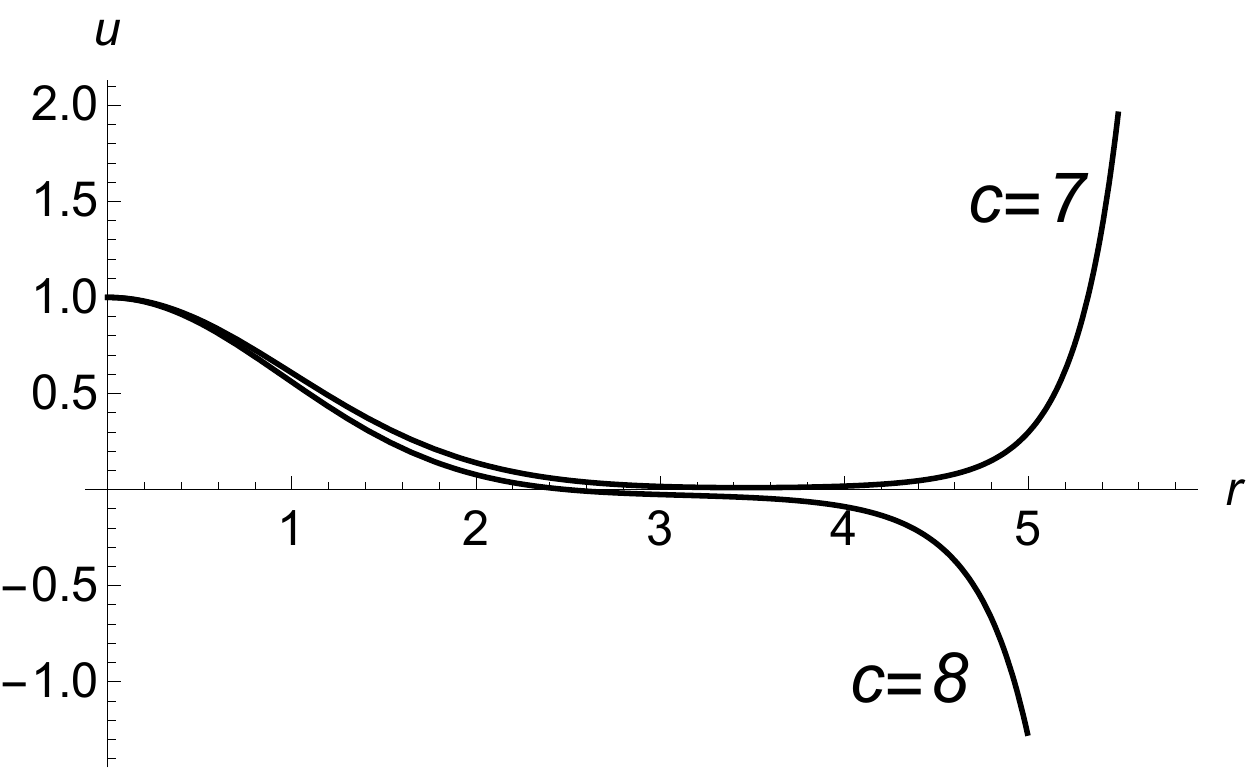}
\caption{}
\label{fig:shoot1a}
\end{subfigure}
~
\begin{subfigure}{0.47\textwidth}
\includegraphics[width=\textwidth]{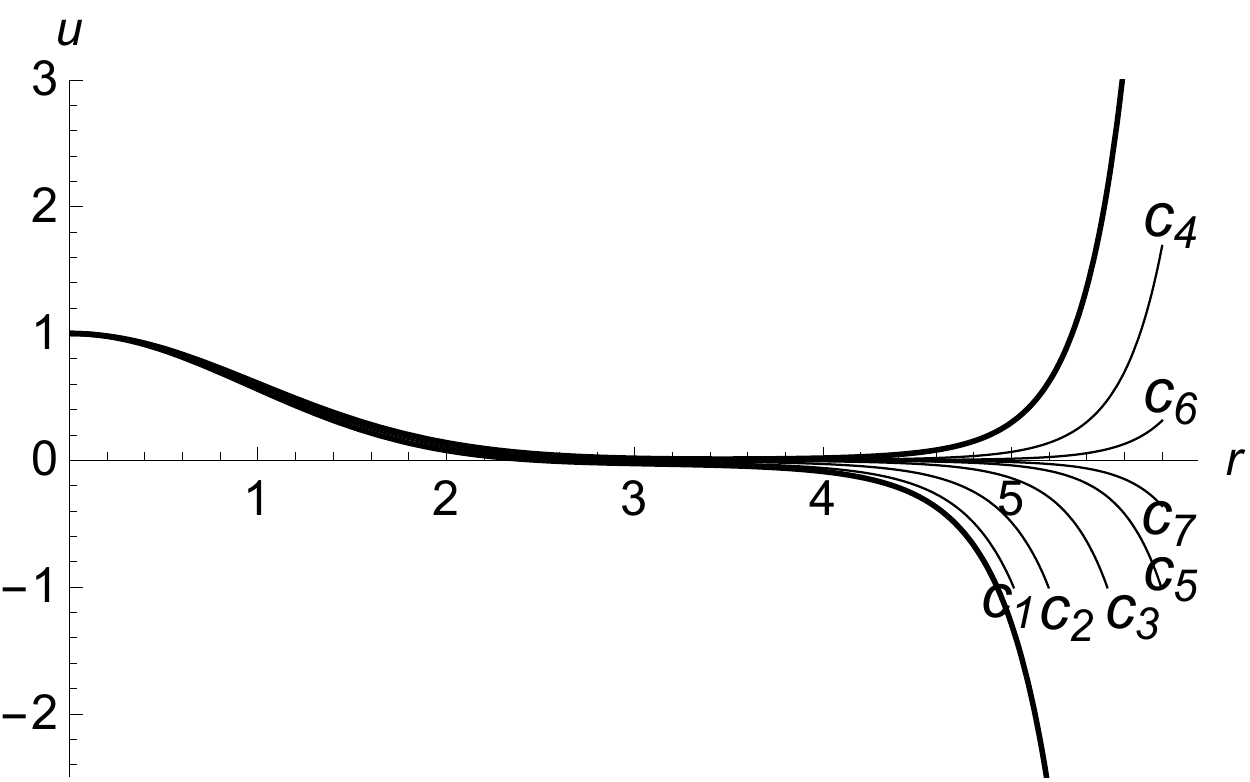}
\caption{}
\label{fig:shoot1b}
\end{subfigure}
\caption{Solutions $u$ to system (\ref{eqn:SNHradh}) with initial conditions (\ref{eqn:SNHradhinitial}) for $d=7$, $b=1$ and various values of $c$. The left plot shows presents solutions for $c=7$ and $c=8$, while the right one shows also solutions for the values of $c$ given by Table \ref{tab:shoot1}.}
\label{fig:shoot1}
\end{figure}
\begin{figure}
\centering
\includegraphics[width=0.65\textwidth]{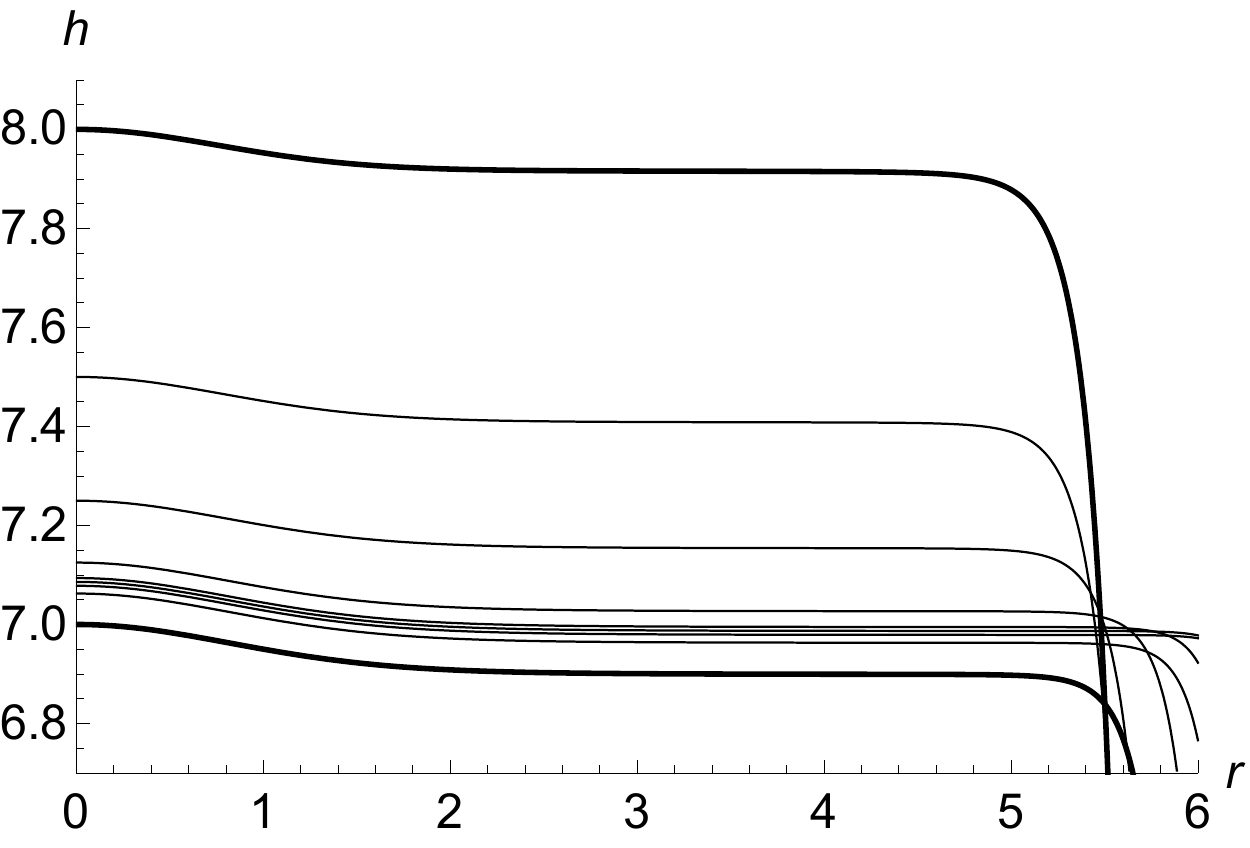}
\caption{Solutions $h$ to system (\ref{eqn:SNHradh}) with initial conditions (\ref{eqn:SNHradhinitial}) for $d=7$, $b=1$ and various values of $c$ given in Table \ref{tab:shoot1}.}
\label{fig:shoot2}
\end{figure}

Now, having established all the necessary details let us perform some numerical experiments. In Fig.\ \ref{fig:shoot1a} we present the sample plots obtained for a SNH system in $d=7$ for $b=1$ and with $c=7$ and $c=8$. We can see that both solutions lean towards the horizontal axis, but then one of them departs staying positive all the time, while the other one crosses zero and falls down. The continuous dependence of the solutions on the value of $c$ then suggests that between $c=7$ and $c=8$ there exist values of $c$ for which the solutions approach zero in larger intervals, i.e.\ are better approximations of the ground state. We may try to find them with the bisection method. The results of the successive steps are presented in Table \ref{tab:shoot1} and Fig.\ \ref{fig:shoot1b}. They let us expect that in case of $d=7$ and $b=1$ there exists a ground state for $c\approx 7.0817$. We can observe this convergence also in Fig.\ (\ref{fig:shoot2}). It shows that the closer we are to the ground state, the longer is the interval on which the solution $v$ is almost constant. By performing even more iterations in the bisection method we see that the proper candidate for $\lim_{r\to\infty} h(r)$ lays near $6.9826$, giving us the frequency of the ground state $\omega\approx 6.9826$.

\begin{table}[h]
\centering
 \begin{tabular}{cc|cc|cc|cc|cc} 

 $c_1$ & 7.5000 & $c_2$ & 7.2500 & $c_3$ & 7.1250 & $c_4$ & 7.0625 & $c_5$ & 7.0938\\
 \hline
 $c_6$ & 7.0781 & $c_7$ & 7.0859 & $c_8$ & 7.0820 & $c_9$ & 7.0801 & $c_{10}$ & 7.0811\\
 \hline
 $c_{11}$ & 7.0818 & $c_{12}$ & 7.0817 & $c_{13}$ & 7.0817 & $c_{14}$ & 7.0818 & $c_{15}$ & 7.0817 \\
  \hline
 $c_{16}$ & 7.0817 & $c_{17}$ & 7.0817 & $c_{18}$ & 7.0817 & $c_{19}$ & 7.0817 & $c_{20}$ & 7.0817
 \end{tabular}
 \caption{Values of the parameter $c$ (up to five significant digits) obtained in the succeeding steps of the bisection method.}
 \label{tab:shoot1}
\end{table}

\begin{figure}[h]
\centering
\includegraphics[width=0.65\textwidth]{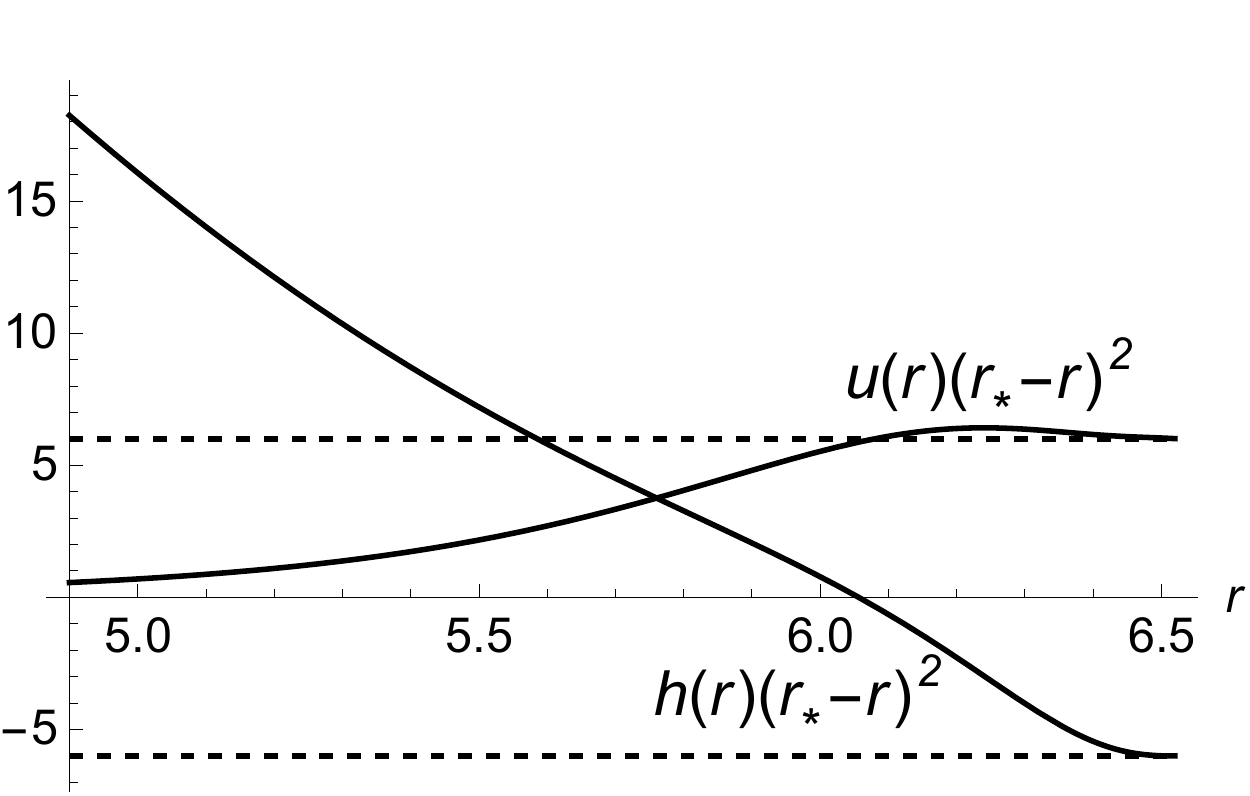}
\caption{Behaviour of the solutions to system (\ref{eqn:SNHradh}) for $d=7$, $b=1$, $c=7$ near the point of the blow-up $r_\ast\approx6.5259$.}
\label{fig:shootblowup}
\end{figure}

It is impossible to find numerically the precise value of $c$ giving the ground state. The solution $u$ always eventually departs from zero and rapidly goes down or up. Moreover, the numerical solver reveals problems with integration at some $r_\ast$, caused by the blow-up of the solutions. We can study the nature of this singularity by assuming that near this $r_\ast$ the solutions behave like $u(r)=A(r_\ast-r)^\alpha$ and $h(r)=B(r_\ast-r)^\beta$, where $A$, $B$, $\alpha$, and $\beta$ are some numerical coefficients to be found. Plugging this ansatz into system (\ref{eqn:SNHradh}) yields
\begin{subequations}
\begin{align*}[left ={ \empheqlbrace}]
A\,\alpha(\alpha-1)(r_\ast-r)^{\alpha-2}- A\,\alpha\frac{d-1}{r}(r_\ast-r)^{\alpha-1}-A\, r^2 (r_\ast-r)^{\alpha}&\\
+ A \,B(r_\ast-r)^{\alpha+\beta}&=0,\\
B\,\beta(\beta-1)(r_\ast-r)^{\beta-2}- B\,\beta\frac{d-1}{r}(r_\ast-r)^{\beta-1}+ A^2(r_\ast-r)^{2\alpha}&= 0.
\end{align*}
\end{subequations}
Near $r=r_\ast$ the main role will be played by the terms of the lowest order in $(r_\ast-r)$, hence we may assume that the first and last term in both equations shall balance. This can happen when $\alpha-2=\alpha+\beta$ and $\beta-2=2\alpha$ giving us $\alpha=\beta=-2$. Then, by comparing the coefficients of these terms we get $B=6$ and $A=\pm 6$ (the sign of $A$ depends on the infinity to which $u$ diverges). One can check numerically (Fig.\ \ref{fig:shootblowup}) that indeed near $r_\ast$ the solutions behave like $u\sim\pm 6\,(r_\ast-r)^{-2}$, $h\sim-6\,(r_\ast-r)^{-2}$. This blow-up behaviour is not important from the point of view of the shooting method, but may pose problems for the dynamical systems approach.

\vspace{\breakFF}
\begin{adjustwidth}{\marwidFF}{\marwidFF}
\small\qquad
In case of systems with local nonlinearities, such as GP, the situation is usually simpler as we start from the level of a single ordinary equation with a single unknown parameter -- the frequency $\omega$. Then we can just pose initial conditions $u(0)=b$ and $u'(0)=0$ and search for the values of $\omega$ giving the best approximations of the ground state. Sample plots for GP equation in $d=5$ with $b=1$ are presented in Fig.\ \ref{fig:shootGP}. In this case the bisection method eventually gives us $\omega\approx 4.8397$ as a candidate for the frequency of the ground state.

\begin{figure}
\centering
\includegraphics[width=0.5\textwidth]{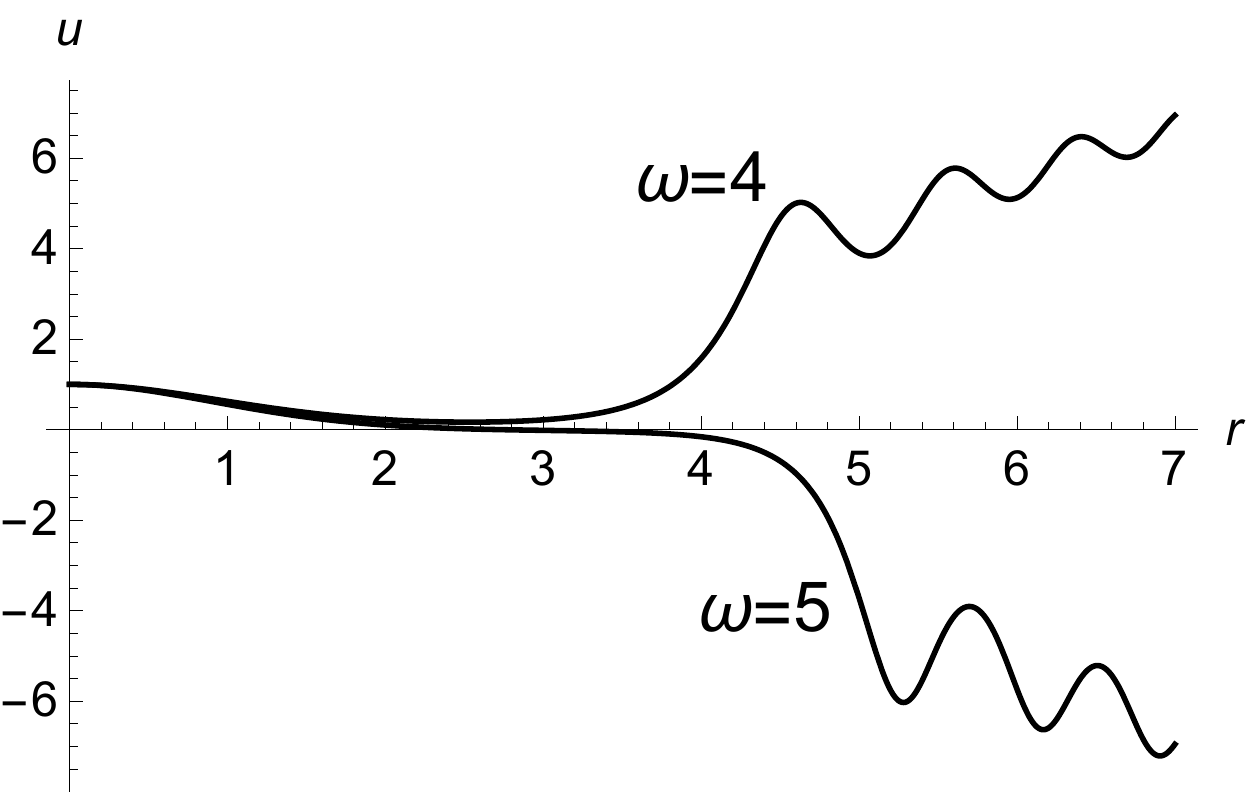}
\captionsetup{width=0.83\textwidth}
\caption{\small Solutions $u$ to Eq.\ (\ref{eqn:4GPrad}) for $d=5$, $b=1$ in cases of $\omega=6$ and $\omega=7$.}
\label{fig:shootGP}
\end{figure}

Interestingly, for this system there is no blow-up and all solutions exist globally. It can be shown by rewriting Eq.\ (\ref{eqn:4GPrad}) in new variables $t=r^2/2$ and $w(t)=u(r)/r$. The idea behind these combinations comes from the fact that after departing from zero the plot of $u$ seems to oscillate around the linear function $u=r$ (see Fig.\ \ref{fig:shootGP}). We also point out that these variables are ill-defined in $r=0$, however we want to discuss here the behaviour of the system for large $r$, so it does not pose any problems. The GP system is then governed by the equation
\begin{align}\label{eqn:GPw}
    \ddot{w}+\frac{d+2}{2t}\dot{w}+w(w^2-1)+\frac{d-1}{4t^2}w+\frac{\omega}{2t}w=0,
\end{align}
with dots denoting derivatives in $t$. Now we may define an energy functional
\begin{align}\label{eqn:GPwE}
    E=\frac{1}{2}\dot{w}^2+\frac{1}{4}w^4-\frac{1}{2}w^2+\frac{d-1}{8t^2}w^2+\frac{\omega}{4t}w^2.
\end{align}
It is bounded from below by $-1/4$ and its differentiation over $t$ gives
\begin{align*}
    \dot{E}=-\frac{d+2}{2t}\dot{w}^2-\frac{d-1}{4t^3}w^2-\frac{\omega}{4t^2}w^2,
\end{align*}
so it decreases with $t$. As a result, $w$ is bounded, hence it exists for all $t>0$. Coming back to $u$, it gives us the global existence of solutions of Eq.\ (\ref{eqn:4GPrad}).
\end{adjustwidth}\vspace{\breakFF}

\subsection{Ground states}
Having seen how the search for the frequency of the ground state looks numerically, we are now ready to cast this procedure into the proof of existence. For this purpose, we again use the formulation (\ref{eqn:SNHradh}) including just one unknown $c$, that from now on will be called the shooting parameter (even though there exist versions of the shooting method dealing with many parameters \cite{Ams, Has}, it is easier to handle a single shooting parameter). In the following we fix some $b>0$ and study how the solutions change as we alter the value of $c$. As the numerical experiments show, when starting with $c$ such that the solution $u$ is positive and eventually blows up, by gradually increasing the value of $c$ at some point we get a solution crossing zero. The idea will be to show that on the border of these two worlds there exists a value of $c$ such that the solution $u$ stays positive and goes to zero instead of blowing up ($u$ is the ground state). 

In the first step we would like to show that there indeed exist different values of $c>0$, such that for one of them the solution $u$ is positive and for the other the solution $u$ crosses zero (as we have seen in numerical results). The former of these results can be achieved for $c=0$ by the method similar to the derivation of the Pohozaev identities (more on them in Section \ref{sec:omegab}). Let us assume that when $c=0$ there exists some point $r=R$ such that $u=0$ for the first time there. Then we may multiply Eq.\ (\ref{eqn:SNHradha}) by $u\, r^{d-1}$ and integrate it over $[0,R]$ getting 
\begin{subequations}\label{eqn:smallc}
\begin{align}
-\int_0^R u'^2 r^{d-1} dr-\int_0^R r^2 u^2 r^{d-1} dr+\int_0^R u^2 h r^{d-1} dr&=0.\label{eqn:smallca}
\end{align}
When we repeat this procedure, but this time multiplying by $u' r^d$ instead, we get
\begin{align}
u'(R)^2 R^d+(d-2)\int_0^R u'^2 r^{d-1} dr-\int_0^R u^2 h' r^d dr&\nonumber\\
+(d+2)\int_0^R r^2 u^2 r^{d-1} dr-d \int_0^R u^2 h r^{d-1} dr&=0.\label{eqn:smallcb}
\end{align}
Analogous strategy can be executed for Eq.\ (\ref{eqn:SNHradhb}), but this time we multiply either by $h r^{d-1}$ or $h' r^d$. It yields
\begin{align}
h'(R)h(R) R^{d-1}-\int_0^R h'^2 r^{d-1} dr+\int_0^R u^2 h r^{d-1} dr&=0,\label{eqn:smallcc}\\
h'(R)^2 R^d+(d-2)\int_0^R h'^2 r^{d-1} dr+2\int_0^R u^2 h' r^d dr&=0.\label{eqn:smallcd}
\end{align}
\end{subequations}
As a result, we got a system of four identities consisting of eight different terms. We can get rid of three of them by taking the combination
 $(d+2)$ $\times$ \eqref{eqn:smallca} + 2 $\times$ \eqref{eqn:smallcb} + $(d-2)$ $\times$ \eqref{eqn:smallcc} + \eqref{eqn:smallcd}, obtaining the final identity
\begin{align*}
(d-6)\int_0^R u'^2 r^{d-1} dr+(d+2)\int_0^R r^2 u^2 r^{d-1} dr+2u'(R)^2 R^d&\\
+h'(R)^2 R^d+(d-2)h(R)h'(R)R^{d-1}&=0.
\end{align*}
All terms, except the last one, are obviously positive. However, one can observe that since Eq.\ (\ref{eqn:SNHradhb}) can be written as
\begin{align*}
    h'(r)=-\frac{1}{r^{d-1}}\int_0^r u(s)^2\,s^{d-1}\, ds,
\end{align*}
$h$ is decreasing in $r$. Since $h(0)=0$, both $h'(R)$ and $h(R)$ are negative, which means that the last term of the identity is also positive. Hence, we end up with a contradiction meaning that for $c=0$ the solution $u$ cannot cross zero so it stays positive. Let us point out that in this argumentation the fact that we are in supercritical (or at least critical) dimensions is essential, as without the assumption $d\geq 6$ the left-hand side of the obtained identity is not necessarily positive and there is no clear contradiction.

\begin{figure}
\centering
\includegraphics[width=0.65\textwidth]{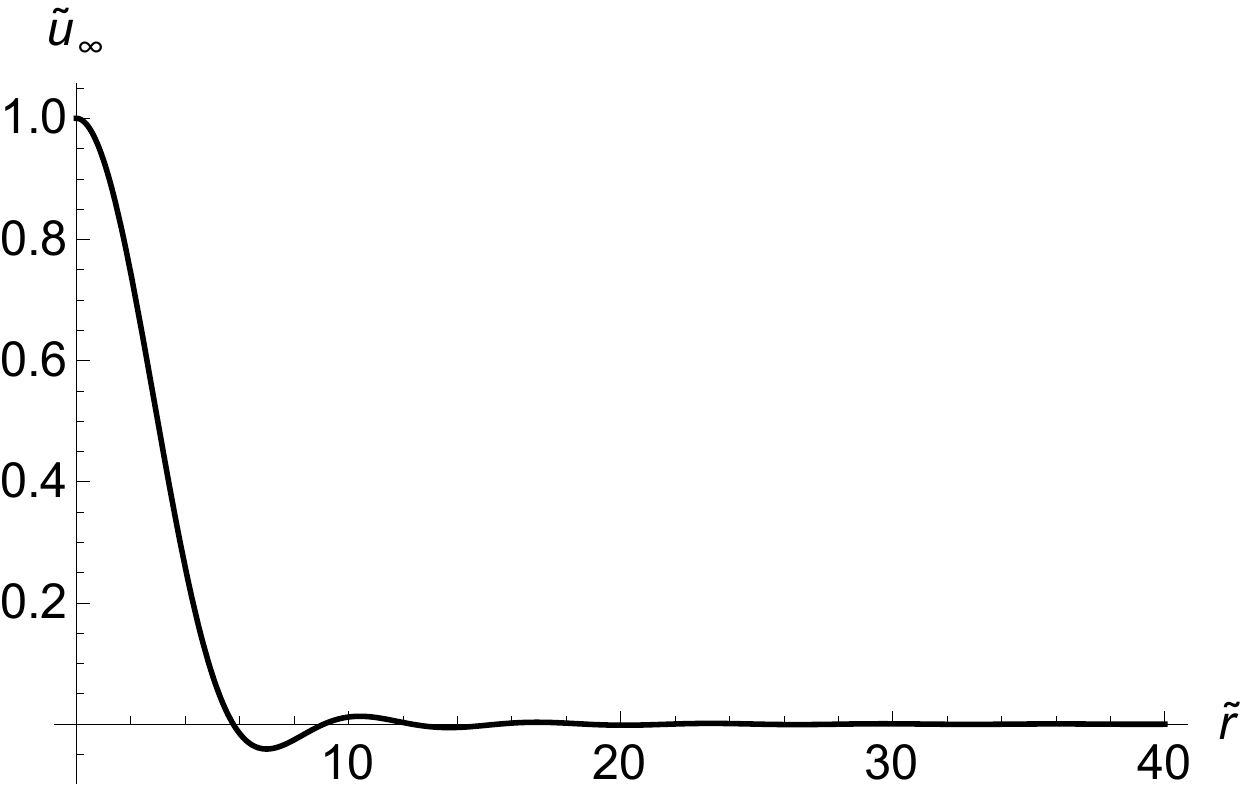}
\caption{Plot of the function $\widetilde{u}_\infty(\widetilde{r})$ given by Eq.\ (\ref{eqn:bessel}) for $d=7$ and $b=1$.}
\label{fig:shootbessel}
\end{figure}

In order to show that there exists such $c$ that the solution $u$ crosses zero, instead of choosing some specific value of $c$, it is easier to see what happens as $c\to\infty$. For this purpose, it is handy to rescale the coordinate $r$ to $\widetilde{r}=\sqrt{c}r$ and the function $h$ to $\widetilde{h}(\widetilde{r})=h(r)/c$. Then Eqs.\ (\ref{eqn:SNHradh}) take the form of
\begin{subequations}\label{eqn:SNHradhbes}
\begin{align}[left ={ \empheqlbrace}]
\widetilde{u}''+\frac{d-1}{\widetilde{r}}\widetilde{u}'-\frac{\widetilde{r}^2}{c^2} \widetilde{u} +\widetilde{h}\widetilde{u}&=0,\label{eqn:SNHradhbesa}\\
\widetilde{h}''+\frac{d-1}{\widetilde{r}}\widetilde{h}'+\frac{1}{c^2}\widetilde{u}^2 &= 0,\label{eqn:SNHradhbesb}
\end{align}
\end{subequations}
where $\widetilde{u}(\widetilde{r})=u(r)$ and all derivatives are with respect to $\widetilde{r}$. The initial conditions now are $\widetilde{u}(0)=b$, $\widetilde{h}(0)=1$, $\widetilde{u}'(0)=\widetilde{h}'(0)=0$. In the limit $c\to\infty$ Eq.\ (\ref{eqn:SNHradhbesb}) becomes just $(\widetilde{h}'\,r^{d-1})'=0$ and has a unique solution $\widetilde{h}_\infty\equiv 1$. As a result, in this limit Eq.\ (\ref{eqn:SNHradhbesa}) is equal to
\begin{align}\label{eqn:SNHbessel}
\widetilde{u}''+\frac{d-1}{\widetilde{r}}\widetilde{u}'+\widetilde{u}=0.
\end{align}
Its solution can be written explicitly in terms of the Bessel function $J_\alpha$:
\begin{align}\label{eqn:bessel}
\widetilde{u}_\infty(\widetilde{r})=b\, \Gamma\left(\frac{d}{2}\right)\frac{2^{\frac{d}{2}-1}}{\widetilde{r}^{\frac{d}{2}-1}}\, J_{\frac{d}{2}-1}(\widetilde{r}).
\end{align}
Since $\widetilde{u}_\infty$ is the limiting solution for $c\to\infty$, on every compact interval $[0,R]$ the solution $\widetilde{u}$ tends to $\widetilde{u}_\infty$ in the supremum norm as $c$ goes to infinity (as we are considering the second-order ODEs, also the first derivative $\widetilde{u}'$ converges to $\widetilde{u}_\infty'$ in this limit). The function $\widetilde{u}_\infty$ oscillates around zero with a decreasing amplitude (see Fig.\ \ref{fig:shootbessel}), so a similar behaviour shall be expected from the solutions $\widetilde{u}$ of system (\ref{eqn:SNHradhbes}) for sufficiently large $c$. The return to the original variables just unfolds the function $u$ along the horizontal axis, not changing its overall shape. Hence, we see that not only there exists $c$ such that solution $u$ of system (\ref{eqn:SNHradh}) crosses zero, but the number of these crossings can be arbitrarily large, provided that $c$ is large enough.

The next step will consist of proving the following trichotomy: if $u$ is a solution of Eq.\ (\ref{eqn:SNHradh}), it either diverges to one of the infinities (we do not specify here whether it happens in finite time through a blow-up or the solution exists globally) or converges to zero at infinity. To show it, and also for further convenience, we need to make some simple observations regarding the behaviour of solutions to system (\ref{eqn:SNHradh}). Let us start by pointing out that since $h$ is a decreasing function, on the whole domain of existence of the solution it holds $h(r)\leq c$. Now assume that the solutions $u$ and $h$ exist globally (otherwise they clearly diverge to one of the infinities). Then Eq.\ (\ref{eqn:SNHradha}) implies that for $r>\sqrt{c}$ the solution $u$ cannot have a positive maximum nor a negative minimum, because in such cases all terms on the left hand side of Eq.\ (\ref{eqn:SNHradha}) would need to have the same sign at the location of such extremum. As a result, for sufficiently large values of $r$ the solution $u$ is monotone (by similar means one can show that among the stationary points there are also no positive decreasing nor negative increasing inflection points). As such, $u$ either diverges to one of the infinities or converges to some finite value $\lim_{r\to\infty}u(r)$. In the latter case, let us write Eq.\ (\ref{eqn:SNHradha}) as
\begin{align*}
    u'(r)=\frac{1}{r^{d-1}}\int_0^r \left[s^2-h(s)\right]u(s)\,s^{d-1}\, ds.
\end{align*}
If the limit $\lim_{r\to\infty}u(r)$ is nonzero, the integral on the right hand side diverges. Then we may use the l'H\^{o}pital's rule to show that $|u'(r)|\to\infty$. It contradicts the convergence of $u$, hence the only possible finite limit point is zero and we obtain the desired trichotomy. 

The last observation is that if the solution $u$ of Eq.\ (\ref{eqn:SNHradh}) satisfies $u(r)=u'(r)=0$ at some $r>0$, then it is just a zero function $u\equiv 0$ (from the uniqueness of the solutions to the Cauchy problem). This trait, known as the non-tangency property, forbids the solution $u$ from acquiring new zeroes at finite values of $r$ as $c$ changes. If under the variations of $c$ new zero of $u$ appears, it must come from infinity.

\begin{figure}
\centering
\includegraphics[width=0.65\textwidth]{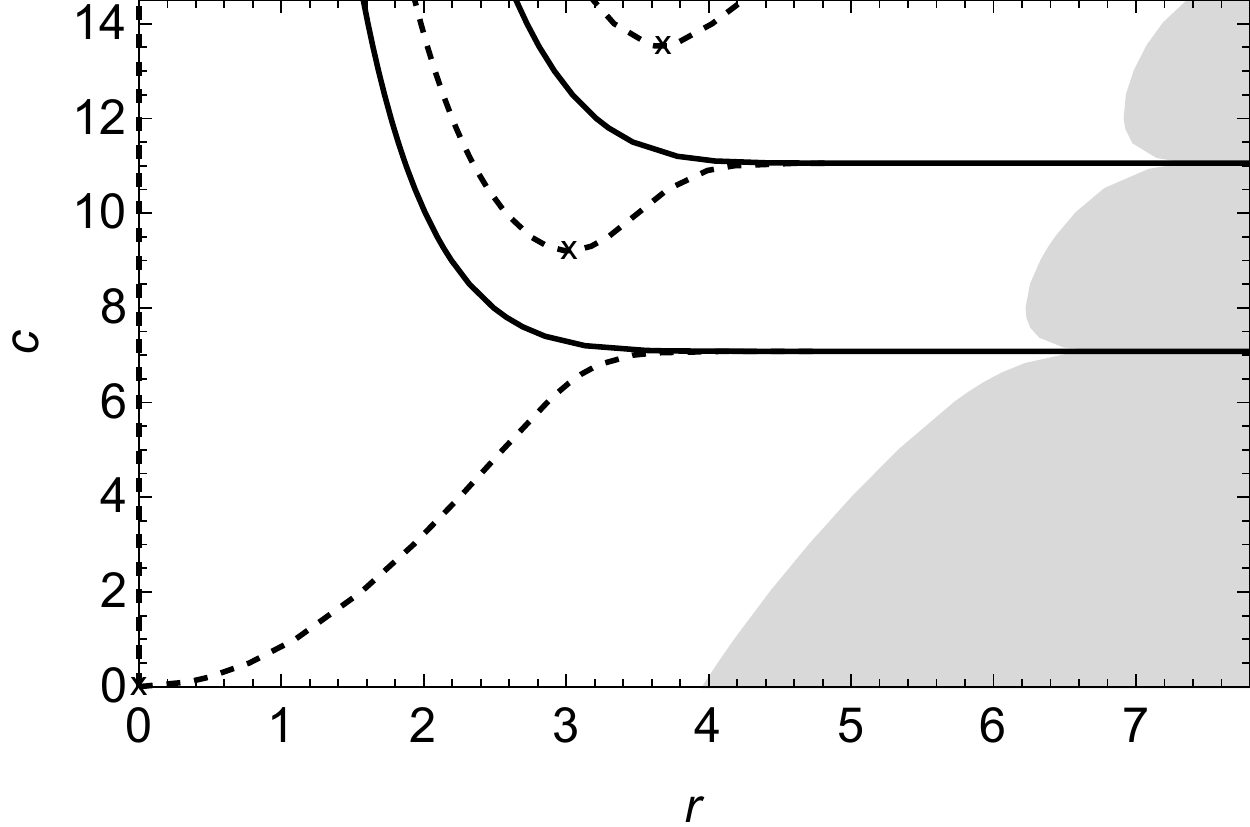}
\caption{"Phase diagram" of solutions $u$ to Eqs.\ (\ref{eqn:SNHradh}) for $d=7$ and $b=1$. The white region denotes the domain of existence of solutions, while the grey region shows the values of $r$ following the blow up. Solid lines denote values of $c$ and $r$ such that $u(r)=0$ and for the dashed lines $u'(r)=0$. The crosses indicate places where $u'(r)=u''(r)=0$.}
\label{fig:shootphase}
\end{figure}

Now we are ready for the final step, where we employ a version of the shooting method. However, before we proceed it is vital to familiarise ourselves with the ways the shape of $u$ can change when $c$ varies. The main tool here is the already mentioned continuous dependence of the solutions on the parameters and initial conditions. As we are dealing with the second-order ODEs, this result applies to both solutions and their derivatives. More precisely, it means that on every compact region of existence of solutions $u$ and $h$, they together with their derivatives change continuously in the supremum norm as one alters $c$. We would like to understand how new zeroes and extrema of $u$ may be created and destroyed in the process of changing $c$. We have already established that new zeroes can emerge only at infinity. We also know that for $r>\sqrt{c}$ there cannot be a positive maximum nor a negative minimum, hence there is a limitation on what kinds of stationary point may come from the infinity. On the other hand, there is no restriction on existence of $r$ such that $u'(r)=u''(r)=0$, so in case of $u'$ new zeroes may also pop up spontaneously. In such instances, we have an inflection point resolving into two extrema. However, the possible types of inflection points restrict this process: when $u$ is positive, a pair maximum -- minimum can appear (in the order of increasing $r$), for negative $u$ this order is reversed. These observations are presented in Fig.\ \ref{fig:shootphase} showing the locations of zeroes of $u$ and $u'$ for $0\leq c\leq 14$ in $d=7$.

\begin{figure}
\centering
\begin{subfigure}{0.47\textwidth}
\includegraphics[width=\textwidth]{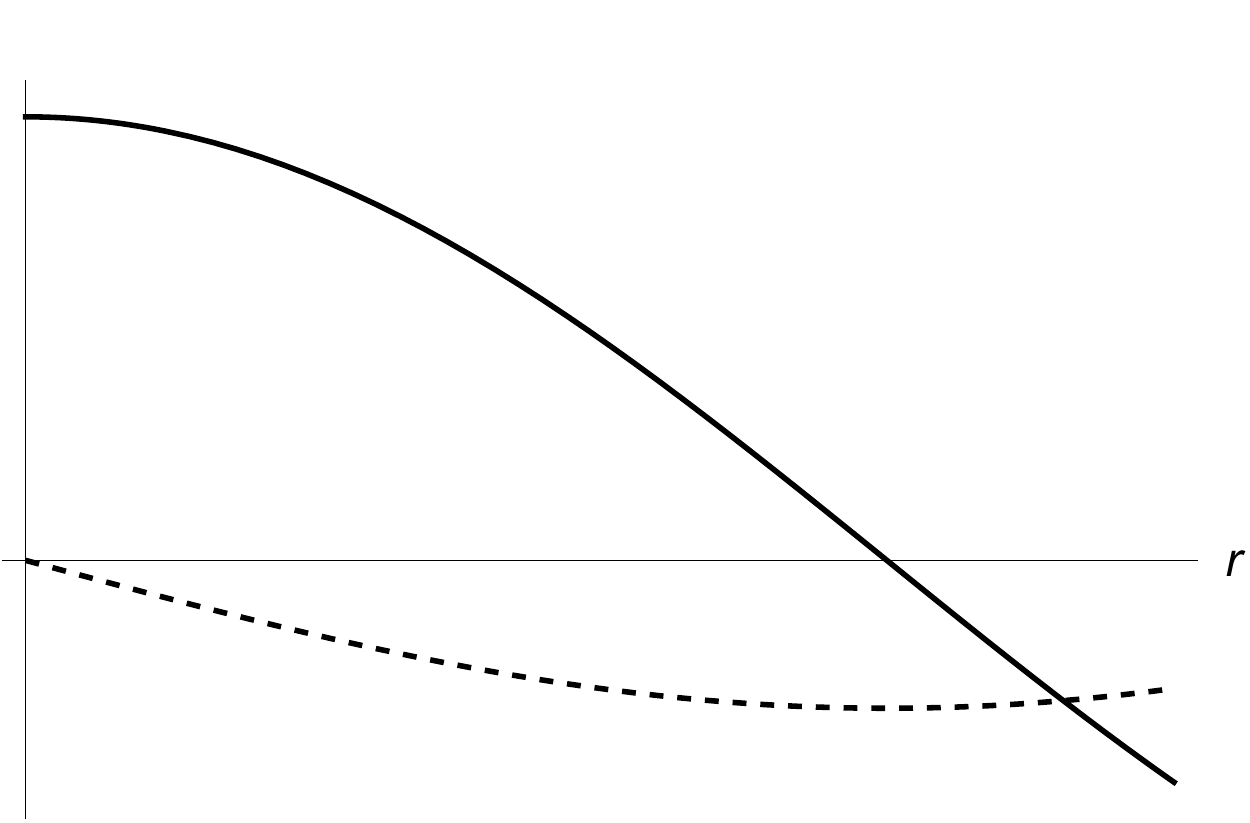}
\caption{}
\label{fig:shootIa}
\end{subfigure}
~
\begin{subfigure}{0.47\textwidth}
\includegraphics[width=\textwidth]{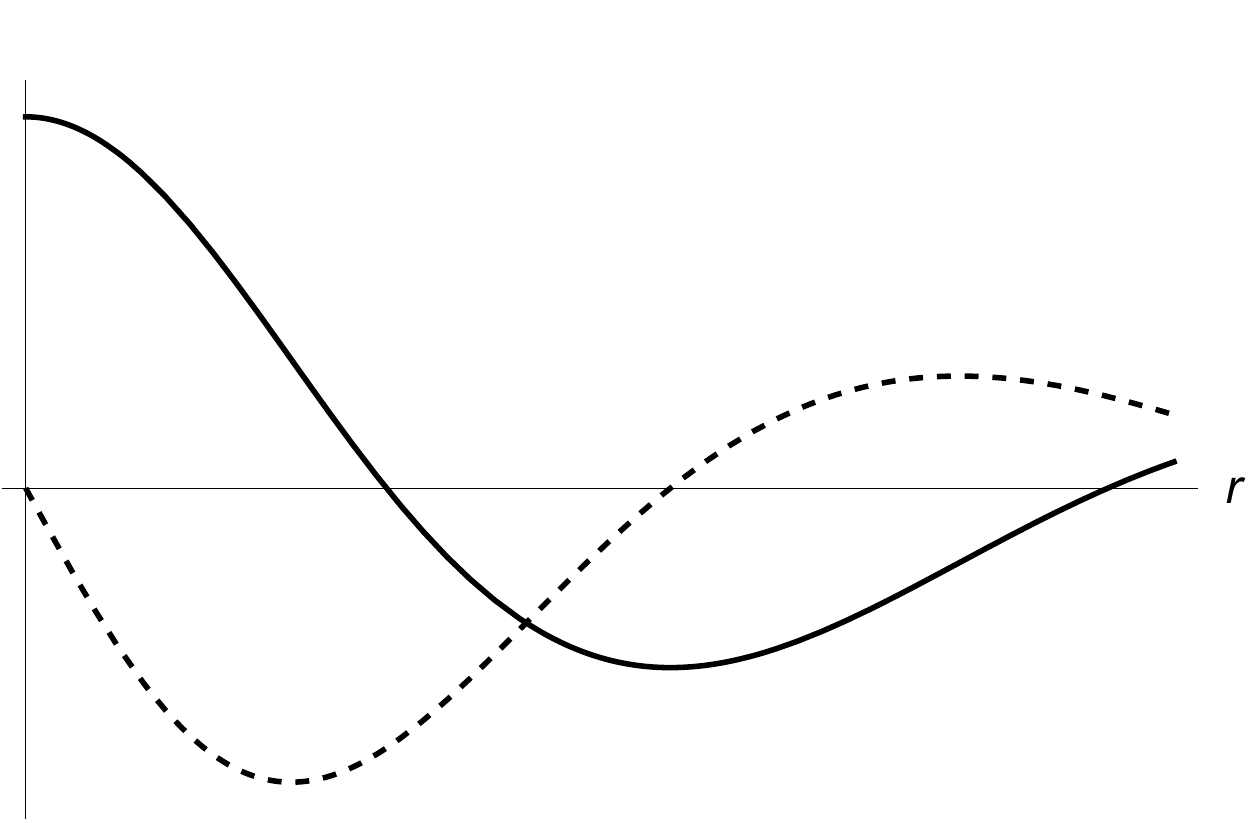}
\caption{}
\label{fig:shootIb}
\end{subfigure}
\caption{Shapes of the solution $u$ (solid line) and its derivative (dashed line) for the values of $c$ belonging to $I_0$ (left) and $I_1$ (right).}
\label{fig:shootI}
\end{figure}

Let us now define the following subset of $\mathbb{R}$:
\begin{align*}
    I_0=\{c\geq 0\, | \, \exists\, r_0 > 0 : u(r_0) = 0 \mbox{ while }
u(r) > 0 \mbox{ and } u'(r) < 0 \mbox{ for } r \in (0,r_0) \},
\end{align*}
where $u$ are the solutions of Eqs.\ (\ref{eqn:SNHradh}) for the given initial value $c$. The shape of a solution $u$ for $c\in I_0$ can be seen in Fig.\ \ref{fig:shootIa}. We already know that $I_0\neq\emptyset$ (because solutions with large enough $c$ are in this set), hence $c_0=\inf I_0$ is finite. We will now show that taking $c=c_0$ gives us the solution $u_0$ that is precisely the ground state. The general idea is to prove that $u_0$ cannot diverge to either infinity, hence by the trichotomy $u_0$ must converge to zero (some additional results obtained along the way show that the result is not just any bound state but the ground state).

A brief look into the initial conditions tells us that for the fixed $b>0$, for sufficiently small $r$ the solution $u$ is separated from zero, but the same cannot be said about its derivative as $u'(0)=0$. Hence, it is a good idea to study the shape of $u$ near zero. From l'H\^{o}pital's rule and Eq.\ (\ref{eqn:SNHradha}) it follows $\lim_{r\to\infty} u'(r)/r=-bc/d$,
leading to $u''(0)=-bc/d$. One can also go further and perform some rather lengthy calculations, including differentiating Eq.\ (\ref{eqn:SNHradha}) twice and using l'H\^{o}pital's rule several times, to find out that (we include here the previous result):
\begin{align*}
    u''(0)=-\frac{b\,c}{d},\qquad u'''(0)=0, \qquad u''''(0)=\frac{3 b \left(b^2+c^2+2 d\right)}{d (d+2)}.
\end{align*}
Alternatively, one can get these formulas just by expanding $u$ and $h$ into power series in $r$ near zero and choosing the coefficients in the way giving cancellations up to the highest possible order. As a result, we see that even though for $c=0$ the solution is initially increasing, when $c$ becomes positive the second-order term starts to dominate and the solution becomes initially decreasing. This domination prevails as $c$ increases, hence no more zeroes of $u'$ can emerge from $r=0$ (see Fig.\ \ref{fig:shootphase}). Additionally, we see here that $c_0\neq 0$, because for sufficiently small positive values of $c$ we have solutions $u$ initially decreasing and then bending up without crossing zero, hence not belonging to $I_0$. Together with $0\not\in I_0$ it means that $c=0$ cannot be the infimum of $I_0$.

All that is left is to establish the profile of $u_0$. Let us assume that $u_0$ crosses zero at $r_0>0$ for the first time. Then, the continuous dependence tells us that under small variations of $c$ the zero in $r_0$ can change a bit its position, but does not vanish (due to the non-tangency property). Depending on whether $u_0$ is decreasing in $(0,r_0)$, $c_0$ belongs to $I_0$ or not. However, the same is true for some small neighborhood of $c_0$, because in the appropriate region no stationary points may vanish nor emerge under the small changes of $c$ (the potentially problematic case of the inflection point is ruled out by the lack of such stationary points for positive, decreasing solutions $u$). Hence, $c_0$ cannot be the infimum of $I_0$ and we get a contradiction. It turns out that $u_0$ keeps positive, therefore $u$ cannot diverge to $-\infty$, as $u(0)=b>0$. Also divergence to $+\infty$ is impossible as it would require from $u_0$ to have a positive minimum. However, for close values of $c$ such minimum may again at most change its position and $c_0$ together with some neighborhood is outside of $I_0$ giving the contradiction. We showed here not only that $u_0$ cannot go to any of the infinities, meaning that it converges to zero and is a bound state, but also that it cannot cross zero, nor have a minimum. Hence, $u_0$ is a positive decreasing solution -- the ground state of Eq.\ (\ref{eqn:SNHradh}). As discussed earlier, we may now retrieve the value of $\omega$ and get the solution to the original problems (\ref{eqn:SNHrad}) and (\ref{eqn:SNHradInt}).

One can try to summarize the procedure we just described in a more general setting. Let us assume that we are investigating NLS equation with a free parameter $c$ and want to show that there exists such a value of $c$ that the solution $u$ (with $u(0)=b$ fixed) is the ground state. To follow the lines of the proof for SNH equation, the considered system needs to satisfy the following points.
\begin{itemize}
    \item For some value of the parameter $c$ the solution $u$ is positive.
    \item For some value of the parameter $c$ the solution $u$ monotonically decreases to zero and crosses it transversally.
    \item There is a trichotomy: the solutions either diverge to one of the infinities (possibly for a finite $r$) or converge to zero.
    \item A non-tangency property: if solution $u$ satisfies $u(r_0)=u'(r_0)=0$ for some $r_0$ then $u\equiv 0$.
    \item There is a sufficient control over the creation of new stationary points of $u$ as $c$ changes (for example, some mechanism excludes positive decreasing inflection points).
\end{itemize}
If these conditions are satisfied, one may define a nonempty $I_0$ as before and easily show that for $c_0=\inf I_0$ the solution does not diverge to any of the infinities, is positive and decreases, hence it is a ground state. Let us point out that even though this procedure does not depend on the dimension of the space, some additional assumptions regarding $d$ may be needed for the conditions above to apply. For example, in case of SNH system we had to assume $d\geq 6$ to show that for $c=0$ the solution $u$ is positive. However, a quick numerical check shows that it should be true also for subcritical dimensions. Hence, the described ground state does exist also in subcritical dimensions.

\vspace{\breakFF}
\begin{adjustwidth}{\marwidFF}{\marwidFF}
\small\qquad
Existence of the ground states of Eq.\ (\ref{eqn:4GPrad}) for any $u(0)=b>0$ has been rigorously proven in \cite{Biz21} using the shooting method combined with techniques of functional analysis and dynamical systems. Here, we would like to show a more heuristic reasoning relaying on some physical intuitions and based on the same approach that we used for SNH system. Since in Eq.\ (\ref{eqn:4GPrad}) the parameter $\omega$ is the only unknown, no reformulation is needed and we can just use frequency as a shooting parameter. 

The first two steps are analogous to SNH system. It turns out that here also taking $\omega=0$ implies that the solution $u$ is positive. To prove this we again assume for contradiction that $u(R)=0$ for some $R>0$. Then multiplying Eq.\ (\ref{eqn:4GPrad}) by $u\, r^{d-1}$ and $u'\, r^{d}$, respectively, and integrating over the interval $[0,R]$ gives
\begin{align*}
-\int_0^R u'^2 r^{d-1} dr-\int_0^R r^2 u^2 r^{d-1} dr+\int_0^R u^4 r^{d-1} dr&=0,\\
u'(R)^2 R^d+(d-2)\int_0^R u'^2 r^{d-1} dr-\frac{d}{2}\int_0^R u^4 r^{d-1} dr&\\
+(d+2)\int_0^R r^2 u^2 r^{d-1} dr&=0.
\end{align*}
This time we obtained two formulae consisting of four terms, so we can get rid of one of them and get the identity
\begin{align*}
2u'(R)^2 R^d+(d-4)\int_0^R u'^2 r^{d-1} dr+(d+4)\int_0^R r^2 u^2 r^{d-1} dr = 0.
\end{align*}
As all terms on its left hand side are positive, it gives a clear contradiction. Once again, this argument works only in critical and supercritical cases ($d\geq 4$), we will return to this observation in Section \ref{sec:omegab}, when discussing the Pohozaev identities. Regarding the limit of large values of $\omega$, this time we introduce $\widetilde{r}=\sqrt{\omega} r$ and $\widetilde{u}(\widetilde{r})=u(r)$. Then Eq.\ (\ref{eqn:4GPrad}) becomes
\begin{align*}
\widetilde{u}''+\frac{d-1}{\widetilde{r}} \widetilde{u}'-\omega^{-2}\widetilde{r}^2 \widetilde{u}+\omega^{-1}\widetilde{u}^3+\, \widetilde{u}=0.\label{eqn:GPrad2}
\end{align*}
Taking the limit $\omega\to\infty$ we arrive precisely at Eq.\ (\ref{eqn:SNHbessel}) and the analysis performed for SNH applies. Also the non-tangency of the solutions is trivial in this case. The only result that requires more work is the trichotomy.

In the proof of the trichotomy for SNH the key observation was that $h$ is a decreasing function of $r$, as it greatly restricted the possible behaviour of the solution $u$. In particular, as an intermediate step we showed that if the solution of SNH exists for sufficiently large $r$, it is monotonic. For GP equation such result is unavailable, as one can see in Fig.\ \ref{fig:shootGP} the solution $u$ for larger $r$ is oscillating rather than monotone. Hence, we need to use another approach. When showing the global existence of the solutions, we were considering GP equation in variables $t=r^2/2$ and $w(t)=u(r)/r$, so the system was described by Eq.\ (\ref{eqn:GPw}). We will now take a moment to study this equation.

Equation (\ref{eqn:GPw}) can be interpreted as an equation describing the motion of a unit mass particle in a time-dependent potential given by
\begin{align*}
    U=\frac{1}{4}w^4-\frac{1}{2}w^2+\frac{d-1}{8t^2}w^2+\frac{\omega}{4t}w^2,
\end{align*}
where $w$ is the position of the particle and $t$ plays the role of time, in the presence of some specific friction. Then the energy of such particle is given by $E=\dot{w}^2/2+U$, i.e.\ Eq.\ (\ref{eqn:GPwE}). Hence, the investigations of this system may be reduced to the problem from classical mechanics. For small values of $t$ the potential in which the particle is moving has a single minimum at $w=0$, but as time goes on, the neighborhood of this point gets flatter and at $t=(\omega+\sqrt{\omega^2+4(d-1)})/4$ we have a transition of this $w=0$ to a maximum. At the same time two new minima emerge at $w=\pm\sqrt{4t^2-2\omega t-(d-1)}/2t$ and the potential $U$ becomes W-shaped (Fig.\ \ref{fig:potentialGP}). As $t\to\infty$, the potential tends to the limit of $w^4/4-w^2/2$ with minima at $w=\pm 1$ (and the whole system reduces to $\ddot{w}+w(w^2-1)=0$). As a result, the physical intuition tells us that $w=\pm 1$ and $w=0$ are the only viable limiting points of $w$ as $t\to\infty$. If the solution $w$ is converging to $0$, it happens at the exponential rate, so in the original variables we also have $u(r)\to 0$. Otherwise, we have $w(t)\to\pm 1$ which translates to $u(r)\to\pm \infty$. Hence, we get the trichotomy.

\begin{figure}
\centering
\captionsetup{font=small, width=.83\linewidth}
\includegraphics[width=0.3\textwidth]{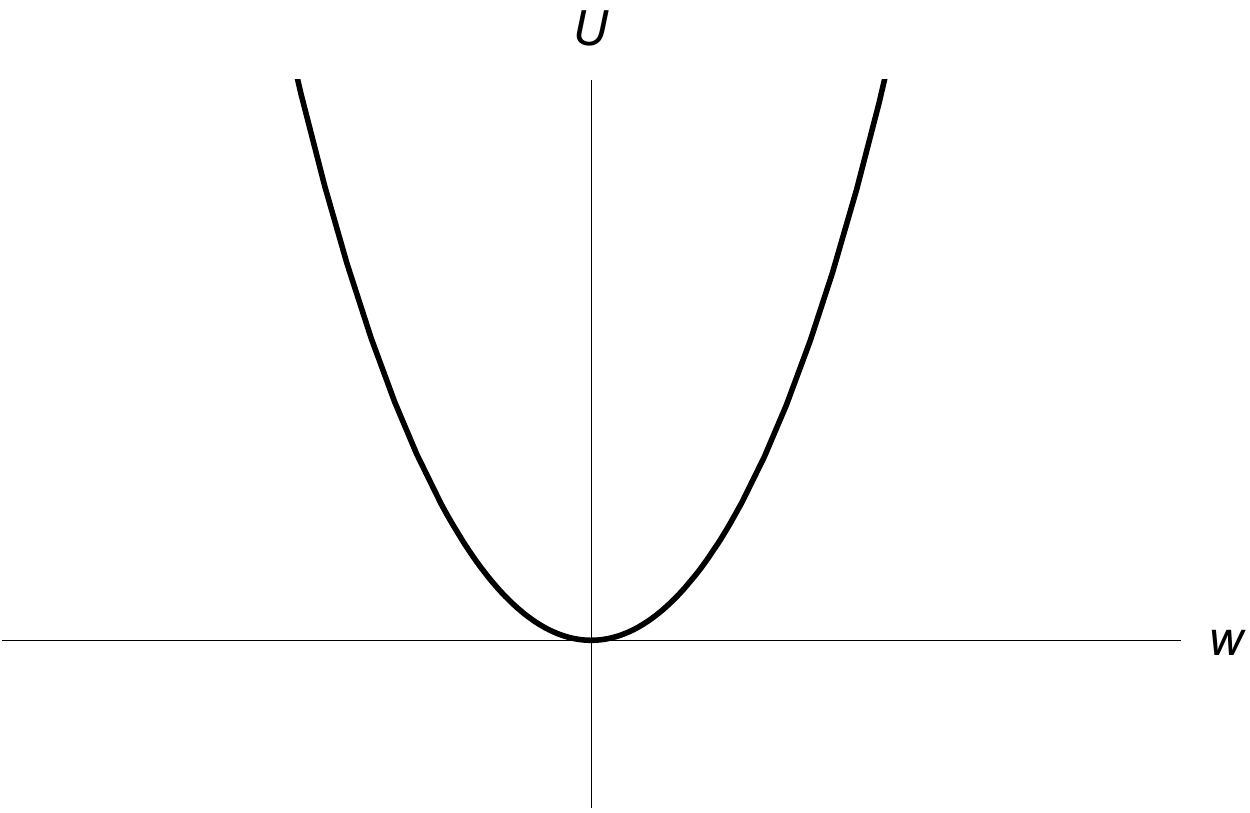}\qquad
\includegraphics[width=0.3\textwidth]{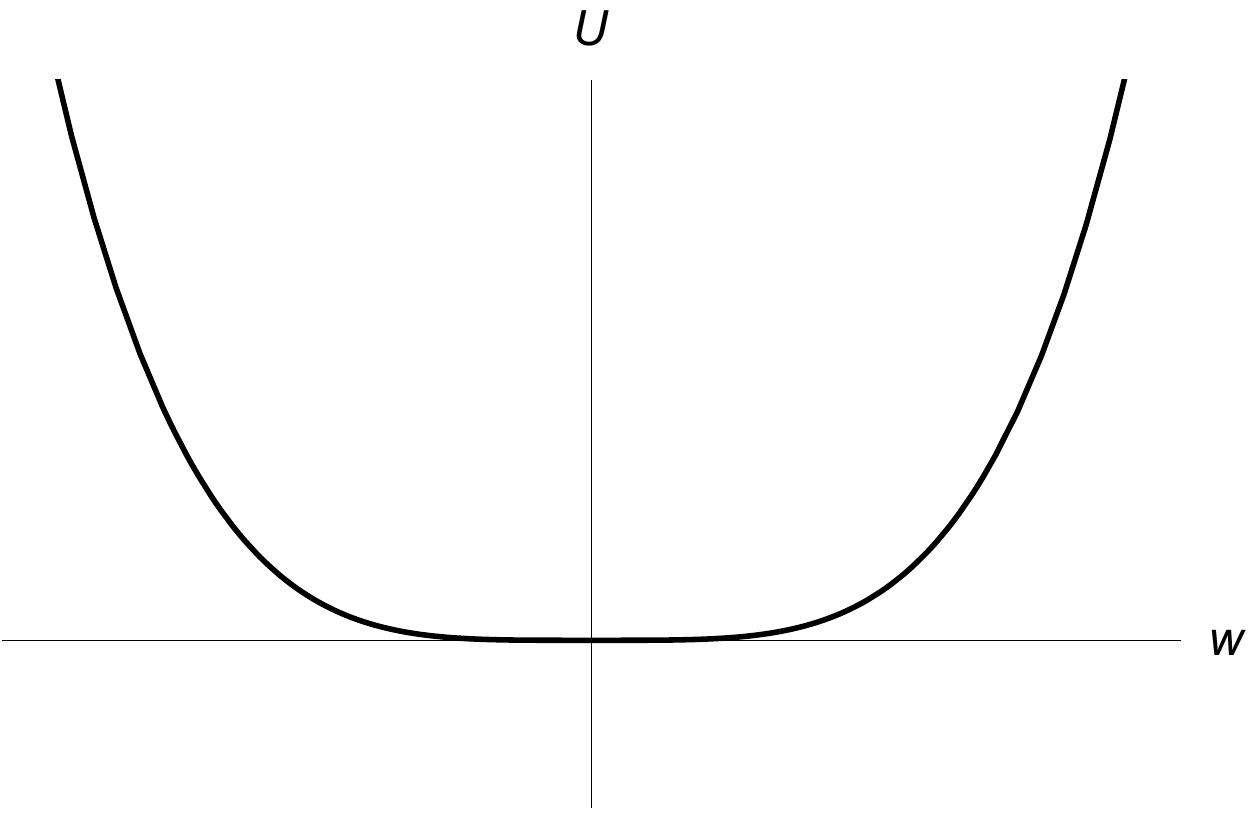}\\
\includegraphics[width=0.3\textwidth]{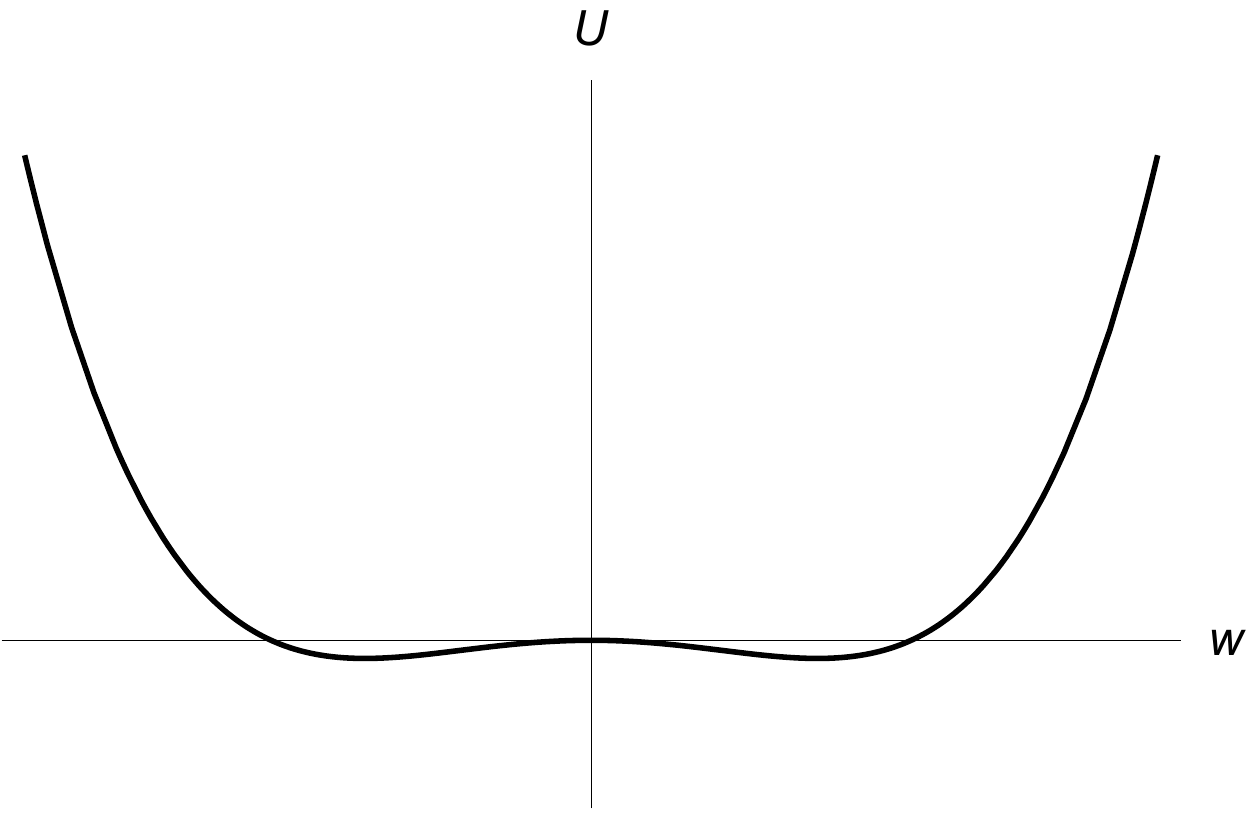}\qquad
\includegraphics[width=0.3\textwidth]{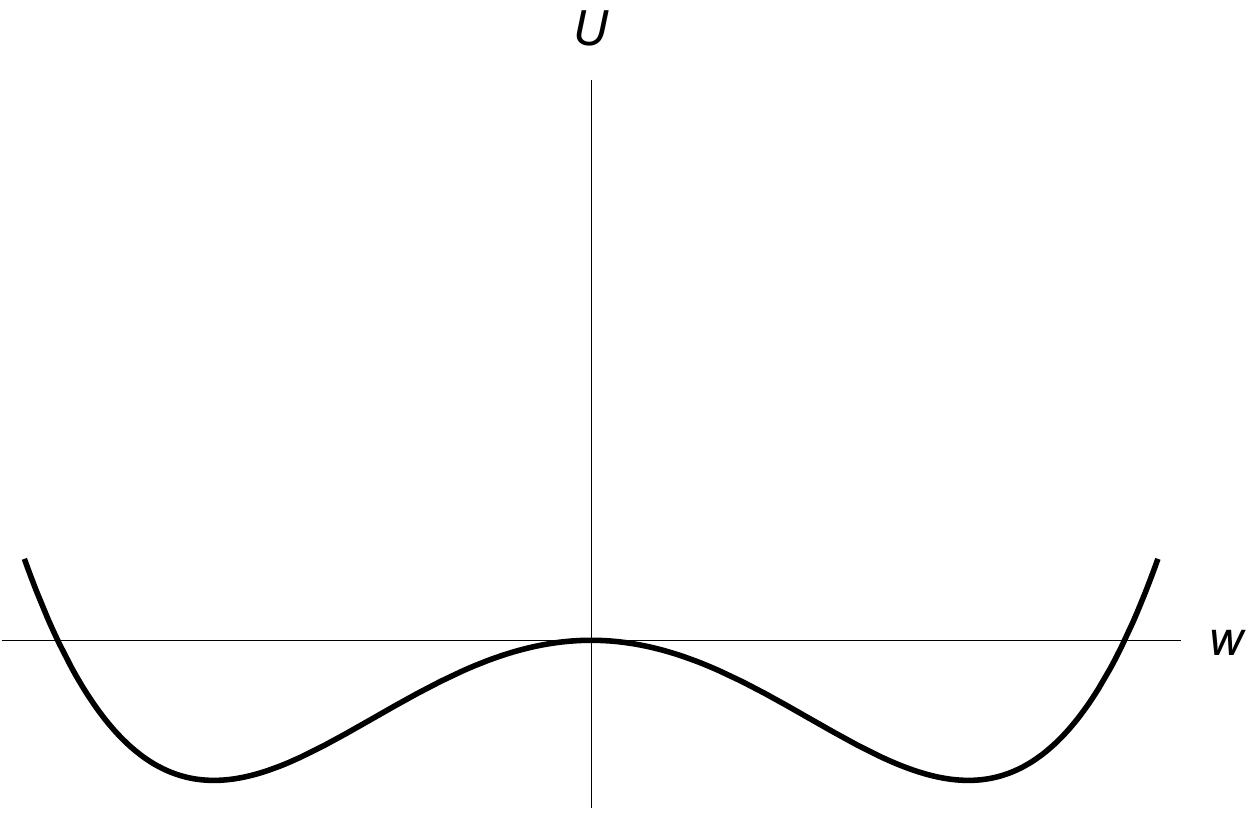}
\caption{Shapes of the potential $U$ as time $t$ increases. The second plot demonstrates the situation at the critical time $t=(\omega+\sqrt{\omega^2+4(d-1)})/4$, while the fourth one shows the limit as $t\to\infty$.}
\label{fig:potentialGP}
\end{figure}

Finally, the observations regarding the possibilities of creation of new stationary points also hold in the case of GP (a sample phase diagram for this case can be seen in Fig.\ \ref{fig:shootphaseGP}). This time we have
\begin{align*}
    u''(0)=-\frac{b\left(b^2+\omega\right)}{d},\, u'''(0)=0, \, u''''(0)=\frac{3 b \left(3 b^4+4 b^2 \omega+\omega ^2 +2 d\right)}{d (d+2)}.
\end{align*}
Therefore, we see that for $\omega>0$ (in Section \ref{sec:omegawindow} we show that this condition holds for stationary states of GP equation in critical and supercritical dimensions) no new stationary point can emerge from $r=0$. Let us now assume that there exists a positive, decreasing inflection point at $R$. Then it holds $[R^2-\omega-u(R)^2]u(R)=0$. Since $u$ is decreasing, for a slightly larger value of $r$ the left hand side of this expression becomes positive. However, then $u''$ and $u'$ are negative and Eq.\ (\ref{eqn:4GPrad}) cannot be satisfied. Hence, we obtained the last needed ingredient and we can conclude that for every $b>0$ there exists a frequency $\omega$ such that the solution $u$ of Eq.\ (\ref{eqn:4GPrad}) is a ground state.
\begin{figure}
\centering
\captionsetup{font=small, width=.83\linewidth}
\includegraphics[width=0.5\textwidth]{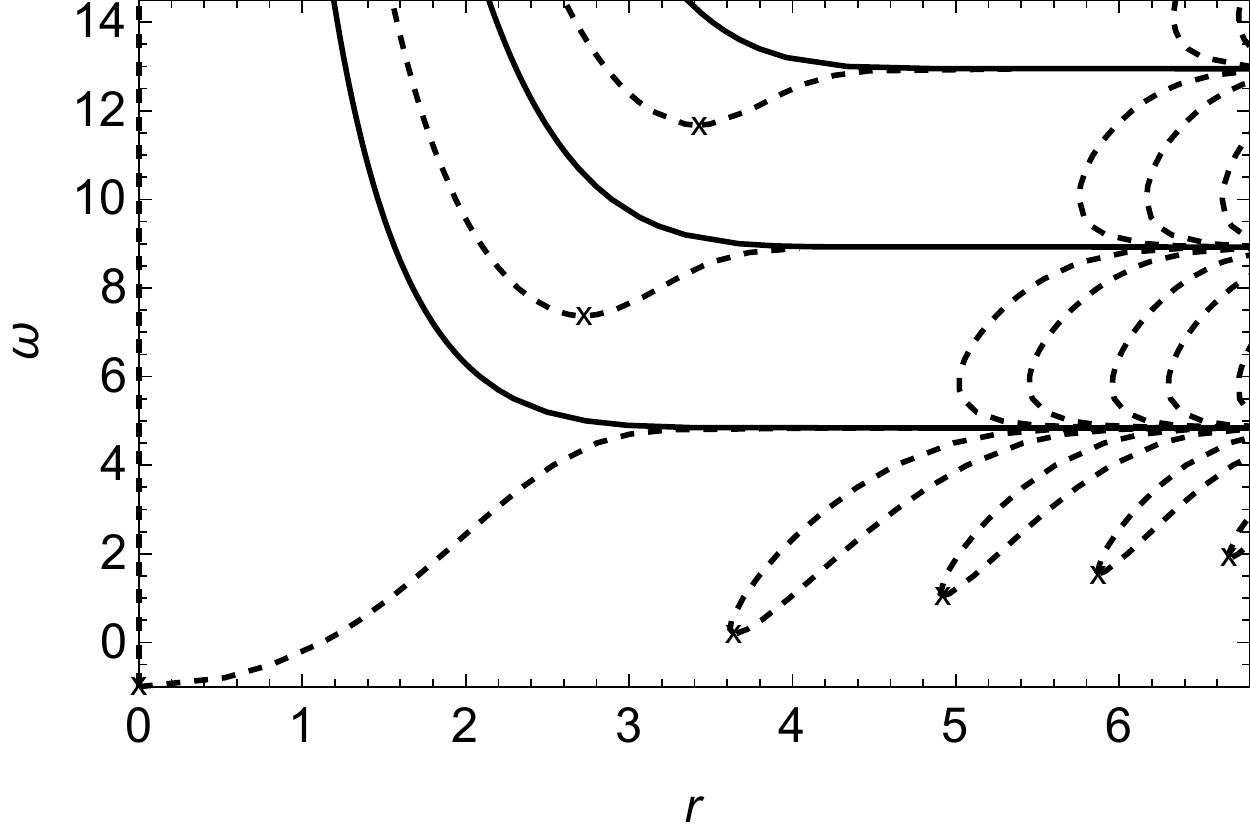}
\caption{"Phase diagram" of solutions $u$ to Eqs.\ (\ref{eqn:4GPrad}) for $d=5$ and $b=1$. Solid lines denote values of $\omega$ and $r$ such that $u(r)=0$ and for the dashed lines $u'(r)=0$. The crosses indicate places where $u'(r)=u''(r)=0$.}
\label{fig:shootphaseGP}
\end{figure}

Similar analysis as for SNH and GP systems can be also done for other nonlinear Schr\"{o}dinger equations. The details regarding some of the assumptions needed for the procedure to work may differ significantly, in some cases requiring much more work. 
\end{adjustwidth}\vspace{\breakFF}

\subsection{Excited states}
The method we have used to construct the ground state can be easily extended to give us also the excited states. We want to show now that for every $b>0$ and $n\in\mathbb{N}_+$ there exists a bound state of Eq.\ (\ref{eqn:SNHradh}) for which $u$ crosses zero exactly $n$ times. The main role here will be played by two facts: 1) in the limit $c\to\infty$ the solution oscillates around zero indefinitely (as seen in Fig.\ \ref{fig:shootbessel}), 2) the solution is absolutely monotone from some point. Then, 1) lets us find for any $n$ a sufficiently large $c$, such that the solution $u$ crosses zero at least $n$ times, while 2) ensures that as $c$ increases and new zeroes of $u$ appear, they do so separately. We will see the details while proving the existence of the first excited state.

Let us define the following set:
\begin{align*}
    I_1=\{&c\geq 0\, | \, \exists\, 0<r_0<\rho_1<r_1 : u(r_0) = u(r_1) = 0 \mbox{ and } u'(\rho_1) = 0 \mbox{ while, }\\
    & u(r) > 0,\, u'(r) < 0 \mbox{ for } r \in (0,r_0), u(r) < 0,\, u'(r) < 0 \mbox{ for } r \in (r_0,\rho_1),\\
    &\mbox{and } u(r) < 0,\, u'(r) > 0 \mbox{ for } r \in (\rho_1,r_1)  \}.
\end{align*}
Even though its definition may look rather complicated, the idea behind it is to simply mimic the shape of the plot of $\tilde{u}_\infty$ from $\tilde{r}=0$ up to the second zero (as shown in Fig.\ \ref{fig:shootIb}).
Then, as for large $c$ the shape of $u$ resembles $\tilde{u}_\infty$ (on some initial interval), we know that $I_1\neq\emptyset$ and we can define $c_1=\inf I_1$. Once again we claim that for $c_1$ the solution $u$, let us call it $u_1$, is the solution we are looking for: the first excited state. To prove it, we need to study the inflection points of solutions to Eq.\ (\ref{eqn:SNHradha}) in a slightly larger detail.

\begin{figure}
\centering
\includegraphics[width=0.55\textwidth]{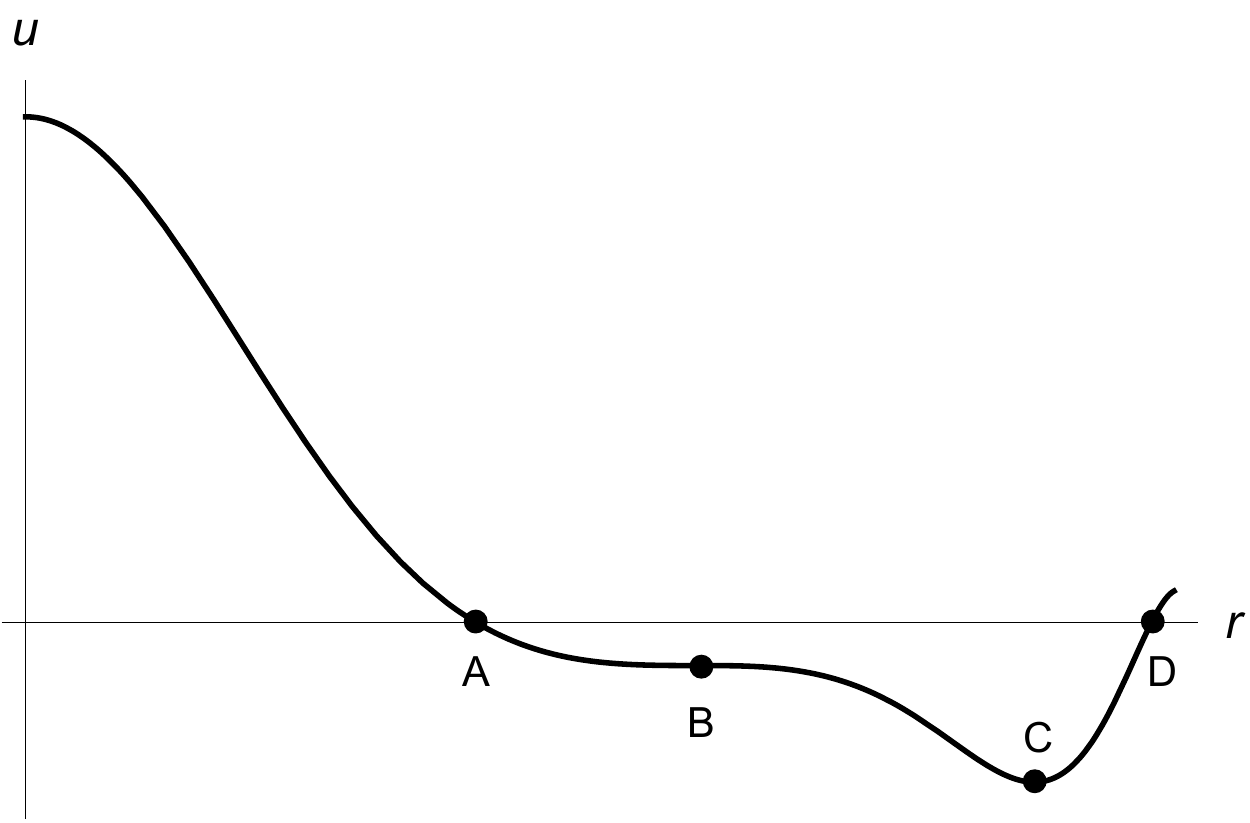}
\caption{Potentially possible shape of the solution $u_1$. Zeroes are here denoted by A and D, a local minimum by C, and an inflection point by B.}
\label{fig:shootinflection}
\end{figure}

We already know that there are no positive decreasing nor negative increasing inflection points, but what about the other possibilities? Let us consider a situation as shown in Fig.\ \ref{fig:shootinflection}, i.e.\ a solution $u$ having a decreasing negative inflection point. Then Eq.\ (\ref{eqn:SNHradha}) tells us that at point B it must hold $h(r)=r^2$. As $h$ is a decreasing function, in the minimum C we have $(h-r^2)u>0$. But there $u'=0$ and $u''>0$ giving us a contradiction. Hence, after a negative decreasing inflection point (and analogously for a positive increasing one) there can be no extremum (and in general no stationary point). This result shows us that $u_1$ cannot have any inflection point, because $c_1=\inf I_1$.

Now we claim that $u_1$ cannot have two zeroes as then, similarly to the case of the ground state, small changes of $c$ would keep the solution $u$ in $I_1$ contradicting $c_1$ being the infimum. It cannot have no zeroes either, as $c_1=\inf I_1$ so some arbitrarily small increase in $c$ would result in the simultaneous creation of two zeroes. However, new zeroes can emerge only from infinity and for $r>\sqrt{c}$ there can be no positive maximum, nor negative minimum. As a result only one zero can appear at a time, giving a contradiction. Hence, $u_1$ must cross the horizontal axis exactly once. It is then obvious that $u_1$ cannot diverge to $+\infty$. Now we assume that it goes to $-\infty$. If this happens monotonically, then small changes of $c$ cannot produce a minimum (there can be no negative minimum of $u$ for $r>\sqrt{c}$) contradicting $c_1=\inf I_1$. If there are some stationary points along the way, small changes of $c$ will not remove them, so there exists a neighborhood of $c_1$ separated from $I_1$ -- contradiction. Hence, by the trichotomy we have $u_1\to 0$ crossing zero exactly once -- the first excited state. As before, we may now calculate the frequency $\omega=\lim_{r\to\infty}h(r)$ and get the first excited state of Eq.\ (\ref{eqn:SNHrad}).

\begin{figure}
\centering
\includegraphics[width=0.75\textwidth]{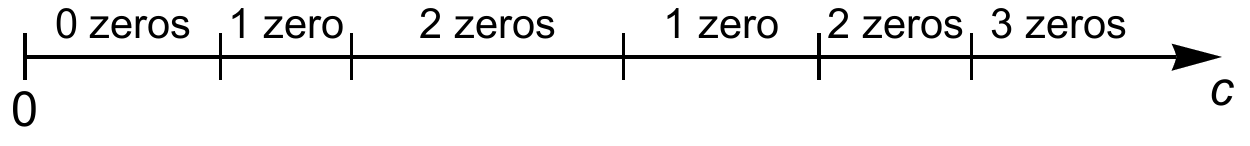}
\caption{Potential number of zeroes of solution $u$ for different values of $c$.}
\label{fig:zeroes}
\end{figure}

This procedure can be repeated for any $n\in\mathbb{N}_+$ implying the existence of the $n-$th excited state. In such case, the subset of $\mathbb{R}$ giving the proper value of $c$ as an infimum can be defined generally as:

\begin{align*}
    I_n=\{  c\geq 0\, |& \, \exists\, 0=\rho_0<r_0<\rho_1<r_1<...<\rho_n<r_n :\\
    &u(r_0) = u(r_1)=...=u(r_n) = 0\\
    &  \mbox{and } u'(\rho_1) =... =  u'(\rho_n) = 0 \mbox{ while between them it holds}\\
& u(r) > 0 \mbox{ and } u'(r) < 0 \mbox{ for } r \in (\rho_{2k},r_{2k}),\\
& u(r) < 0 \mbox{ and } u'(r) < 0 \mbox{ for } r \in (r_{2k},\rho_{2k+1}),\\
& u(r) < 0 \mbox{ and } u'(r) > 0 \mbox{ for } r \in (\rho_{2k+1},r_{2k+1}),\\
& u(r) > 0 \mbox{ and } u'(r) > 0 \mbox{ for } r \in (r_{2k+1},\rho_{2k+2})\}.
\end{align*}
Once again, this rather complicated definition just describes the initial shape of $\tilde{u}_\infty$, this time up to $(n+1)$-th zero. Hence, we are able to define the whole ladder of excited states given by $c_n=\inf I_n$, and then a whole ladder of solutions to Eqs.\ (\ref{eqn:SNHrad}), being excited states with frequencies $\omega_n$. Let us point out here that even though the method we presented is constructive, we do not know at this point whether we have found all spherically symmetric bound states. In principle, it is possible that there exists a whole interval of $c$ giving some bound state, or there may be numerous separated values $c$ giving bound states with the same number of zeroes (such situation might happen if the number of zeroes of $u$ changes with $c$ as shown in Fig.\ \ref{fig:zeroes}). We discuss this matter further in Section \ref{sec:uniqueness}.

As in the case of the ground state, the features needed to perform the reasoning above can be easily summarized. Let us have a NLS equation with some free parameter $c$ and satisfying:
\begin{itemize}
    \item For some value of the parameter $c$ the solution $u$ is positive.
    \item There exists a (possibly limiting) solution that oscillates indefinitely.
    \item There is a trichotomy: solutions of the equation either diverge to one of the infinities (possibly in a finite time) or converge to zero.
    \item The equation has a non-tangency property, i.e.\ if a solution $u$ satisfies $u(r_0)=u'(r_0)=0$ for some $r_0$ then $u\equiv 0$.
    \item There is some mechanism letting us to control the creation of new stationary points as $c$ changes.
    \item As $c$ changes, new zeroes of $u$ come from infinity separately.
\end{itemize}
Then one can repeat the presented logic and show the existence of spherically symmetric bound states with any prescribed number of zeroes. When comparing these conditions with the ones sufficient for the ground state, we see that here we need a stronger control on the creation of new stationary states, additional control on the emergence of new zeroes and a way to obtain solutions with arbitrarily many oscillations.

\vspace{\breakFF}
\begin{adjustwidth}{\marwidFF}{\marwidFF}
\small\qquad
Just as in the case of the ground state, the mechanisms behind these assumptions may vary. For GP one can reason heuristically and try to use the \textit{point mass in the potential well} picture. Then the $n$-th excited state translates to the solution where the point mass settles to zero after making $n$ oscillations. In this picture it is rather intuitive that the additional conditions needed to show the existence of excited states hold. In principle, it should be possible to give a strict proof of this fact using just simple ODE methods, in a manner similar to \cite{Coh21}.
\end{adjustwidth}\vspace{\breakFF}

\subsection{Singular solutions}\label{sec:exsingular}
So far we have been studying solutions with a fixed finite $b>0$. One may wonder what happens if the value of $b$ gets larger and larger. The limiting case of such procedure would be a singular solution -- solution to Eqs.\ (\ref{eqn:SNHrad}) satisfying $\lim_{r\to 0} u(r)=\infty$. We would like to end this section with a short discussion of the existence of ground and excited singular states.

As before, we will be using the formulation given by Eq.\ (\ref{eqn:SNHradh}). The first step is to factor out the singular behaviour of the solution near zero. To this end, let us introduce the rescaled variables for some finite $b>0$:
\begin{align*}
    \rho=\sqrt{b}\, r,\qquad U(\rho)=\frac{u(r)}{b}, \qquad H(\rho)=\frac{h(r)}{b}.
\end{align*}
Then Eqs. (\ref{eqn:SNHradh}) can be rewritten as
\begin{subequations}\label{eqn:SNHbinf}
\begin{align}[left ={ \empheqlbrace}]
U'' + \frac{d-1}{\rho} U' - b^{-2} \rho^2 U + U H&= 0,\label{eqn:SNHbinfa}\\
H''+\frac{d-1}{\rho} H' +U^2&= 0.\label{eqn:SNHbinfb}
\end{align}
\end{subequations}
In the limit $b\to\infty$ this system can be reduced to a single equation if we assume $H=U$ (such solution is called synchronised):
\begin{align}\label{eqn:laneemden}
U''+\frac{d-1}{\rho} U' +U^2&= 0.
\end{align}
This is a $d$-dimensional Lane-Emden equation with quadratic nonlinearity. It has a singular solution $U(\rho)=2(d-4)/\rho^2$. Hence, we would like to factor out such behaviour by defining new functions $\widetilde{u}$ and $\widetilde{h}$ such that:
\begin{align*}
u(r)=\frac{2(d-4)}{r^2} \tilde{u}(r),\qquad h(r)=\frac{2(d-4)}{r^2} \tilde{h}(r).
\end{align*}
Then Eqs.\ (\ref{eqn:SNHradh}) become 
\begin{subequations}\label{eqn:utilde}
\begin{align}[left ={ \empheqlbrace}]
\widetilde{u}'' + \frac{d-5}{r}\widetilde{u}' +\frac{2(d-4)}{r^2}(\widetilde{u}\widetilde{h}-\widetilde{u})-r^2 \widetilde{u}&= 0,\label{eqn:utildea}\\
\widetilde{h}''+\frac{d-5}{r} \widetilde{h}'+ \frac{2(d-4)}{r^2}(\widetilde{u}^2-\widetilde{h})&= 0.\label{eqn:utildeb}
\end{align}
\end{subequations}

For small values of $r$ the harmonic term in Eq.\ (\ref{eqn:SNHbinfa}) may be neglected and the solutions $u$ and $h$ are expected to behave like $2(d-4)/r^2$. It suggests that the proper initial conditions for the solutions are $\widetilde{u}(0)=\widetilde{h}(0)=1$. At first sight it is not evident how to use the shooting method, because there is no apparent shooting parameter here. However, due to the presence of singular coefficients in Eqs.\ (\ref{eqn:utilde}), $\widetilde{u}(r)=\widetilde{h}(r)=1$ and $\widetilde{u}'(r)=\widetilde{h}'(r)=0$ are not sufficient as initial conditions. One can show that in supercritical dimensions ($d>6$), for every $c\in\mathbb{R}$ local solutions to system (\ref{eqn:utilde}) behave as follows:
\begin{align}\label{eqn:asymptilde}
    \widetilde{u}(r) = 1 - c\,r^\lambda + \mathcal{O}(r^4), \qquad \widetilde{h}(r) = 1 + 2c\,r^\lambda + \mathcal{O}(r^4)
\end{align}
where
\begin{align*}
    \lambda=\frac{-d+6+ \sqrt{d^2 + 4d - 28}}{2}
\end{align*}
(for the future reference let us point out that in supercritical dimensions $3\leq \lambda<4$). A rather technical derivation of this fact, based on the analysis of the unstable manifold of an appropriate dynamical system and a series of estimates, can be found in Appendix \ref{sec:appsingular} (it is a slightly extended version of the proof published in \cite{Fic21}). Parameter $c$ describing the asymptotic behaviour near $r=0$ is our desired shooting parameter. Now we will briefly argue that in supercritical dimensions system (\ref{eqn:utilde}) together with initial conditions given by Eqs.\ (\ref{eqn:asymptilde}) satisfies all the assumptions needed to show that the whole ladder of excited states exists, similarly to the regular case.

One may easily show that for $c<0$ the solution $\widetilde{u}$ is positive. To see it, let us observe that then there exists a neighborhood of zero where $\widetilde{u}>1$ is larger than 1 and $\widetilde{h}<1$. If we assume that $\widetilde{u}$ has at some point maximum, Eq.\ (\ref{eqn:utildea}) tells us that then $\widetilde{h}>1$. However, from Eq.\ (\ref{eqn:utildeb}) we know that $\widetilde{h}$ cannot have a minimum for $\widetilde{u}<1$, so we have a contradiction, hence $\widetilde{u}$ is a positive, increasing function. From the continuous dependence on the initial conditions we conclude the same for $c=0$.

To study the limit of large values of $c$ it is convenient to introduce a new variable: $s=\ln c^{1/\lambda} r$ (unfolding the radial variable into the whole line). Taking $c\to\infty$ then changes system (\ref{eqn:utilde}) into
\begin{subequations}\label{eqn:utilde2}
\begin{align}[left ={ \empheqlbrace}]
\ddot{\widetilde{u}} + (d-6)\dot{\widetilde{u}} +2(d-4)\left(\widetilde{u}\widetilde{h}-\widetilde{u}\right)&= 0,\label{eqn:utilde2a}\\
\ddot{\widetilde{h}}+(d-6) \dot{\widetilde{h}}+ 2(d-4)\left(\widetilde{u}^2-\widetilde{h}\right)&= 0,\label{eqn:utilde2b}
\end{align}
\end{subequations}
where dots denote derivative with respect to $s$. The asymptotic behaviour (\ref{eqn:asymptilde}) translates to
\begin{align*}
\widetilde{u}(s)=1- e^{\lambda s} + \mathcal{O} (e^{4s}),\qquad \widetilde{h}(s)=1+2e^{\lambda s} -\mathcal{O} (e^{4 s})
\end{align*}
as $s\to-\infty$. Let us define a functional $E$:
\begin{align*}
E= \dot{\widetilde{u}}^2+\frac{1}{2}\dot{\widetilde{h}}^2-2(d-4)\widetilde{u}^2-(d-4)\widetilde{h}^2+2(d-4)\widetilde{h}\tilde{u}^2.
\end{align*}
Then it holds $\dot{E}=-2(d-6)\dot{\widetilde{u}}^2-(d-6)\dot{\widetilde{h}}^2<0$ and $\lim_{s\to -\infty} E(s)=-(d-4)$. Hence, for any $s$ we have
\begin{align*}
    -(d-4)=\lim_{s\to-\infty}E(s)>E>(d-4)(-2\tilde{u}^2-\tilde{h}^2+2\tilde{h}\tilde{u}^2)
\end{align*} giving $(\tilde{h}-1)(\tilde{h}+1-2\tilde{u}^2)>0$. The asymptotic behaviour of $\widetilde{u}$ and $\widetilde{h}$ near $-\infty$ tells us that there exists $s_0$ such that $\widetilde{h}+1>2\widetilde{u}^2$ holds for $s<s_0$. Then we have $\widetilde{u}^2-\widetilde{h}<\frac12 (1-\widetilde{h})$ and the last term in the left hand side of Eq.\ (\ref{eqn:utilde2b}) is negative for $s<s_0$. It means that $\widetilde{h}$ cannot have a maximum in this interval and hence it is increasing. Then, also in $(-\infty,s_0)$, Eq.\ (\ref{eqn:utilde2a}) is a damped harmonic oscillator with increasing frequency. It means that we can indefinitely extend the interval on which $\widetilde{h}+1>2\widetilde{u}^2$ holds. As a result, $\widetilde{u}$ is an oscillating function with decreasing amplitude over the whole half-line $[0,\infty)$.

The other property shared by regular and singular solutions to SNH equation are constraints on stationary points. To show them, it is convenient to consider the function $h$ instead of $\widetilde{h}$. From Eq.\ (\ref{eqn:asymptilde}) we know that near $r=0$ its derivative behaves like
\begin{align}
    h'(r)=-\frac{4(d-4)}{r^3}-4c\,(d-4)(\lambda-2)r^{\lambda-3}+\mathcal{O}(r).
\end{align}
Hence, the limit of $h' r^{d-1}$ as $r\to 0$ exists and is equal to zero.  We then have
\begin{align}
    h'(r)=-\frac{1}{r^{d-1}}\int_0^r u^2(s)\, s^{d-1}\, ds,
\end{align}
so $h$ is a decreasing function of $r$ also in the singular case. It means that $\widetilde{h}/r^2=h/2(d-4)$ is bounded from above in every interval $(r,\infty)$ with $r>0$. As a result, Eq.\ (\ref{eqn:utildea}) can be analysed as the regular case: it turns out $\widetilde{u}$ cannot have positive maxima, nor negative minima for sufficiently large $r$. Also there can be no stationary points after a negative decreasing (or positive increasing) inflection point. Some other key conclusions coming from these results are the monotonicity of $\widetilde{u}$ from some point on and the fact that new zeroes of $\widetilde{u}$ appear separately as $c$ changes. We also point out that Eq.\ (\ref{eqn:asymptilde}) tells us that no new stationary points of $\widetilde{u}$ can emerge from zero.

The remaining assumptions are easy to check. As there are no singularities for $r>0$ in Eqs.\ (\ref{eqn:utilde}), the non-tangency property is obvious. Also the trichotomy property is simple as the solution $\widetilde{u}$ eventually becomes monotone. Then it either diverges to one of the infinities or converges to a finite value. Regarding the second case, if we rewrite Eq.\ (\ref{eqn:utildeb}) in the form
\begin{align*}
    \widetilde{u}'(r)=\frac{1}{r^{d-5}}\int_0^r \left[s^2+\frac{2(d-4)}{s^2}-\frac{2(d-4)}{s^2}\widetilde{h}(s)\right]\widetilde{u}(s)s^{d-5}\, ds
\end{align*}
and use l'H\^{o}pital's rule it becomes apparent that $\lim_{r\to\infty}\widetilde{u}(r)=0$ as otherwise we get a contradiction, just as in the regular case.

Having all these results, the same argument as in the regular case gives the singular solutions: the ground state together with the whole ladder of the excited states. Even though such solutions are usually deemed unphysical \cite{Ban02}, they will prove useful in Section \ref{sec:omegab} as limits of regular solutions when $b\to\infty$. Let us point out that here we were assuming the supercritical dimensions $d\geq 7$. This is caused by the fact that in lower dimensions the character of the eigenvalues of a linear system appearing in the proof of asymptotic behaviour (\ref{eqn:asymptilde}) changes and the proof is no longer valid (see Appendix \ref{sec:appsingular} for the details). In fact, in the critical dimension as $b\to\infty$, the mass of the solution goes to zero.

\vspace{\breakFF}
\begin{adjustwidth}{\marwidFF}{\marwidFF}
\small\qquad
For other NLS equations the analysis of the singular solutions may differ in a significant way from what we presented here. For example, in case of GP equation (\ref{eqn:4GPrad}), after we factor out the singular part by introducing $\widetilde{u}$ such that $u=\sqrt{d-3}\widetilde{u}/r$, the study of the behaviour near zero shows that there is no unstable manifold. This is a significant difference when compared to SNH system, however it poses no problem as $\omega$ may still play the role of the shooting parameter for singular GP equation. For detailed proofs of the existence of singular solutions of GP equation in supercritical dimensions (although considering only the ground states) we refer to \cite{Biz21} and \cite{Sel13} (see also \cite{Mer91}).
\end{adjustwidth}\vspace{\breakFF}

\section{Uniqueness}\label{sec:uniqueness}
Usually, the next natural question that one poses after showing the existence of some solutions is the matter of their uniqueness. We would like to know whether the values of the shooting parameter $c$ (and consequently frequencies $\omega$) giving $n$-th stationary state are uniquely defined. 

We start by considering the regular ground states of SNH. Then the uniqueness can be easily proved by the same method as presented in \cite{Cho08}. For two positive solutions $u_1$ and $u_2$ one can define a Wro\'{n}skian $W$ and then with some minor changes it is possible to repeat the proof of Lemma 3.1 in \cite{Cho08} resulting in the monotonicity of $r^{d-1} W(r)$. Since this combination is equal to zero at $r=0$ and the fast decay of the solution ensures that it also vanishes in infinity, it must be zero indefinitely, giving us the uniqueness.

However, we would like to present here in a greater detail another proof. This one is inspired by Proposition 1.1 in \cite{Gal11} and can be found in \cite{Fic21}). We prove by contradiction and assume that for some $b>0$ there exist two different initial values, $c_1$ and $c_2$, giving the solutions $u_1$, $h_1$ and $u_2$, $h_2$ that are the ground states. Without loss of generality one may assume that $c_1>c_2$. Since we are considering the ground states, $u_1$ and $u_2$ are positive and the function $\rho(r):=u_1(r)/u_2(r)$ is well-defined. Of course, it satisfies $\rho(0)=1$ and $\rho'(0)=0$. To analyze the monotonicity of this function, it is convenient to introduce $\delta(r):=h_2(r)-h_1(r)$ and $\mu(r):=r^{d-1}\, u_2(r)^2\,\rho'(r)$. They satisfy $\mu(0)=\mu'(0)=0$, $\delta(0)<0$, $\delta'(0)=0$, and are solutions to equations
\begin{align}\label{eqn:uniqueSNH1}
    \mu'(r)=r^{d-1}u_2(r)^2\,\rho(r)\,\delta(r)
\end{align}
and
\begin{align}\label{eqn:uniqueSNH2}
    \left(r^{d-1}\delta'(r)\right)'=-r^{d-1}u_2(r)^2\left[1-\rho(r)^2\right].
\end{align}
In some neighborhood of $r=0$ it holds $\mu<0$ implying that also $\rho'<0$ there. It means that initially $\rho$ is decreasing. This fact together with Eq.\ (\ref{eqn:uniqueSNH2}) leads to the conclusion that $\delta$ is initially decreasing and stays so as long as $\rho<1$.

Now we show that $\rho'<0$ indefinitely. Assume otherwise and let $r_0$ be the lowest argument for which $\rho'=0$. Then, in $(0,r_0)$ we have $0<\rho<1$, $\delta<0$, and $\mu'<0$. But $\mu(r_0)=0$ giving us a contradiction. As a result $\rho'<0$ and $\mu$ is a decreasing function. The latter lets us to write for $r>1$: $r^{d-1} u_2(r)^2 \rho'(r) < u_2(1)^2 \rho'(1)<0$ giving
\begin{align*}\label{eqn:uniqueSNH3}
    \rho'(r)<\frac{u_2(1)^2\rho'(1)}{r^{d-1}u_2(r)^2}<0.
\end{align*}
Using $0<\rho<1$ once again we can integrate this inequality getting
\begin{align*}\label{eqn:uniqueSNH4}
    -1<\lim_{r\to\infty} \rho(r)-\rho(1)=\int_1^\infty \rho'(r)\,dr< u_2(1)^2\rho'(1)\int_1^\infty \frac{dr}{r^{d-1}u_2(r)^2}<0
\end{align*}
As a result, the integral on the right-hand side is convergent. Now we can use the fact that $u_2$ decays exponentially and the Cauchy-Schwarz inequality yields the following contradiction:
\begin{align*}
    \infty =& \int_1^\infty dr = \int_1^\infty \left(r^{d-1}u_2(r)^2\right)^{1/2}\, \left(\frac{1}{r^{d-1}u_2(r)^2}\right)^{1/2}\,dr\\
    \leq& \left(\int_1^\infty r^{d-1}u_2(r)^2 \, dr\right)^{1/2} \, \left(\int_1^\infty \frac{1}{r^{d-1}u_2(r)^2}\, dr\right)^{1/2}<\infty.
\end{align*}
The uniqueness of $c$ for the ground state of Eqs.\ (\ref{eqn:SNHradh}) for a fixed $b$ translates to the uniqueness of $\omega$ for the ground state of Eqs.\ (\ref{eqn:SNHrad}). Hence, in a fixed dimension one can define a function $\omega(b)$ being the frequency of a ground state for any $b>0$. In the next section we investigate this function using both analytical and numerical methods. In particular, we study how its qualitative behaviour changes with the dimension.

\vspace{\breakFF}
\begin{adjustwidth}{\marwidFF}{\marwidFF}
\small\qquad
Unfortunately, the presented approach employs some specific features of SNH, such as the possibility of getting rid of $\omega$, and in principle may be impossible to repeat in case of other nonlinear Schr\"{o}dinger equations. Even for a relatively simple GP, this line of reasoning seems to fail. To see it, let us say that for Eq.\ (\ref{eqn:4GPrad}) we want to show the uniqueness of the frequency $\omega$ giving the ground state for some fixed $b>0$. In this case the analogue of Eq.\ (\ref{eqn:uniqueSNH1}) takes the form of
\begin{align}\label{eqn:uniqueGP}
    \mu '=r^{d-1}u_2^2\,\rho\,\left(\omega_2 u_2+u_2^3-\omega_1 u_1 -u_1^3\right).
\end{align}
The more complicated form of the right-hand side of this formula obstructs the further analysis in the presented spirit. In particular, let us assume that $\omega_1>\omega_2$. Then near zero it holds $u_1<u_2$, because
\begin{align*}
    u''(0)=-\frac{b^3+\omega b}{d}.
\end{align*}
If we assume $\omega_1<\omega_2$, we get $u_1>u_2$. As a result, the initial sign of the expression on the right hand side of Eq.\ (\ref{eqn:uniqueGP}) cannot be established as seen for SNH. The same obstacle emerges when trying to adapt the uniqueness proof from \cite{Cho08} to GP equation.
\end{adjustwidth}\vspace{\breakFF}

The reasoning giving the uniqueness of regular ground states of SNH can be applied, with only minor changes, to show that the singular ground state is also unique. This time we also assume that there exist two different positive solutions $\widetilde{u}_1$, $\widetilde{u}_2$ to Eqs.\ (\ref{eqn:utilde}). They behave near zero like Eq.\ (\ref{eqn:asymptilde}) with different parameters $c_1$ and $c_2$, respectively. Without loss of generality we assume $c_1>c_2$. We can again define $\rho$ as the ratio $\widetilde{u}_1/\widetilde{u}_2$ and $\delta$ as the difference $\widetilde{h}_2-\widetilde{h}_1$. Then the asymptotic behaviours tell us that near zero $\rho<1$, while the equation satisfied by $\delta$
\begin{align*}
    \left(r^{d-5}\delta'(r)\right)'=2(d-4)r^{d-7}\left[ \widetilde{u}_2(r)^2\left(\rho(r)^2-1\right)+\delta(r)\right],
\end{align*}
gives us $\delta<0$ there. Defining $\mu=r^{d-5} \widetilde{u}_2^2 \rho'$ yields
\begin{align*}
    \mu'(r)=2(d-4)r^{d-7} \widetilde{u}_2(r)^2\delta(r).
\end{align*}
Now, one may use these auxiliary functions in exactly the same way as in the regular case to get a contradiction (the only non-obvious point is the observation that $\widetilde{u}_2\in L(\mathbb{R}^d)$, but it is analogous to the regular case). Let us also point out that, similarly to the proof of the existence of singular solutions, here we also assume $d\geq 7$. As a result, we get a unique singular ground state for every such dimension. Its frequency will be denoted by $\omega_\infty$ in the following.

So far, this section was concentrated only on the ground states. The reason for it is simple: the uniqueness of excited states of nonlinear elliptic equations is a very challenging and in most cases open problem \cite{Has}. The existing results are extremely scarce and apply to rather specific cases, assuming a specific form of the nonlinearity \cite{Tro05} or imposing very strong growth conditions \cite{Cor09}. Even with these restrictions, they focus only on the first excited (the authors \cite{Tro05} claim that extension of their method to higher states should be possible, although technically challenging). Recently, a new approach emerged \cite{Coh21}: to give a computer-assisted proof using the methods of validated numerics \cite{Tuc}. With this method, the authors of the cited paper were able to show that the equation $-\Delta u+u-u^3=0$ has a unique set of initial conditions $u(0)=b$, $u'(0)=0$ giving a spherically-symmetric bound state with exactly $n$ zeroes, as long as $n\leq 20$. Unfortunately, also this technique does not seem to be applicable in our case -- the use of validated numerics restricts us to single values of $b$, we are not able to prove the uniqueness for all $b$ at one time, as we did for the ground states. In principle, it should be possible to use the continuous dependence of solutions on $b$ to extend this result to some finite range of $b$, but all $b>0$ seem to be out of reach. However, the numerical observations suggest that the excited states (at least the lower ones) are also unique, in the same sense as the ground states. It also means that the situation presented in Fig.\ \ref{fig:zeroes} does not take place.

\section{Behaviour of \texorpdfstring{$\omega(b)$}{ω(b)}}\label{sec:omegab}
Observations regarding uniqueness (based either on rigorous arguments or numerical results) let us define a function $\omega_n(b)$ being the unique frequency of the $n$-th state for some central value $b>0$. In this section we study its properties in various dimensions. We focus especially on its behaviour for small and large values of $b$. We also present numerical results for SNH and other relevant systems.

\subsection{Preliminaries}\label{sec:omegawindow}
First, we would like to introduce the notation that will be used in this section and Chapter \ref{sec:dynamics}. In the following it will be handy to work with a scalar product given by
\begin{align*}\label{eqn:scalardot}
    \langle f, g \rangle = \int_0^\infty \bar{f}(r) g(r) \, r^{d-1}\, dr.
\end{align*}
The norm associated with this scalar product will be denoted simply by $\Vert\cdot\Vert$. Let us also point out that even though the presence of the complex conjugation is meaningless here, as we consider only real solutions, it will become relevant in the next chapter where the stability is discussed. The other useful notion regards solutions of the linearized Eq.\ (\ref{eqn:SNHstatInt}). By $e_n$ we will denote $n$-th radially symmetric eigenfunction (enumerated by the number of nodes, so $e_0$ is a ground state) of $-\Delta+r^2$. The eigenvalue to the function $e_n$ will be denoted by $\Omega_n$, so one has $-\Delta e_n + r^2 e_n=\Omega_n e_n$. We additionally impose the normalization with respect to $\Vert\cdot\Vert$ on $e_n$, so explicitly it holds
\begin{align}\label{eqn:freqlin}
    e_n(r)=\sqrt{\frac{2n!}{\Gamma\left(n+\frac{d}{2}\right)}}L^{\left(\frac{d}{2}-1\right)}_n(r^2) e^{-\frac{r^2}{2}},\qquad
    \Omega_n=d+4n,
\end{align}
with $L_n^{(\alpha)}$ being the generalised Laguerre polynomials. Even though we focus here on SNH equation, in this section and the next chapter it will be sometimes handy to use $e_n$ and $\Omega_n$ instead of the explicit expressions. Then, the obtained formulas can be easily adapted to describe other NLS equations. In the same spirit it will be often beneficial to use Eq.\ (\ref{eqn:4NLS}) instead of Eq.\ (\ref{eqn:SNHstatInt}), namely, hide the exact forms of potential and nonlinearity behind $V$ and $F$.

\vspace{\breakFF}
\begin{adjustwidth}{\marwidFF}{\marwidFF}
\small\qquad
Since Eq.\ (\ref{eqn:4GP}) has the same linear part as Eq.\ (\ref{eqn:SNHstatInt}), GP has exactly the same eigenfunctions $e_n$ and eigenvalues $\Omega_n$ as SNH. In the case of general Eq.\ (\ref{eqn:4NLS}), these quantities need to be replaced by the eigenfunctions and eigenvalues of $-\Delta+V$ (i.e.\ satisfying $-\Delta e_n+V e_n=\Omega_n e_n$).
\end{adjustwidth}\vspace{\breakFF}

For a moment, let us focus on the ground states and let $u$ be a positive solution of Eq.\ (\ref{eqn:SNHstatInt}) with some frequency $\omega_0$. As mentioned, we may write it as Eq.\ (\ref{eqn:4NLS}) with $V=|x|^2$ and $F=A_d (|u|^2 \ast |x|^{-(d-2)})\, u$. Then $\omega_0 u=-\Delta u+V u-F$ and
\begin{align*}
    0&=\left\langle e_0,-\Delta u+ V u-F-\omega_0 u\right\rangle
    =\left\langle e_0, (-\Delta+V)u\right\rangle-\left\langle e_0,F\right\rangle-\omega_0\langle e_0, u\rangle\\
    &=\langle(-\Delta+V) e_0,u\rangle-\left\langle e_0,F\right\rangle-\omega_0\langle e_0, u\rangle
    =\Omega_0\langle e_0, u\rangle-\left\langle e_0,F\right\rangle-\omega_0\langle e_0, u\rangle\\
    &=(\Omega_0-\omega_0)\langle e_0, u\rangle-\left\langle e_0,F\right\rangle.
\end{align*}
In the third equality we used the self-adjointness of $-\Delta+V$ with respect to our scalar product. Since both $u$ and $e_0$ are positive, it follows that $\langle e_0,u\rangle>0$. Also $F$ is positive, so for $\omega_0>\Omega_0$ one gets a contradiction. It means that $\omega_0\leq\Omega_0=d$, giving us the upper bound for the ground state frequency of SNH. Of course, this line of argumentation ceases to work for the excited states, as $u$ is no longer a positive function.

\vspace{\breakFF}
\begin{adjustwidth}{\marwidFF}{\marwidFF}
\small\qquad
Clearly this reasoning is very general and works for any NLS with focusing nonlinearity, i.e.\ such equations that $F$ is positive. Hence, also for GP it holds $\omega_0\leq\Omega_0=d$. Of course, for other trapping potentials, the value of $\Omega_0$ changes accordingly.
\end{adjustwidth}\vspace{\breakFF}

The lower bounds for $\omega_n$ are given by the so-called Pohozaev identities \cite{Poh65}. We may obtain them in a similar way as we showed that for $c=0$ the solution $u$ to Eqs.\ (\ref{eqn:SNHradh}) is positive, but this time we use Eqs.\ (\ref{eqn:SNHrad}). Let us assume that $u$ and $v$ describe a bound state of this system (it may be either the ground state or any excited state). Then Eq.\ (\ref{eqn:SNHrada}) can be multiplied either by $u\, r^{d-1}$ or $u'\, r^d$ and integrated over $(0,\infty)$ in each of these cases. An analogous operation can be done for Eq.\ (\ref{eqn:SNHradb}), but with $v\, r^{d-1}$ or $v'\, r^d$. Since the solutions decay sufficiently fast in infinity, some terms may be integrated by parts with the boundary terms discarded. As a result, we get the following four identities:
\begin{subequations}\label{eqn:poh1}
\begin{align}
-\left\Vert u' \right\Vert^2-\left\Vert r u \right\Vert^2+\omega \left\Vert u \right\Vert^2 - \int_0^\infty u^2 v\, r^{d-1}\, dr&=0,\label{eqn:poh1a}\\
(d-2) \left\Vert u' \right\Vert^2+(d+2)\left\Vert r u \right\Vert^2-\omega d \left\Vert u \right\Vert^2+\int_0^\infty u^2 v' \, r^{d-1}\, dr&\nonumber\\
+d\int_0^\infty u^2 v\, r^{d-1}\, dr&=0,\\
\left\Vert v' \right\Vert^2+\int_0^\infty u^2 v\, r^{d-1}\, dr&=0,\\
(d-2)\left\Vert v' \right\Vert^2-2\int_0^\infty u^2 v'\, r^{d-1}\, dr&=0.
\end{align}
\end{subequations}
Once again, they can be combined in such a way that some of the terms disappear giving us, for example,
\begin{align*}
(d-6)\left\Vert u'\right\Vert^2 +(d+2)\left\Vert r u\right\Vert^2= \omega (d-2)\left\Vert u \right\Vert^2.\label{eqn:pohSNH}
\end{align*}
For $d\geq 6$, i.e.\ for critical and supercritical dimensions, both terms on the left-hand side are positive, hence, we must have $\omega\geq 0$. Let us recall that we did not assume anything about the nature of the bounds state, hence for every $n$ we have $\omega_n\geq 0$. Even though during these calculations we were utilizing the spherical symmetry of our solutions, they can be repeated in full generality without this assumption (see \cite{Van18}).

Interestingly, if we restrict ourselves to the supercritical dimensions $(d>6)$, this bound may be improved. Combining Eqs.\ (\ref{eqn:poh1}) in a slightly different manner it is possible to obtain
\begin{align}\label{eqn:poh2}
8 \left\Vert r u\right\Vert^2-(d-6) \int_0^\infty u^2 v\, r^{d-1}\, dr = 4\,\omega \left\Vert u \right\Vert^2.
\end{align}
Now we recall that $d$ is the smallest eigenvalue of the linear operator $-\Delta + r^2$. As $e_n$ constitute the Schauder basis of $L^2(0,\infty)$, any solution $u$ can be represented in this basis. Then, it is obvious that $\Vert u' \Vert^2+ \Vert r u \Vert^2\geq d \Vert u \Vert^2$. Using this inequality together with Eqs.\ (\ref{eqn:poh1a}) and (\ref{eqn:poh2}), we get
\begin{align*}
    d \Vert u \Vert^2 \leq \Vert u' \Vert^2+ \Vert r u \Vert^2&= \omega \left\Vert u \right\Vert^2 - \int_0^\infty u^2 v\, r^{d-1}\, dr \\
    &= \omega \left\Vert u \right\Vert^2 +\frac{4\omega}{d-6}\Vert u\Vert^2 -\frac{8}{d-6}\Vert r u\Vert^2,
\end{align*}
which for $d>6$ gives the improved lower bound
\begin{align*}
    \omega \geq d -\frac{4}{d-2}+\frac{8}{d-2}\frac{\Vert ru\Vert^2}{\Vert u\Vert^2} \geq d -\frac{4}{d-2} = \frac{d-6}{d-2}d.
\end{align*}

\vspace{\breakFF}
\begin{adjustwidth}{\marwidFF}{\marwidFF}
\small\qquad
Analogous calculations for GP lead to the Pohozaev identity
\begin{align*}
(d-4)\left\Vert u'\right\Vert^2 +(d+4)\left\Vert r u\right\Vert^2= \omega d\left\Vert u \right\Vert^2.
\end{align*}
Here also the criticality of the dimension is crucial, as only for $d\geq 4$ we get the constraint $\omega_n\geq 0$. In fact, this observation can be made also for other nonlinearities: the Pohozaev identities forbid our systems to have solutions with negative frequency in critical and higher dimensions. As an example, let us take Eq.\ (\ref{eqn:4NLS}) with $V=|x|^n$ and $F=u^p$, then one gets
\begin{align*}
\left(d+\frac{p+1}{p-1}n\right)\Vert r^{n/2} u\Vert^2+ \left(d-2\frac{p+1}{p-1}\right)\Vert u'\Vert^2= \omega d \Vert u\Vert^2.
\end{align*}
In such case for this argumentation to work (of course, assuming that all present integrals are convergent), one needs to assume $d\geq2(p+1)/(p-1)$ being precisely the condition for critical and supercritical dimensions.

If the considered system is in a supercritical dimension, also the improved bound can be obtained. Calculations analogous to the SNH case give then
\begin{align*}
    \omega \geq \Omega_0 -\frac{2(p+1)}{(p-1)d}+\frac{(p+1)(n+2)}{(p-1)d}\frac{\Vert r^{n/2}u\Vert^2}{\Vert u\Vert^2} \geq \Omega_0 -\frac{2(p+1)}{(p-1)d}
\end{align*}
For GP it means that $\omega\geq d-4/d$.

\end{adjustwidth}\vspace{\breakFF}

\subsection{Small and large values of  \texorpdfstring{$b$}{b}}\label{sec:smalllargeb}
The precise shape of $\omega_n (b)$ for any $n$ cannot be obtained analytically and can be studied only numerically, as we do in the next subsection. However, for very small and very large values of $b$ it is possible to get some approximate results, as we proceed to show here.

Solutions of SNH with very small $b$ can be investigated with the aid of the bifurcation theory. For brevity, let us use again Eq.\ (\ref{eqn:4NLS}) with $V$ and $F$ given by (\ref{eqn:SNHstatInt}). Then we may rewrite it as $\mathcal{A}(\omega,u)=0$, where $\mathcal{A}$ is a functional on $\mathbb{R}\times\Sigma$ ($\Sigma$ is a function space defined in Section \ref{sec:subcritical}) given by
\begin{align}
    \mathcal{A}(\omega,u)=-\Delta u+V u-F(u)-\omega u.
\end{align}
Since $\mathcal{A}(\omega, 0) \equiv 0$ we can see that zero solutions satisfy our equation for every $\omega$ giving us a line in $\mathbb{R}\times\Sigma$. We would like to find the values of $\omega$ for which new solutions bifurcate from this line. The implicit function theorem tells us that it is possible only when the linear operator
\begin{align}
    \mathcal{A}_u(\omega,0)[w]=-\Delta w+V w-\omega w
\end{align}
is not invertible. It means that $\omega$ is the eigenvalue, hence it is equal to one of $\Omega_n$. Then the standard local bifurcation theory \cite{Amb} says that from the line $u\equiv 0$ in $\Omega_n$ there emerges a pair of solutions given by
\begin{subequations}\label{eqn:bifur}
\begin{align}\label{eqn:bifura}
    u=\pm\sqrt{(\Omega_n-\omega)a_n} e_n+\mathcal{O}(|\Omega_n-\omega|),
\end{align}
where 
\begin{align}\label{eqn:bifub}
a_{n}=6\frac{\langle e_n,\mathcal{A}_{\omega,u}(\Omega_n,0)[e_n]\rangle}{\langle e_n,\mathcal{A}_{u,u,u}(\Omega_n,0)[e_n]^3\rangle}
\end{align}
\end{subequations}
Obviously $\mathcal{A}_{\omega,u}(\Omega_n,0)[e_n]=-e_n$, so the numerator is equal to $-1$ (because our functions $e_n$ are normalized). However, to calculate the denominator we need to bring back the explicit formula for $F$. Then the calculation gives
\begin{align*}
\mathcal{A}_{u,u,u}(\Omega_n,0)[e_n]^3=-6A_d\left(\int_{\mathbb{R}^d} \frac{|e_n(y)|^2}{|x-y|^{d-2}} d y\right) e_n.
\end{align*}
This expression can be simplified with the use of the Newton formula (\ref{eqn:newton}), giving
\begin{align*}
\langle e_n, \mathcal{A}_{u,u,u}(\Omega_n,0)[e_n]^3\rangle=-\frac{6}{d-2}\int_0^\infty \int_0^\infty \frac{e_n(r)^2 e_n(s)^2\, r^{d-1} s^{d-1}}{\max\{r, s\}^{d-2}} dr\, ds.
\end{align*}
The double integral on the right-hand side does not have an explicit form that is valid for any $d$ and $n$, so for now we will denote it as $S_{nnnn}$. Then $a_n=(d-2)/S_{nnnn}$ and we get an expression for small solutions $u$ with frequencies close to $\Omega_n=d+4n$:
\begin{align*}
    u_n=\sqrt{\frac{d-2}{S_{nnnn}}} (d+4n-\omega)^{1/2} e_n+\mathcal{O}(|d+4n-\omega|).
\end{align*}
We have restricted ourselves here only to the positive branch. Remembering that $u(0)=b$, the approximate formula for $\omega(b)$ follows:
\begin{align*}
    \omega_n(b)=d+4n-\frac{S_{nnnn}}{(d-2)e_n(0)}b^2+\mathcal{O}(b^3).
\end{align*}
For the ground state $S_{0000}=2^{1-\frac{d}{2}}/\Gamma(d/2)$ and $e_0(0)=\sqrt{2/\Gamma(d/2)}$, hence
\begin{subequations}\label{eqn:bifursnh}
\begin{align}
    u_0=\sqrt{2^{\frac{d}{2}-1}(d-2)\Gamma\left(d/2\right)} (d-\omega)^{1/2} e_0+\mathcal{O}(|d-\omega|),\label{eqn:bifursnha}\\
    \omega_0(b)=d-\frac{b^2}{2^{d/2}(d-2)}+\mathcal{O}(b^3).\label{eqn:bifursnhb}
\end{align}
\end{subequations}

\vspace{\breakFF}
\begin{adjustwidth}{\marwidFF}{\marwidFF}
\small\qquad
The expressions (\ref{eqn:bifur}) can be applied to other NLS equations with cubic nonlinearities. For GP the coefficient $a_0$ takes the form
\begin{align*}
    a_n=\left(\int_0^\infty e_n(r)^4 r^{d-1} \,dr\right)^{-1}.
\end{align*}
Once again, there is no compact expression for $a_n$, but for the ground states it can be simplified to $a_0=2^{\frac{d}{2}-1}\Gamma(d/2)$. Then
\begin{align*}
    u_0=\sqrt{2^{\frac{d}{2}-1}\Gamma\left(d/2\right)} (d-\omega)^{1/2} e_0+\mathcal{O}(|d-\omega|),\\
    \omega_0(b)=d-\frac{b^2}{2^{d/2}}+\mathcal{O}(b^3).
\end{align*}

If the nonlinearity in Eq.\ (\ref{eqn:4NLS}) is of higher order than cubic, the formulae above do not work anymore and a separate treatment is needed. As an example, for $V=|x|^2$ and $F=|u|^4 u$ the bifurcation analysis shows that the small ground states are given by
\begin{align*}
u_0=3^{d/8}(d-\omega)^{1/4} e_0+\mathcal{O}(|\Omega_n-\omega|^{1/2}),\\ \omega_0(b)=d-\frac{b^4}{3^{d/2}}+\mathcal{O}(b^5).
\end{align*}

Let us point out that the behaviour of $\omega(b)$ for small $b$ is determined by both the potential $V$ and nonlinearity $F$. The latter determines the bifurcation scheme and so impacts the shape of the branch. However, the former not only establishes locations of the bifurcation points (via linear part eigenvalues) but also influences the shape of the branch through the eigenfunctions present in the relevant expressions.
\end{adjustwidth}\vspace{\breakFF}

Having established the behaviour of $\omega(b)$ for small central values $b$, we now move to the opposite case and investigate the behaviour for large $b$. This time, instead of the bifurcation theory, we will use some simple tools from the field of dynamical systems. Let us recall that we have already encountered a similar problem when studying the singular solution. We return to Eq.\ (\ref{eqn:laneemden}) obtained as a synchronised solution in the limit $b\to\infty$. Its analysis can be simplified with the help of the Emden-Fowler transformation, reducing it to an autonomous equation. Hence, let us introduce $s=\ln \rho$ and $w(s)=\rho^2 U(\rho)$ getting
\begin{align*}
    w''+(d-6)w'-2(d-4)w+w^2=0.
\end{align*}
This system has fixed points at $w=0$ and $w=2(d-4)$. We are interested in the nontrivial one, since it corresponds to the singular solution $U=2(d-4)/\rho^2$. As we want to understand the behaviour of the system near this point, we perform a linearization there. To do so, we introduce such function $\nu$ that $w(s) = 2(d-4)+ \nu(s)$ and then drop all terms of orders higher than linear in $\nu$. This gives
\begin{align*}
    \nu''+(d-6)\nu'+2(d-4)\nu=0.
\end{align*}
The same equation was already encountered in the study of the singular problem in Appendix \ref{sec:appsingular} as Eq.\ (\ref{eqn:Betaxia}). The nature of its eigenvalues depends on the dimension $d$. If we introduce $\beta=-\frac{d}{2}+3$ and $\alpha_1=\sqrt{|d^2-20d+68|}$, we may write them for $7\leq d\leq 15$ as $\beta\pm i\alpha_1$, while for $d\geq 16$ as $\beta\pm \alpha_1$. The change of their character takes place at $d=10+4\sqrt{2}\approx 15.66$. The solutions of this linear system are
\begin{align*}
    \nu(s)= \begin{cases}A e^{\beta s}\sin(\alpha_1s+\delta) & \mbox{ for $7\leq d\leq 15$,}
    \\ A e^{(\beta-\alpha_1) s }+ B e^{(\beta+\alpha_1) s }& \mbox{ for $d\geq 16$,}
    \end{cases}
\end{align*}
where $A$, $B$, and $\delta$ are some real constants (we do not need their exact values, so we use them here as some dummy variables -- they may differ between the following equations). Since this system is hyperbolic, one can use these formulas to approximate solutions $w$:
\begin{align*}
    w(s)\approx \begin{cases}2(d-4)\left[ 1+ A e^{\beta s}\sin(\alpha_1s+\delta)\right]& \mbox{ for $7\leq d\leq 15$,}
    \\ 2(d-4)\left[1+ A e^{(\beta-\alpha_1) s }+ B e^{(\beta+\alpha_1) s }\right]& \mbox{ for $d\geq 16$}.
    \end{cases}
\end{align*}
Finally, returning to the original variables, i.e.\ $u(r)=b\, U(\sqrt{b} r)$, yields
\begin{align}\label{eqn:uapprox}
    u(r)\approx \begin{cases}\frac{2(d-4)}{r^2}\left[ 1+ A (\sqrt{b} r)^{\beta}\sin(\alpha_1 \ln \sqrt{b}r+\delta)\right]& \mbox{ for $7\leq d\leq 15$,}
    \\ \frac{2(d-4)}{r^2}\left[1+ A (\sqrt{b}r)^{\beta-\alpha_1}+ B (\sqrt{b}r)^{\beta+\alpha_1}\right]& \mbox{ for $d\geq 16$}.
    \end{cases}
\end{align}
We may expect this to be a good approximation of the solution in the intermediate range $1/b\ll r\ll b$. For smaller values of $r$ it obviously cannot be right, since it is a singular approximation of a regular solution. For larger values of $r$ the harmonic term we have omitted in the derivation of Eq.\ (\ref{eqn:laneemden}) starts to dominate and the solution behaves like Eq.\ (\ref{eqn:solas}). This observation can be used to get a formula for the behaviour of $\omega(b)$ for large $b$. Let us assume that at some $r_0$ both Eq.\ (\ref{eqn:uapprox}) and (\ref{eqn:solas}) give good enough approximations to the actual solution $u$. Both of these expressions depend on $b$, however the former does it explicitly, while the latter through $\omega$. Hence, for this fixed $r_0$ we can make an expansion $u(r_0)\approx C_0 + C_1 (\omega-\omega_\infty)$, where $\omega_\infty$ was the frequency of the singular solution. Now, we compare this formula with Eq.\ (\ref{eqn:uapprox}) keeping only terms non-constant in $b$. As a result, we get
\begin{align}\label{eqn:snhlargeb}
    \omega(b)\approx \begin{cases} A\, b^{\beta/2} \sin(\frac{\alpha_1}{2}\ln b +\delta)& \mbox{ for $7\leq d\leq 15$,}
    \\ b^{\beta/2} \left(A\, b^{\alpha_1/2}+B\, b^{-\alpha_1/2} \right)& \mbox{ for $d\geq 16$}.
    \end{cases}
\end{align}
Thus, in dimensions $7\leq d \leq 15$ the function $\omega(b)$ oscillates, while for $d\geq 16$ it is monotonic. This fact has many interesting implications, for instance, when $7\leq d \leq 15$ there exist an infinite number of ground states with frequency $\omega_\infty$. The situation changes drastically for $d\geq 16$, where we get uniqueness in the sense that for each $\omega$ there exists at most one ground state with this frequency. Since our considerations were not assuming the positivity of the solution at any place, the same shall hold for the excited states. This change of behaviour also leads to some interesting observations regarding the stability of the ground states, as we discuss in the next chapter.

\vspace{\breakFF}
\begin{adjustwidth}{\marwidFF}{\marwidFF}
\small\qquad
The same considerations can be repeated for GP resulting in almost the same outcome with one major difference: in this case the change of behaviour happens between $d=12$ and $d=13$. This fact has been noticed numerically in \cite{Sel11}. A rigorous proof of this result can be found in \cite{Biz21}. Similar effect has also been observed earlier in nonlinear systems with compact domain, such as a ball \cite{Bud87, Bud89, Dol07, Jos73}. 
\end{adjustwidth}\vspace{\breakFF}

In the end of this analysis we want to point out an interesting feature characteristic for SNH system. For small values of $b$ it behaves like NLS equation with a cubic nonlinearity (as can be seen during the bifurcation analysis in the beginning of this subsection), while for large values of $b$ it is well described by NLS equation with quadratic nonlinearity (as shown in Section \ref{sec:exsingular} and then used in the present subsection). Such property is present also for other nonlocal NLS equations, for example, when considering the more general version of the Choquard equation \cite{Mor17} (SNH system is recovered by taking $p=2$):
\begin{align*}
 -\Delta u+V\, u=\left(\int_{\mathbb{R}^d} \frac{|u(y)|^p}{|x-y|^{d-2}}\, dy\right) |u|^{p-2}u,
\end{align*}
similar analysis shows that for small $b$ the system behaves as it had nonlinearity of the form $u^{2p-1}$, while for large $b$ it is rather $u^p$.

\subsection{Numerical results}
\begin{figure}
\centering
\includegraphics[width=0.65\textwidth]{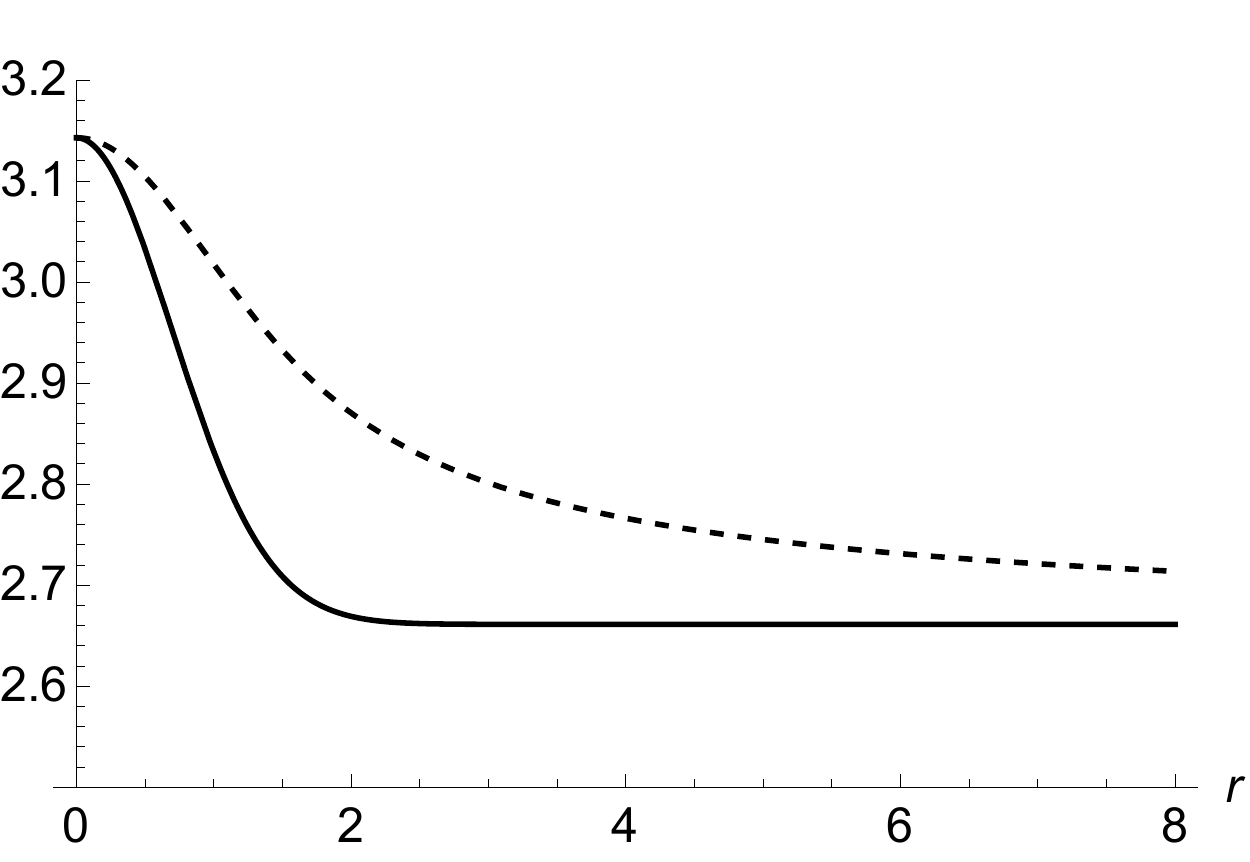}
\caption{Plots of functions $h(r)$ (dashed) and $h(r)+r\,h'(r)/(d-2)$ (solid) for the ground state of SNH in $d=3$ with $b=1$.}
\label{fig:hconv}
\end{figure}

Finally, we would like to verify numerically the results we got in this section. We obtain plots of functions $\omega(b)$ by the means described in Section \ref{sec:4pre}. The only additional improvement is the way we extract the values of $\omega$. As they are equal to $\lim_{r\to\infty} h(r)$, one needs to evaluate such limits. The convergence of $h(r)$ to $\omega$ is slow (see Fig.\ \ref{fig:hconv}), especially in lower dimensions, therefore to calculate this limit we proceed as follows. We integrate Eq.\ (\ref{eqn:SNHradhb}) to
\begin{align*}
    h'(r)\, r^{d-1} = -\int_0^r f(s)^2 s^{d-1} \, ds.
\end{align*}
For large values of $r$ the right hand is almost constant and equals the mass of the solution (divided by the area of a $(d-1)$-sphere), let us call it $M$. Then for large $r$ it holds $h'(r)\approx - M/r^{d-1}$. By integrating this expression once more, this time over the interval $(r,\infty)$, we get
\begin{align*}
    \omega \approx h(r)-\frac{M}{(d-2)r^{d-2}}\approx h(r)+\frac{r h'(r)}{d-2}
\end{align*}
for large $r$. This expression converges much faster then $h$, see Fig.\ \ref{fig:hconv}, and we use it to find the value of $\omega$.

\begin{figure}
\centering
\begin{subfigure}{0.45\textwidth}
\includegraphics[width=\textwidth]{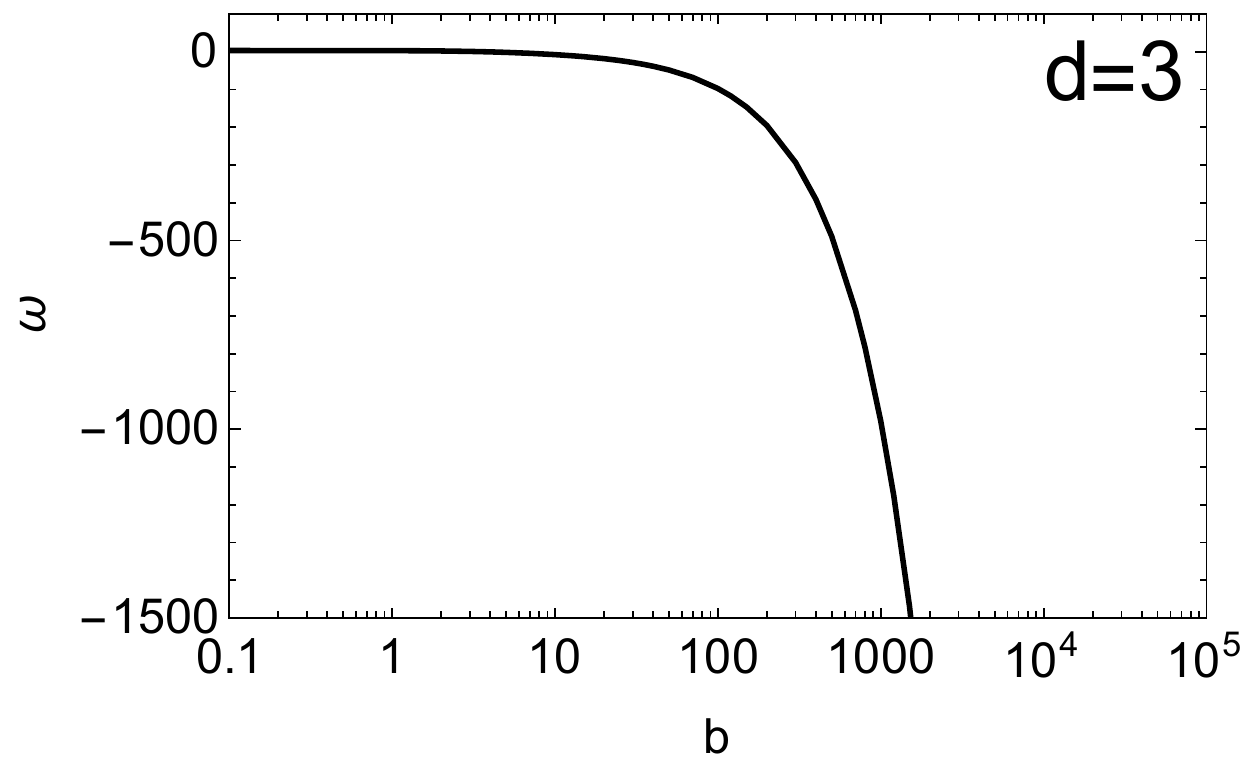}
\end{subfigure}
~
\begin{subfigure}{0.45\textwidth}
\includegraphics[width=\textwidth]{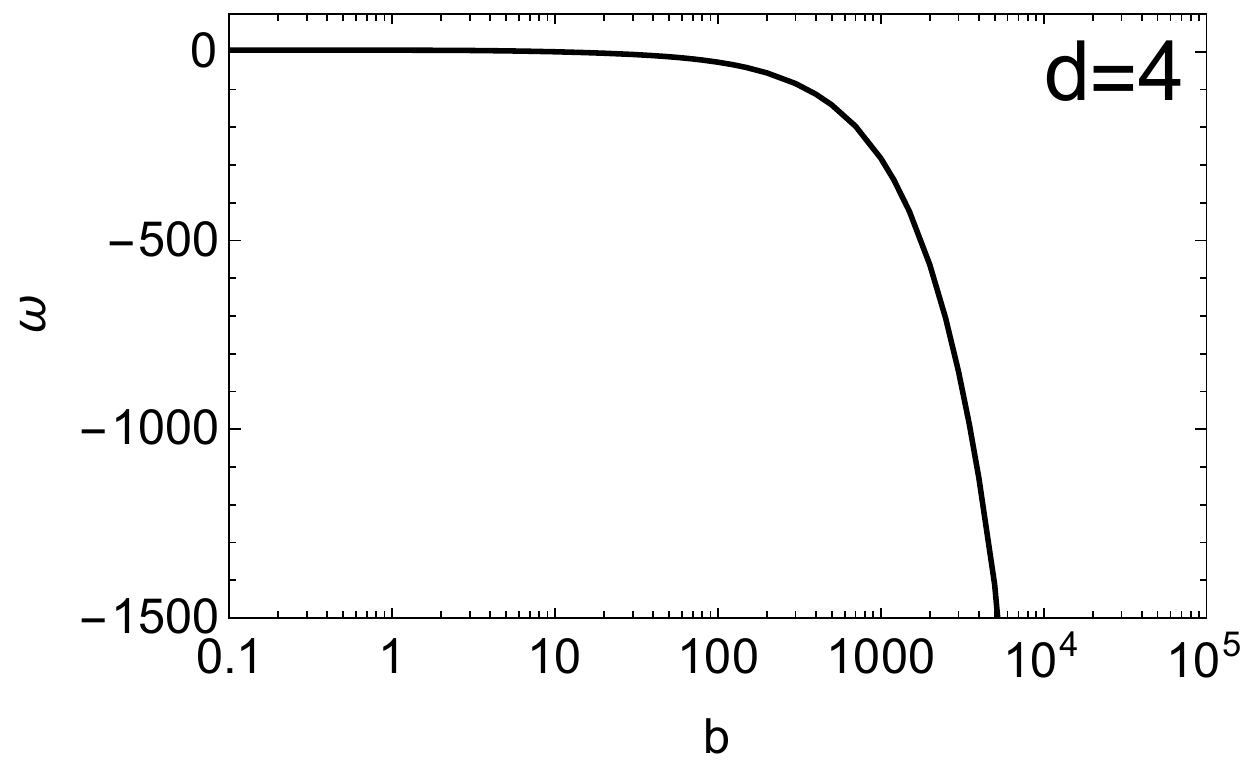}
\end{subfigure}
\\
\begin{subfigure}{0.45\textwidth}
\includegraphics[width=\textwidth]{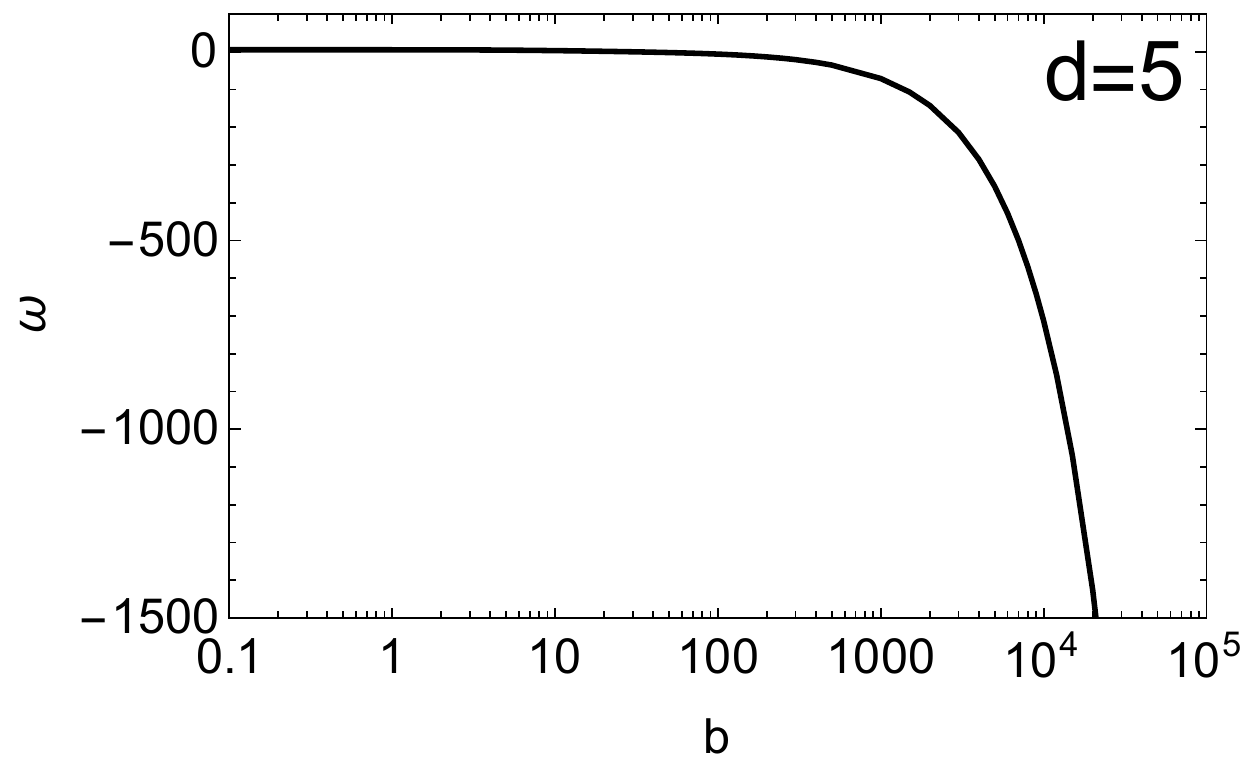}
\end{subfigure}
~
\begin{subfigure}{0.45\textwidth}
\includegraphics[width=\textwidth]{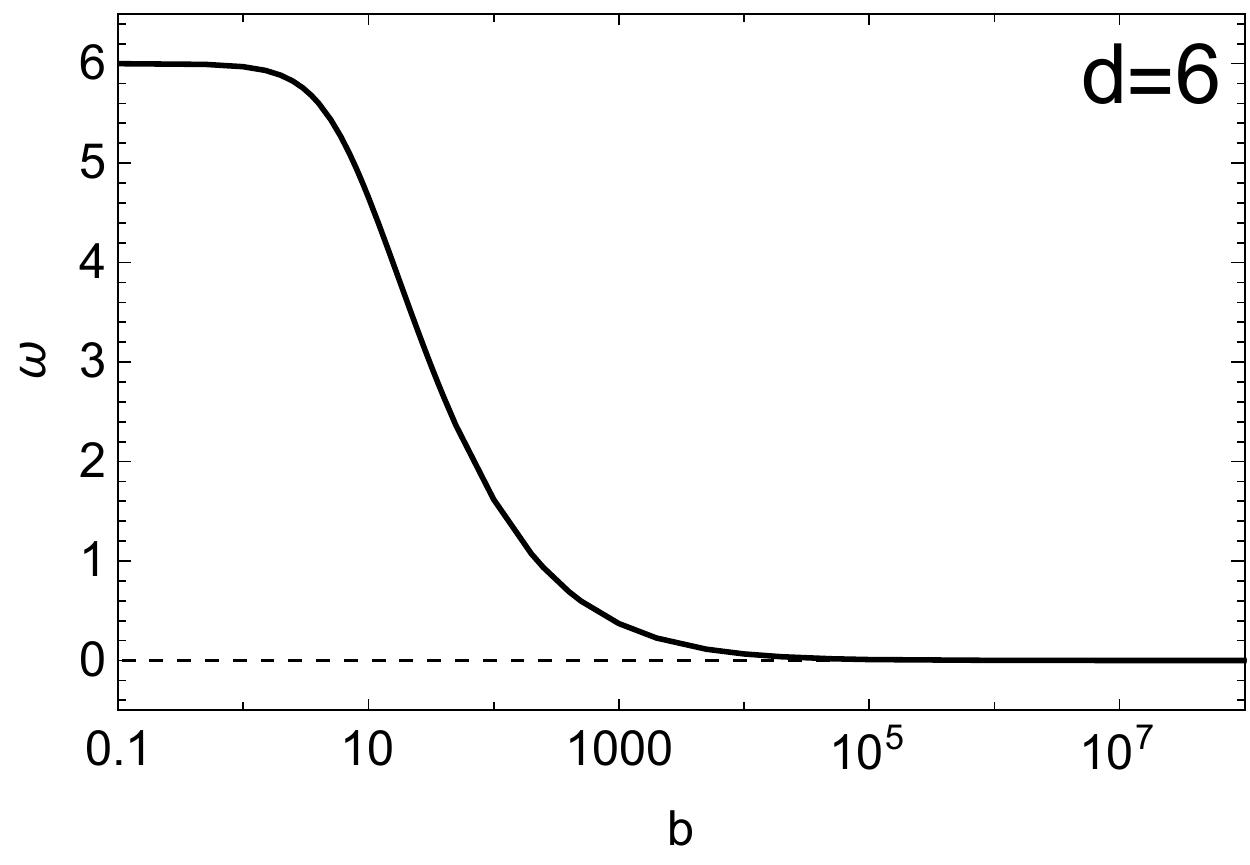}
\end{subfigure}
\\
\begin{subfigure}{0.45\textwidth}
\includegraphics[width=\textwidth]{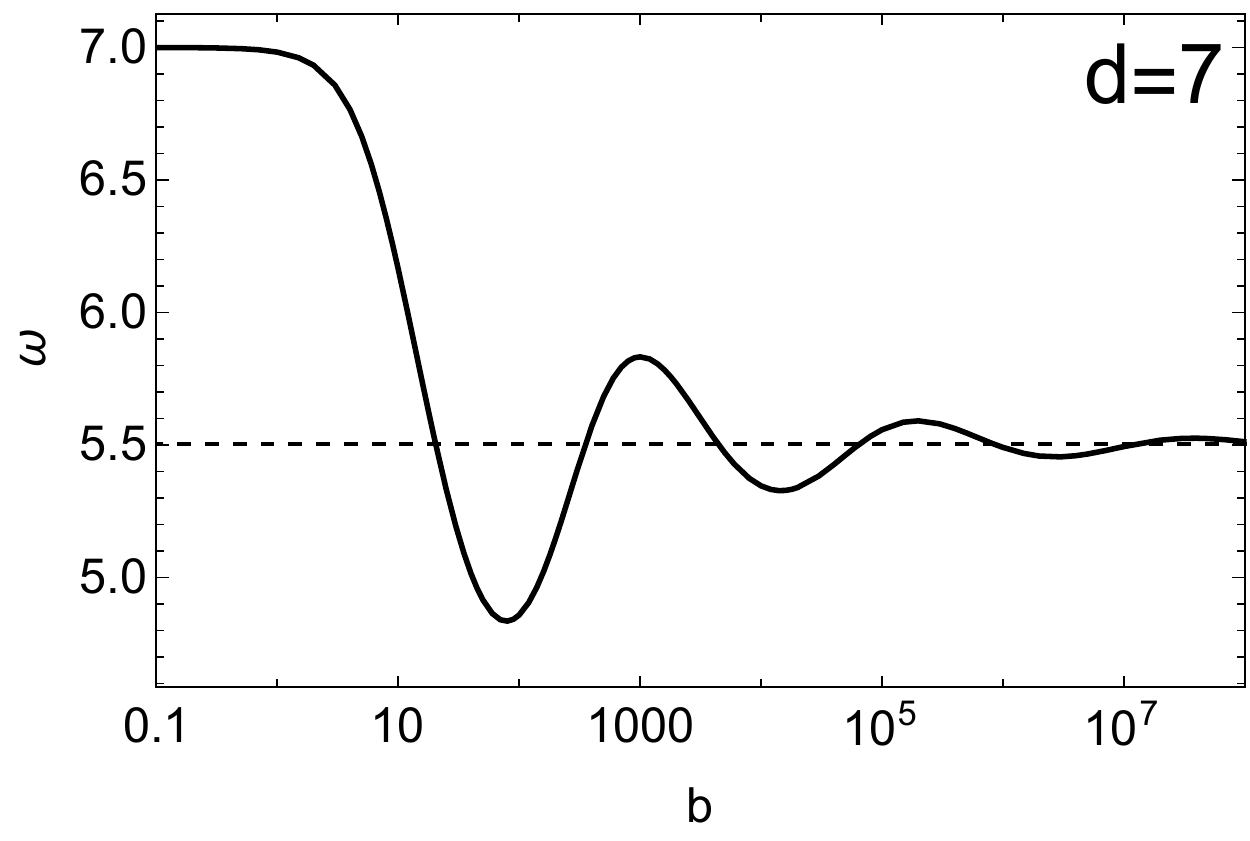}
\end{subfigure}
~
\begin{subfigure}{0.45\textwidth}
\includegraphics[width=\textwidth]{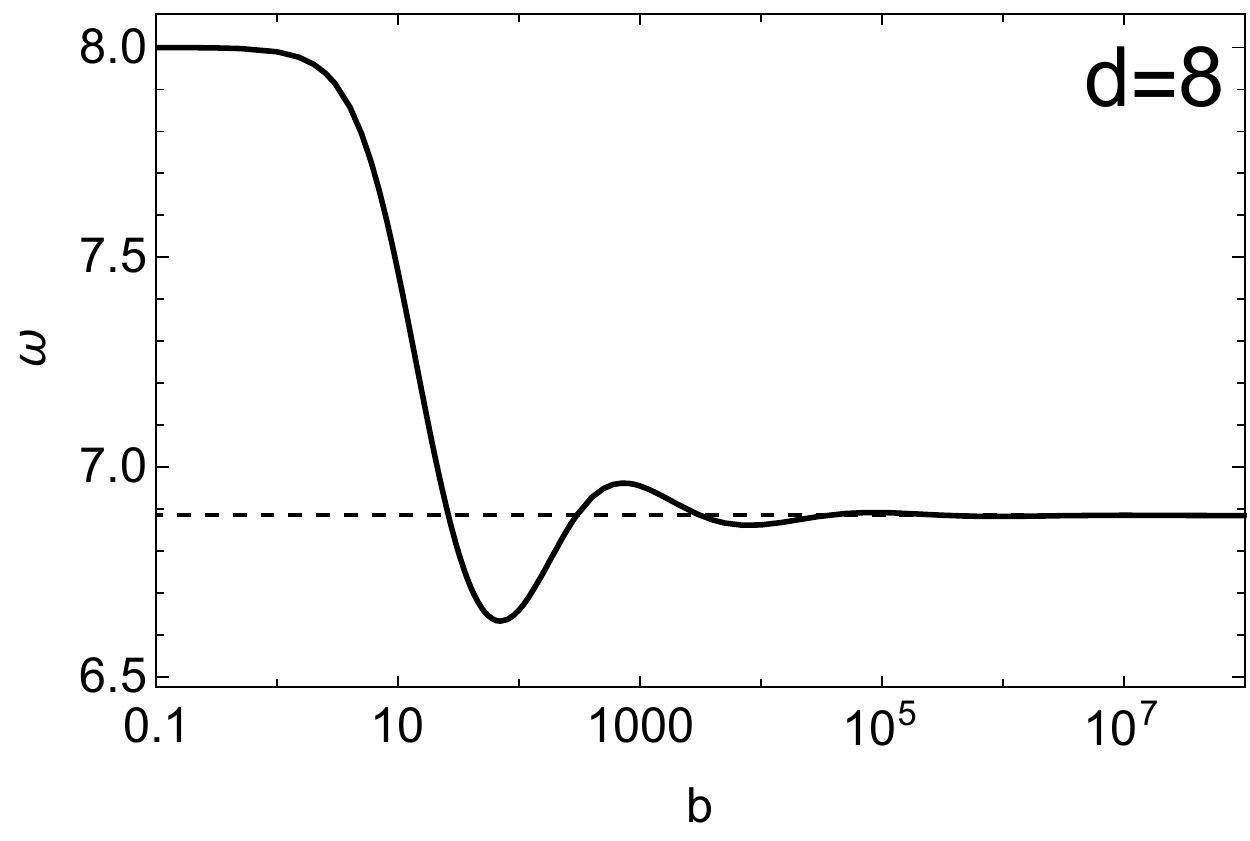}
\end{subfigure}
\\
\begin{subfigure}{0.45\textwidth}
\includegraphics[width=\textwidth]{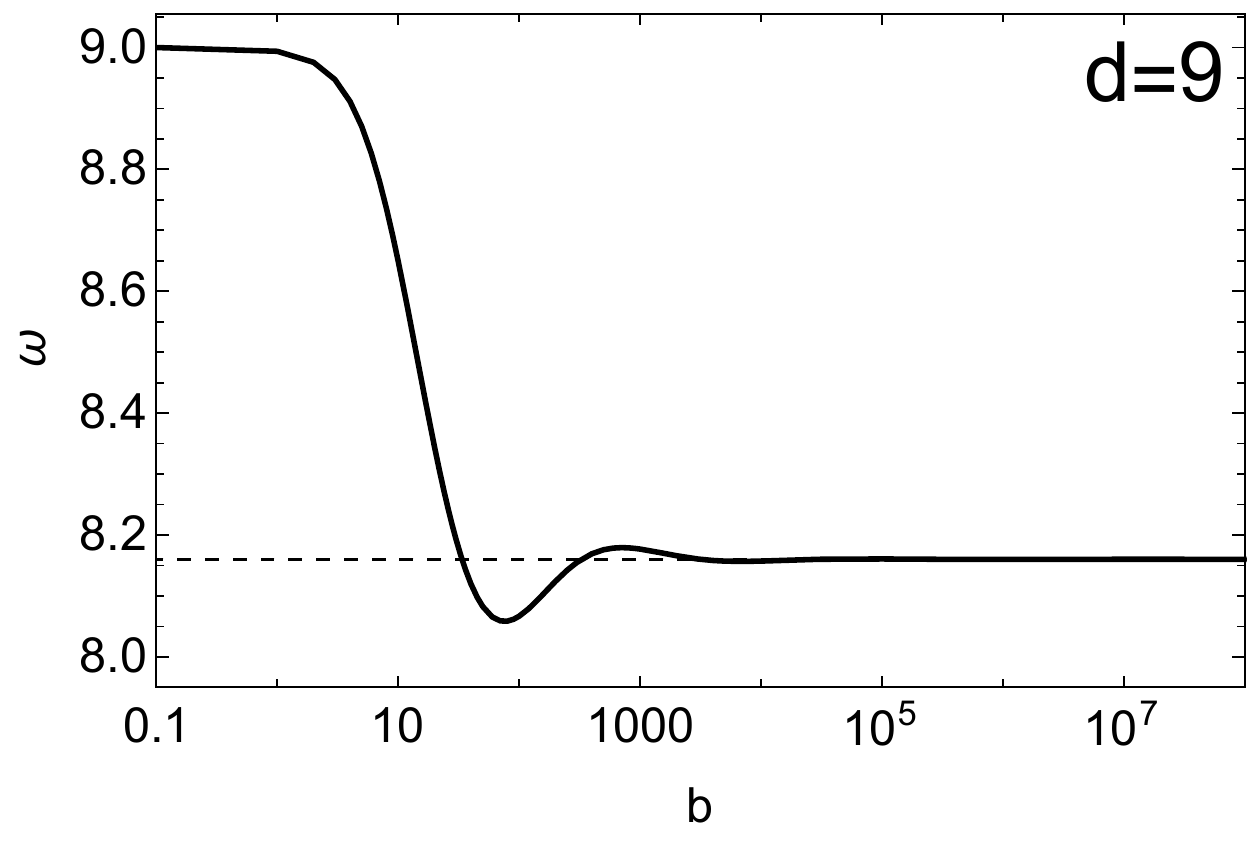}
\end{subfigure}
~
\begin{subfigure}{0.45\textwidth}
\includegraphics[width=\textwidth]{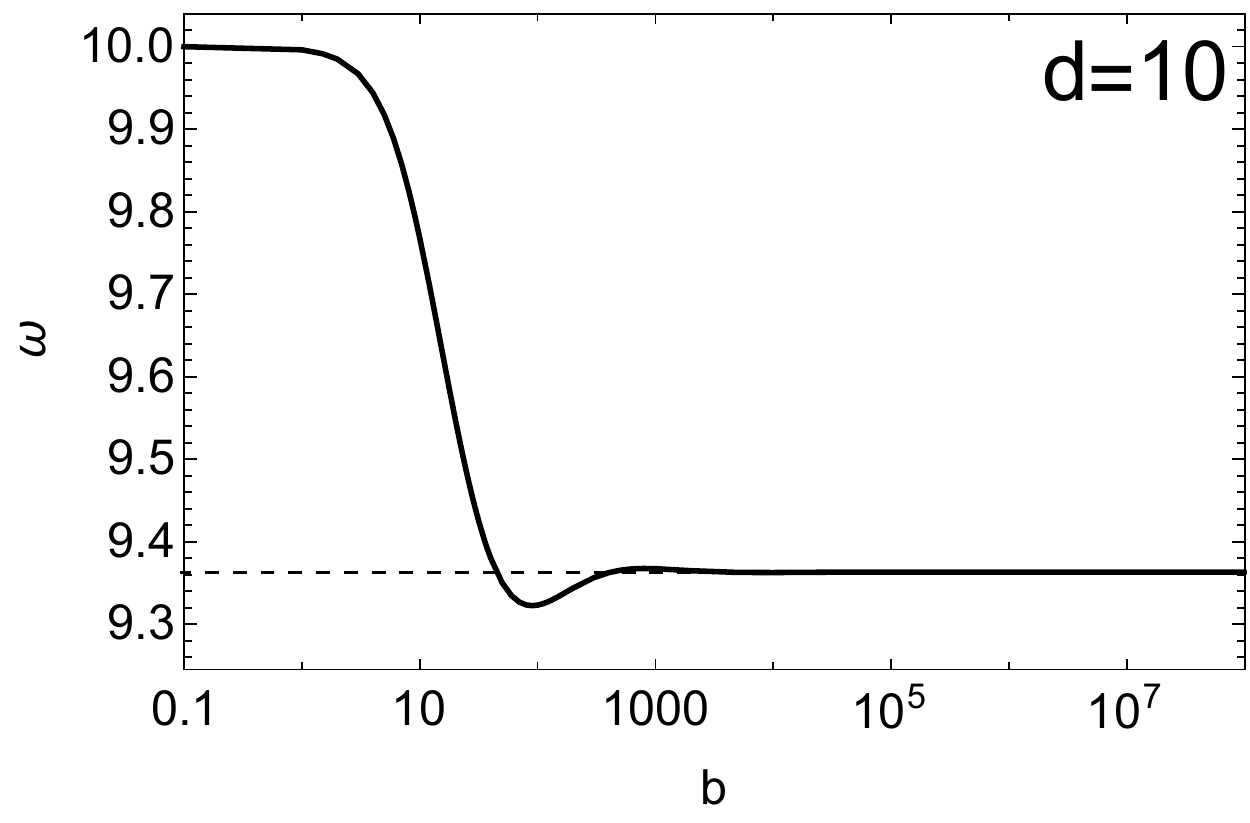}
\end{subfigure}
\end{figure}

\begin{figure}
\ContinuedFloat
\centering
\begin{subfigure}{0.45\textwidth}
\includegraphics[width=\textwidth]{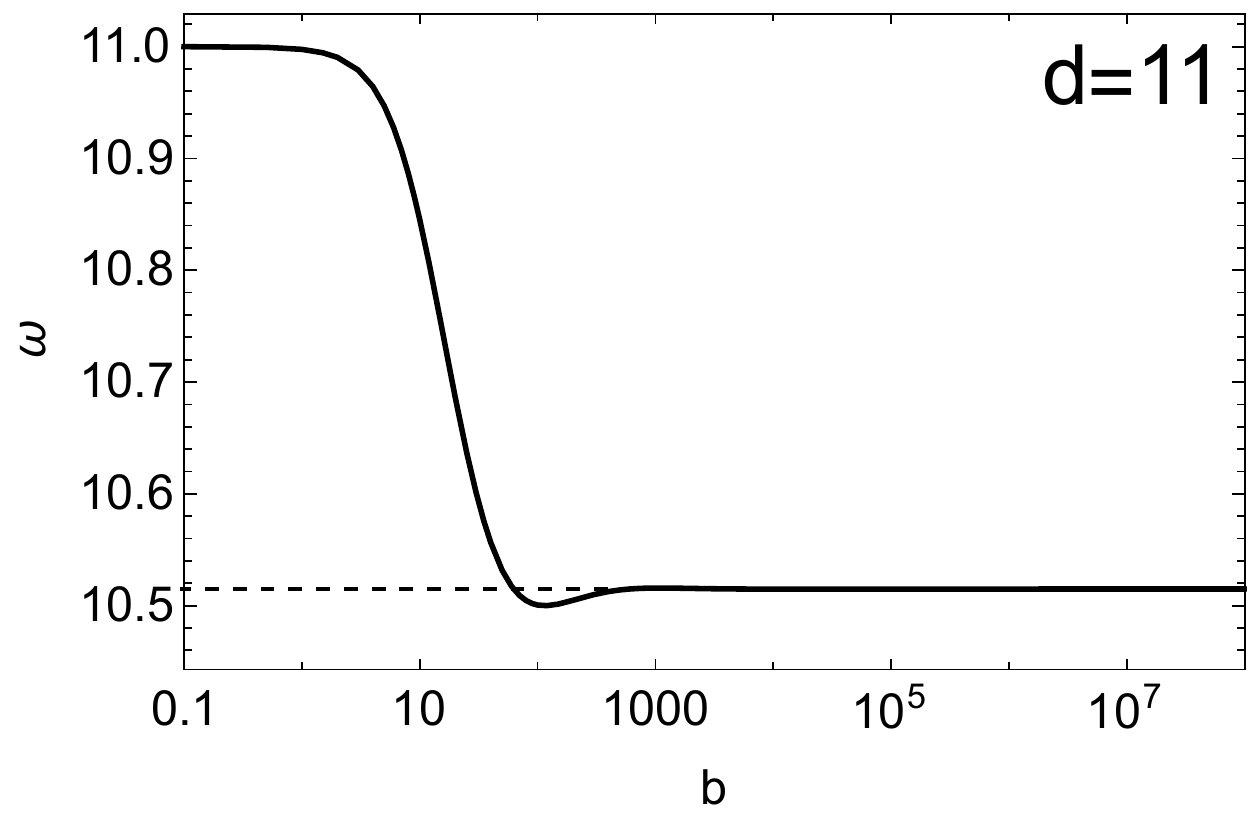}
\end{subfigure}
~
\begin{subfigure}{0.45\textwidth}
\includegraphics[width=\textwidth]{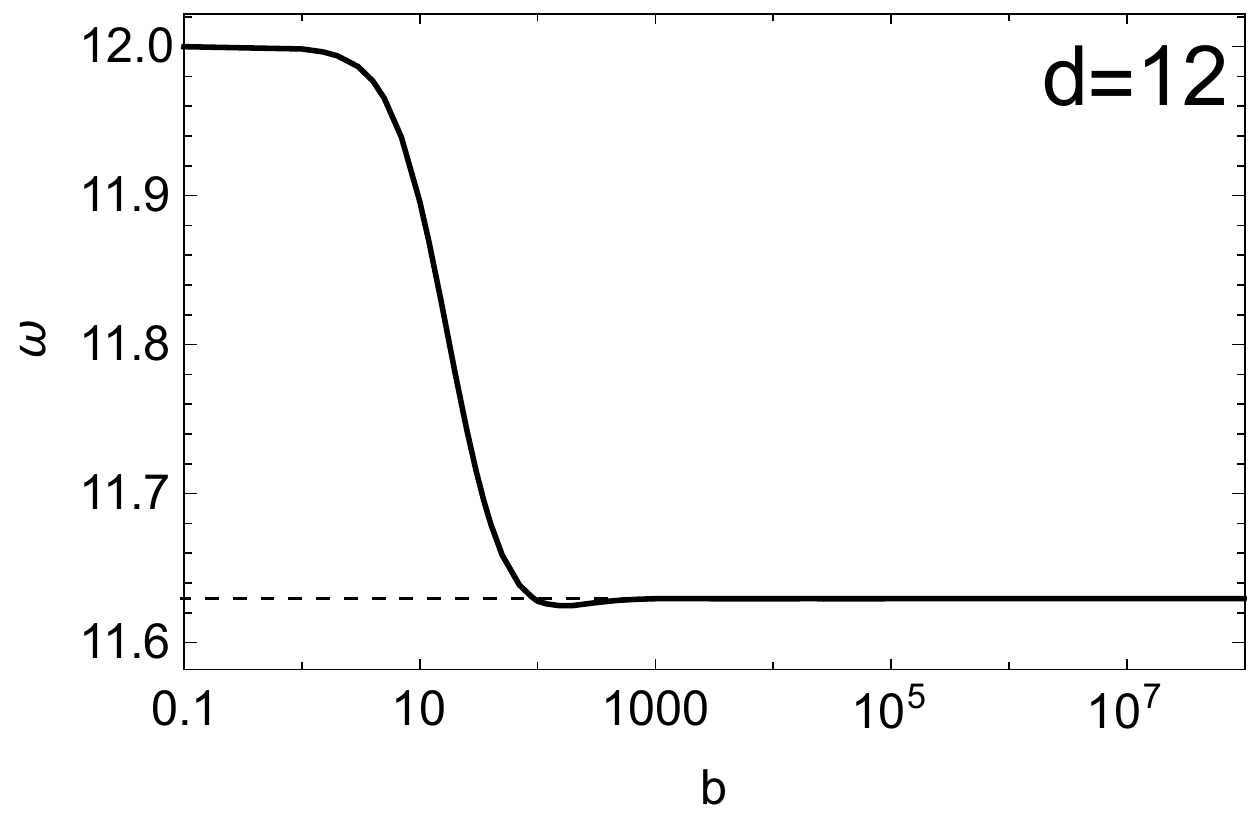}
\end{subfigure}
\\
\begin{subfigure}{0.45\textwidth}
\includegraphics[width=\textwidth]{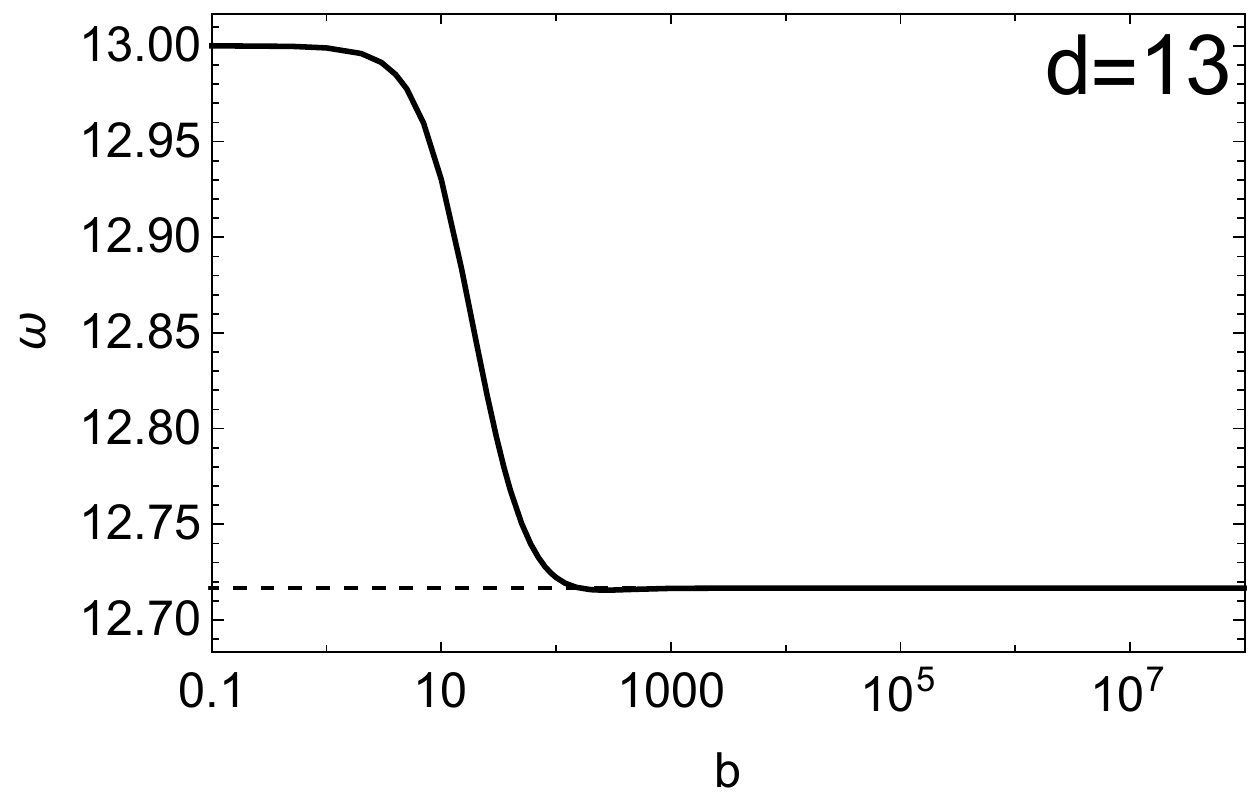}
\end{subfigure}
~
\begin{subfigure}{0.45\textwidth}
\includegraphics[width=\textwidth]{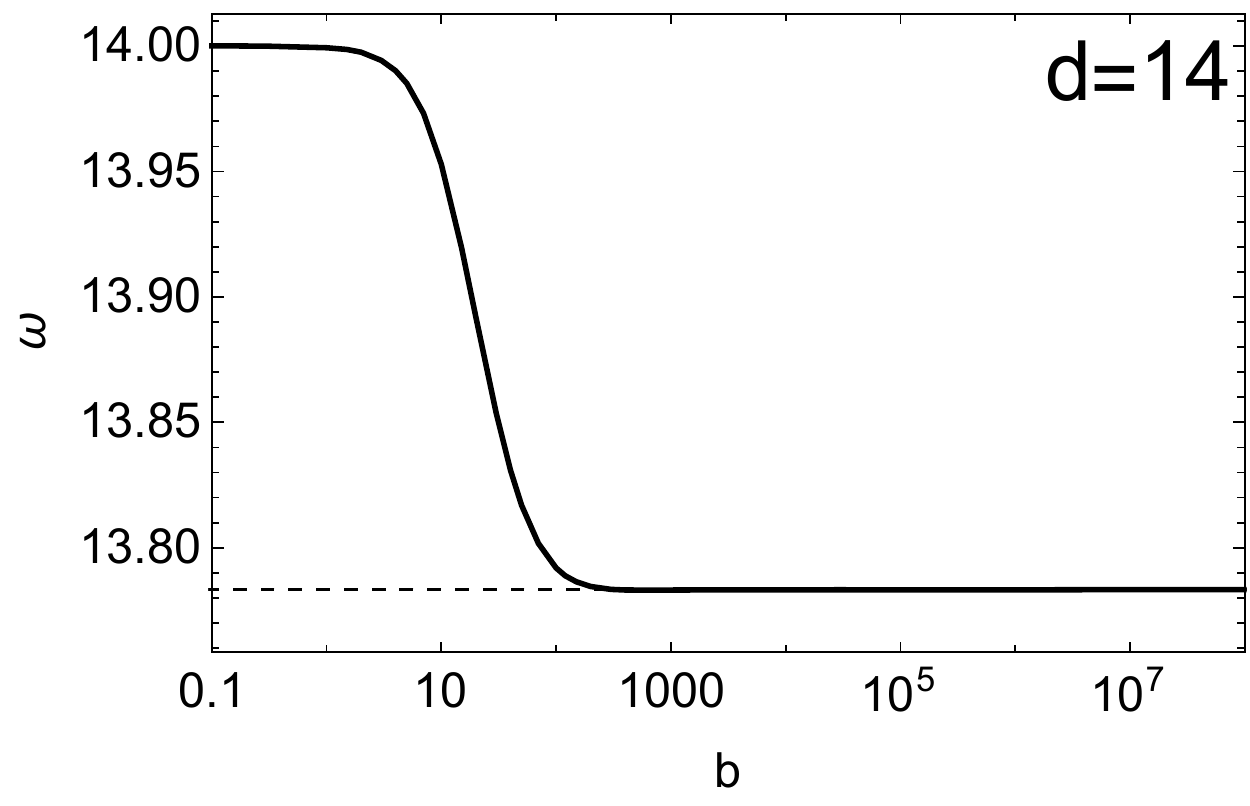}
\end{subfigure}
\\
\begin{subfigure}{0.45\textwidth}
\includegraphics[width=\textwidth]{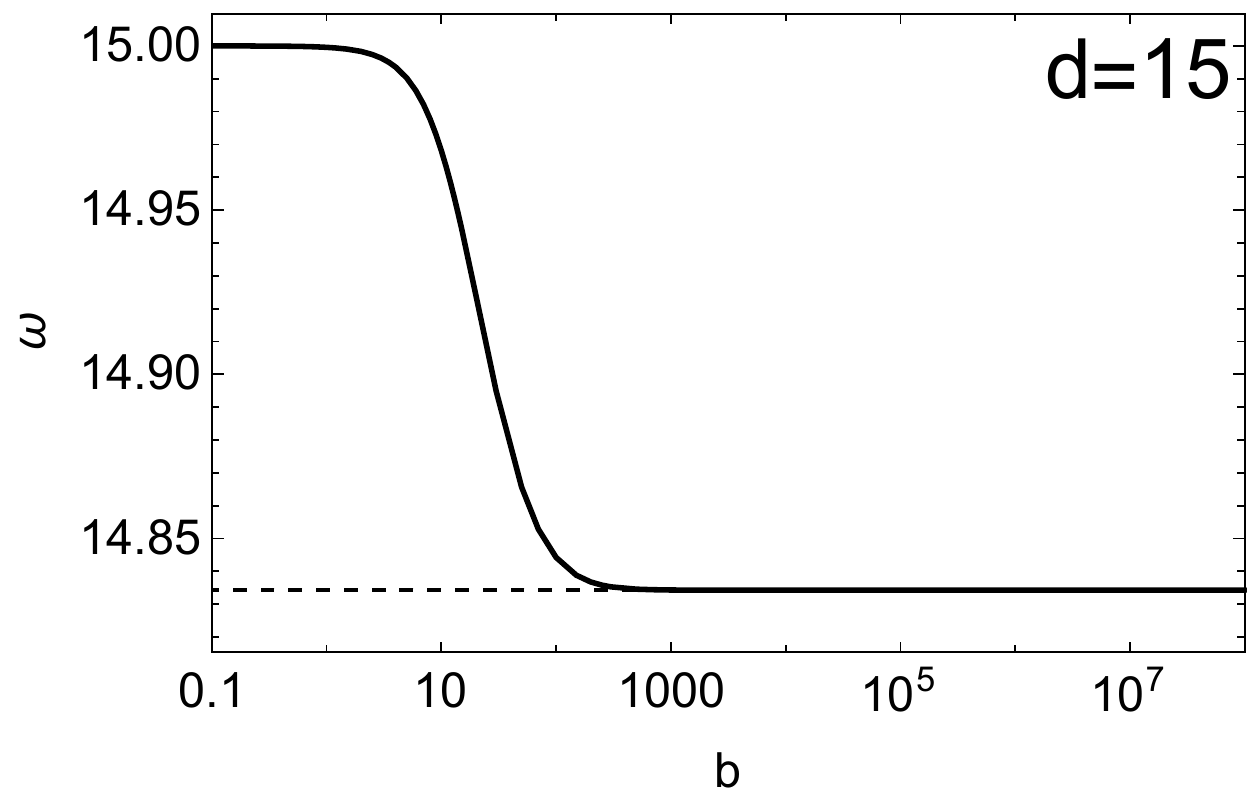}
\end{subfigure}
~
\begin{subfigure}{0.45\textwidth}
\includegraphics[width=\textwidth]{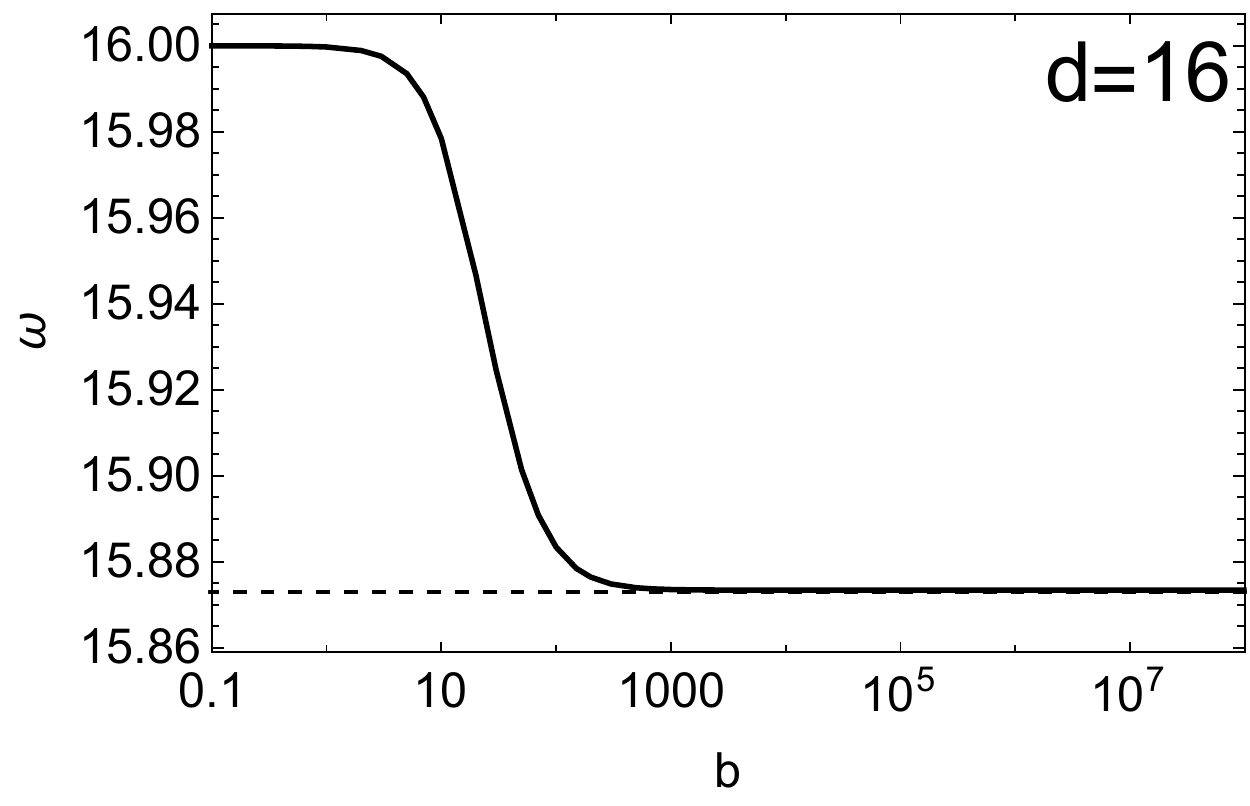}
\end{subfigure}
\\
\begin{subfigure}{0.45\textwidth}
\includegraphics[width=\textwidth]{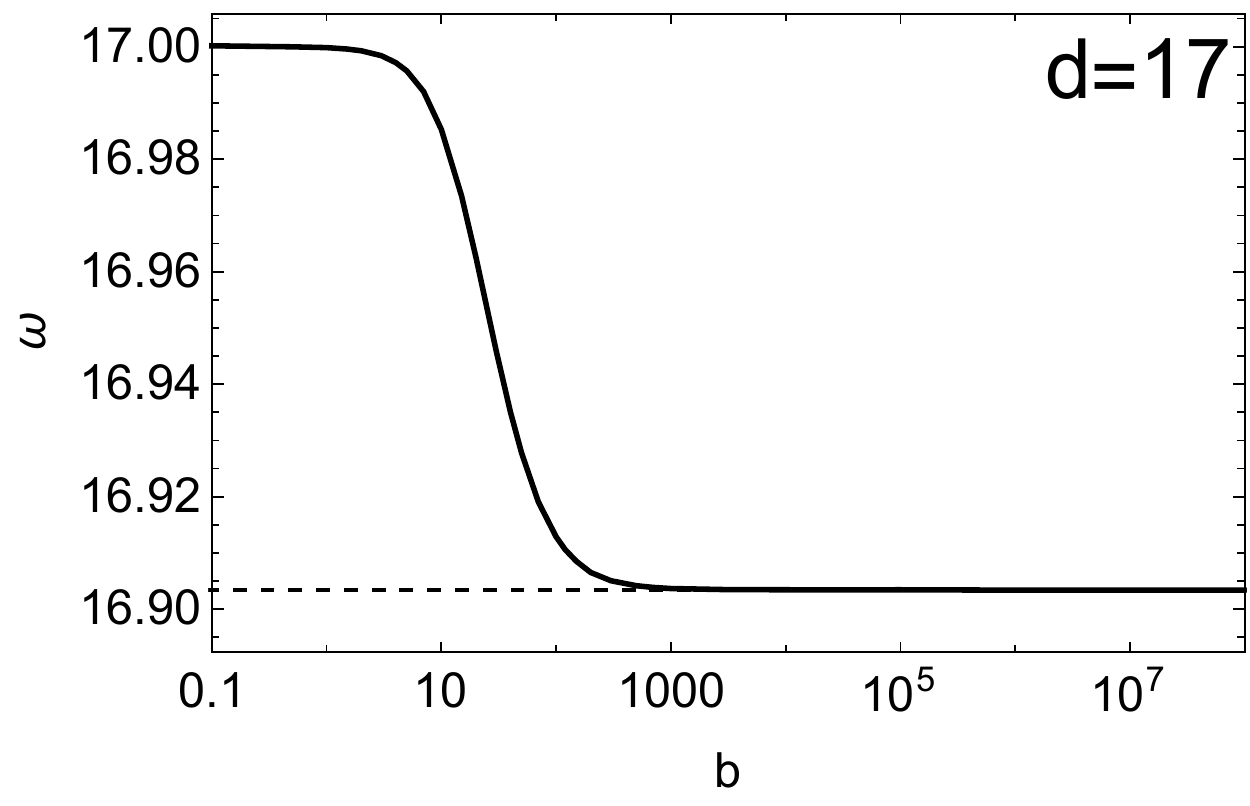}
\end{subfigure}
~
\begin{subfigure}{0.45\textwidth}
\includegraphics[width=\textwidth]{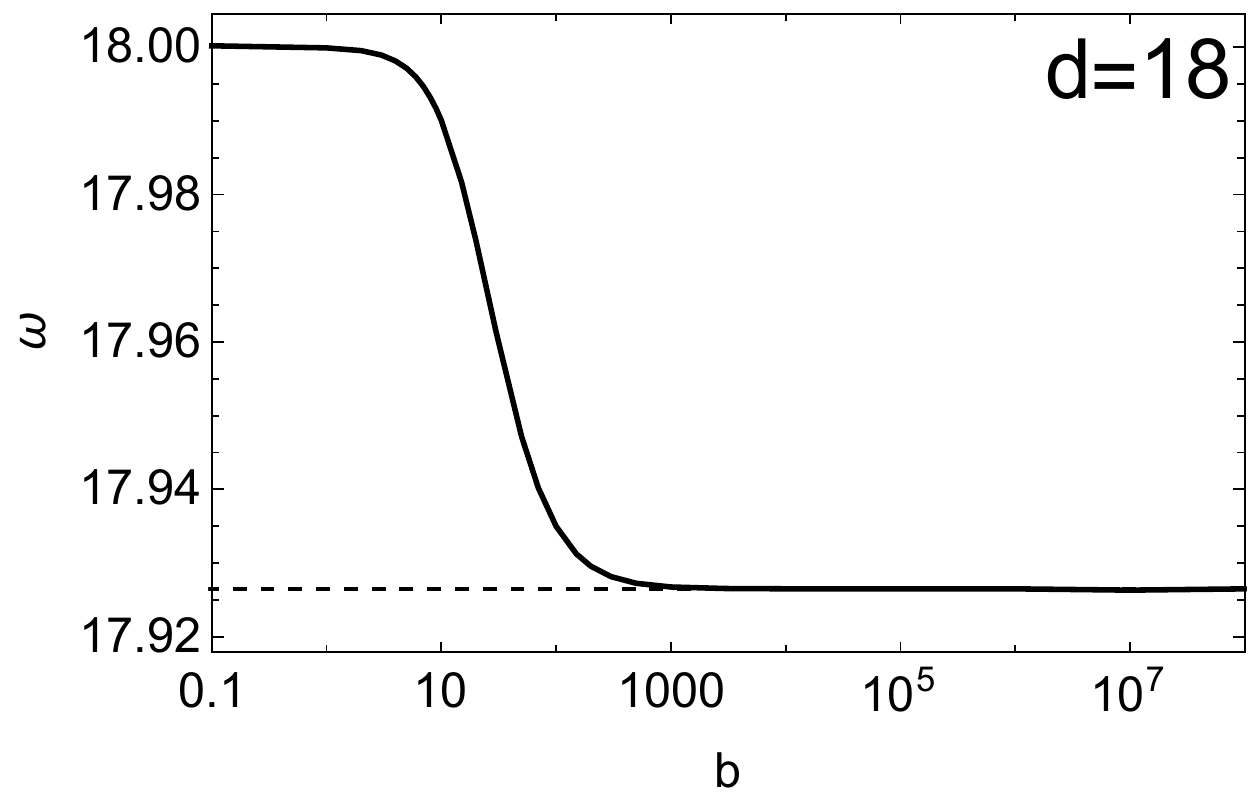}
\end{subfigure}
\caption{Plots of function $\omega (b)$ for the ground states of SNH in various dimensions. Horizontal dashed lines correspond to the position of the limiting frequency $\omega_\infty$.}
\label{fig:snhnum}
\end{figure}

\begin{figure}
\centering
\includegraphics[width=0.65\textwidth]{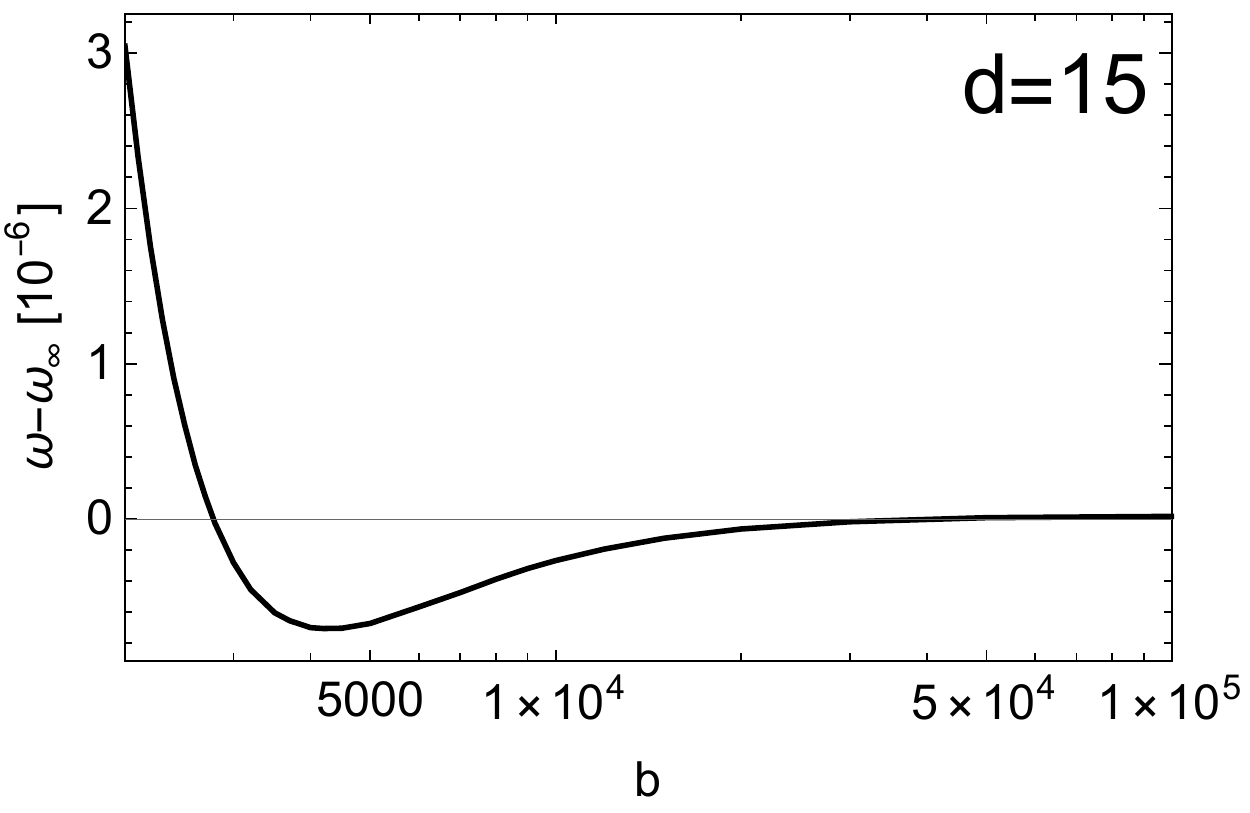}
\caption{The close-up of the plot for $d=15$ from Fig.\ \ref{fig:snhnum} showing the beginning of the oscillations.}
\label{fig:snh15zoom}
\end{figure}

The plots of functions $\omega(b)$ for the ground states in various dimensions are presented in Fig.\ \ref{fig:snhnum}. We start with the subcritical case $d=3$ (we do not include the lower dimensions, since there the gravitational interpretation of SNH system ceases to work). As pointed out in Section \ref{sec:subcritical}, in subcritical dimensions the curve $\omega(b)$ bifurcates from $d$ and then decreases indefinitely. Indeed, such behaviour agrees with the plots for $d=3$, $4$, and $5$. The situation becomes more interesting in the critical dimension $d=6$. As we showed in Section \ref{sec:omegawindow}, our function then must be contained in the interval $[0,6]$. Numerical evidence shows us that it is indeed the case and that $\omega(b)$ saturates these bounds (it is also consistent with the fact that the improved bounds we described can be obtained only for the supercritical dimension). We also point out that in this case the function $\omega(b)$ is monotonically decreasing. This behaviour changes in the supercritical dimensions, as for $d=7$ we can see clear oscillations around $\omega_\infty$. The amplitude of these oscillations decreases as the dimension increases, and at some point they become invisible in these plots. We zoom the relevant fragment of $\omega(b)$ plot for $d=15$ in Fig.\ (\ref{fig:snh15zoom}). In dimension $d=16$ this structure completely vanishes. Even though the lack of oscillations is impossible to show in Fig.\ \ref{fig:snhnum}, in the next chapter we give some other numerical evidence supporting this result. This situation remains in higher dimensions, as we get plots of $\omega(b)$ that are monotonically decreasing down to $\omega_\infty$.

\begin{figure}
\centering
\includegraphics[width=0.65\textwidth]{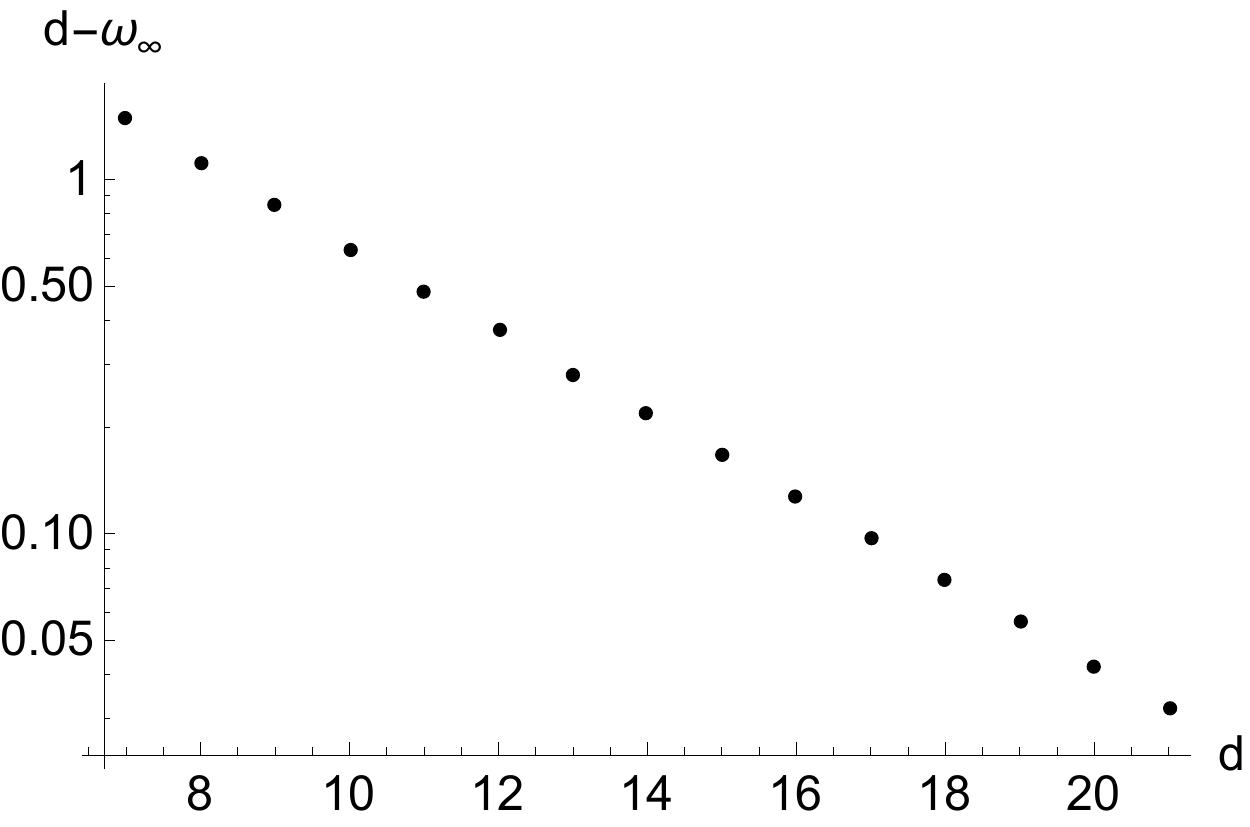}
\caption{Dependence of $d-\omega_\infty$ on the dimension $d$.}
\label{fig:omegainfty}
\end{figure}

Another interesting observation coming from these plots are the values of $\omega_\infty$ in different dimensions. They are presented in Table \ref{tab:omegainfty}. These results show that as $d$ increases, the value of $\omega_\infty$ approaches $d$ (implying that the range of possible frequencies shrinks). Additionally, as Fig.\ \ref{fig:omegainfty} shows, this dependence seems to be exponential. Its approximate form is $d-\omega_\infty(d)= A\, e^{-\gamma d}$, where $A\approx9.85$ and $\gamma\approx2.29$.

\begin{table}[]
    \centering
    \begin{tabular}{c|cccccccc}
        $d$ & 6 & 7 & 8 & 9 & 10 & 11 & 12 & 13 \\
        $\omega_\infty$ & 0.000 & 5.504 & 6.885 & 8.161 & 9.363 & 10.515 & 11.623 & 12.717 \\ \hline
        $d$& 14 & 15 & 16 & 17 & 18 & 19 & 20 & 21 
        \\ $\omega_\infty$& 13.783 & 14.834 & 15.873 & 16.903 & 17.926 & 18.944 & 19.958 & 20.968
    \end{tabular}
    \caption{Limiting frequencies as $b\to\infty$ in various dimensions $d$ for the ground states of SNH.}
    \label{tab:omegainfty}
\end{table}

\begin{figure}
\centering
\includegraphics[width=0.65\textwidth]{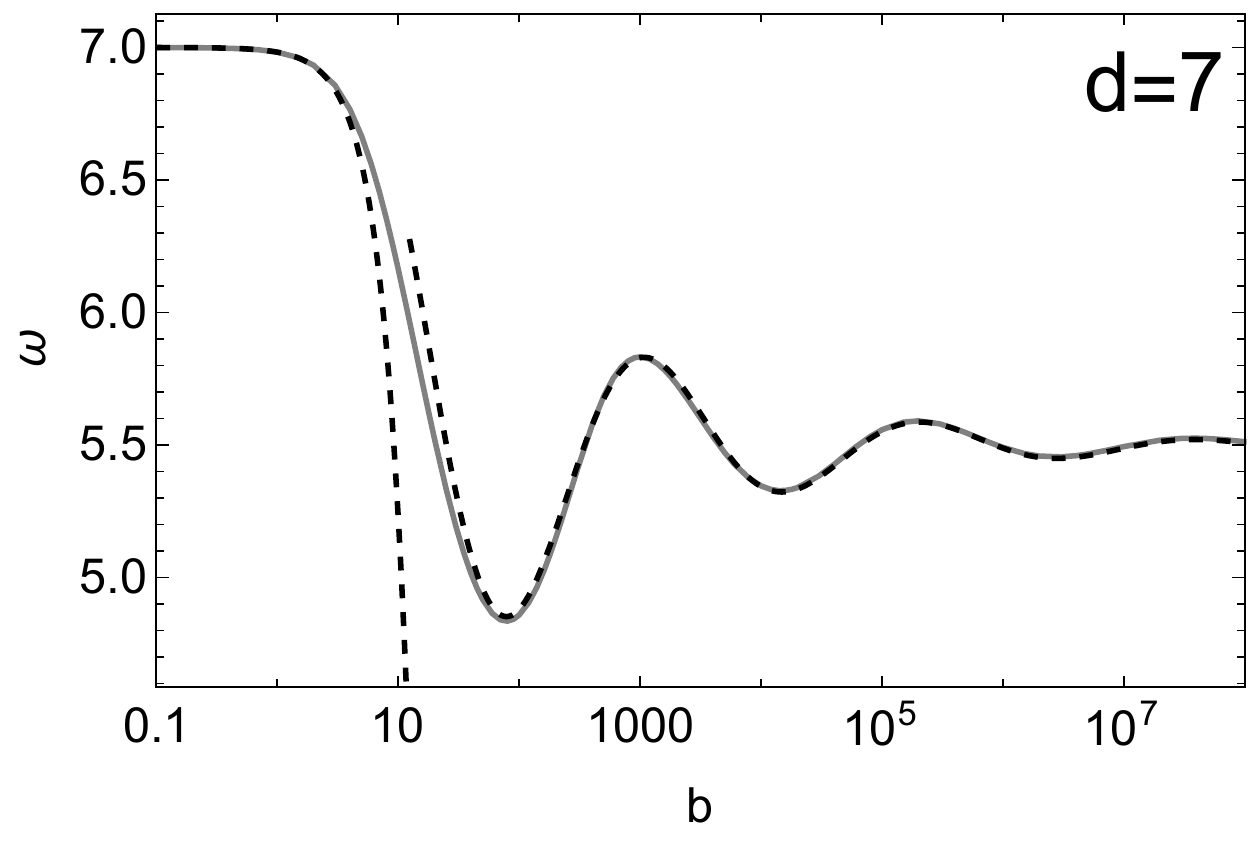}
\caption{Function $\omega(b)$ for a ground state in $d=6$ (solid gray line) together with approximations given by Eqs.\ (\ref{eqn:bifursnhb}) and (\ref{eqn:snhlargeb}) (dashed black lines).}
\label{fig:snhapprox}
\end{figure}

We also can verify the behaviours of $\omega(b)$ for small and large values of $b$ derived in Section \ref{sec:smalllargeb}. However, while Eq.\ (\ref{eqn:bifursnhb}) gives the approximate shape of $\omega(b)$ for $b$ close to zero explicitly, Eq.\ (\ref{eqn:snhlargeb}) describing it for large $b$ contains some unknown constants and their values need to be fitted numerically. We present both of these approximations for SNH in $d=7$ in Fig.\ \ref{fig:snhapprox}.

\begin{figure}
\centering
\begin{subfigure}{0.47\textwidth}
\includegraphics[width=\textwidth]{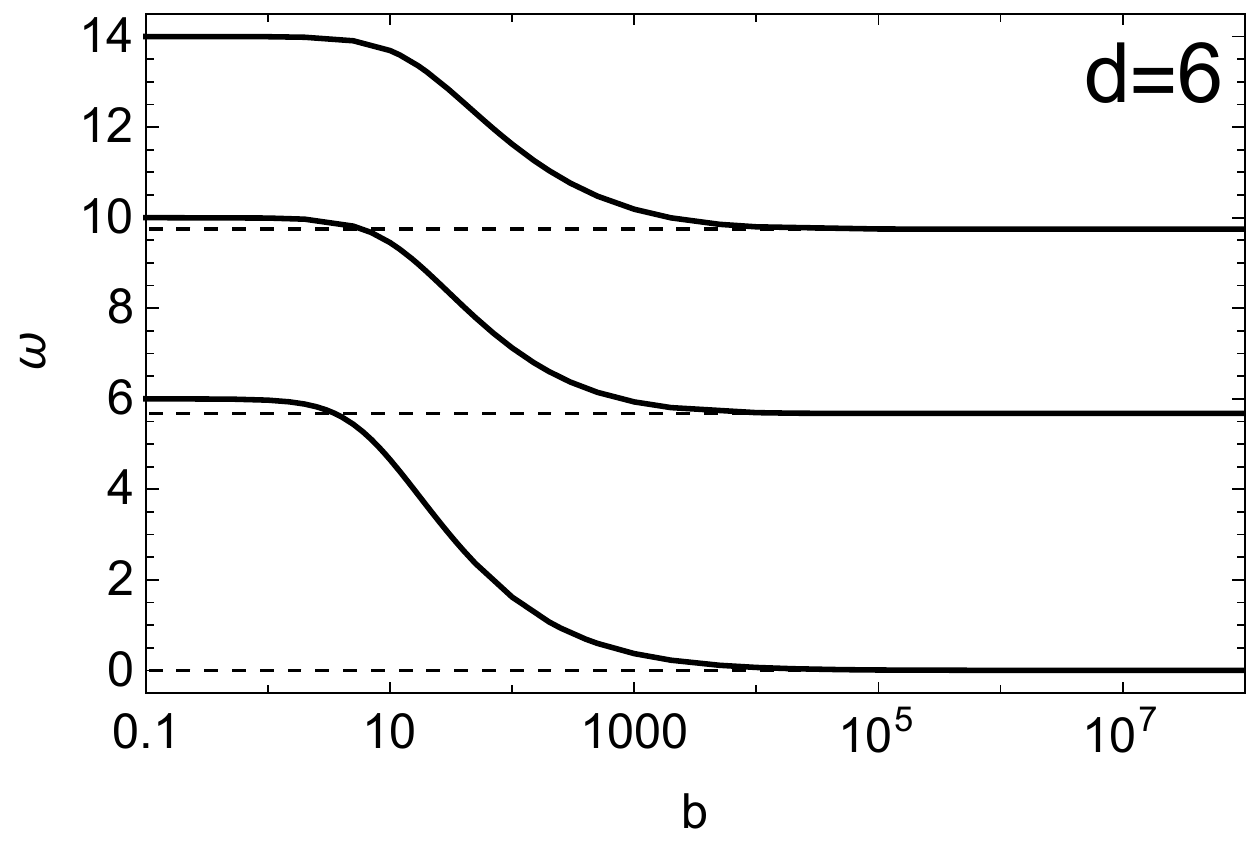}
\label{fig:snhexciteda}
\end{subfigure}
~
\begin{subfigure}{0.47\textwidth}
\includegraphics[width=\textwidth]{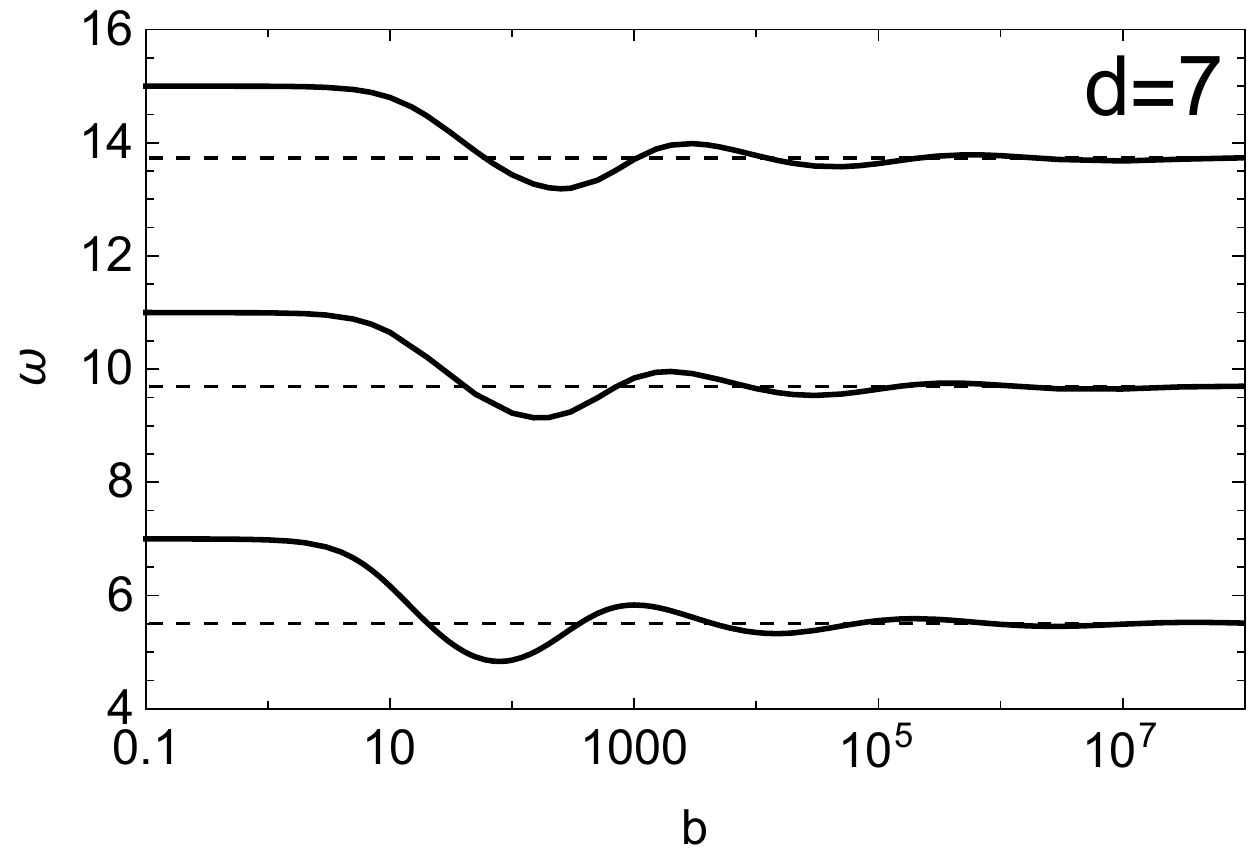}
\label{fig:snhexcitedb}
\end{subfigure}
\caption{Plots of functions $\omega(b)$ for the ground state and two lowest excited states of SNH in critical and first supercritical dimension. The horizontal dashed lines denote the positions of the limiting frequencies $\omega_\infty$.}
\label{fig:snhexcited}
\end{figure}

All the numerical results discussed so far regarded the ground state. Of course, one may produce the analogous plots the for excited states. In Fig.\ \ref{fig:snhexcited} we show functions $\omega(b)$ for the ground state and the two lowest excited states in the most interesting cases of $d=6$ and $d=7$. We can see that they are qualitatively identical. This observation propagates to higher dimensions, where the oscillations for the excited states also weaken as $d$ increases and finally vanish completely for $d\geq16$.

\vspace{\breakFF}
\begin{adjustwidth}{\marwidFF}{\marwidFF}
\small\qquad
We present the plots of $\omega(b)$ for the ground state of GP in chosen dimensions in Fig.\ \ref{fig:gpnum}. Comparing them with Fig.\ \ref{fig:snhnum} shows almost perfect qualitative similarity. The only major difference between these results are the dimensions in which certain behaviours can be observed. The oscillations appear above the critical dimension, which this time is $d=4$, and vanish between $d=12$ and $d=13$, as dictated by the analysis of the eigenvalues of the appropriate dynamical system. A rigorous proof of this result can be found in \cite{Biz21}. The plots also agree with the bounds on possible frequencies we obtained in the previous subsection.
\begin{figure}
\centering
\captionsetup{font=small, width=.82\linewidth}
\includegraphics[width=0.35\textwidth]{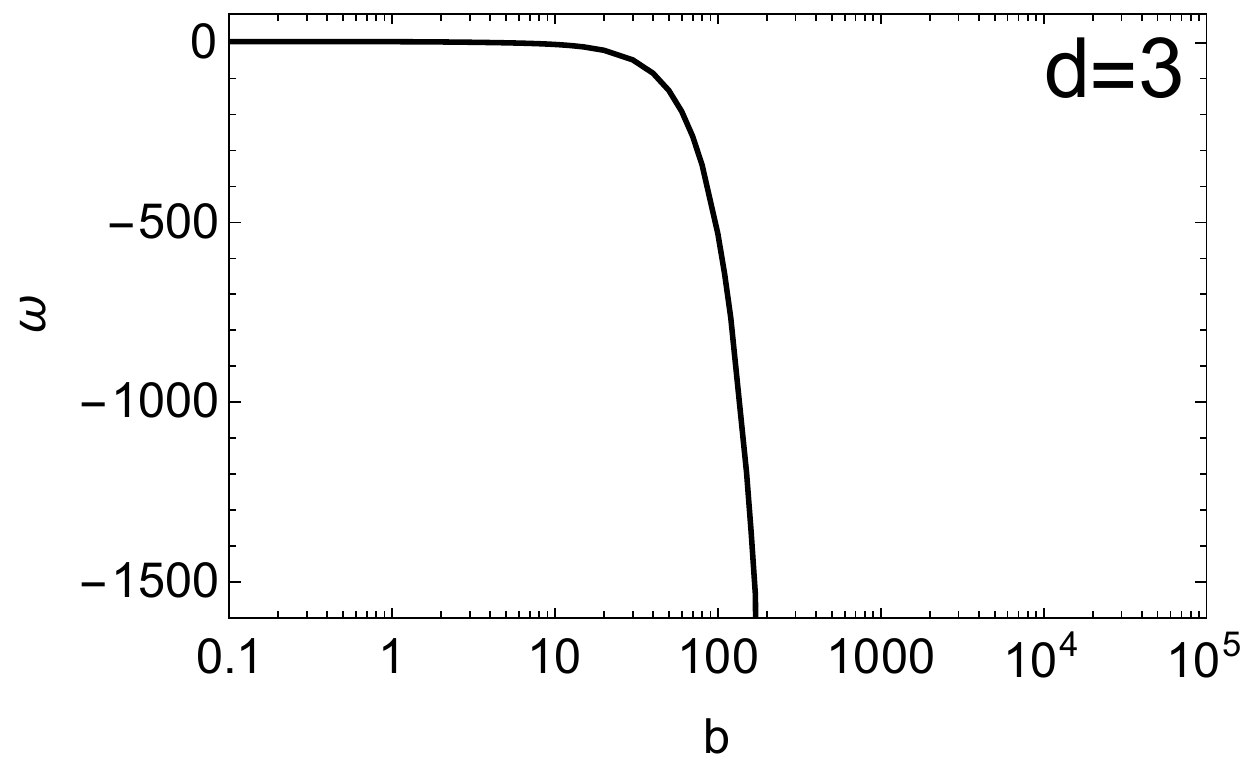}\qquad
\includegraphics[width=0.35\textwidth]{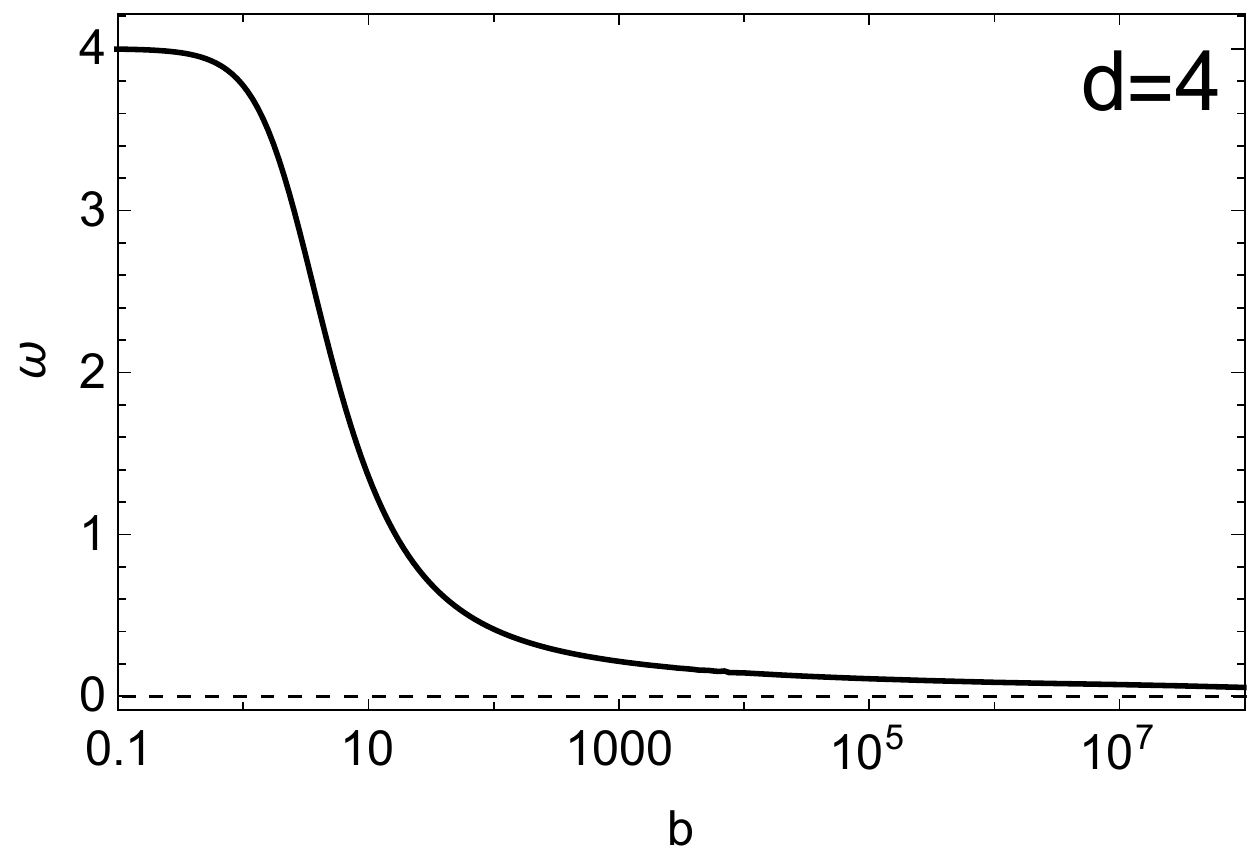}\\
\includegraphics[width=0.35\textwidth]{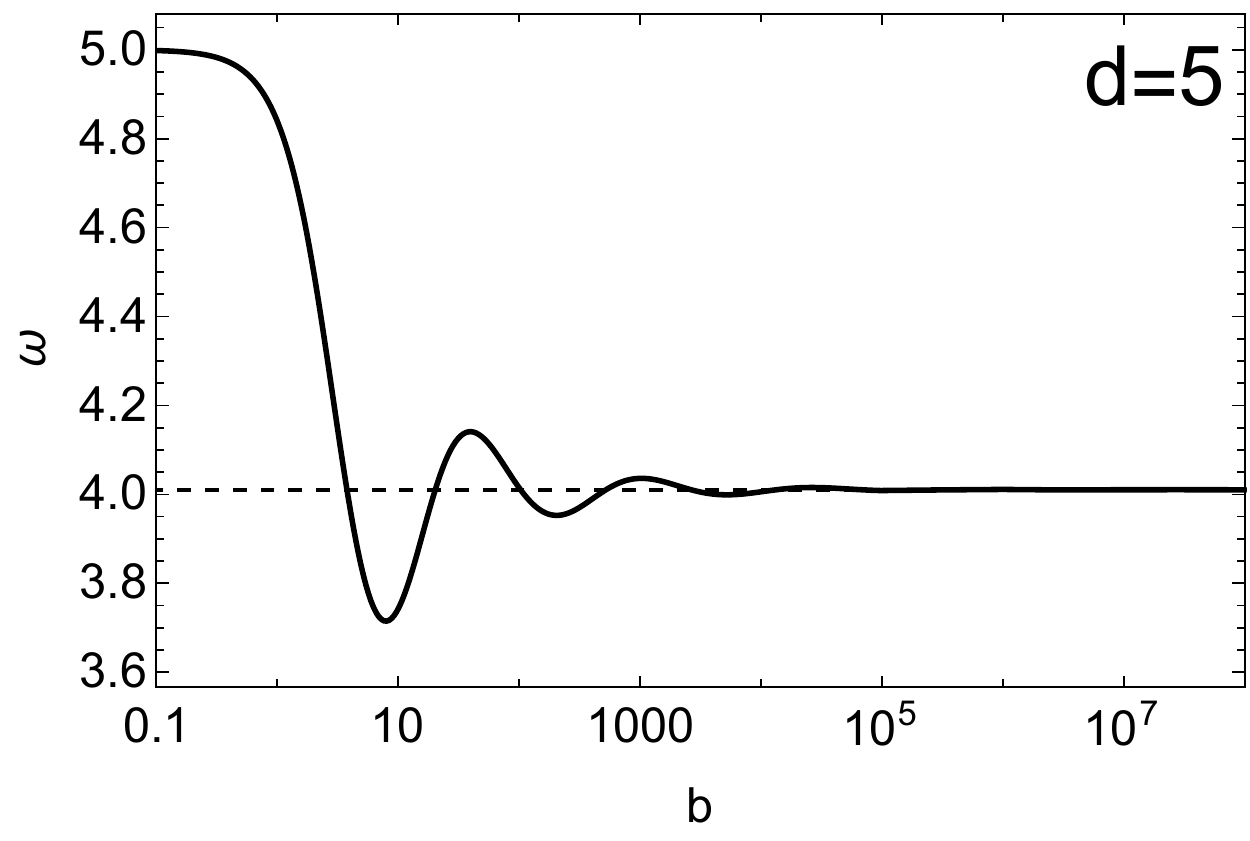}\qquad
\includegraphics[width=0.35\textwidth]{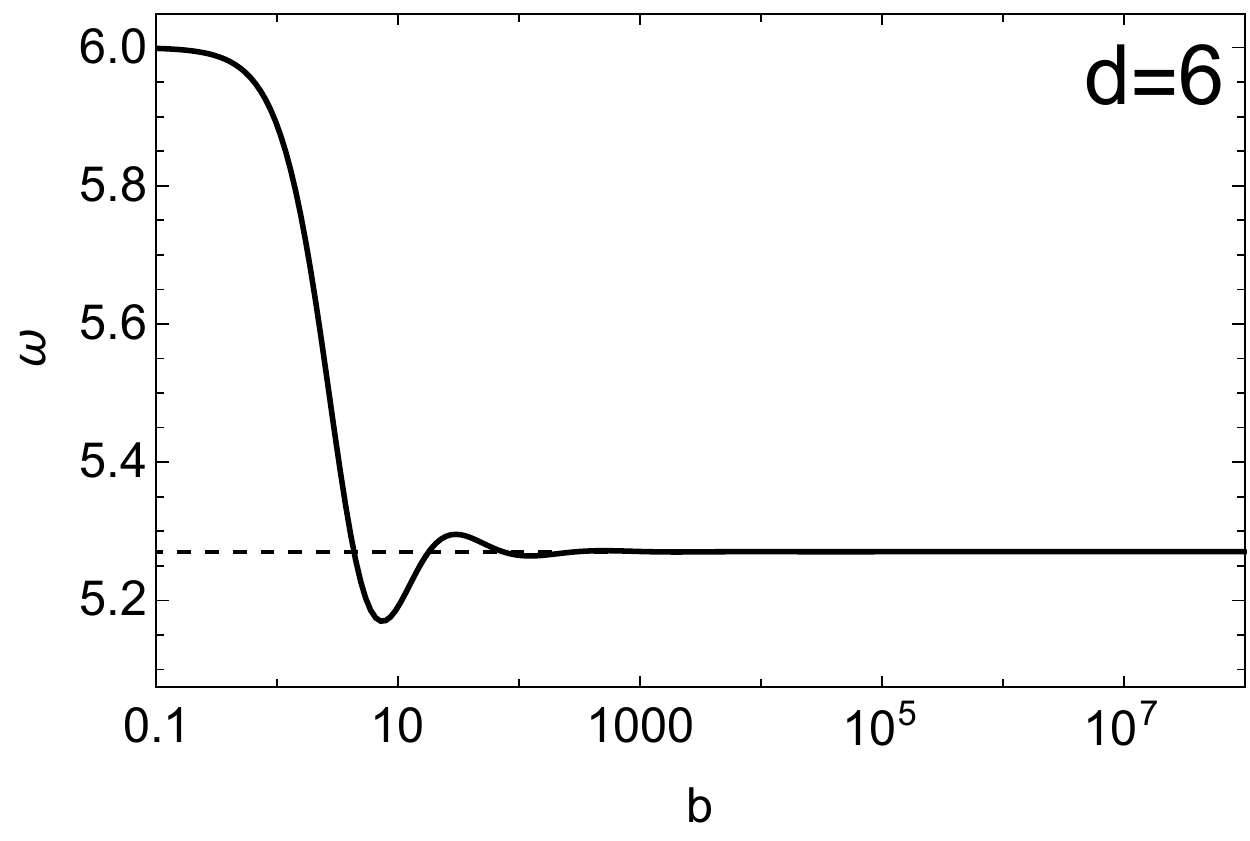}\\
\includegraphics[width=0.35\textwidth]{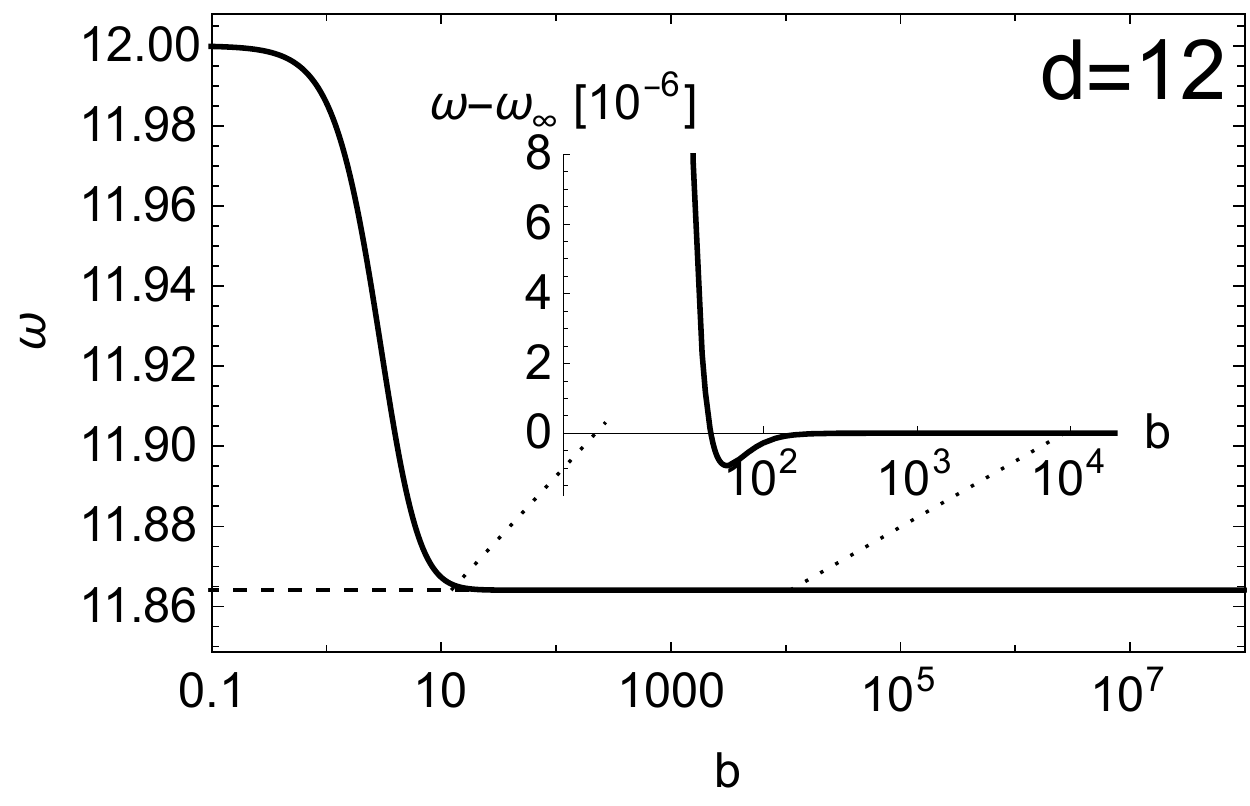}\qquad
\includegraphics[width=0.35\textwidth]{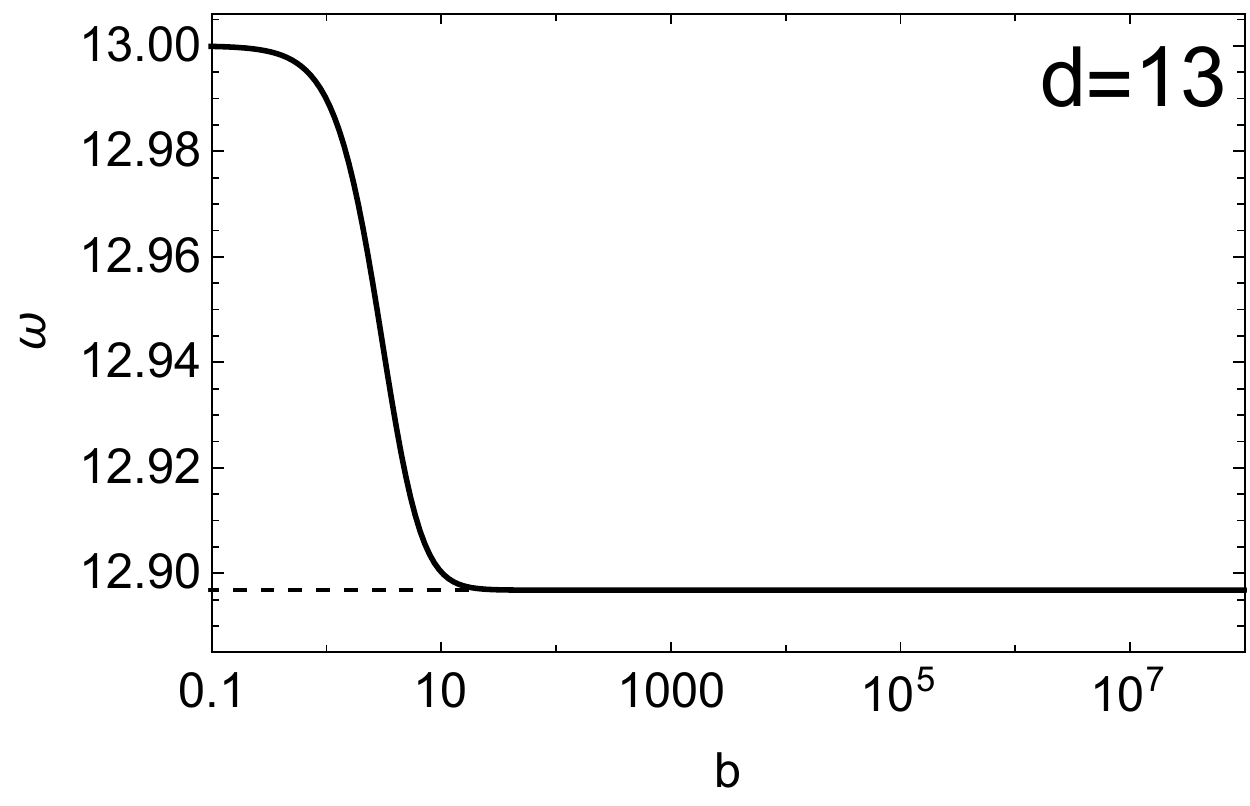}
\caption{Plots of function $\omega (b)$ for the ground states of GP in various dimensions. In relevant cases, the horizontal dashed lines denote the position of the limiting frequency $\omega_\infty$.}
\label{fig:gpnum}
\end{figure}

One can investigate in a similar way other systems, for example, Eq.\ (\ref{eqn:4NLS}) with harmonic potential and quintic nonlinearity $F=u^5$. Then the critical dimension is $d=3$ and the plots of $\omega(b)$ for the ground states in the corresponding dimensions look very similar to GP and SNH, as can be seen in Fig.\ \ref{fig:nls5num}. There is, however, one interesting difference: while SNH and GP in their critical dimensions saturate the bounds on possible frequencies $\omega$ enforced by the Pohozaev identity, this is not the case here, as $\omega\in[1,3]$ instead of $[0,3]$. One may want to investigate this matter further by analysing other simple models with nonlinearities $F=u^p$ for some $p$ in their critical dimensions. However, the only cases that seem to be physically relevant are the ones where both $p$ and $d$ are natural numbers (one usually also wants $p$ to be odd), hence we are left with $p=2$ (in many ways similar to SNH), $p=3$ (GP), and $p=5$ (qunitinc NLS) as the only possibilities. On the other hand, one can increase the number of available options by making two simple observations. First, changing the nonlinearity to $F=|u|^{p-1}u$ deals with potential problems caused by taking the roots of negative numbers and lets us use $p$ that are not necessarily integer. Let us also point out that since we are studying spherically symmetric solutions, the dimension $d$ just plays the role of a parameter in our equation. Hence, any value can be imposed on it and we can easily consider also fractional dimensions. Then, for each $p>1$ we look at the ground states of Eq.\ (\ref{eqn:4NLS}) with harmonic potential and $F=|u|^{p-1}u$. In particular, it can be done in critical "dimensions" $d=2(p+1)/(p-1)$ obtaining plots $\omega(b)$ of the same shape as before, but with various ranges of $\omega$. These ranges are marked in Fig.\ (\ref{fig:nlscritical}). The case of GP seems to be the border case, as for the systems with higher critical dimensions we get the saturated range $\omega\in[0,d]$, while for $2<d\leq 4$ it holds $\omega\in[0,4-d]$.

\begin{figure}
\centering
\captionsetup{font=small, width=.78\linewidth}
\includegraphics[width=0.35\textwidth]{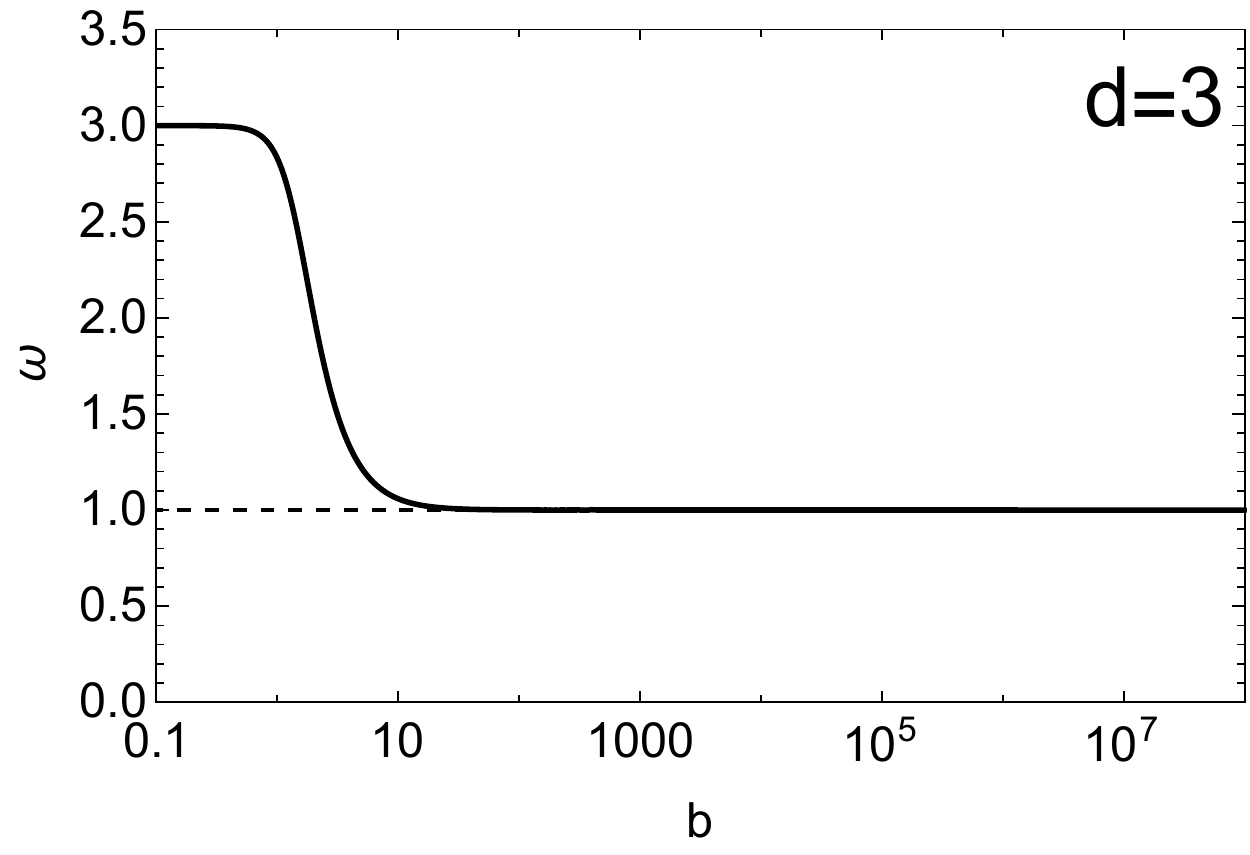}\qquad
\includegraphics[width=0.35\textwidth]{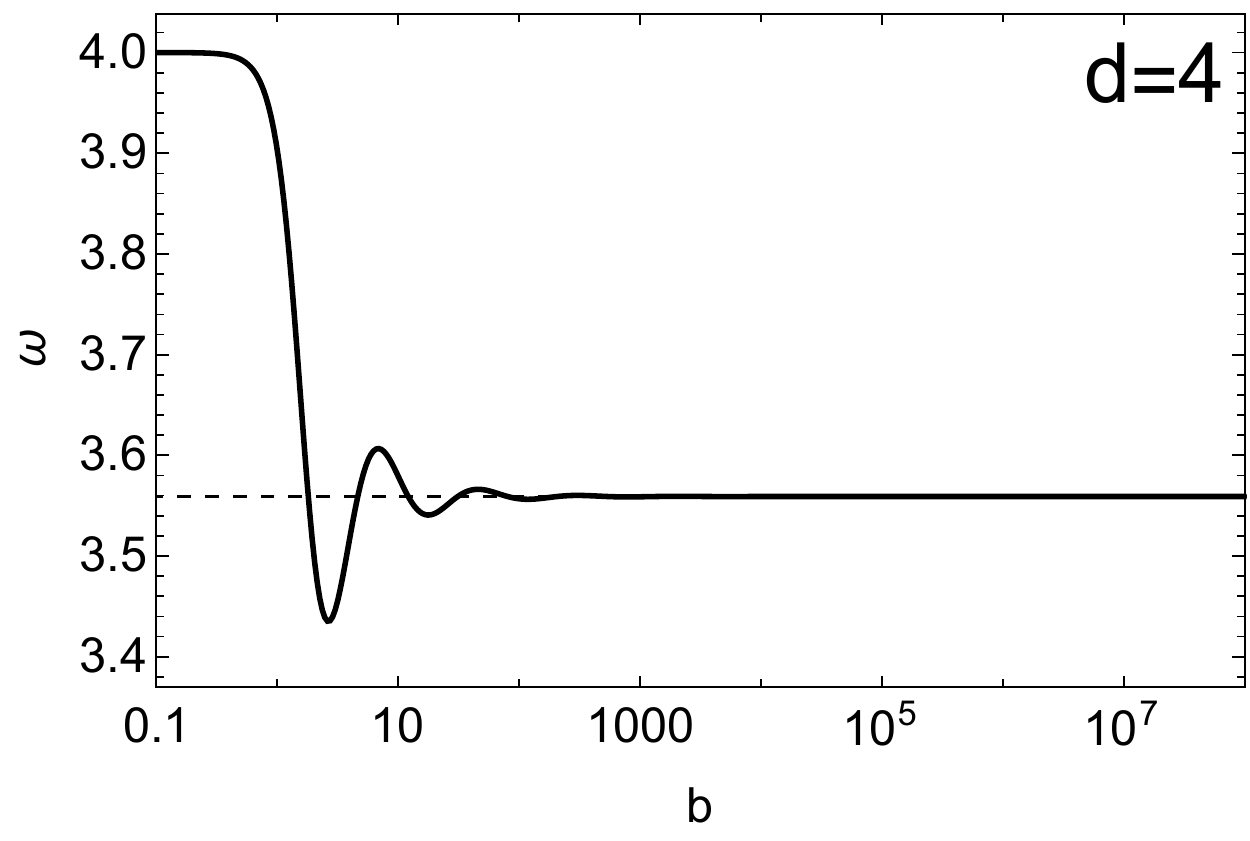}
\caption{Plots of function $\omega (b)$ for the ground states of NLS with quintic nonlinearity in critical and supercritical dimensions. The horizontal dashed lines correspond to the limiting frequencies $\omega_\infty$.}
\label{fig:nls5num}
\end{figure}

\begin{figure}
\centering
\captionsetup{font=small, width=.82\linewidth}
\includegraphics[width=0.5\textwidth]{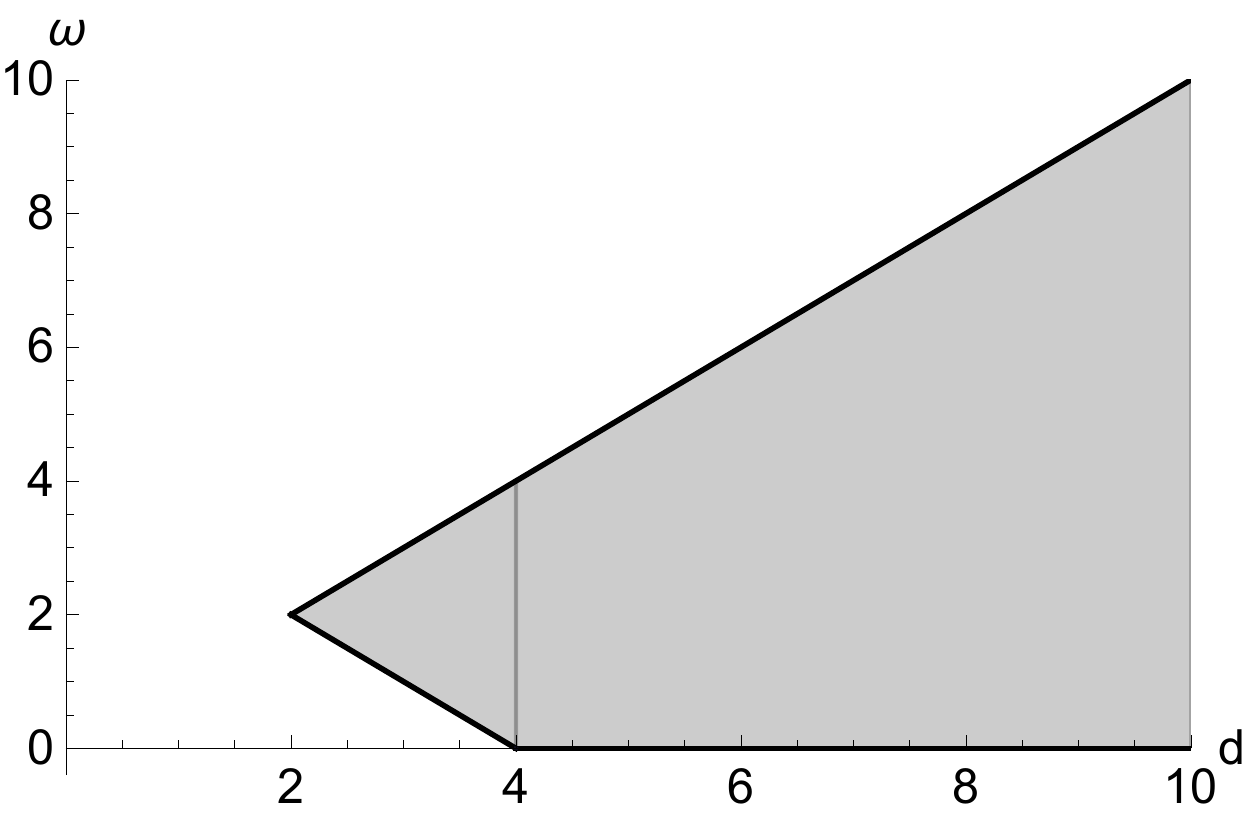}
\caption{Image of the function $\omega(b)$ for ground states of Eq.\ (\ref{eqn:4NLS}) with harmonic potential and $F=|u|^{p-1}u$ in (not necessarily integer) critical dimensions $d$. The relation between $p$ and $d$ here is given by $p=(d+2)/(d-2)$.}
\label{fig:nlscritical}
\end{figure}

In most of our reasonings the explicit form of the potential did not play any role -- we just were using the facts that it is trapping and ensures fast enough decay of the solutions at infinity. To check this claim, in Fig.\ \ref{fig:potentials} we show the plots of function $\omega(b)$ for the ground state of Eq.\ (\ref{eqn:4NLS}) with $F=u^3$ and various trapping potentials $V$ that are increasing faster (quartic potential $V=r^4$) and slower (linear potential $V=r$) than the harmonic one. We present plots for the most interesting cases of the critical and the lowest supercritical dimensions in Fig.\ \ref{fig:potentials}. From the qualitative point of view, the results are the same as before. The situation is more interesting for the Coulomb potential $V=-1/r$. As this time the potential is negative, one cannot repeat the reasoning giving us the lower bound from the Pohozaev identity. It is consistent with the observation that now the bound states of the linear part have negative frequencies. In particular, in the critical dimension the branch $\omega(b)$ bifurcating from such negative frequency decreases indefinitely, as seen in Fig.\ \ref{fig:potentials}. In supercritical dimension the situation seems to be similar to the previous cases. In the end, we study a system with potential $V=(r^2-1)^2$ which is trapping, but not monotonic. Interestingly, for such choice the function $\omega(b)$ in the critical dimension is no longer decreasing. In all these cases, for $d=5$ we eventually get similar oscillating plots. It is probably caused by the fact that the large $b$ behaviour is mainly controlled by the nonlinear term, while the potential term is dominating for small values of $b$.

\begin{figure}
\centering
\captionsetup{font=small, width=.78\linewidth}
\includegraphics[width=0.35\textwidth]{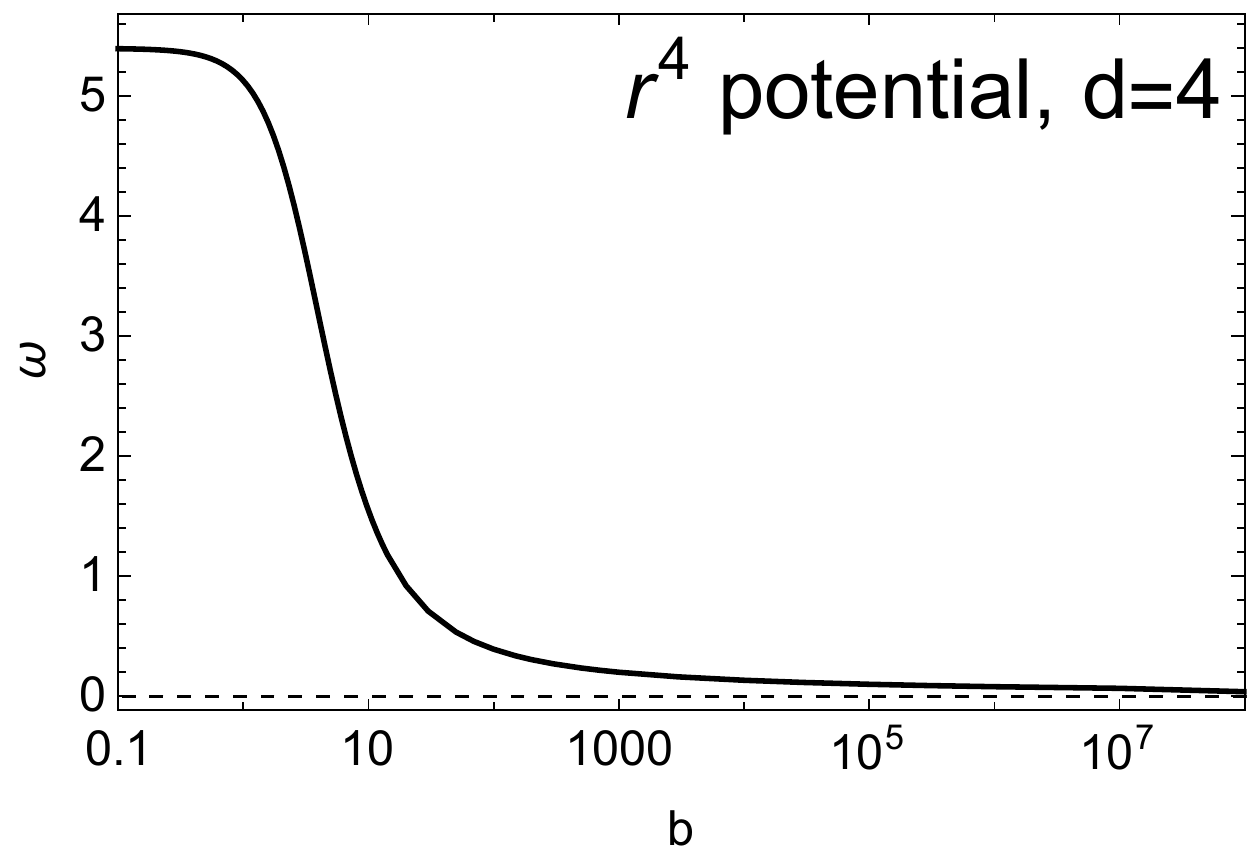}\qquad
\includegraphics[width=0.35\textwidth]{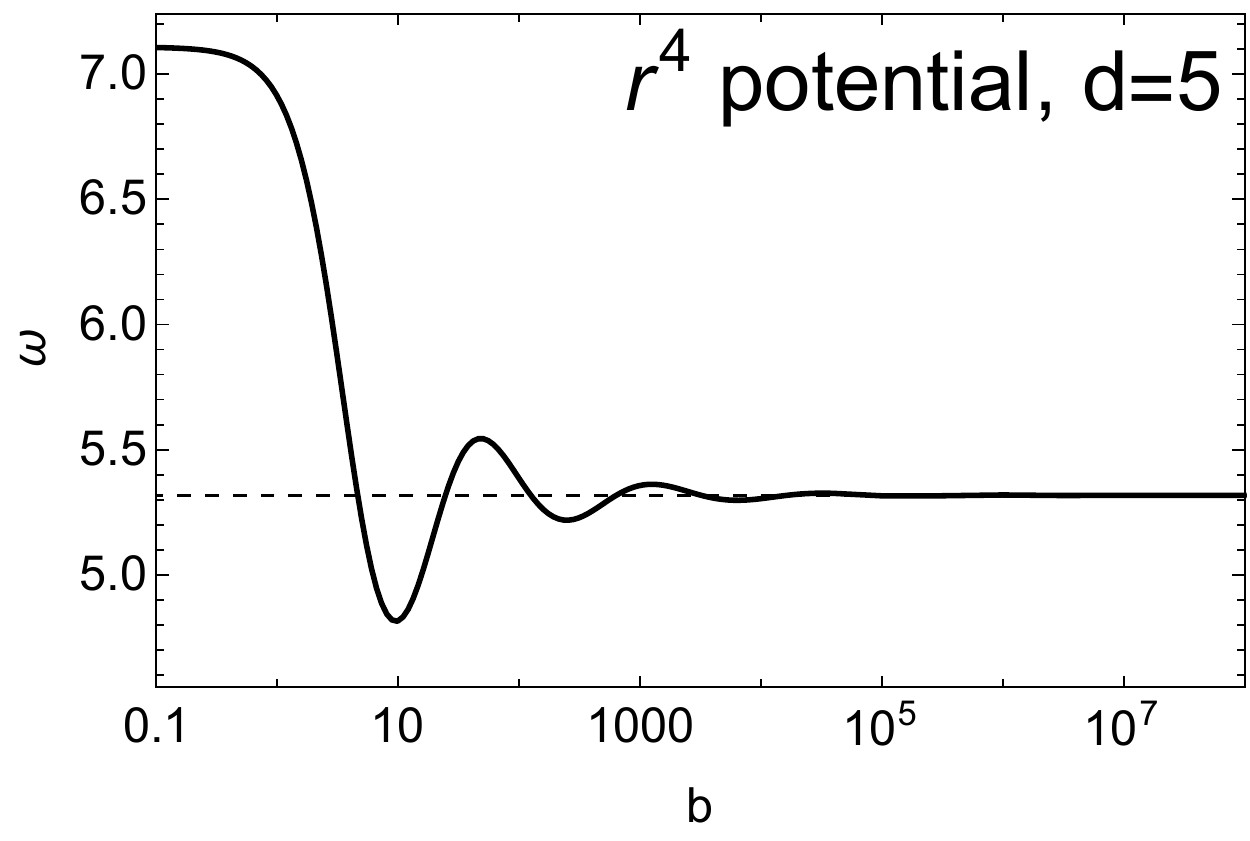}\\
\includegraphics[width=0.35\textwidth]{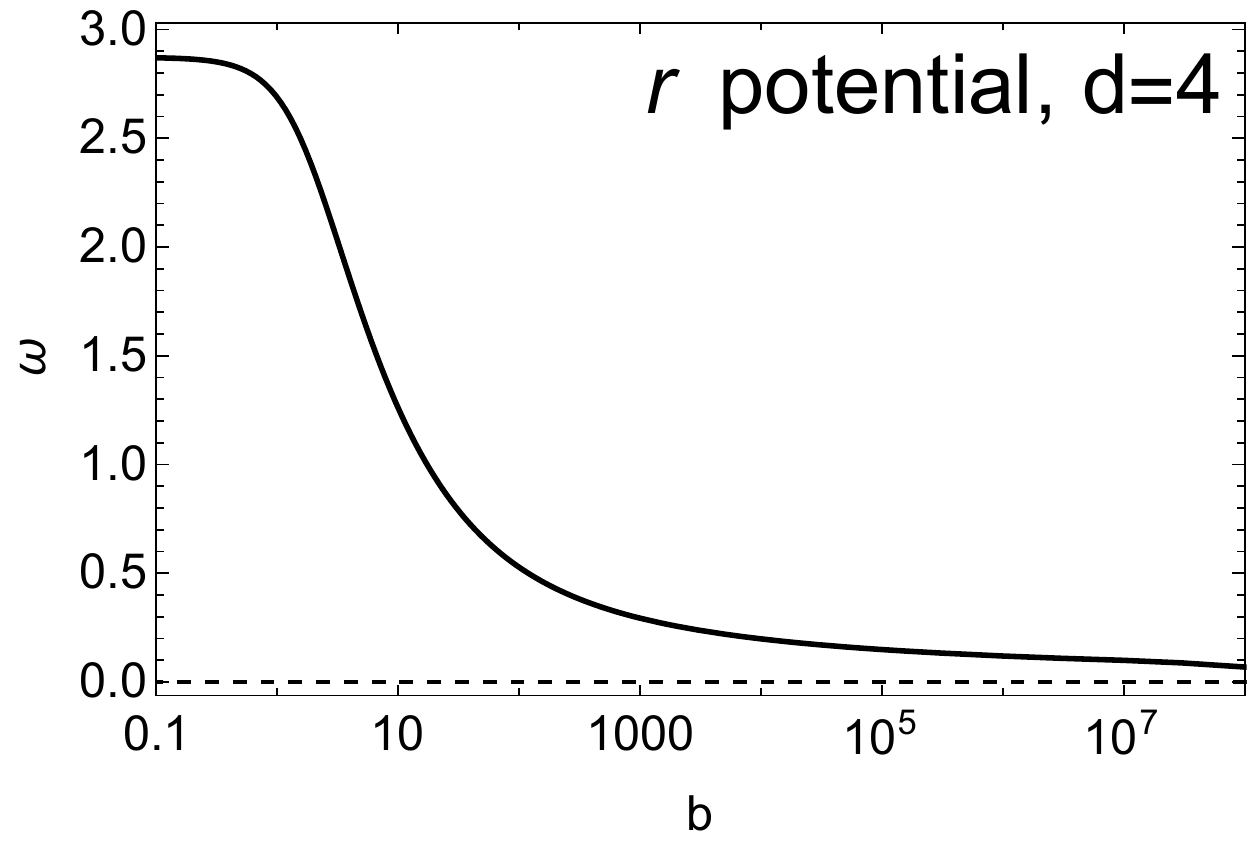}\qquad
\includegraphics[width=0.35\textwidth]{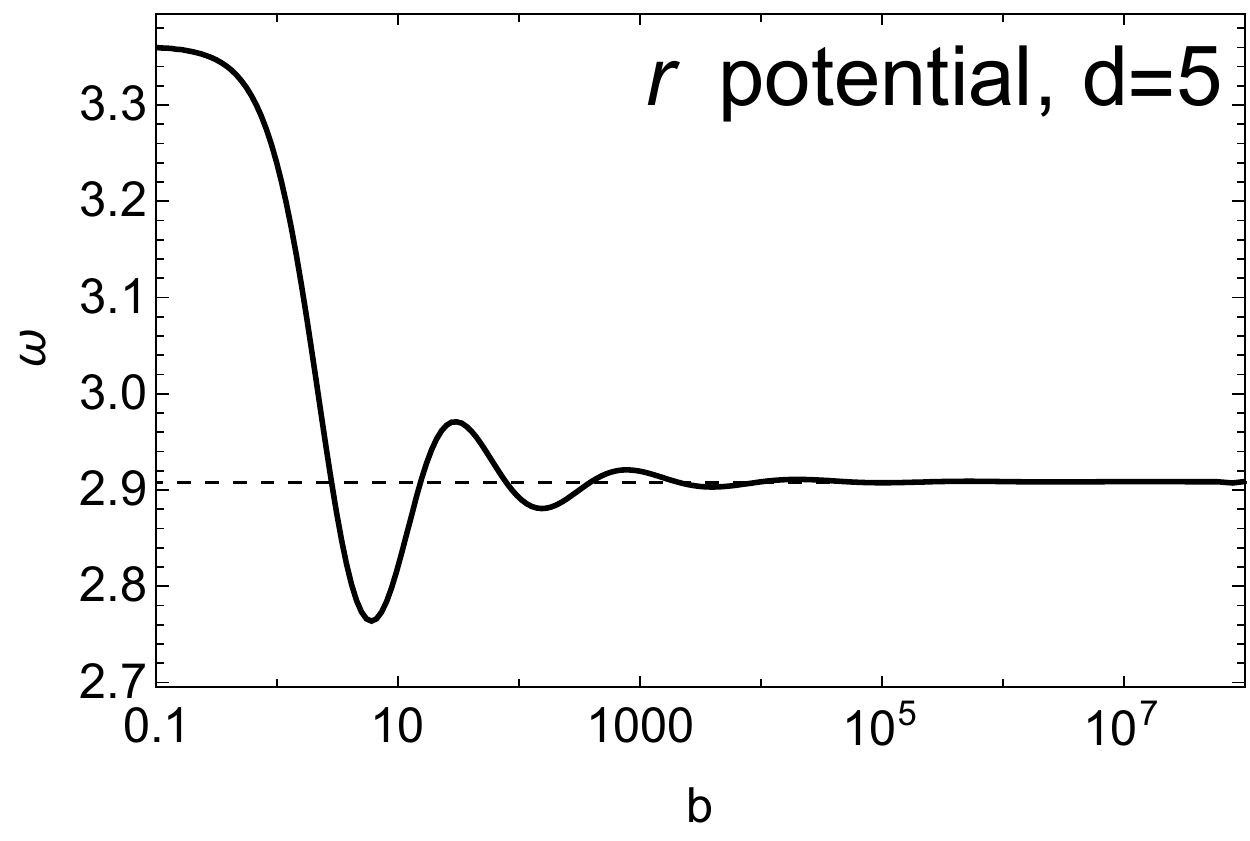}\\
\includegraphics[width=0.35\textwidth]{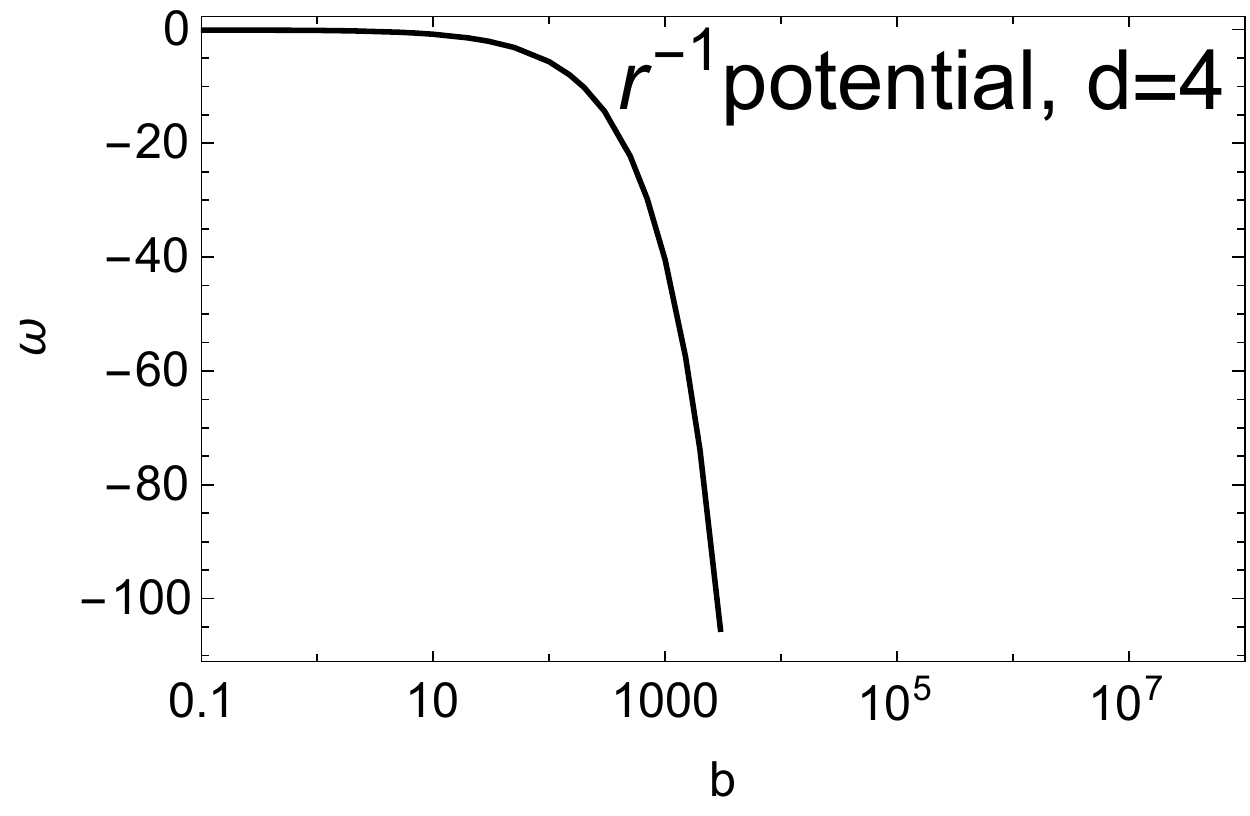}\qquad
\includegraphics[width=0.35\textwidth]{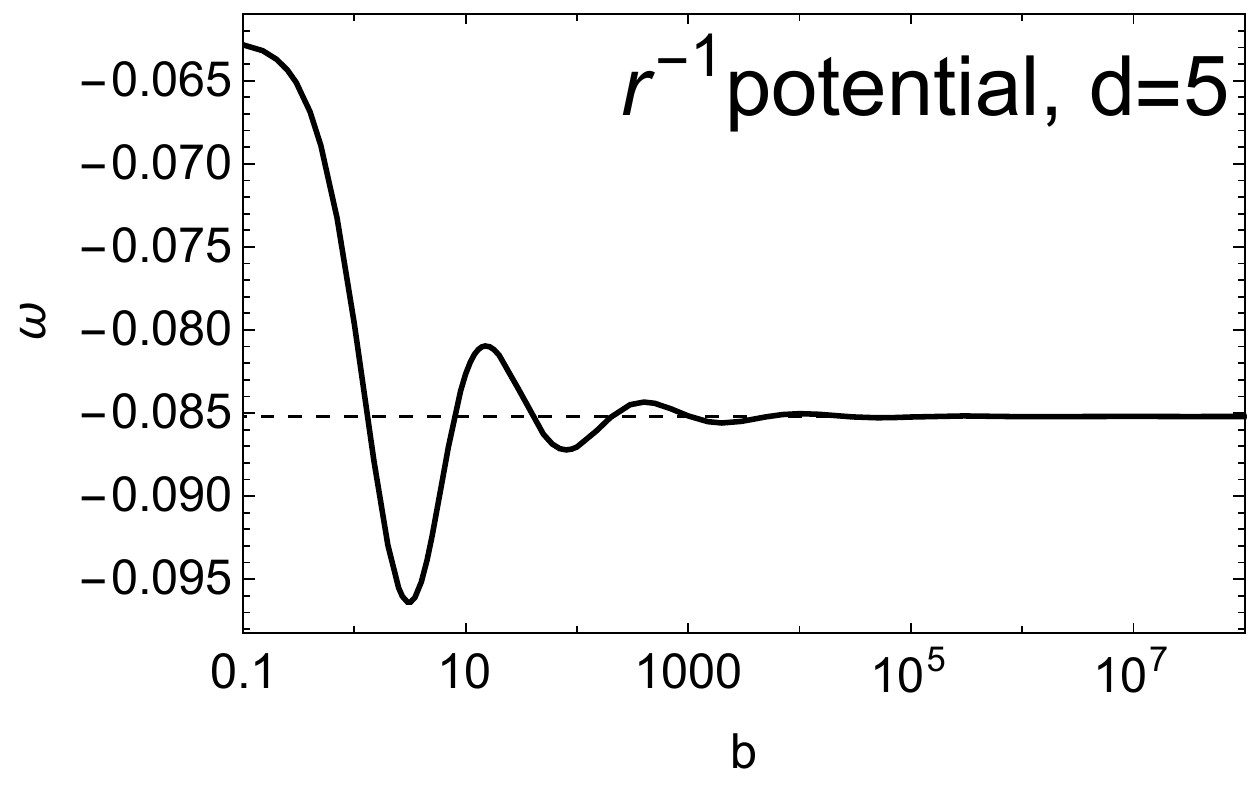}\\
\includegraphics[width=0.35\textwidth]{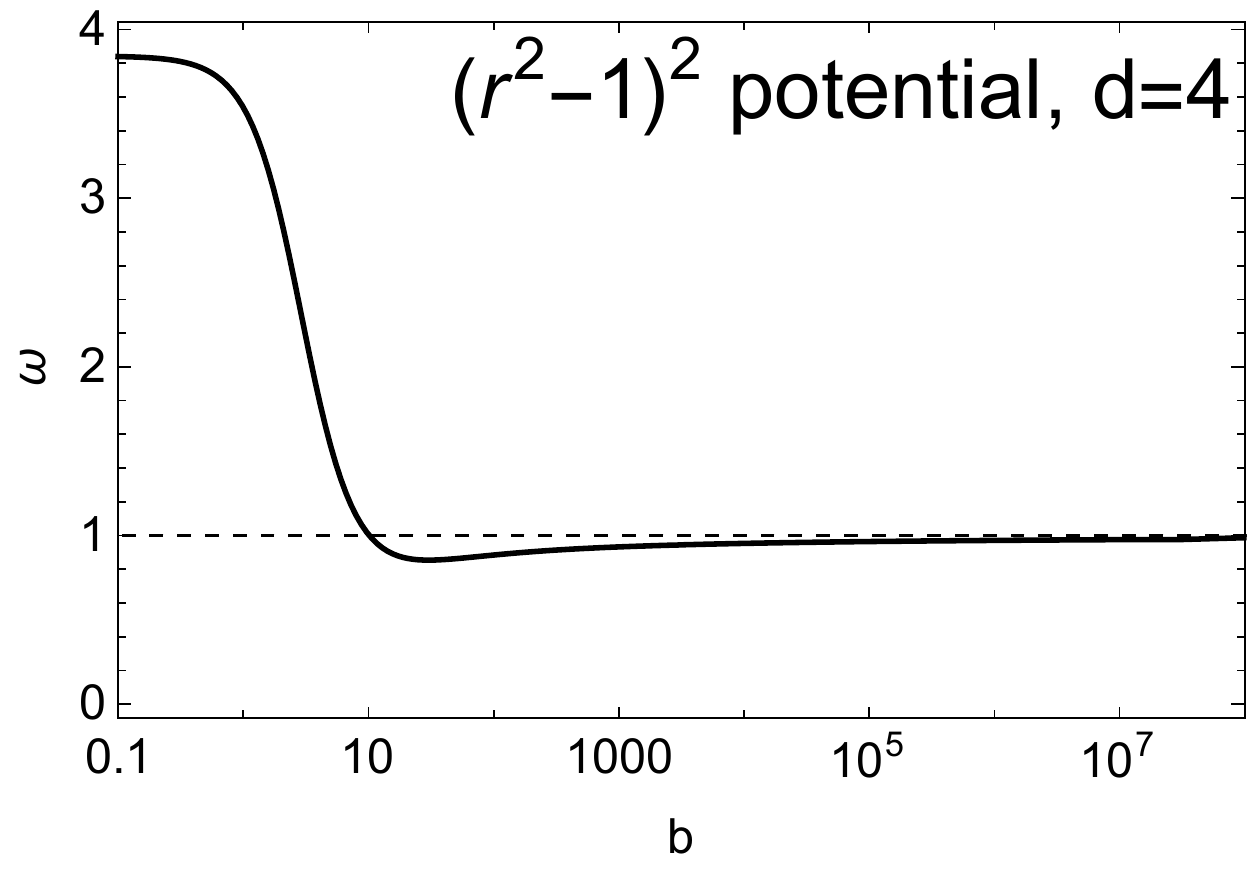}\qquad
\includegraphics[width=0.35\textwidth]{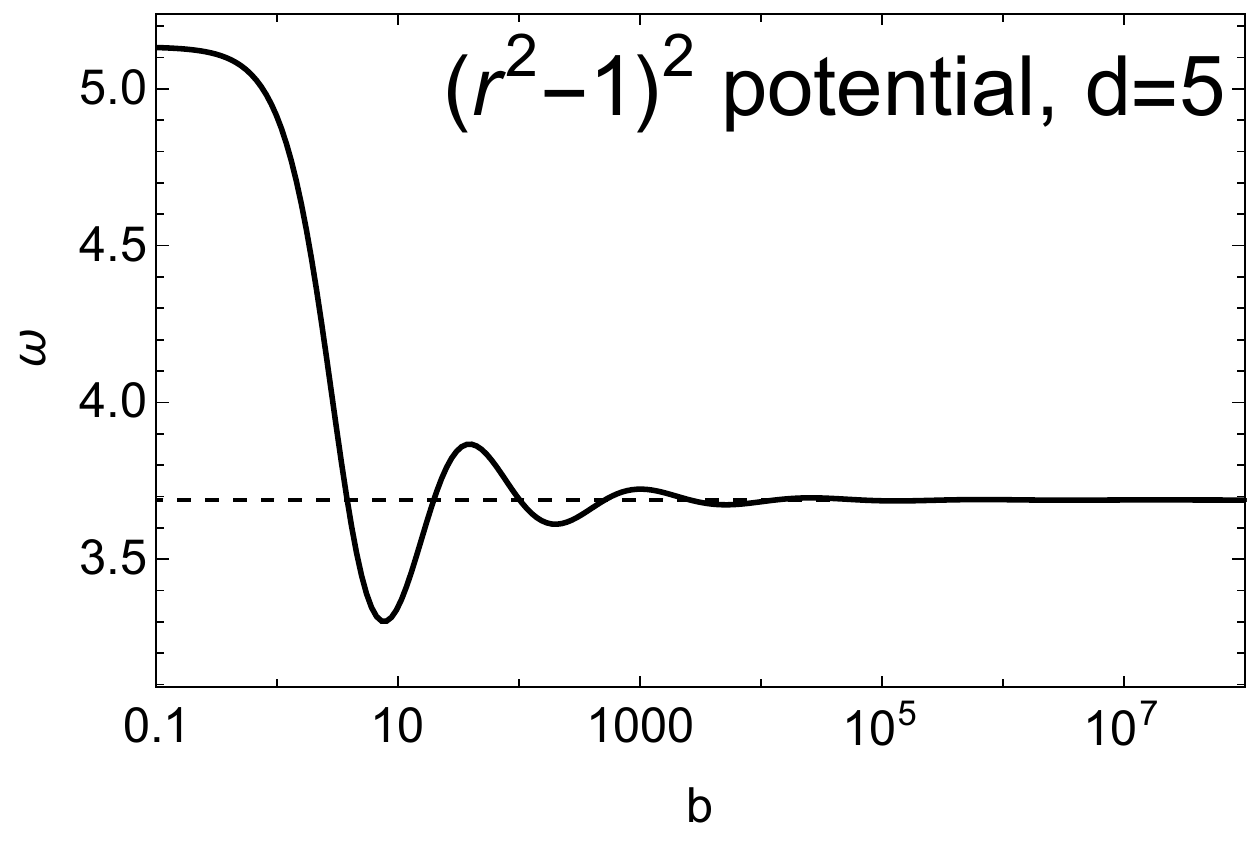}
\caption{Plots of function $\omega (b)$ for the ground states of GP with various potentials in critical and lowest supercritical dimensions.}
\label{fig:potentials}
\end{figure}

\end{adjustwidth}\vspace{\breakFF}

\chapter{Dynamics}\label{sec:dynamics}
Now we return to Eq.\ (\ref{eqn:SNHt}) to briefly discuss the dynamical properties of SNH system. We begin by investigating the stability of the stationary states obtained in the previous chapter. This topic is covered in Section \ref{sec:stability}. It turns out that some observations made earlier, such as the change of behaviour of $\omega(b)$ between $d=15$ and $d=16$, have important implications also here. 

The next natural step would be the study of a full evolution of Eq.\ (\ref{eqn:SNHt}). However, potentially interesting phenomena are expected to occur for very large times. Performing numerical simulations with adequate precision for such long times seems to be rather challenging and is a topic that needs further work, as we hint in Section \ref{sec:evolution}. In return, in Section \ref{sec:resonant} we focus on the simpler version of the problem given by the resonant approximation. This approach not only simplifies the numerical calculations, but also lets us get some interesting analytical results.

Most of the results presented in this chapter (except for the part regarding the resonant system) are in some sense preliminary. Further work is needed to achieve a better understanding of the dynamical behaviour of SNH equation and similar NLS systems in supercritical dimensions. 

\section{Stability}\label{sec:stability}
To study the linear stability of the stationary states we need to introduce an appropriate framework. Similarly to Section \ref{sec:smalllargeb}, it is more convenient to work here with general Eq.\ (\ref{eqn:introNLS}), instead of a specific realisation such as SNH equation. By $u$ we will denote here the stationary state, i.e.\ the solution to Eq.\ (\ref{eqn:4NLS}) that the stability we are investigating, while $\omega$ is its frequency. Small perturbations of this solution can be described with the help of the following ansatz
\begin{equation}\label{eqn:5ansatzstab}
    \psi(t,x)=e^{-i\omega t}\left( u(x)+f(t,x)+i g(t,x)\right).
\end{equation}
Functions $f$ and $g$ are assumed here to be small (in comparison with $|u|$) and real. It is also convenient to assume that the nonlinearity $F$ can be factored out to $F(\psi)=G(\psi)\,\psi$, where $G(\psi)=G(|\psi|)$ (for SNH system $G=A_d \,|\psi|^2\ast |x|^{-(d-2)}$). Then, by inserting this ansatz into the nonlinear term we get
\begin{align*}
    F(\psi)&=G(\psi)\cdot\psi=e^{-i\omega t}\left(G(u+f+ig)\cdot(u+f+ig)\right)\\
    &\approx e^{-i\omega t} \left( u\,G(u)+(f+ig)G(u)+ u\,G_u(u)[f]\right),
\end{align*}
where we have restricted ourselves to terms at most linear in $f$ and $g$. Now we may plug ansatz (\ref{eqn:5ansatzstab}), together with this expansion of $F(\psi)$, to the dynamical equation (\ref{eqn:introNLS}). Using the fact that $u$, $f$, and $g$ are real, we split the resulting equation into real and imaginary parts:
\begin{align*}
    \partial_t f =& -\Delta g+V g-\omega g-G(u)\, g,\\
    \partial_t g =& \Delta f-V f+\omega f+G(u)\, f+ u\,G_u(u)[f].
\end{align*}
These equations can be written conveniently in a matrix form
\begin{align}\label{eqn:matL}
    \partial_t \left(\begin{array}{c}f\\g\end{array}\right)=
    \left(\begin{array}{cc}0&L_-\\-L_+ & 0\end{array}\right)\left(\begin{array}{c}f\\g\end{array}\right),
\end{align}
where the linear operators in the antidiagonal are given by
\begin{subequations}\label{eqn:Lpmoperators}
\begin{align}
    L_-&:=-\Delta+V-\omega-G(u),\\
    L_+&:=-\Delta+V-\omega-G(u)-u\, G_u(u).
\end{align}
\end{subequations}
The full linear operator present in Eq.\ (\ref{eqn:matL}) will be denoted by $\mathcal{L}$.

Now let us assume that the vector $(f, g)$ can be written as $(\alpha, \beta)\, e^{\lambda t}$, where $\alpha$ and $\beta$ do not depend on time. Such ansatz transforms Eq.\ (\ref{eqn:matL}) to an eigenproblem. Hence, the nature of the eigenvalues of $\mathcal{L}$ determines the stability of our stationary states: if any of the eigenvalues has a positive real part, the solution is linearly unstable. We can begin the study of this matter with the following simple observations.  Even though the operators $L_\pm$ are obviously self-adjoint with respect to the scalar product introduced in Section \ref{sec:omegawindow}, the operator $\mathcal{L}$ is clearly not. However, it has another interesting property. Let $\alpha$ and $\beta$ be functions such that for some $\lambda\in\mathbb{C}$
\begin{align}\label{eqn:eigenL}
    \mathcal{L}\left(\begin{array}{c}\alpha\\ \beta\end{array}\right)= \left(\begin{array}{r}L_-\beta\\ -L_+\alpha\end{array}\right) = \lambda \left(\begin{array}{c}\alpha\\ \beta\end{array}\right),
\end{align}
i.e.\ $\lambda$ is an eigenvalue. Then one easily checks that
\begin{align*}
    \mathcal{L}\left(\begin{array}{c}\bar{\alpha}\\ \bar{\beta}\end{array}\right)&= \overline{\mathcal{L}\left(\begin{array}{c}\alpha\\ \beta\end{array}\right)}= \overline{\lambda \left(\begin{array}{c}\alpha\\ \beta\end{array}\right)}= \bar{\lambda}\left(\begin{array}{c}\bar{\alpha}\\ \bar{\beta}\end{array}\right),\\
    \mathcal{L}\left(\begin{array}{c}\alpha\\ -\beta\end{array}\right)&=\left(\begin{array}{c}-L_-\beta\\ -L_+\alpha\end{array}\right)=\left(\begin{array}{r}-\lambda\alpha\\ \lambda\beta\end{array}\right)=-\lambda \left(\begin{array}{c}\alpha\\ -\beta\end{array}\right),\\
    \mathcal{L}\left(\begin{array}{c}\bar{\alpha}\\ -\bar{\beta}\end{array}\right)&= \overline{ \mathcal{L}\left(\begin{array}{c}\alpha\\ -\beta\end{array}\right)}= \overline{-\lambda \left(\begin{array}{c}\alpha\\ -\beta\end{array}\right)}= -\bar{\lambda} \left(\begin{array}{c}\bar{\alpha}\\ -\bar{\beta}\end{array}\right).
\end{align*}
Hence, $\bar{\lambda}$, $-\lambda$, and $-\bar{\lambda}$ are also the eigenvalues of $\mathcal{L}$. In a more geometric picture, if some point in the complex plane represents an eigenvalue of $\mathcal{L}$, then so do its reflections with regard to the real and imaginary axes. It also means that our system is unstable if $\mbox{Re}\,\lambda\neq 0$.

\subsection{Ground states}
Since operator $\mathcal{L}$ is not self-adjoint, investigating its eigenvalues may in general be difficult (as we will see in the next subsection). However, for the ground states the situation simplifies greatly as one can decide the stability by using the Vakhitov-Kolokolov criterion \cite{Vak73, Pel}. To formulate it, let us go back to the whole branch of ground states discussed in \ref{sec:omegab}. In Fig.\ \ref{fig:snhnum} we presented the ground states in plots $(b,\omega)$. However, one can draw this branch in coordinates $(\omega, \mathcal{M})$, where $\mathcal{M}$ is the mass of the solution (as before, for convenience we divide it by the area of a $d-1$-sphere). Such plots, being just curves parameterized by $b$, for the ground states in various dimensions are presented in Fig.\ \ref{fig:massSNH} (it can be seen that only in $d=3$ there exist ground states with arbitrarily large mass $\mathcal{M}$, which agrees with the results mentioned in Section \ref{sec:subcritical}). Let us point out that in general $\mathcal{M}(\omega)$ is not a function, since for $7\leq d\leq 15$ there are values of $\omega$ characterising multiple ground states. Nevertheless, in almost all points the function $\mathcal{M}(\omega)$ can be introduced locally, telling us how the mass changes with $\omega$ as we move along the curve. Then, the Vakhitov-Kolokolov criterion tells us that the ground state $u$ is stable if $\mathcal{M}'(\omega)<0$, $L_+$ has exactly one negative eigenvalue and $L_-$ is a non-negative operator.

\begin{figure}
\centering
\begin{subfigure}{0.45\textwidth}
\includegraphics[width=\textwidth]{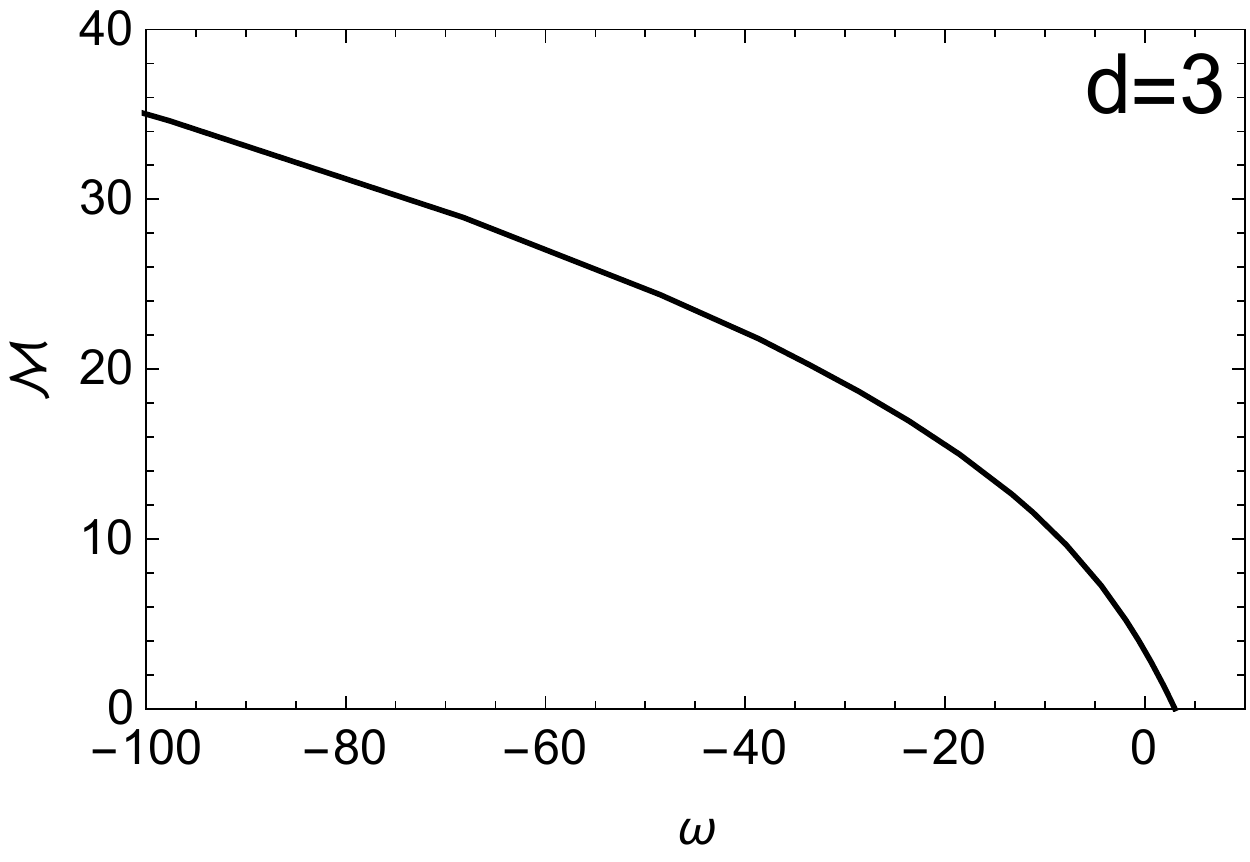}
\end{subfigure}
~
\begin{subfigure}{0.45\textwidth}
\includegraphics[width=\textwidth]{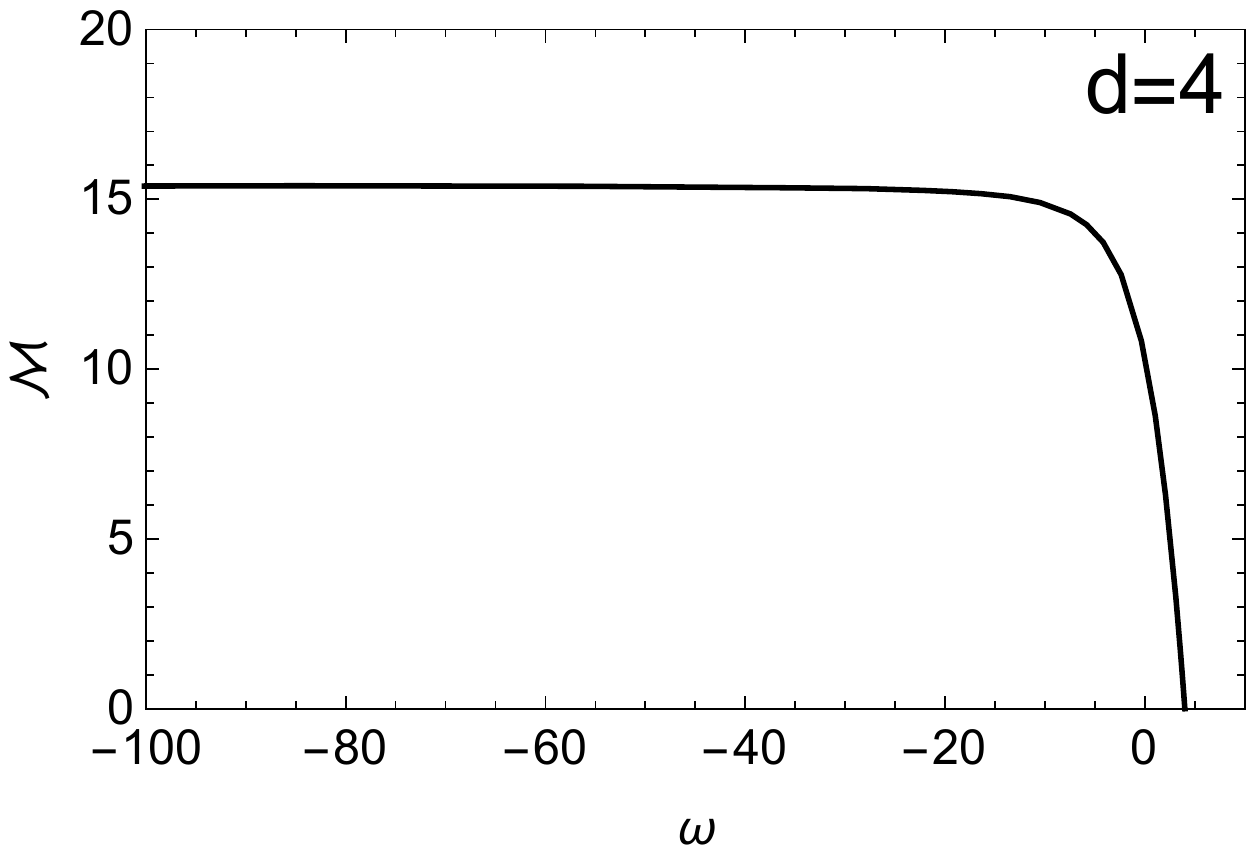}
\end{subfigure}
\\
\begin{subfigure}{0.45\textwidth}
\includegraphics[width=\textwidth]{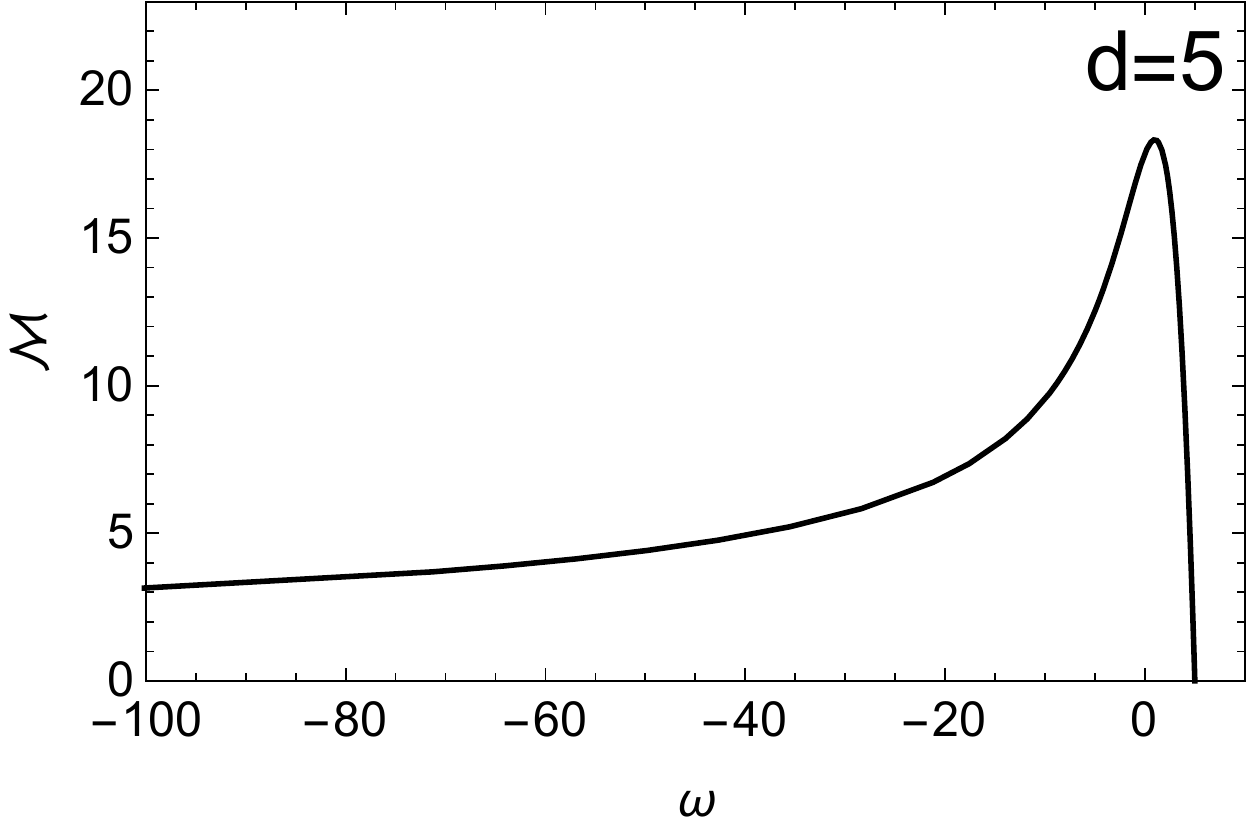}
\end{subfigure}
~
\begin{subfigure}{0.45\textwidth}
\includegraphics[width=\textwidth]{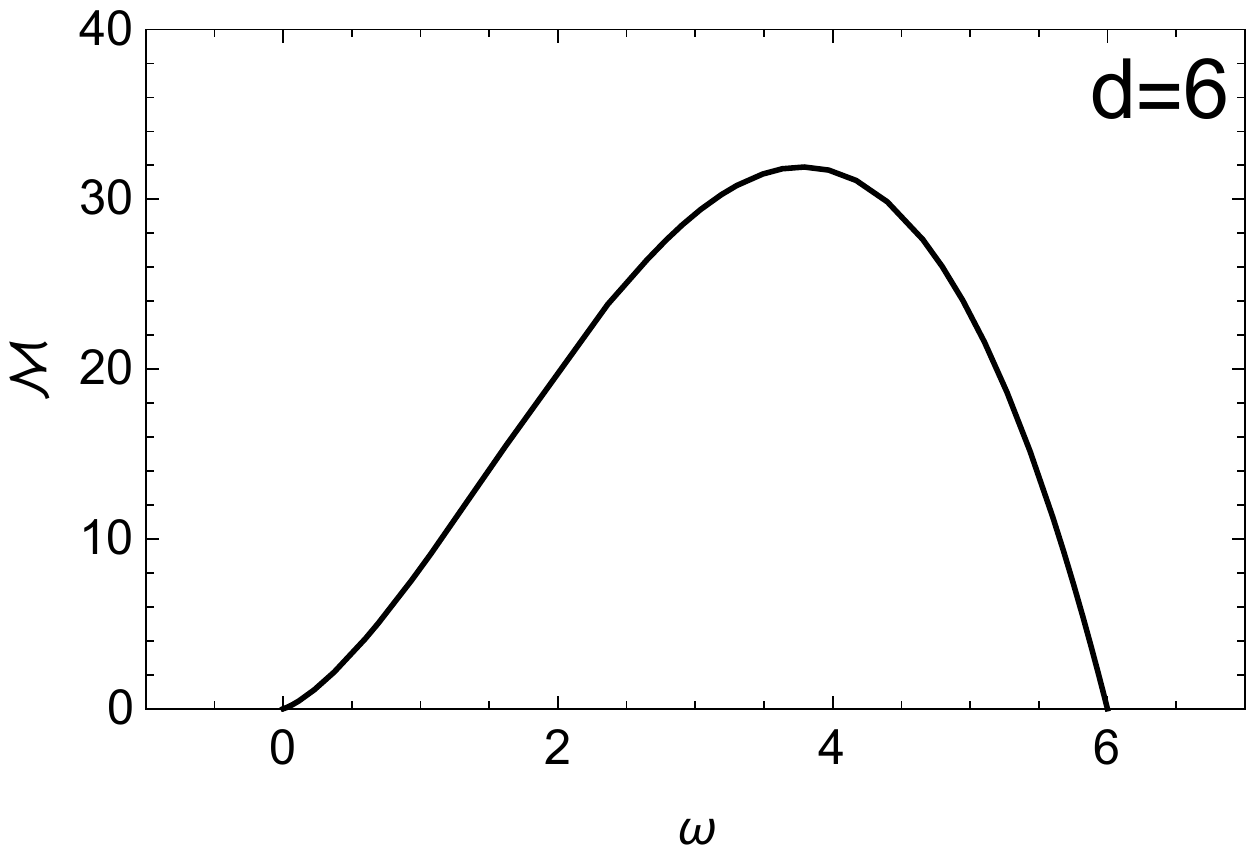}
\end{subfigure}
\\
\begin{subfigure}{0.45\textwidth}
\includegraphics[width=\textwidth]{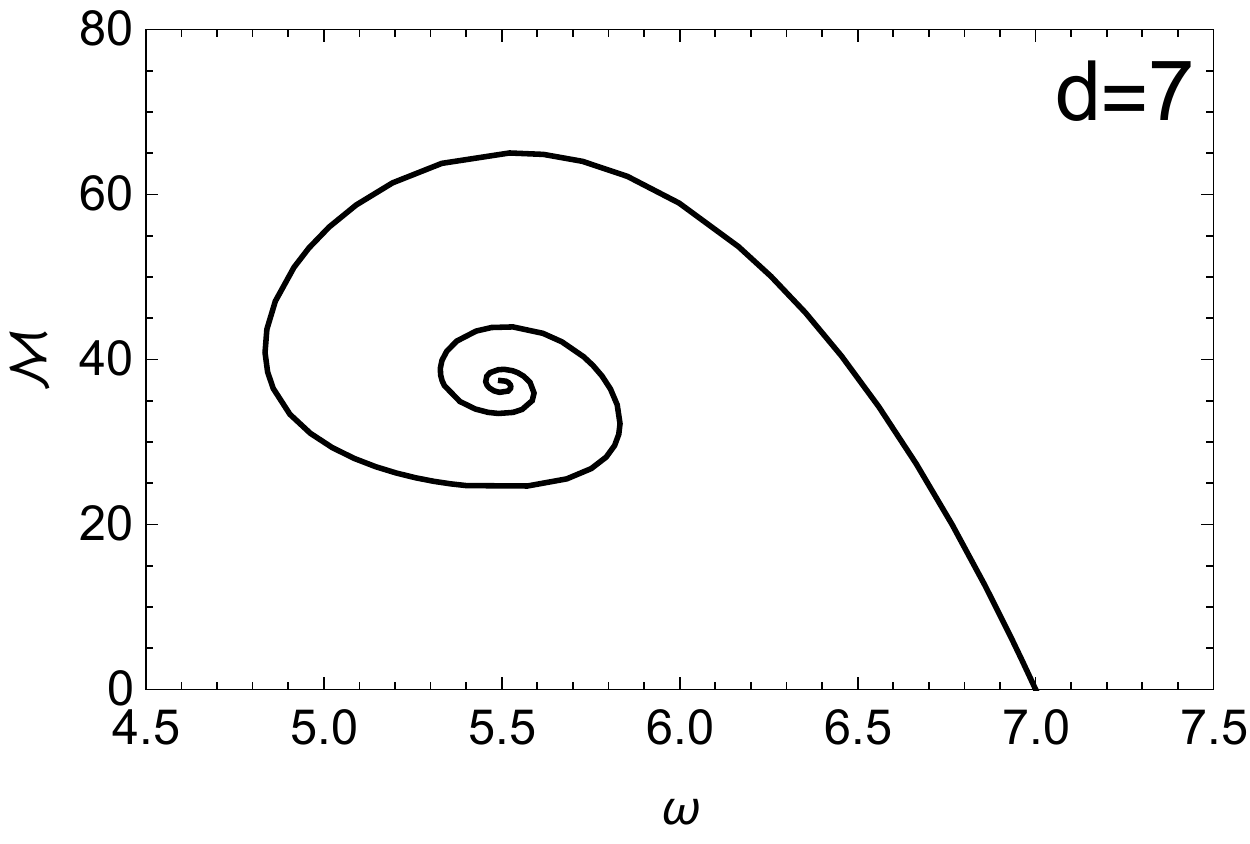}
\end{subfigure}
~
\begin{subfigure}{0.45\textwidth}
\includegraphics[width=\textwidth]{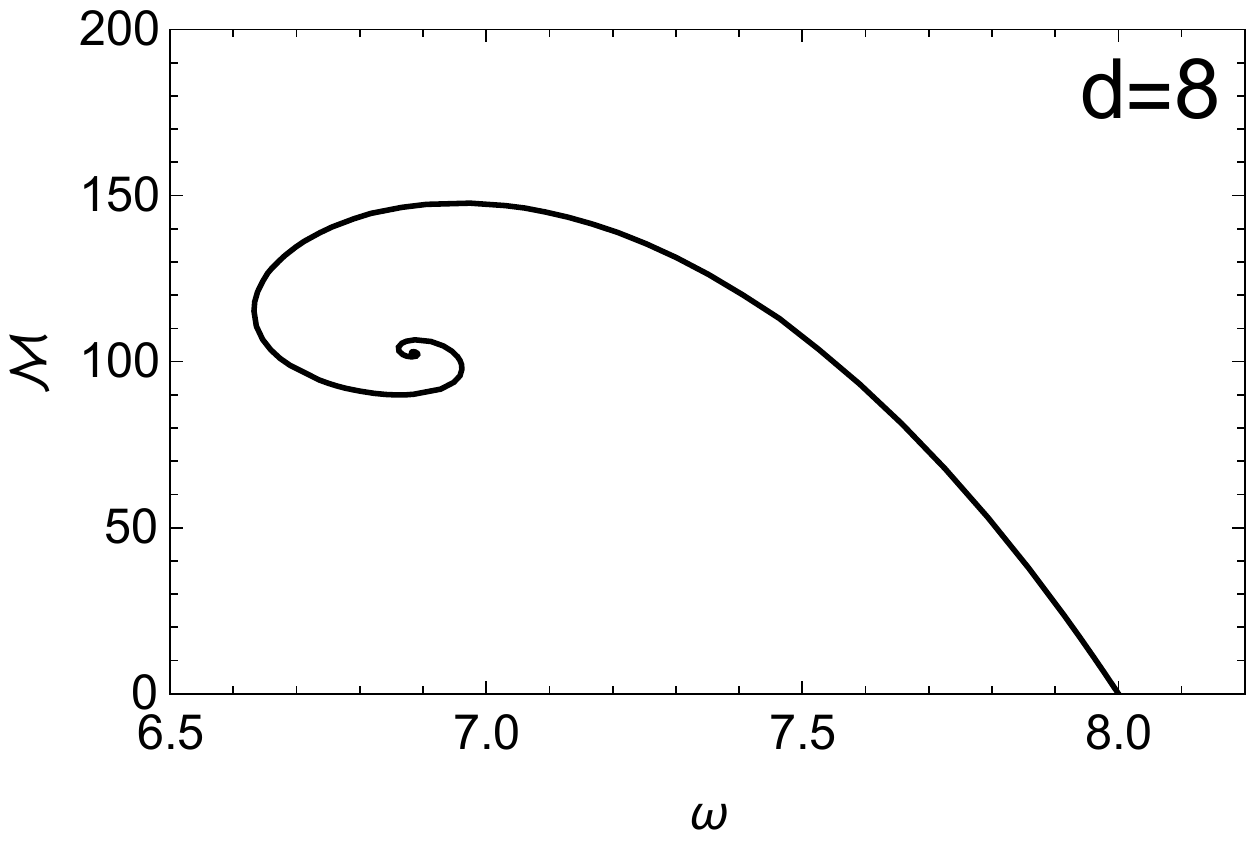}
\end{subfigure}
\\
\begin{subfigure}{0.45\textwidth}
\includegraphics[width=\textwidth]{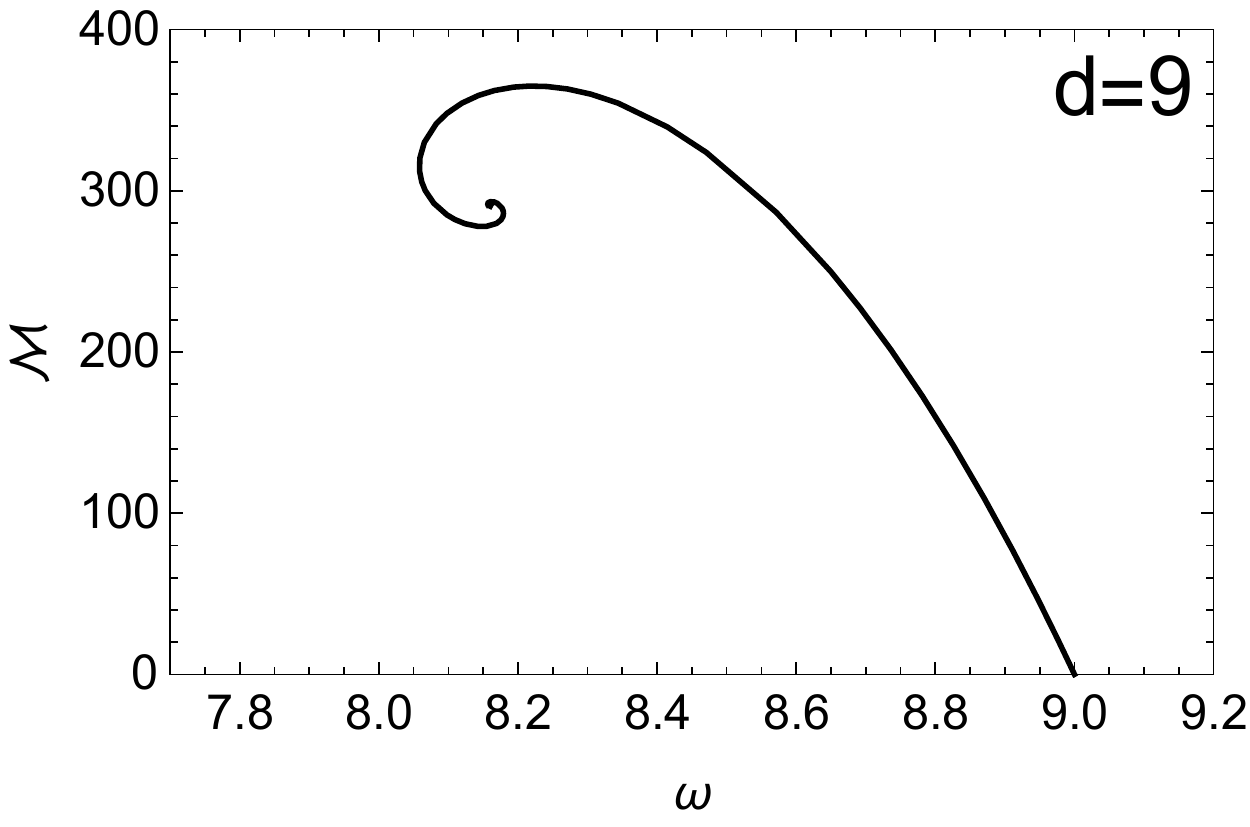}
\end{subfigure}
~
\begin{subfigure}{0.45\textwidth}
\includegraphics[width=\textwidth]{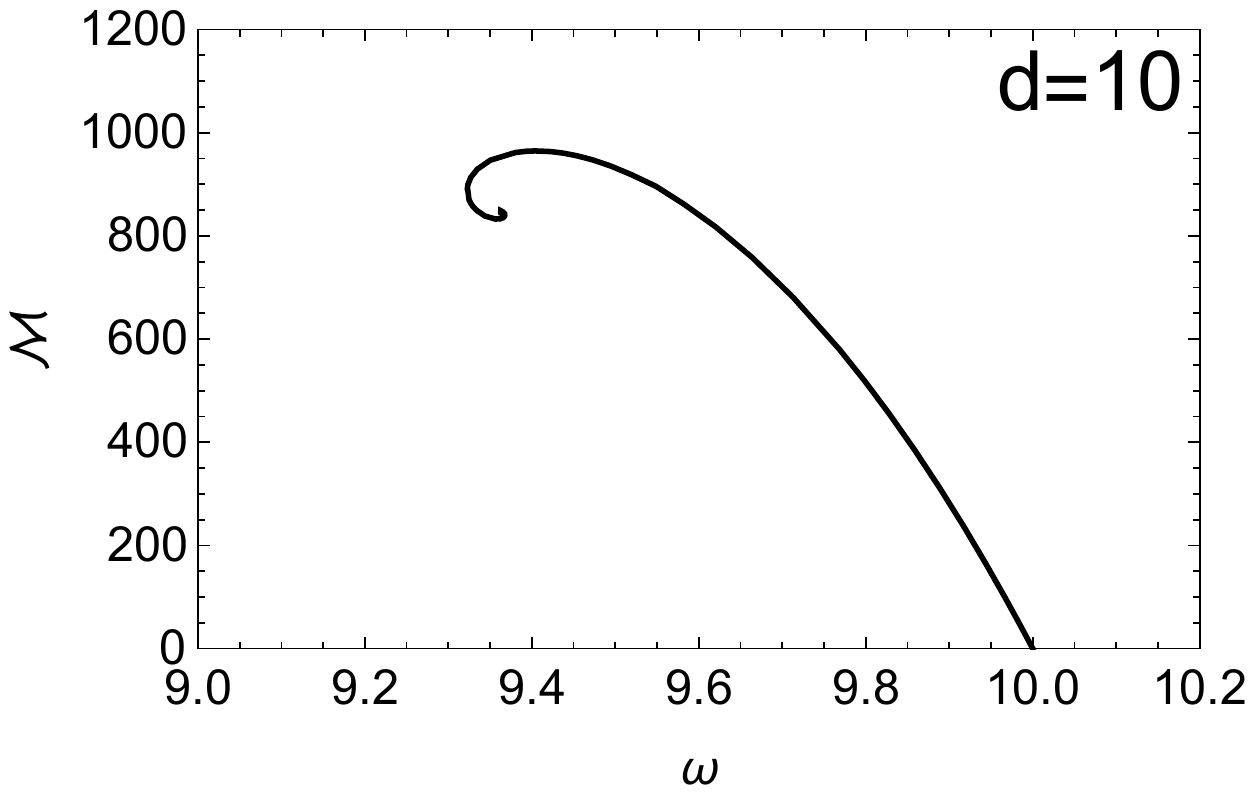}
\end{subfigure}
\end{figure}

\begin{figure}
\ContinuedFloat
\centering
\begin{subfigure}{0.45\textwidth}
\includegraphics[width=\textwidth]{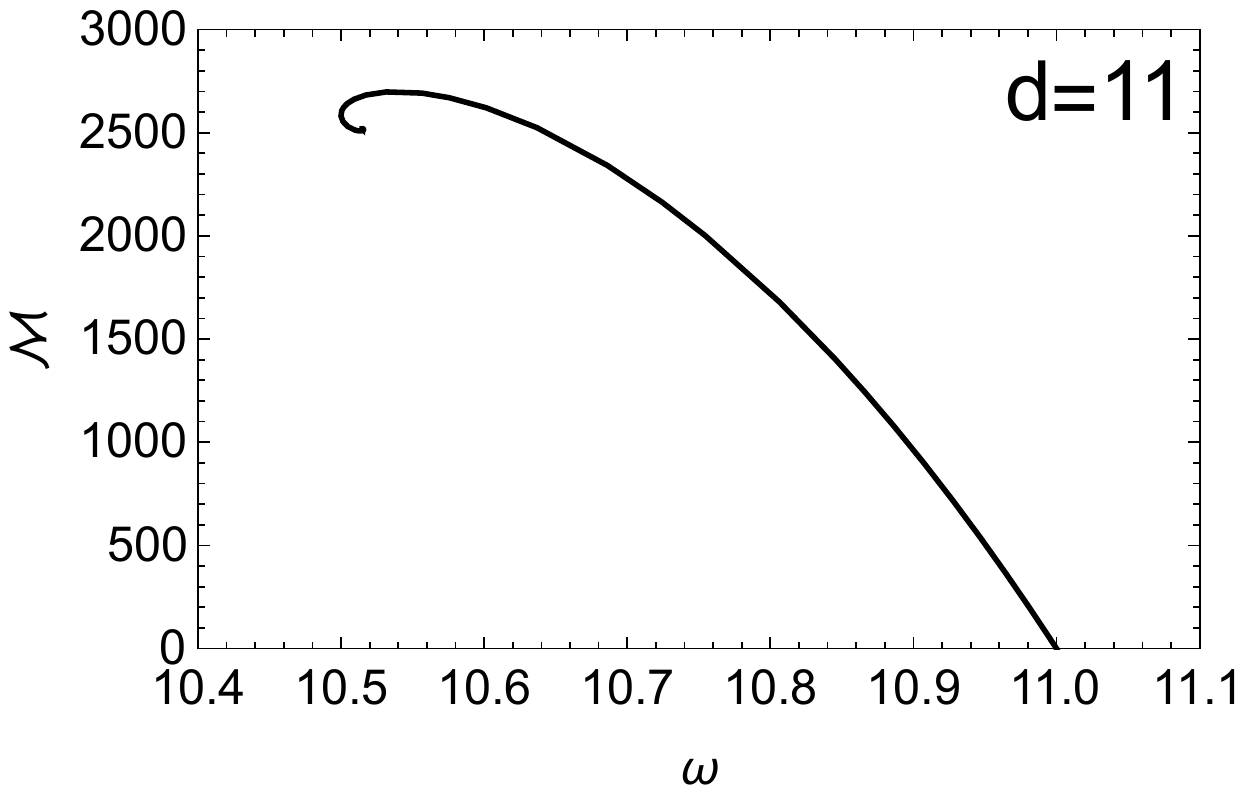}
\end{subfigure}
~
\begin{subfigure}{0.45\textwidth}
\includegraphics[width=\textwidth]{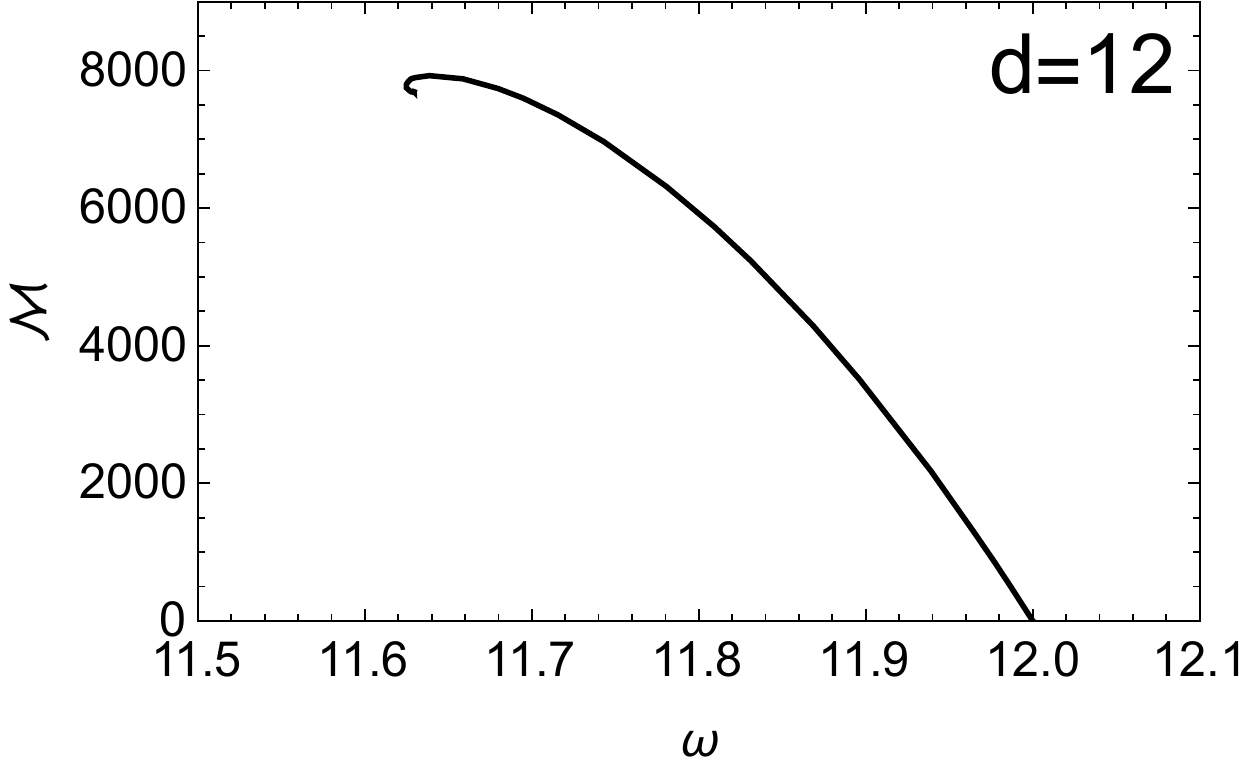}
\end{subfigure}
\\
\begin{subfigure}{0.45\textwidth}
\includegraphics[width=\textwidth]{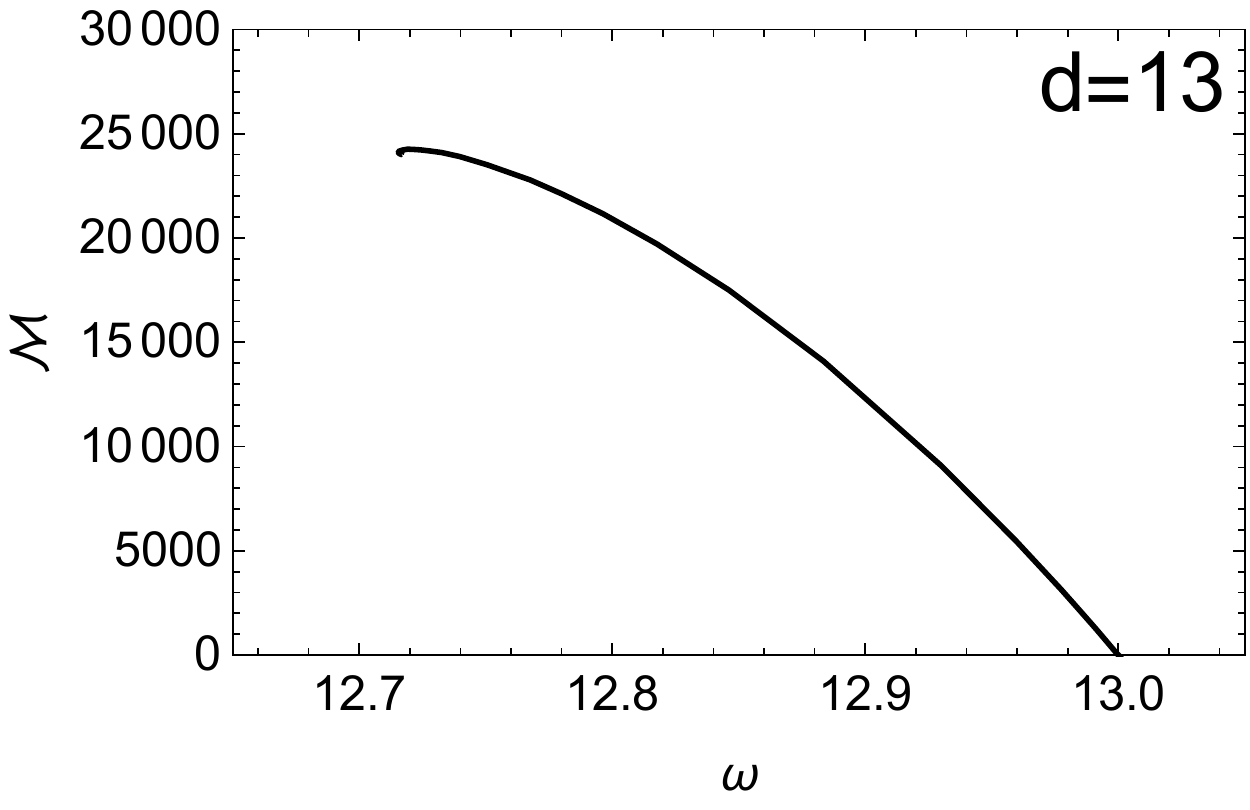}
\end{subfigure}
~
\begin{subfigure}{0.45\textwidth}
\includegraphics[width=\textwidth]{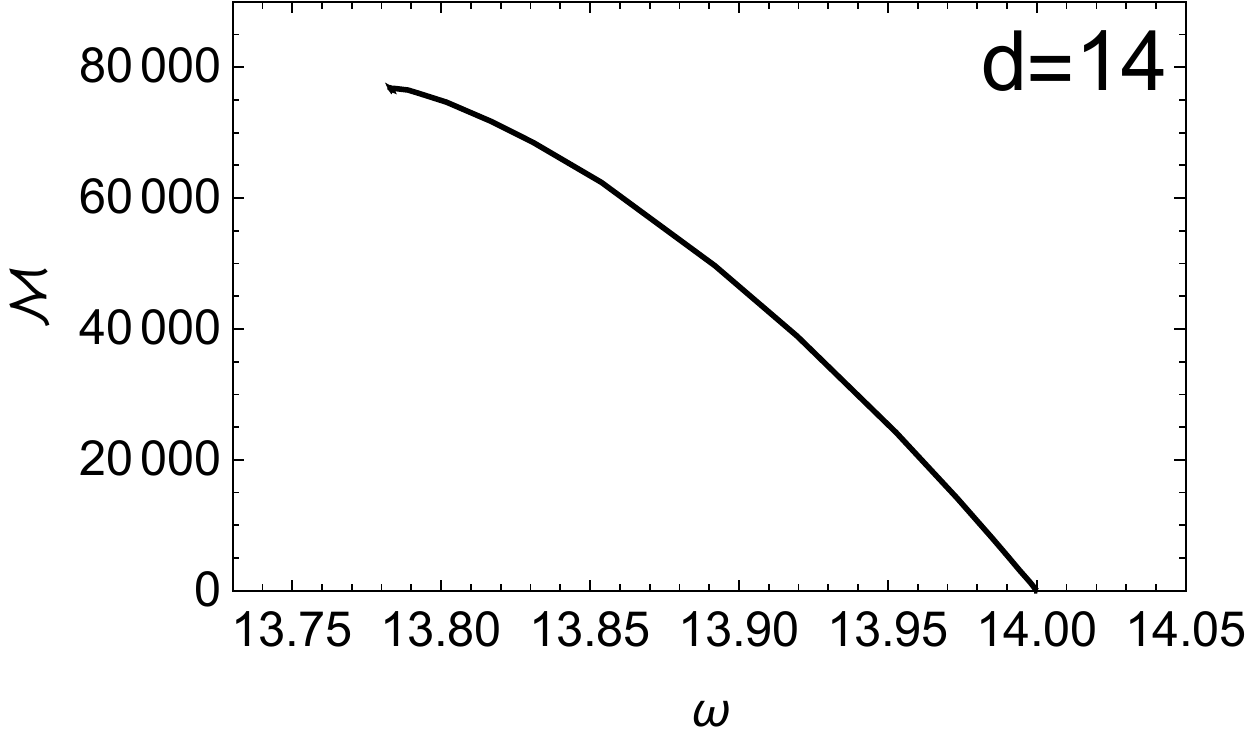}
\end{subfigure}
\\
\begin{subfigure}{0.45\textwidth}
\includegraphics[width=\textwidth]{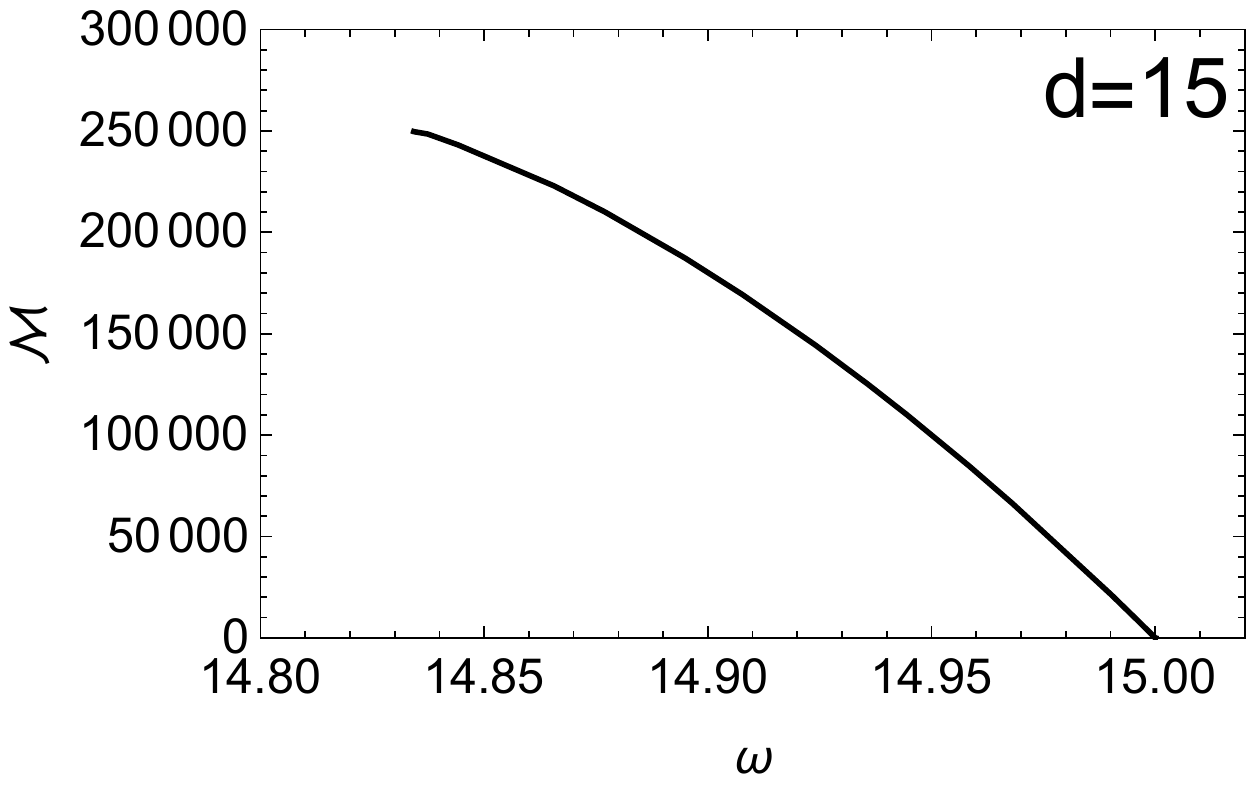}
\end{subfigure}
~
\begin{subfigure}{0.45\textwidth}
\includegraphics[width=\textwidth]{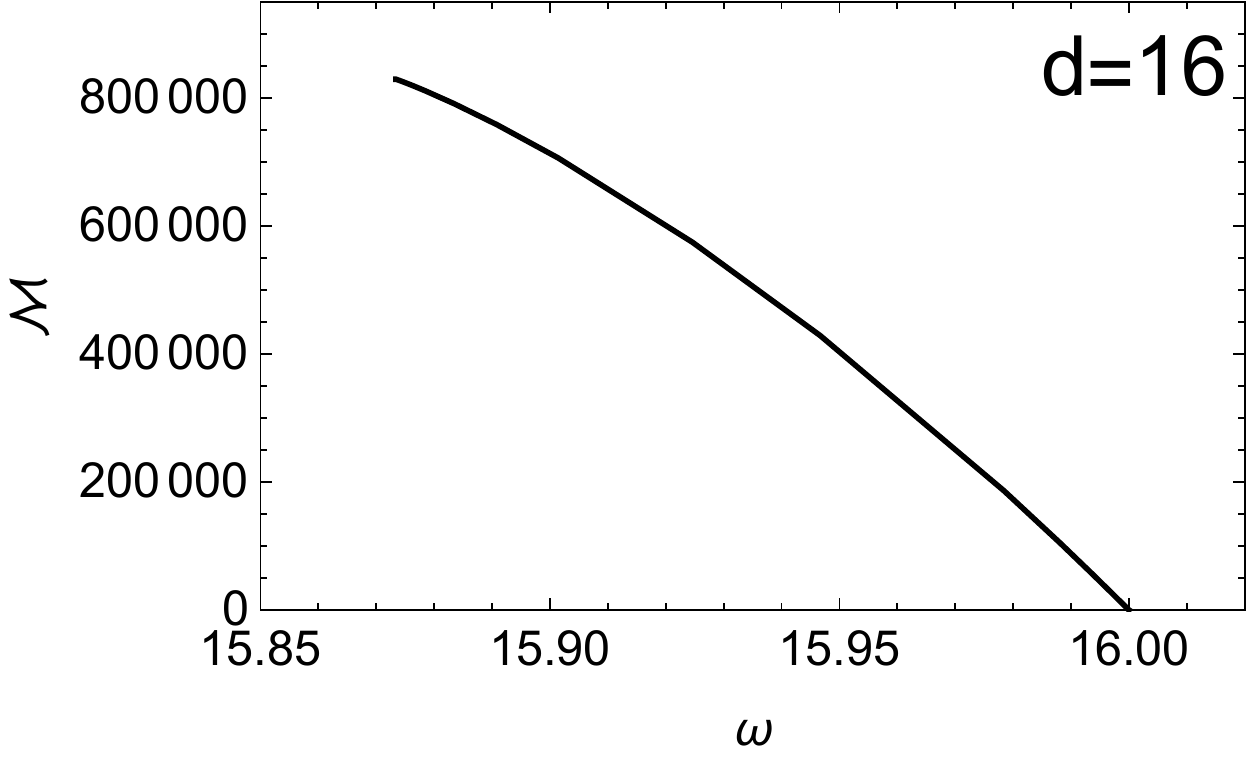}
\end{subfigure}
\\
\begin{subfigure}{0.45\textwidth}
\includegraphics[width=\textwidth]{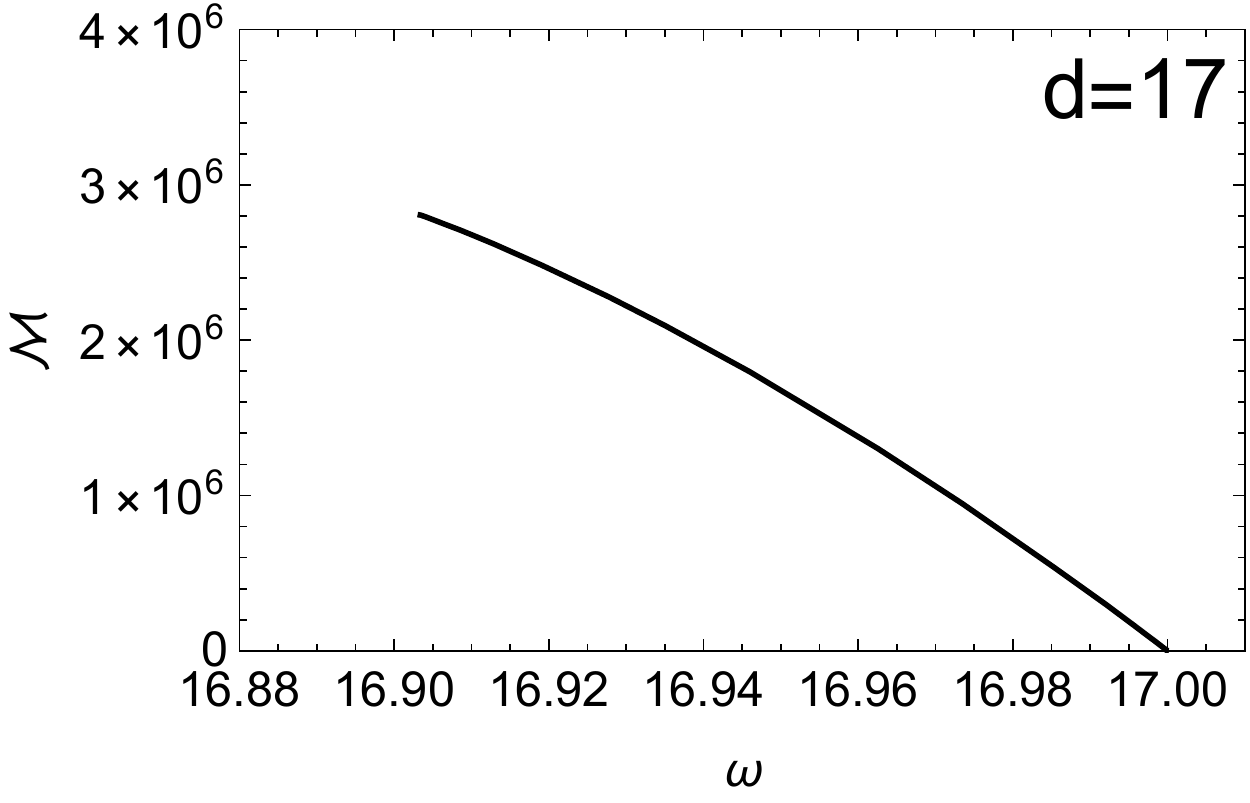}
\end{subfigure}
~
\begin{subfigure}{0.45\textwidth}
\includegraphics[width=\textwidth]{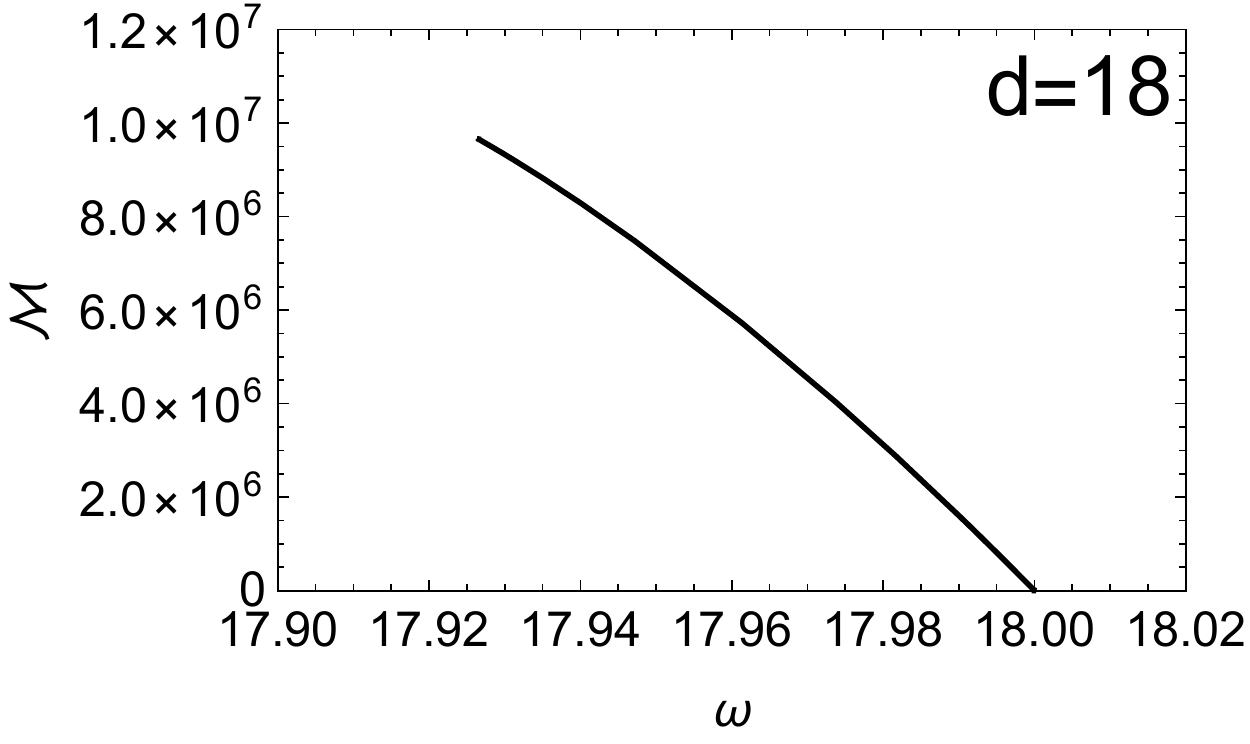}
\end{subfigure}
\caption{Plots of $\mathcal{M} (\omega)$ for the ground states of SNH in various dimensions.}
\label{fig:massSNH}
\end{figure}

One can immediately observe that for the positive solution $u$ it holds $L_- u=0$. Hence, $u$ is an eigenfunction to zero eigenvalue meaning $L_-$ is non-negative and one of the conditions is satisfied. For $L_+$ such a simple analysis will not work, however, one can differentiate Eq.\ (\ref{eqn:4NLS}) over parameter $b$ to show that $L_+\,\partial_b u=\omega'(b) u$. This means that the extrema of $\omega(b)$ correspond to $L_+$ having an  eigenvalue equal to zero. This is just a sufficient condition, but it lets us expect that the stability properties of the system change together with the change of the behaviour of $\omega(b)$. However, in general one still needs to check numerically whether the conditions on $L_+$ and $\mathcal{M}'(\omega)<0$ hold. The explicit calculations are possible only in special cases, such as when $b$ is small so one can use the results of the bifurcation theory, as we now show.

We impose an additional restriction on the nonlinearity $F(\psi)$: let it be cubic in $\psi$ (this holds for SNH and GP systems). Then the calculations made in Section \ref{sec:smalllargeb} apply and also for every $\mu>0$ it is $G(\mu u)=\mu^2 G(u)$. Equations (\ref{eqn:bifursnh}) together with the condition $u(0)=b$ give us general expressions for $u_0$ and $\omega_0$ for small $b$:
\begin{align}\label{eqn:bifurgeneral}
    u_0(r)=b\, \frac{e_0(r)}{e_0(0)}+\mathcal{O}(b^2),\qquad \omega_0(b)=\Omega_0-\frac{b^2}{a_0 e_0(0)^2}+\mathcal{O}(b^3).
\end{align}
Thanks to them, we may write operators $L_\pm$ as
\begin{align*}
    L_-&=-\Delta+V-\Omega_0+\frac{b^2}{e_0(0)^2}\left(\frac{1}{a_0}- G(e_0)\right)+\mathcal{O}(b^3),\\
    L_+&=-\Delta+V-\Omega_0+\frac{b^2}{e_0(0)^2}\left(\frac{1}{a_0}-G(e_0)- e_0 \, G_u(e_0)\right)+\mathcal{O}(b^3).
\end{align*}
These operators can be decomposed into $L_\pm=H+b^2 U_\pm/e_0(0)^2+\mathcal{O}(b^3)$, where
\begin{align*}\label{eqn:HUU}
    H=-\Delta+V-\Omega_0, \qquad
    U_- = \frac{1}{a_0}- G(e_0),\qquad
    U_+ = \frac{1}{a_0}-G(e_0)-  e_0 \,G_u(e_0).
\end{align*}
Hence, for small $b$ we can treat $L_\pm$ as perturbations of the operator $H$ and study the eigenvalues of these operators using the perturbation theory. For potentials $V$ such as the trapping potential there is no degeneracy and these eigenvalues are simply given by
\begin{align*}\label{eqn:5lambda0}
\lambda_{k,\pm}=(\Omega_k-\Omega_0)+\frac{b^2}{e_0(0)^2}\langle e_k, U_\pm e_k \rangle +\mathcal{O}(b^3).
\end{align*}
Regarding the function $\mathcal{M}(\omega)$, obviously $\mathcal{M}(\omega(b))=b^2/e_0(0)^2+\mathcal{O}(b^3)$. Then Eq.\ (\ref{eqn:bifurgeneral}) leads to
\begin{align*}\label{eqn:5mass}
	\mathcal{M}'(\omega)=\frac{d\mathcal{M}(b)}{db}\left(\frac{d\omega(b)}{db}\right)^{-1}=-a_0+\mathcal{O}(b).
\end{align*}
The definition of $a_0$ suggests that in the focusing case it is positive, hence $\mathcal{M}'(\omega)$ is negative for small $b$.

Now we can use these formulas to investigate the stability of small ground states of SNH system (\ref{eqn:SNHt}). Then $e_n$ and $\Omega_n$ are given by Eq.\ (\ref{eqn:freqlin}). We also introduce
\begin{equation}\label{eqn:apSSNH}
	S_{klmn}=\int_0^\infty \int_0^\infty \frac{e_k(r)e_l(r)e_m(s)e_n(s)}{\max\{r,s\}^{d-2}} r^{d-1} s^{d-1}\,dr \, ds,
\end{equation}
agreeing with the previously defined $S_{nnnn}$, so the terms coming from the nonlinearity can be written as
\begin{align*}
	\langle e_k, G(e_0)\, e_k \rangle =\frac{1}{d-2} S_{kk00},\qquad \langle e_k, G_u(e_0)\, e_k \rangle =\frac{2}{d-2} S_{k0k0}.
\end{align*}
All of these lead to 
\begin{subequations}
\begin{align*}
    \lambda_{k,-} &= 4k+\frac{b^2}{2(d-2)}\Gamma\left(\frac{d}{2}\right) \left(S_{0000}-S_{kk00}\right)+\mathcal{O}(b^3)\\
    \lambda_{k,+} &= 4k+\frac{b^2}{2(d-2)}\Gamma\left(\frac{d}{2}\right) \left(S_{0000}-S_{kk00}-2S_{k0k0}\right)+\mathcal{O}(b^3).
\end{align*}
\end{subequations}
In principle, one could try to simplify these expressions using the fact that
\begin{align*}
    S_{k0k0}=\frac{\Gamma\left(\frac{d}{2}+2k-1\right)}{(d-2)k!\, 2^{\frac{d}{2}+2k}\Gamma\left(\frac{d}{2}\right)\Gamma\left(\frac{d}{2}+k\right)},
\end{align*}
however we believe there is no similar closed formula for $S_{kk00}$, so we leave $S_{klmn}$ coefficients implicitly. Recall that $S_{0000}=2^{1-d/2}/\Gamma(d/2)$, so 
\begin{align*}
    \lambda_{0,-}=\mathcal{O}(b^3)\, \qquad \lambda_{0,+}=-\frac{b^2}{2^{d/2-1}(d-2)}+\mathcal{O}(b^3).
\end{align*}
The fact that $\lambda_{0,-}$ is zero up to the examined order is in agreement with previous considerations regarding the positivity of $L_-$. On the other hand, for small values of $b$ we have $\lambda_{0,+}<0$, while $\lambda_{1,+}>0$ (as it bifurcates from $\Omega_1-\Omega_0=4$), so $L_+$ has a single negative eigenvalue. In the end, we check that indeed
\begin{align*}
	\mathcal{M}'(\omega)=-2^{d/2}(d-2)\,\Gamma\left(\frac{d}{2}\right)+\mathcal{O}(b)<0
\end{align*}
for small values of $b$. In conclusion, thanks to the Vakhitov-Kolokolov criterion, there exists such interval $(0,b_0)$ that for central field values $b\in(0,b_0)$, the corresponding unique ground state is spectrally stable. 

\begin{figure}
\centering
\includegraphics[width=0.3\textwidth]{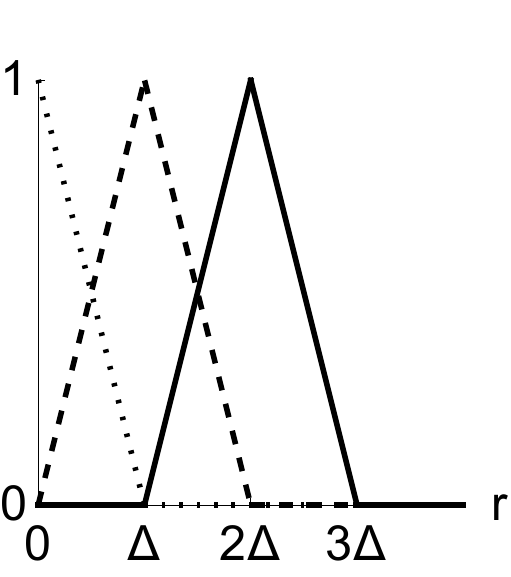}
\caption{Three first hat functions: $\chi_0$ (dotted), $\chi_1$ (dashed), and $\chi_2$ (solid).}
\label{fig:hatfunctions}
\end{figure}

For larger values of $b$ we calculate $\lambda_\pm$ and $\mathcal{M}$ numerically. Probably the simplest way of doing it is by discretization with the use of the hat functions. Since the stationary solutions we found converge to zero with $r\to\infty$ rather quickly, let us truncate their domain to some interval $[0,R]$. Now we can divide it into $N$ smaller intervals of length $\Delta=R/N$. Then we define the family of $N+1$ hat functions $\chi_n$ such that $\chi_n$ is a triangular function with a support $[(n-1)\Delta,(n+1)\Delta]$ and a maximum at $n\Delta$ (see Fig.\ \ref{fig:hatfunctions}). Finally, for any stationary state $u$ one can calculate the matrix elements $\langle \chi_n L_\pm \chi_m \rangle$. The matrices obtained this way can be then diagonalised with respect to the matrix $\langle \chi_n,\chi_m\rangle$, giving us the approximate eigenvalues of $L_\pm$ operators. Let us point out, that the biggest numerical cost in this procedure is introduced by the term $u\, G_u(u)$ in $L_+$ since all other terms give tri-diagonal matrices in the hat function basis.

\begin{figure}
\centering
\includegraphics[width=0.45\textwidth]{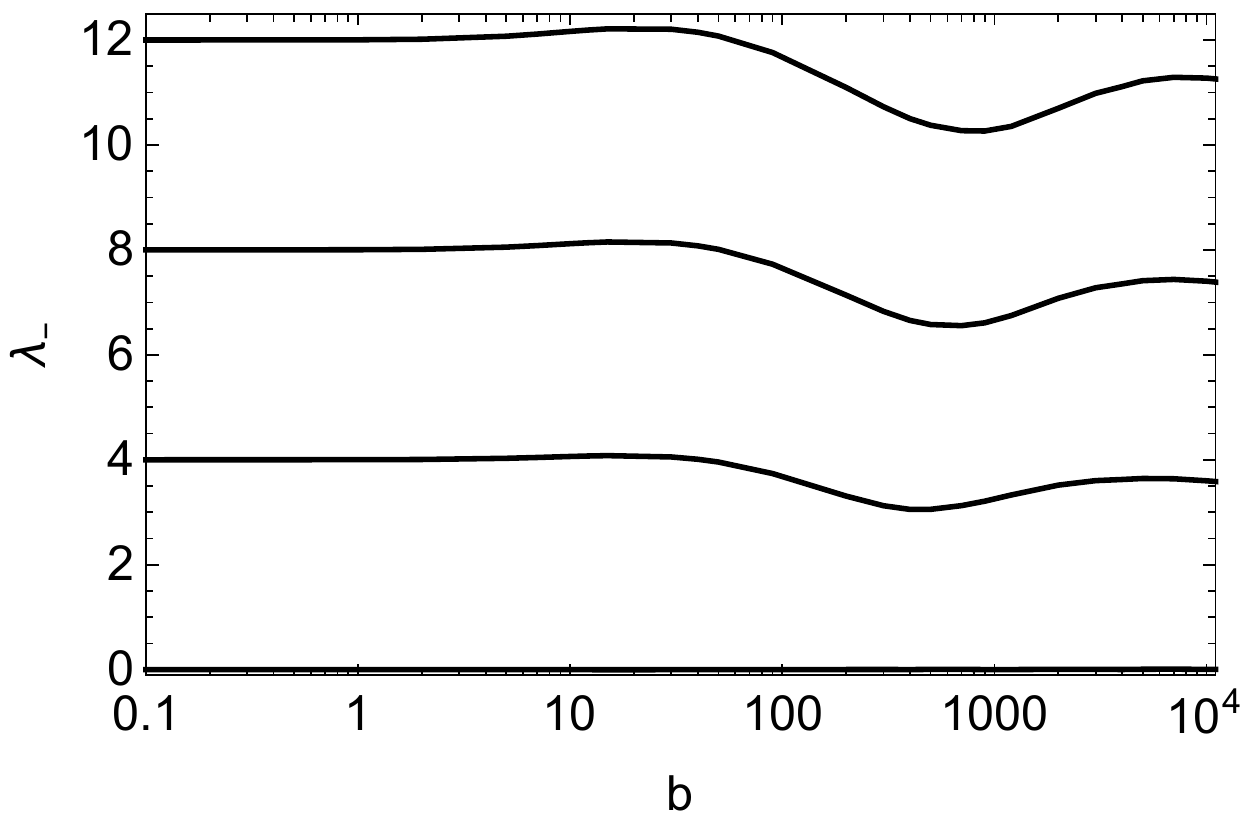}\,
\includegraphics[width=0.45\textwidth]{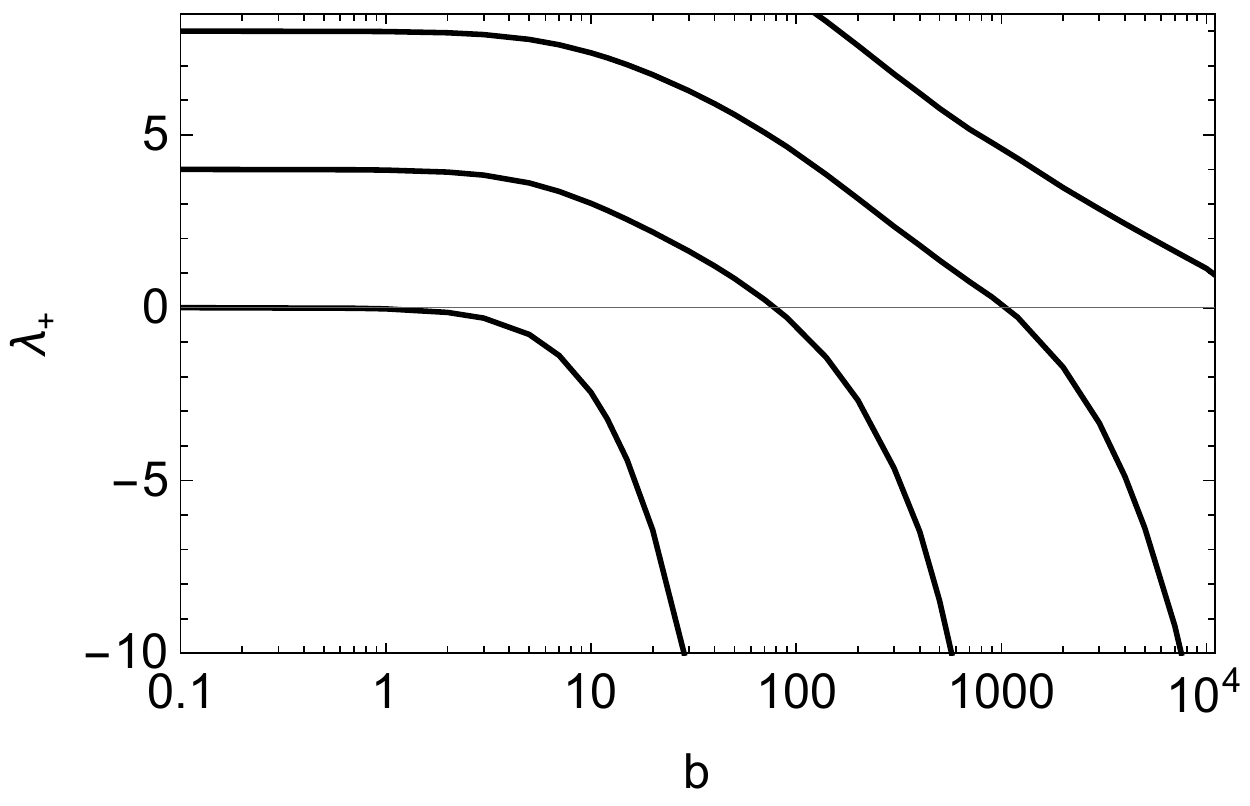}\\
\includegraphics[width=0.45\textwidth]{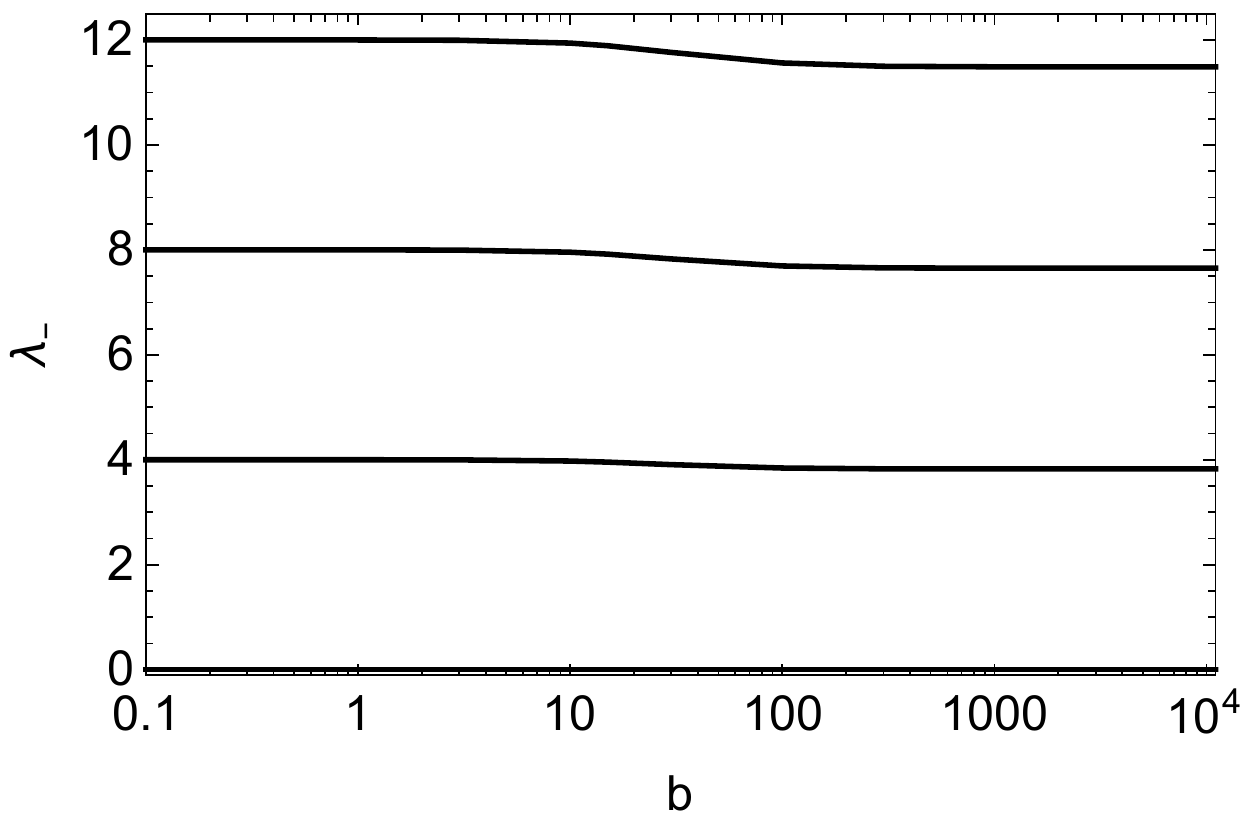}
\includegraphics[width=0.45\textwidth]{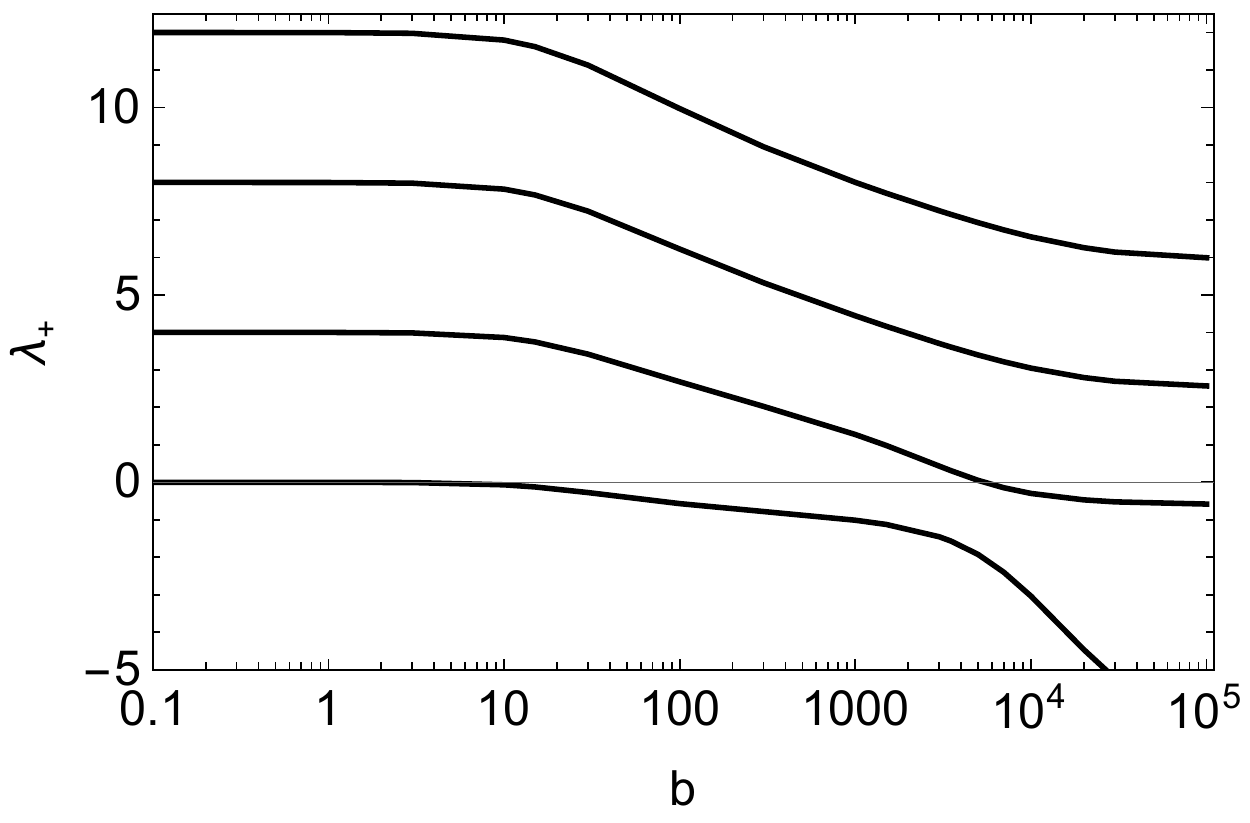}\\
\includegraphics[width=0.45\textwidth]{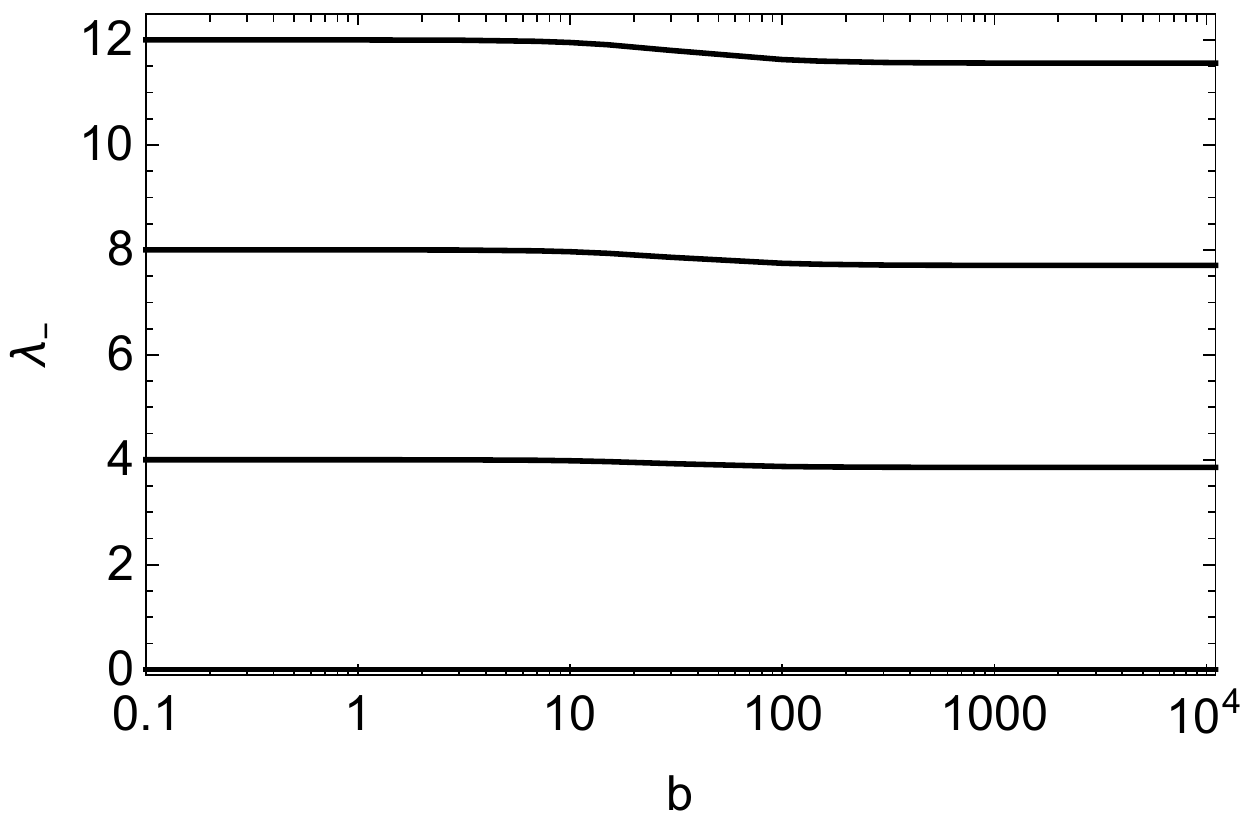}
\includegraphics[width=0.45\textwidth]{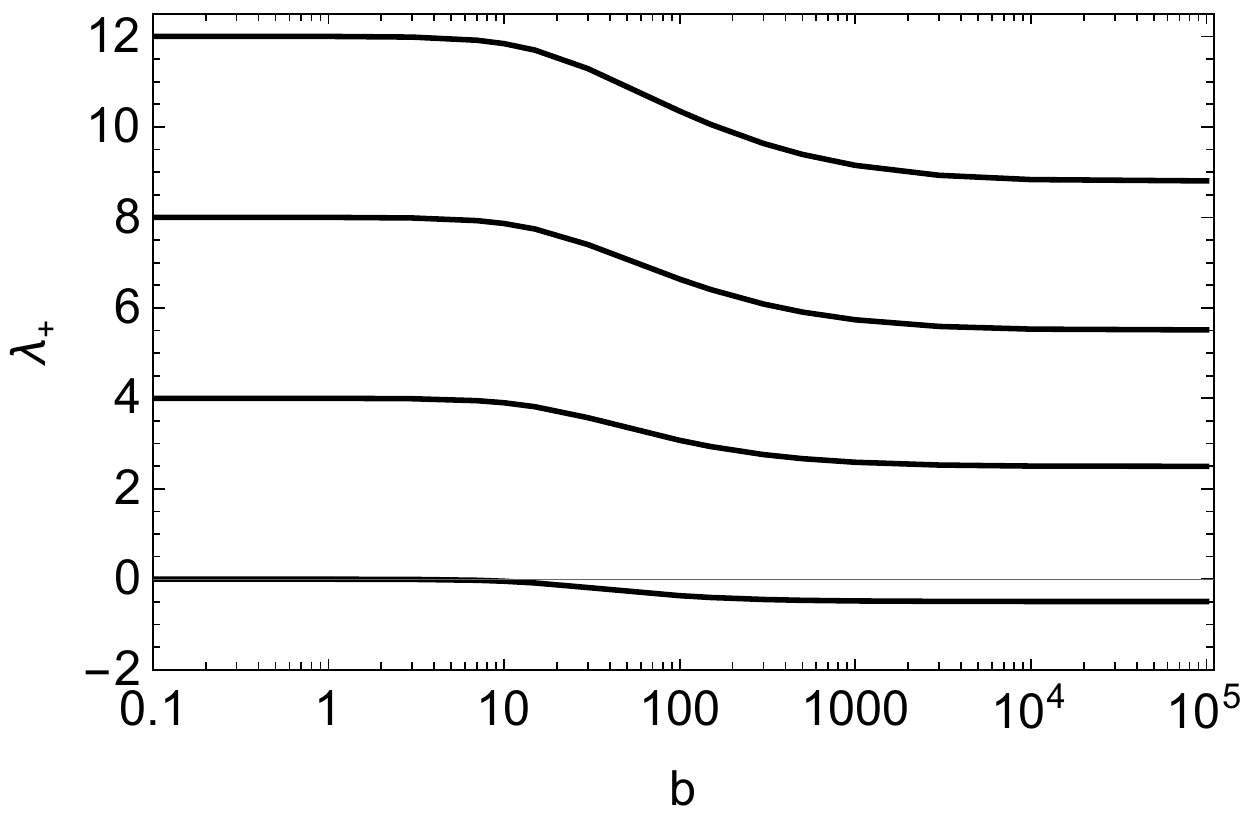}
\caption{Eigenvalues of the operators $L_-$ and $L_+$ for ground states of SNH in $d=7$ (top row) $d=15$ (medium row) and $d=16$ (bottom row).}
\label{fig:LpmSNH}
\end{figure}

\begin{figure}
\centering
\includegraphics[width=0.47\textwidth]{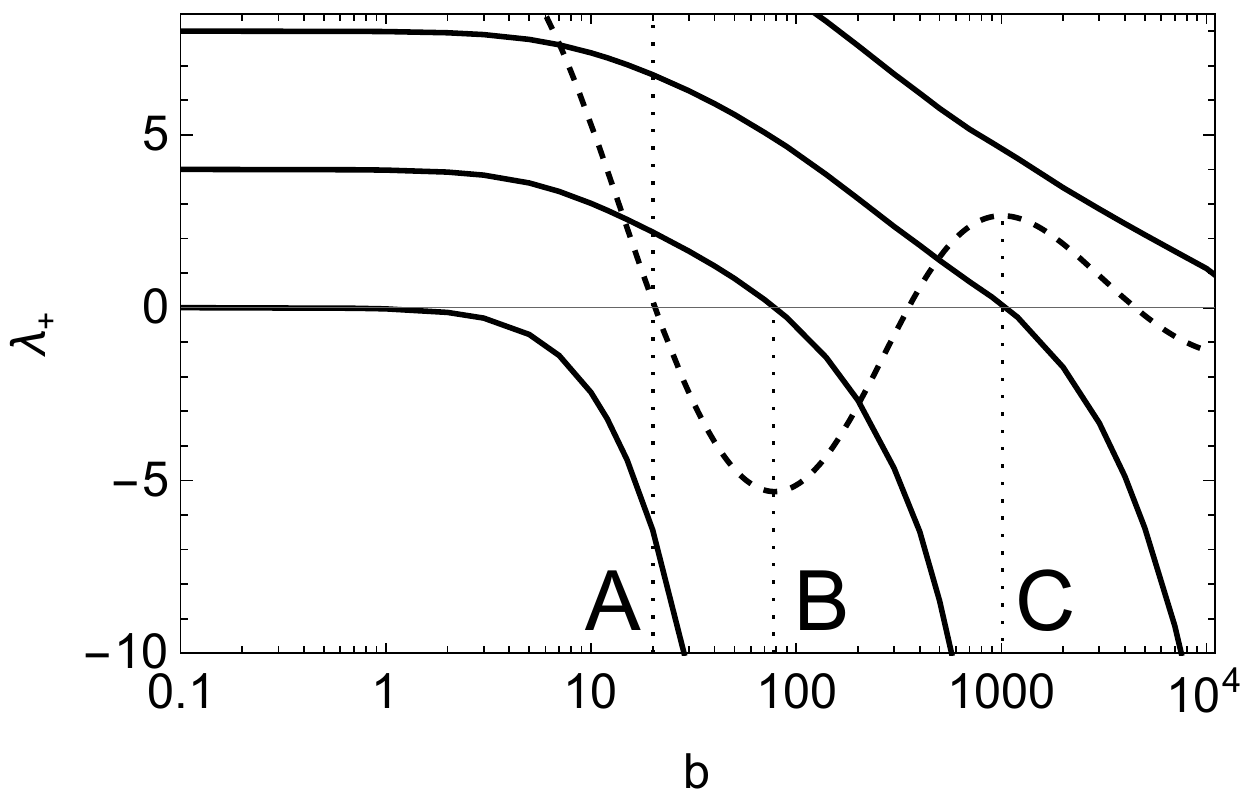}\,
\includegraphics[width=0.47\textwidth]{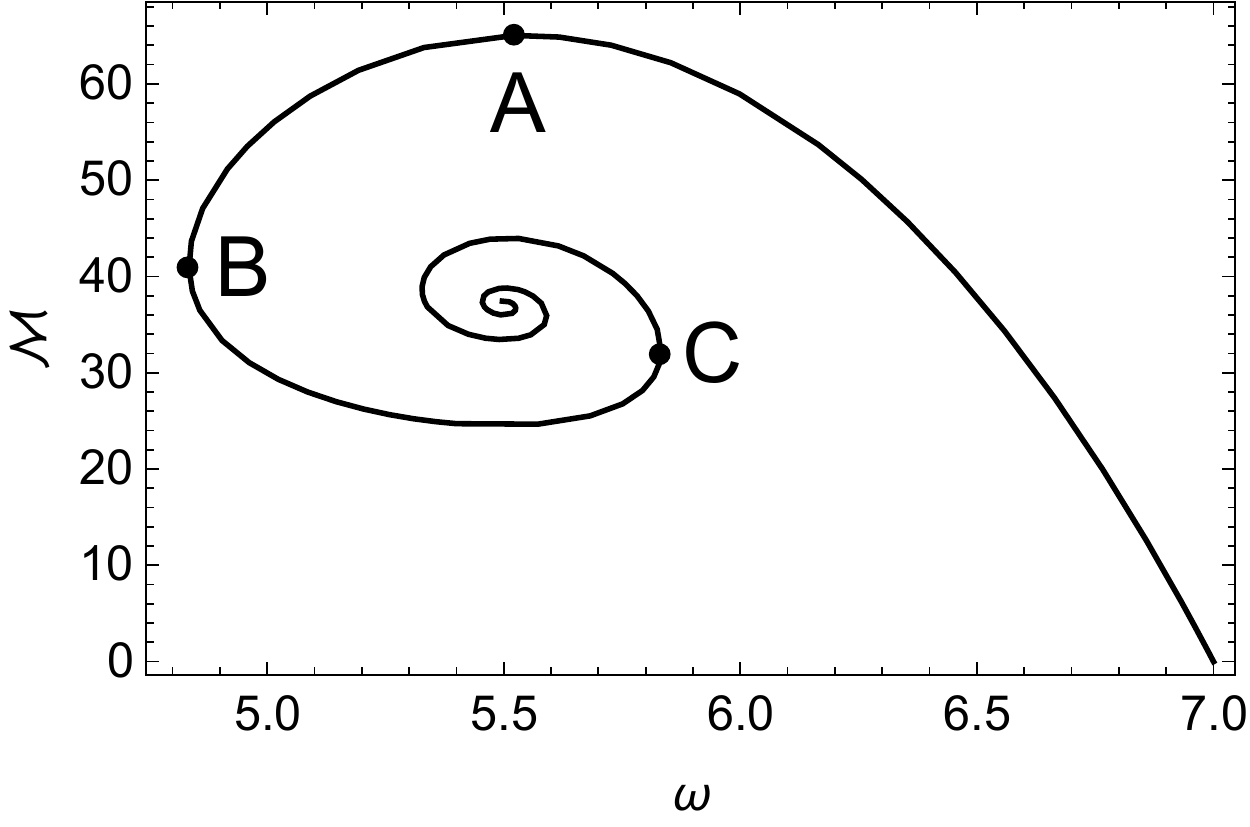}
\caption{Eigenvalues of the operator $L_+$ together with the plot of $\omega(b)$ (solid and dashed lines, respectively, in left figure) and plot of $\mathcal{M}(\omega)$ (right figure) for ground states of SNH in $d=7$. By A we mark the loss of stability of the ground state, while B and C denotes places where the next eigenvalue of $L_+$ becomes negative.}
\label{fig:stabABCD}
\end{figure}

We present eigenvalues obtained with this method in Fig.\ \ref{fig:LpmSNH}. Together with the results shown in Fig. \ref{fig:massSNH} they let us conclude that in $7\leq d\leq 15$, where $\omega(b)$ is an oscillating function, the stability is lost at the first maximum of $\mathcal{M}(\omega)$. Even though eventually $\mathcal{M}'(\omega)<0$ again, in the mean time an additional eigenvalue of $L_+$ irreversibly loses its positivity (see Fig.\ \ref{fig:stabABCD}) so the solution remains unstable. In $d\geq 16$ there is no such effect and the ground states are stable indefinitely.

\vspace{\breakFF}
\begin{adjustwidth}{\marwidFF}{\marwidFF}
\small\qquad
For GP system one can apply the same framework, although this time the coefficient $1/(d-2)$ is absent, so one gets $\langle e_k, G(e_0)\, e_k \rangle=S_{kk00}$ and $\langle e_k, G(e_0)\, e_k \rangle=S_{k0k0}$ where
\begin{equation}\label{eqn:SGP}
	S_{klmn}=\int_0^\infty e_k(r)e_l(r)e_m(r)e_n(r)\,  r^{d-1} \,dr.
\end{equation}
This expression is much simpler than in the case of SNH system, as it is completely symmetric in its indices and in some cases it can be evaluated explicitly, for example
\begin{equation*}
	S_{nn00}=\frac{\Gamma\left(\frac{d}{2}+2n\right)}{n!\, 2^{\frac{d}{2}+2n-1}\Gamma\left(\frac{d}{2}\right)\Gamma\left(\frac{d}{2}+n\right)}.
\end{equation*}
Hence, one obtains
\begin{subequations}
\begin{align*}
    \lambda_{k,-} &= 4k+\frac{b^2}{2^{\frac{d}{2}}}\left(1-\frac{\Gamma\left(\frac{d}{2}+2k\right)}{4^k\, k!\, \Gamma\left(\frac{d}{2}+k\right)}\right)+\mathcal{O}(b^3)\\
    \lambda_{k,+} &= 4k+\frac{b^2}{2^{\frac{d}{2}}}\left(1-\frac{3\,\Gamma\left(\frac{d}{2}+2k\right)}{4^k\, k!\, \Gamma\left(\frac{d}{2}+k\right)}\right)+\mathcal{O}(b^3).
\end{align*}
\end{subequations}
Also
\begin{align*}
	\mathcal{M}'(\omega)=-2^{d/2}\Gamma\left(\frac{d}{2}\right)+\mathcal{O}(b),
\end{align*}
so the ground states are spectrally stable for small values of $b$. Numerical tests for larger values of $b$ give a similar conclusion as for SNH system: the ground states are stable for small $b$, but in dimensions where $\omega(b)$ is oscillating ($5\leq d\leq 12$) at some point they lose stability, while in $d\geq 13$ they stay stable indefinitely (this case was recently investigated in a more rigorous way in \cite{PelIP}).

For SNH and GP systems the interesting effects, such as the change of the profile of $\omega(b)$ and the difference in the stability for large $b$, happen in higher dimensions. Hence, one could ask whether there exists a similar system for which such changes take place between dimensions 2 and 3 and can be observed experimentally. Unfortunately, it seems that such an effect is impossible there (at least based on the same mechanism), as we will try to argue now.

We began our analysis of the large $b$ behaviour with the investigation of the Lane-Emden equation
\begin{equation*}
    U''+\frac{d-1}{r}U'+U^p=0.
\end{equation*}
This equation with $p=2$ was an approximation for the behaviour of SNH equation with large $b$ (and $p=3$ would give such approximation for GP equation). Similarly, by introducing $\rho=b^a r$ and $U(\rho)=u(r)/b$ with some appropriate $a$ one should be able to take the limit $b\to\infty$ and obtain the same equation with some exponent $p>1$ for other NLS equations of the form Eq.\ (\ref{eqn:4NLS}). Then performing the Emden-Fowler transformation yields an autonomous system that can be linearized around its nontrivial fixed point, giving an equation
\begin{equation}\label{eqn:LameEmden2}
    \ddot{\nu}+\left(d-2-\frac{4}{p-1}\right)\dot{\nu}+2\left(d-2-\frac{2}{p-1}\right)\nu=0.
\end{equation}
We are interested in the situation in which the eigenvalues of this equation change from real to complex and vice versa, hence we examine the discriminant of its characteristic equation being
\begin{multline*}
    \left(d-\frac{4}{p-1}-2\right)^2-8 \left(d-\frac{2}{p-1}-2\right)\\
    =d^2-\left(12+\frac{8}{p-1}\right)d+4\frac{5p^2-2p+1}{(p-1)^2}.
\end{multline*}
This expression is zero if $d$ and $p$ satisfy one of these two conditions (plotted in Fig.\ \ref{fig:LaneEmden})
\begin{equation*}
    d = 2 \left(3+\frac{2}{p-1}\pm 2\sqrt{\frac{p}{p-1}}\right).
\end{equation*}
Larger of these numbers corresponds to the gain of stability observed in SNH and GP systems. In the limit $p\to\infty$ it converges to $d=10$, hence these effects cannot be observed in lower dimensions. For $d\geq 11$ the corresponding value of $p$ is called the Joseph-Lundgren exponent \cite{Jos73, Dol07, Far07}.
\begin{figure}
\centering
\captionsetup{font=small, width=.78\linewidth}
\includegraphics[width=0.5\textwidth]{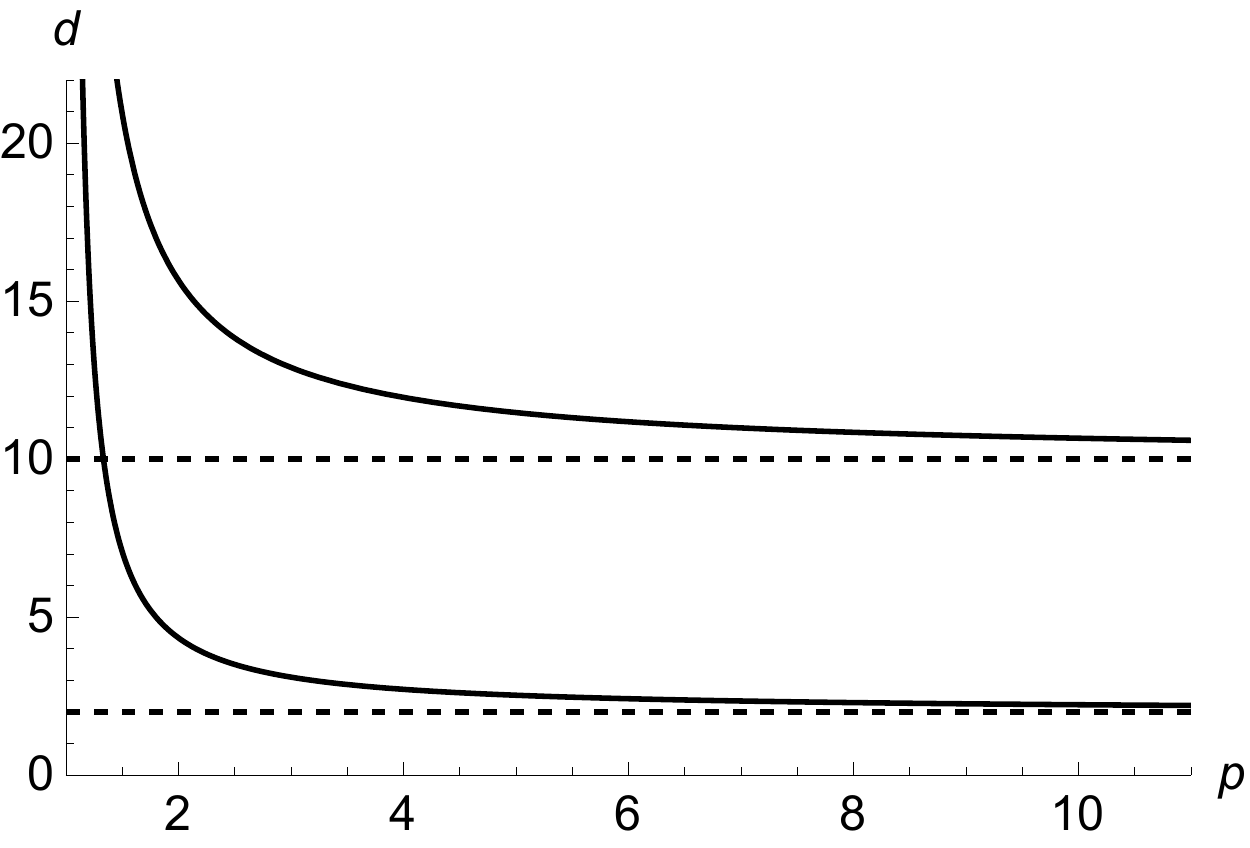}
\caption{Dimensions $d$ in which discriminant of the characteristic equation for Eq.\ (\ref{eqn:LameEmden2}) is zero. The horizontal dashed lines indicate the limits as $p\to\infty$.}
\label{fig:LaneEmden}
\end{figure}
\end{adjustwidth}\vspace{\breakFF}

\subsection{Excited states}
For excited states there is no tool similar to Vakhitov-Kolokolov criterion, so one has to investigate the stability by studying the spectral properties of $\mathcal{L}$. In general, it can be done numerically, but for small values of $b$ one can again perform an analysis based on the bifurcation results from Section \ref{sec:smalllargeb}.

Let us focus on the $n$-th excited state (with $n=0$ being the ground state). Then we may introduce operators analogous to the ones defined for the ground state:
\begin{align*}\label{eqn:HUU}
    H=-\Delta+V-\Omega_n, \quad
    U_- = \frac{1}{a_n}- G(e_n),\quad
    U_+ = \frac{1}{a_n}-G(e_n)-  e_n \,G_u(e_n).
\end{align*}
They can be used to decompose the operator $\mathcal{L}$. By discarding all terms of higher order in $b$ than quadratic we get:
\begin{align*}
    \mathcal{L}= \left(\begin{array}{cc}0&L_-\\-L_+ & 0\end{array}\right)\approx\left(\begin{array}{cc}0&H\\-H & 0\end{array}\right)+\frac{b^2}{e_n(0)^2} \left(\begin{array}{cc}0&U_-\\-U_+ & 0\end{array}\right)=:\mathcal{H}+b^2\mathcal{U}.
\end{align*}
We want to treat this problem using the perturbation method, with $\mathcal{U}$ being the perturbation of $\mathcal{H}$. However, $\mathcal{H}$ and $\mathcal{U}$ are not self-adjoint operators, hence we must be careful during the implementation. Since it holds $H e_k=(\Omega_k-\Omega_n)e_k$, the eigenvectors of $\mathcal{H}$ are given by
\begin{align*}\label{eqn:5epsilonstate}
    |\varepsilon_{k,\pm}^{(0)}\rangle = \frac{1}{\sqrt{2}} \left(\begin{array}{c}e_k\\\pm i e_k\end{array}\right),
\end{align*}
and they satisfy $\mathcal{H}|\varepsilon_{k,\pm}^{(0)}\rangle=\pm i(\Omega_k-\Omega_n)|\varepsilon_{k,\pm}^{(0)}\rangle $. Thus, the eigenvalues of $\mathcal{L}$ at $b=0$ are $\lambda^{(0)}_{k,\pm}=\pm i(\Omega_k-\Omega_n)$. Our goal is to get corrections of the order $b^2$ to these values. We consider an eigenproblem $\mathcal{L}|\varepsilon_{k,\pm}\rangle = \lambda |\varepsilon_{k,\pm}\rangle$ and expand the objects present there in formal series in the perturbation parameter $b^2$: $\lambda_{k,\pm}=\lambda_{k,\pm}^{(0)}+b^2\lambda_{k,\pm}^{(1)}+...$ and $|\varepsilon_{k,\pm}\rangle=|\varepsilon_{k,\pm}^{(0)}\rangle+b^2 |\varepsilon_{k,\pm}^{(1)}\rangle+...$ with the zeroth orders already known. Then at the first order in $b^2$ our eigenproblem is
\begin{align*}
    \mathcal{H}|\varepsilon_{k,\pm}^{(1)}\rangle + \mathcal{U}|\varepsilon_{k,\pm}^{(0)}\rangle =  \lambda_{k,\pm}^{(0)}|\varepsilon_{k,\pm}^{(1)}\rangle +  \lambda_{k,\pm}^{(1)}|\varepsilon_{k,\pm}^{(0)}\rangle.
\end{align*}
This equation can be contracted from the left side with $\langle\varepsilon_{k,\pm}^{(0)}|$ giving
\begin{align}\label{eqn:5perturbderiv}
    \langle\varepsilon_{k,\pm}^{(0)}|\mathcal{H}|\varepsilon_{k,\pm}^{(1)}\rangle + \langle\varepsilon_{k,\pm}^{(0)}|\mathcal{U}|\varepsilon_{k,\pm}^{(0)}\rangle =  \lambda_{k,\pm}^{(0)}\langle\varepsilon_{k,\pm}^{(0)}|\varepsilon_{k,\pm}^{(1)}\rangle +  \lambda_{k,\pm}^{(1)}.
\end{align}
If $\mathcal{H}$ was self-adjoint, the first terms on both sides of this equality would be the same and we would get a simple expression for $\lambda_{k,\pm}^{(1)}$, as in the standard perturbation theory. Performing an explicit calculation shows us, however, that even though $\mathcal{H}$ is not self-adjoint, its special structure also leads to $\langle\varepsilon_{k,\pm}^{(0)}|\mathcal{H}|\varepsilon_{k,\pm}^{(1)}\rangle=\lambda_{k,\pm}^{(0)}\langle\varepsilon_{k,\pm}^{(0)}|\varepsilon_{k,\pm}^{(1)}\rangle$. To see that, let us write $|\varepsilon_{k,\pm}^{(1)}\rangle$ as $\binom{\alpha}{\beta}$ for a moment. Then
\begin{align*}
    \langle\varepsilon_{k,\pm}^{(0)}|\mathcal{H}|\varepsilon_{k,\pm}^{(1)}\rangle&= \int \frac{1}{\sqrt{2}}\left(\begin{array}{c}e_k\\ \pm i e_k\end{array}\right)^\dagger \left(\begin{array}{cc}0&H\\-H & 0\end{array}\right)\left(\begin{array}{c}\alpha \\ \beta \end{array}\right)\\
    &=\int \frac{1}{\sqrt{2}} (e_k H \beta \pm i  e_k H \alpha)=\int \frac{1}{\sqrt{2}} (\beta H e_k \pm i \alpha H e_k)\\
    &= (\Omega_k-\Omega_n) \int\frac{1}{\sqrt{2}}(\beta e_k\pm i \alpha e_k)\\
    &= \pm i (\Omega_k-\Omega_n) \int \frac{1}{\sqrt{2}} ( e_k \alpha \mp i e_k \beta)\\
    &=\pm i (\Omega_k-\Omega_n) \int \frac{1}{\sqrt{2}} \left(\begin{array}{c}e_k\\ \pm i e_k\end{array}\right)^\dagger \left(\begin{array}{c}\alpha\\ \beta\end{array}\right)=\lambda_{k,\pm}^{(0)}\langle\varepsilon_{k,\pm}^{(0)}|\varepsilon_{k,\pm}^{(1)}\rangle
\end{align*}
For brevity, we have slightly simplified the notation in this derivation: all integrations should be understood as $\int f=\int_0^\infty f(r) r^{d-1}\,dr$. As a result, Eq.\ (\ref{eqn:5perturbderiv}) can be reduced to
\begin{align}\label{eqn:5eUe}
    \lambda_{k,\pm}^{(1)}&= \langle\varepsilon_{k,\pm}^{(0)}|\mathcal{U}|\varepsilon_{k,\pm}^{(0)}\rangle\nonumber \\
    &=\int \frac{1}{2e_n(0)^2}\left(\begin{array}{c}e_k\\\pm i e_k\end{array}\right)^\dagger \left(\begin{array}{cc}0&U_-\\-U_+ & 0\end{array}\right)\left(\begin{array}{c}e_k\\\pm i e_k\end{array}\right)\nonumber\\
    &=\pm \frac{i}{2e_n(0)^2} \left(\langle e_k , U_- e_k\rangle+\langle e_k, U_+ e_k\rangle\right).
\end{align}

The presented derivation applies only to the nondegenerate cases, when all eigenvalues of $\mathcal{H}$ are simple. Even though it is true for the considered operators $H$, it is possible for $\mathcal{H}$ to have eigenvalues of higher multiplicity, as it may happen that there exist $k\neq l$ such that $(\Omega_k-\Omega_n)=-(\Omega_l-\Omega_n)$. Whether it is the case or not depends on the potential $V$. Since the following calculations depend heavily on the choice of $V$, we fix it here to be the harmonic potential $V=r^2$. Then $\Omega_k=d+4k$ and $\lambda_{k,\pm}^{(0)}=\pm 4 i (k-n)$. In such case, for a ground state both vectors $|\varepsilon_{0,+}^{(0)}\rangle$ and $|\varepsilon_{0,-}^{(0)}\rangle$ are eigenvectors of $\mathcal{H}$ to zero. For $n=1$, $|\varepsilon_{1,+}^{(0)}\rangle$ and $|\varepsilon_{1,-}^{(0)}\rangle$ are both eigenvectors to zero, but additionally $|\varepsilon_{2,+}^{(0)}\rangle$ and $|\varepsilon_{0,-}^{(0)}\rangle$ are eigenvectors to $4i$ (and their complex conjugates are eigenvectors to $-4i$). Going further, we see that as $n$ increases, the operator $\mathcal{H}$ has more and more double eigenvalues. For a more general description of this problem, let us assume that $\lambda^{(0)}$ is a double eigenvalue of $\mathcal{H}$ with $|\epsilon^{(0)}_{k,+}\rangle$ and $|\epsilon^{(0)}_{l,-}\rangle$ as eigenvectors. As in standard degenerate perturbation theory, we are looking for a combination of these two vectors adjusted to our perturbation $\mathcal{U}$:
\begin{align*}
    |\psi^{(0)}\rangle=c_+ |\epsilon^{(0)}_{k,+}\rangle+c_- |\epsilon^{(0)}_{l,-}\rangle.
\end{align*}
This leads to the system of equations for $c_+$ and $c_-$ \cite{Lan}:
\begin{align*}
    \lambda^{(1)} c_+ &= c_+ \langle \epsilon^{(0)}_{k,+} |\mathcal{U}| \epsilon^{(0)}_{k,+} \rangle + c_- \langle \epsilon^{(0)}_{k,+} |\mathcal{U}| \epsilon^{(0)}_{l,-} \rangle\\
    \lambda^{(1)} c_- &= c_+ \langle \epsilon^{(0)}_{l,-} |\mathcal{U}| \epsilon^{(0)}_{k,+} \rangle + c_- \langle \epsilon^{(0)}_{l,-} |\mathcal{U}| \epsilon^{(0)}_{l,-} \rangle.
\end{align*}
This linear system has nontrivial solutions only if
\begin{align}\label{eqn:lambdadouble}
&\lambda^{(1)}_\pm=\frac{1}{2}\Biggl[\langle \epsilon^{(0)}_{k,+} |\mathcal{U}| \epsilon^{(0)}_{k,+} \rangle + \langle \epsilon^{(0)}_{l,-} |\mathcal{U}| \epsilon^{(0)}_{l,-}\rangle\nonumber\\
&\pm \sqrt{\left(\langle \epsilon^{(0)}_{k,+} |\mathcal{U}| \epsilon^{(0)}_{k,+} \rangle - \langle \epsilon^{(0)}_{l,-} |\mathcal{U}| \epsilon^{(0)}_{l,-} \rangle\right)^2+4 \langle \epsilon^{(0)}_{l,-} |\mathcal{U}| \epsilon^{(0)}_{k,+} \rangle \langle \epsilon^{(0)}_{k,+} |\mathcal{U}| \epsilon^{(0)}_{l,-} \rangle }\Biggr].
\end{align}
Hence, we obtained the formulas for $\lambda_\pm^{(1)}$. Expressions for the symmetric matrix elements $\langle\varepsilon_{k,\pm}^{(0)}|\mathcal{U}|\varepsilon_{k,\pm}^{(0)}\rangle$ are given in Eq.\ (\ref{eqn:5eUe}), while for the mixed elements the analogous calculations lead to
\begin{align}\label{eqn:5epUem}
\langle\varepsilon_{k,+}^{(0)}|\mathcal{U}|\varepsilon_{l,-}^{(0)}\rangle=\frac{i}{2e_n(0)^2} \left(-\langle e_k , U_- e_l\rangle+\langle e_k, U_+ e_l\rangle\right).
\end{align}
To calculate $\langle\varepsilon_{l,-}^{(0)}|\mathcal{U}|\varepsilon_{k,+}^{(0)}\rangle$ it is sufficient to take a complex conjugation of the expression from Eq.\ (\ref{eqn:5epUem}).

These results may be used to give eigenvalues of $\mathcal{L}$ for $n$-th excited state of SNH system with small value of $b$. For this system it holds
\begin{align*}
    \langle e_k , U_- e_l\rangle&=\frac{1}{d-2}\left(S_{nnnn}\,\delta_{kl}-S_{klnn}\right),\\
    \langle e_k , U_+ e_l\rangle&=\frac{1}{d-2}\left(S_{nnnn}\,\delta_{kl}-S_{klnn}-2S_{knln}\right),
\end{align*}
    with $\delta_{kl}$ denoting the Kronecker delta. As the framework we developed works only for small values of $b$, where nothing interesting happens when going from $d=15$ to $d=16$, we focus here only on the case $d=7$. Then, let us check the first excited state ($n=1$). For $b=0$ it has double eigenvalues in $0$ and $4i$, as we already mentioned. The lowest eigenvalues of $\mathcal{L}$ are then equal to
\begin{align*}
    \lambda_0&=\mathcal{O}(b^3),\\
    \lambda_{1,+}&=4i+\left(\frac{\sqrt{45145}+811}{602112 \sqrt{2}}i\right)b^2+\mathcal{O}(b^3)\approx 4i+0.001202 i\, b^2+\mathcal{O}(b^3),\\
    \lambda_{1,-}&=4i+\left(\frac{\sqrt{45145}-811}{602112 \sqrt{2}}i\right)b^2+\mathcal{O}(b^3)\approx 4i-0.000703 i\, b^2+\mathcal{O}(b^3),\\
    \lambda_2&=8i-\left(\frac{43105 }{8830976 \sqrt{2}}i\right)\, b^2+\mathcal{O}(b^3)\approx 8i-0.003451i\, b^2+\mathcal{O}(b^3),\\
    \lambda_3&=12i-\left(\frac{13590411}{3673686016 \sqrt{2}}i\right) b^2+\mathcal{O}(b^3)\approx 12i-0.002616i\, b^2 +\mathcal{O}(b^3).
\end{align*}
For brevity, we did not write eigenvalues that can be obtained as conjugates or negatives of the listed ones. As one can see, all corrections of order $b^2$ are imaginary. In case of the branches starting in simple eigenvalues of $\mathcal{H}$ it is obvious, as they are driven by Eq.\ (\ref{eqn:5eUe}). However, for double eigenvalues of $\mathcal{H}$ the first-order corrections are given by Eq.\ (\ref{eqn:lambdadouble}). The presence of a square root in this expression allows both real and imaginary results. It turns out that here these corrections remain purely imaginary for the first excited state of SNH in $d=7$. Of course, it does not mean the stability of such states with small $b$, as in principle it is possible that somewhere at higher orders of the bifurcation theory there appear corrections with non-zero real part. Even though, it would mean that for small $b$ the real parts of the eigenvalues are very small, hence the initial evolution of the slightly perturbed state shall look stable and instabilities can be seen only after a sufficiently long time. 
    
Similarly, we perform calculations for the second excited state $(n=2)$. There we know that $0$, $4i$, and $8i$ are double eigenvalues for $b=0$: the first of them has eigenvectors $|\varepsilon_{2,+}^{(0)}\rangle$ and $|\varepsilon_{2,-}^{(0)}\rangle$, the second one $|\varepsilon_{3,+}^{(0)}\rangle$ and $|\varepsilon_{1,-}^{(0)}\rangle$, while the last one $|\varepsilon_{4,+}^{(0)}\rangle$ and $|\varepsilon_{0,-}^{(0)}\rangle$. Now using the derived formulas we get
\begin{align*}
    \lambda_0&=\mathcal{O}(b^3),\\
    \lambda_{1,\pm}&=4i+\left(\frac{\sqrt{180684295679}\pm 933785 i}{1907490816 \sqrt{2}}\right)b^2+\mathcal{O}(b^3)\\
    &\approx 4i+(0.000158 \pm 0.000346 i)\, b^2+\mathcal{O}(b^3),\\
    \lambda_{2,\pm}&=8i+\left(\frac{\sqrt{121105311620579831}\pm 1025827843 i}{793516179456 \sqrt{2}}\right)\, b^2+\mathcal{O}(b^3)\\
    &\approx 8i+(0.00031 \pm 0.000914 i)\, b^2+\mathcal{O}(b^3),\\
    \lambda_3&=12i+\left(\frac{888172171}{352673857536 \sqrt{2}}i\right) b^2+\mathcal{O}(b^3)\approx 12i+0.001781i\, b^2 +\mathcal{O}(b^3).
\end{align*}
In contrary to the first excited state, here two branches of eigenvalues admit real corrections at the order $b^2$. It means that such solutions are unstable for small values of $b$.

This analysis gives us hints about the stability of excited states for small values of $b$, but for larger values it is necessary to resort to numerical methods. One can do it similarly as when looking for the spectrum of $L_\pm$, by discretizing $\mathcal{L}$ with the use of hat functions. However, for this operator the precision we were able to achieve with this approach within the reasonable range of parameters was not satisfactory. Hence, we restrict the discussion of the stability of excited states of SNH to the bifurcation analysis above, leaving the numerical studies for the future publication.


\vspace{\breakFF}
\begin{adjustwidth}{\marwidFF}{\marwidFF}
\small\qquad
The same bifurcation analysis can be repeated for other NLS equations. In case of GP system it is especially easy, as it also has a harmonic potential and cubic nonlinearity. Then, the obtained formulas are very similar with the only differences laying in the interaction coefficients, now given by Eq.\ (\ref{eqn:SGP}), and the matrix elements, this time lacking the factor $1/(d-2)$:
\begin{align*}
    \langle e_k , U_- e_l\rangle=S_{nnnn}\,\delta_{kl}-S_{klnn},\\
    \langle e_k , U_+ e_l\rangle=S_{nnnn}\,\delta_{kl}-S_{klnn}-2S_{knln}.
\end{align*}
Then for small $b$ the first excited state in $d=5$ yields $\mathcal{L}$ with the eigenvalues given by
\begin{align*}
    \lambda_0&=\mathcal{O}(b^3),\\
    \lambda_{1,\pm}&=4i+\left(\frac{\pm\sqrt{388239}-1449 i}{51200 \sqrt{2}}\right)b^2+\mathcal{O}(b^3)\\
    &\approx 4i+(\pm 0.008605 - 0.020012 i)\, b^2+\mathcal{O}(b^3),\\
    \lambda_2&=8i-\left(\frac{21217}{204800 \sqrt{2}}i\right)b^2+\mathcal{O}(b^3)\approx 8i-0.073255i\, b^2+\mathcal{O}(b^3),\\
    \lambda_3&=12i-\left(\frac{599081}{6553600 \sqrt{2}}i\right)b^2+\mathcal{O}(b^3)\approx 12i-0.064638i\, b^2+\mathcal{O}(b^3),
\end{align*}
As before, further eigenvalues can be obtained by taking negative values and complex conjugations of these expressions. In contrary to SNH system in $d=7$ we can see that for the first excited state the terms of order $b^2$ in the eigenvalues have nonzero real part, meaning that for small values of $b$ the first excited state of GP equation in $d=5$ is unstable. Interestingly, for this system Eq.\ (\ref{eqn:lambdadouble}) can be evaluated to a explicit formula for $\lambda_{1,\pm}^{(1)}$ containing only $d$ (however, it is rather complicated, so we do not present it here). A quick analysis of this expression leads to the conclusion that for any natural dimension other than $5$, $\lambda_{1,\pm}^{(1)}$ is purely imaginary. For example, in $d=6$, the first excited state gives
\begin{align*}
    \lambda_0&=\mathcal{O}(b^3),\\
    \lambda_{1,\pm}&=4i-\left(\frac{1}{576} \left(15\pm \sqrt{3}\right)i\right)b^2+\mathcal{O}(b^3)\\
    &\approx 4i-(0.026042\pm 0.003007 )i\, b^2+\mathcal{O}(b^3),\\
    \lambda_2&=8i-\left(\frac{7}{96}i\right)b^2+\mathcal{O}(b^3)\approx 8i-0.072917i\, b^2+\mathcal{O}(b^3),\\
    \lambda_3&=12i-\left(\frac{107}{1536}i\right)b^2+\mathcal{O}(b^3)\approx 12i-0.069662i\, b^2+\mathcal{O}(b^3).
\end{align*}

The observation of a special character of $d=5$ can be pushed even further with the numerical results. In case of GP system, due to the lack of a nonlocal term in the nonlinearity, we were able to investigate the eigenvalues of $\mathcal{L}$ with the use of the discretization described in the discussion of SNH system. The plots are presented in Fig.\ \ref{fig:LGP1st}. The first interesting observation is the fact that there is an agreement with the bifurcation analysis of the first order: for small $b$ the first excited states of GP system are stable in $d=6$ but unstable in $d=5$. The other peculiar finding is the window of stability present for $d=5$: even though for small $b$ the bound state is unstable, there exists a range of $b$ in which it becomes stable.
\begin{figure}
\centering
\captionsetup{font=small, width=.78\linewidth}
\includegraphics[width=0.45\textwidth]{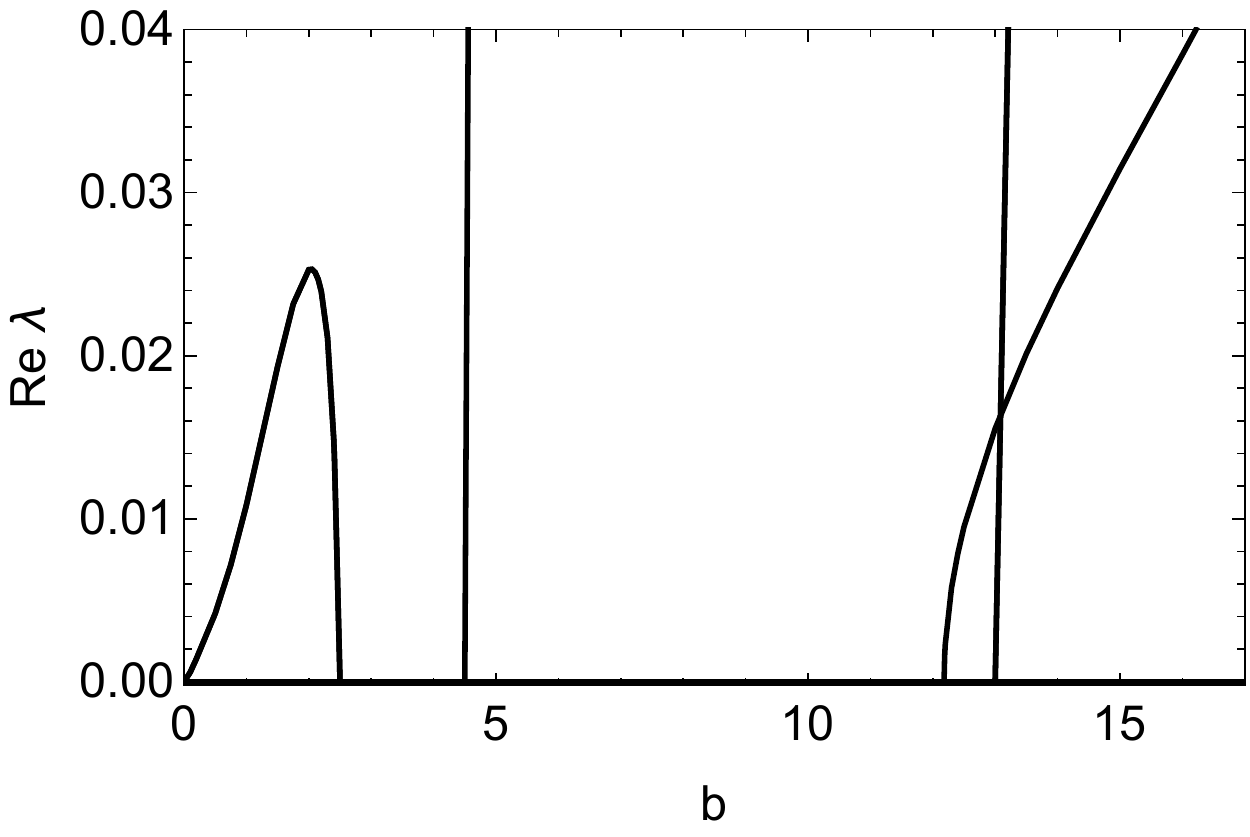}
\includegraphics[width=0.45\textwidth]{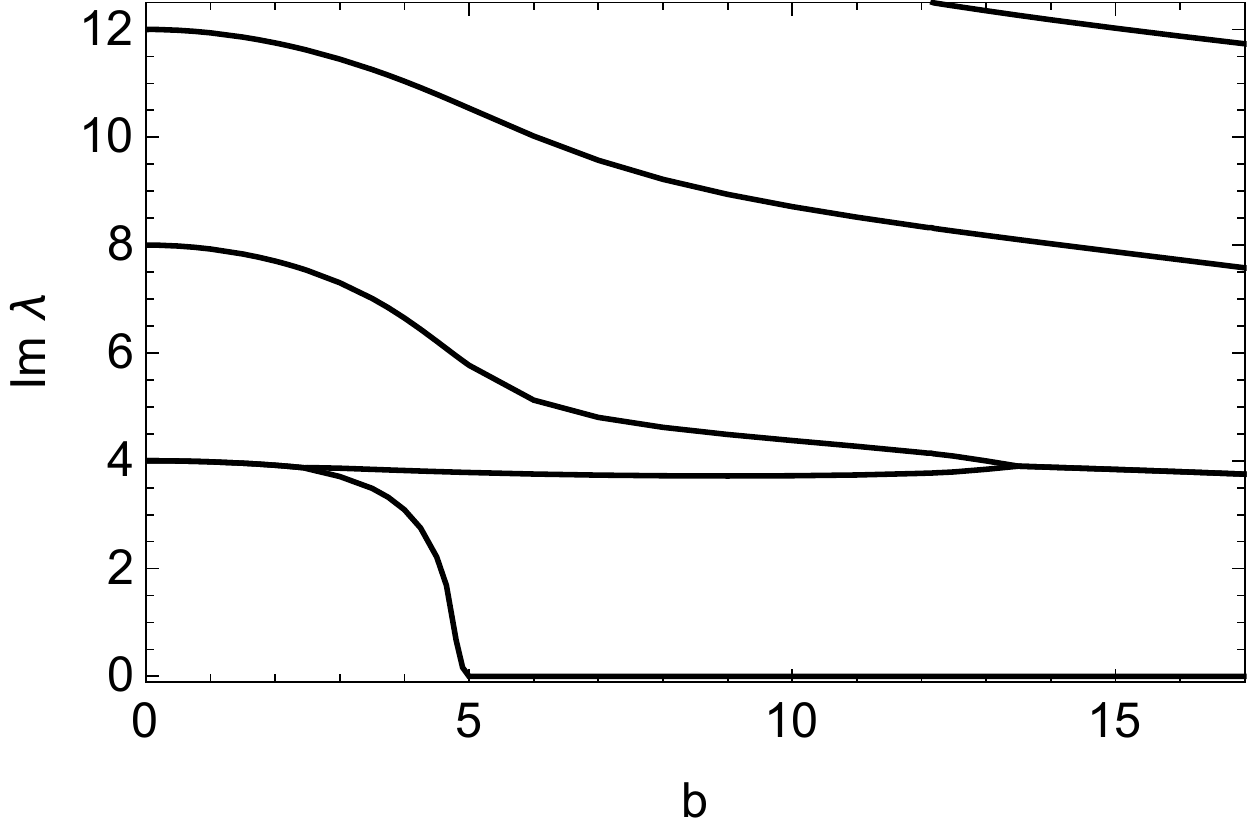}
\includegraphics[width=0.45\textwidth]{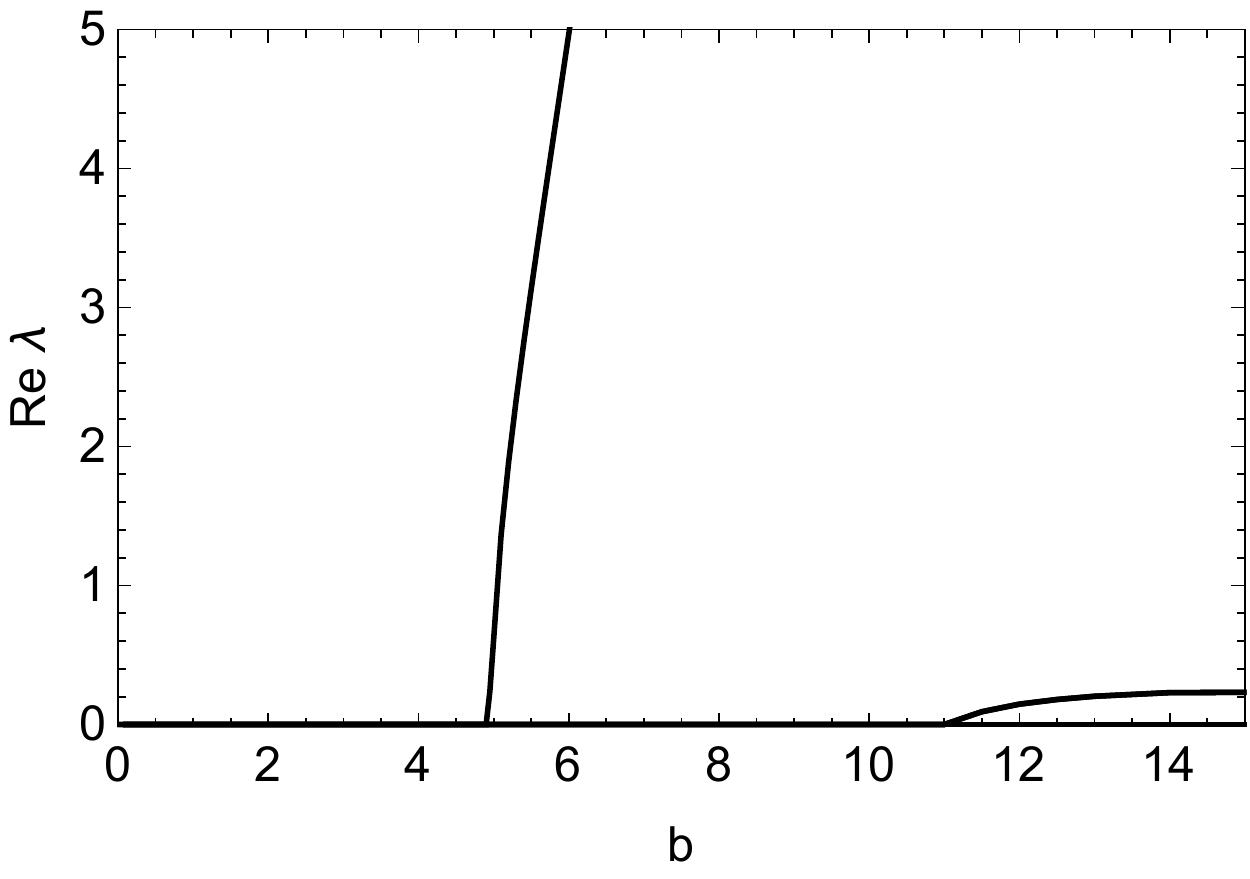}
\includegraphics[width=0.45\textwidth]{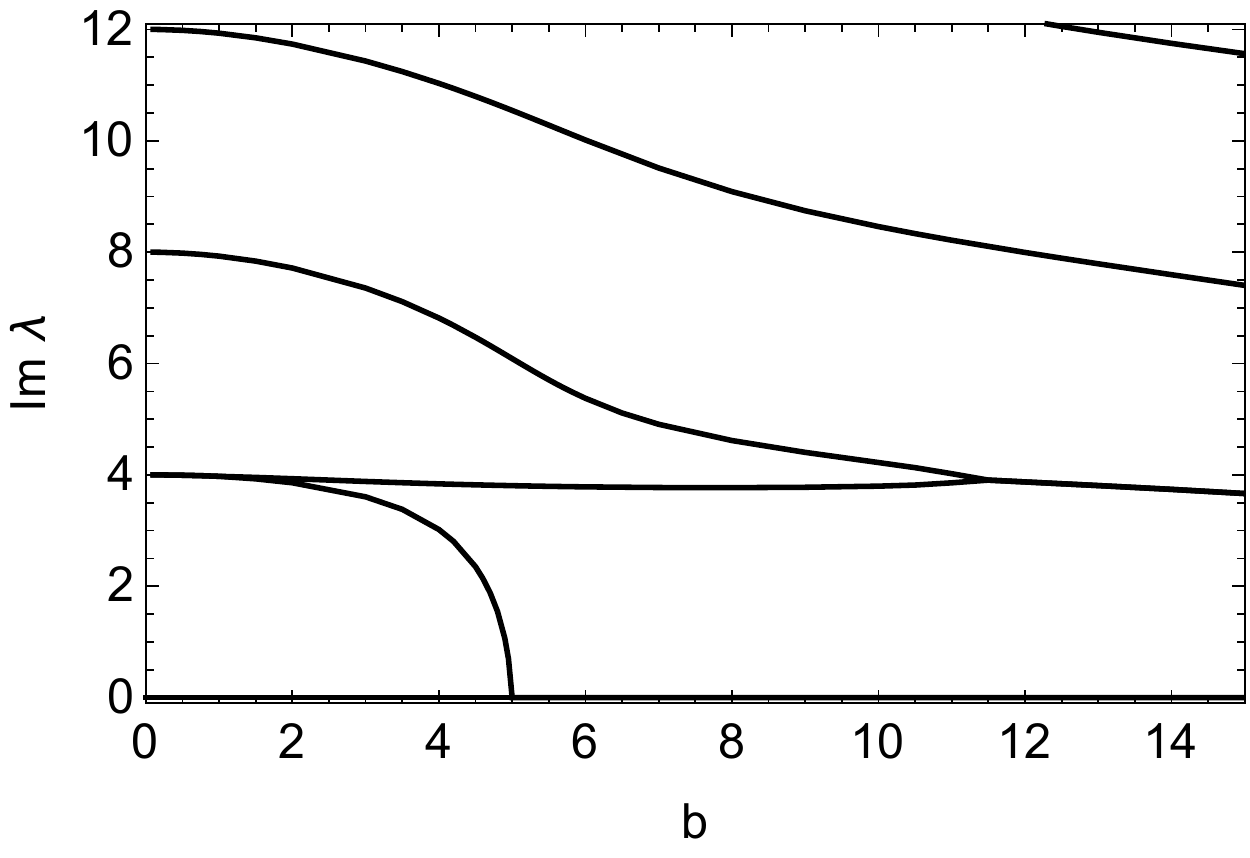}
\caption{Plots of real and imaginary parts of eigenvalues of $\mathcal{L}$ for the first excited state of GP system in $d=5$ (upper plots) and $d=6$ (lower plots).}
\label{fig:LGP1st}
\end{figure}

As before, we repeat some calculations for the second excited states. In $d=5$ the bifurcation analysis gives
\begin{align*}
    \lambda_0&=\mathcal{O}(b^3),\\
    \lambda_{1,\pm}&=4i+\left(\frac{\pm\sqrt{1315188538188839}-84322459 i}{2569011200 \sqrt{2}}\right)b^2+\mathcal{O}(b^3)\\
    &\approx 4i+(\pm 0.009982 - 0.023209 i)\, b^2+\mathcal{O}(b^3),\\
    \lambda_{2,\pm}&=8i+\left(\frac{\pm\sqrt{355202931527}-1009411 i}{80281600 \sqrt{2}}\right)b^2+\mathcal{O}(b^3)\\
    &\approx 8i+(\pm 0.005249 -0.008891 i)\, b^2+\mathcal{O}(b^3),\\
    \lambda_3&=12i-\left(\frac{734233553}{10276044800 \sqrt{2}}i\right)b^2+\mathcal{O}(b^3)\approx 12i- 0.050524 i\, b^2+\mathcal{O}(b^3).
\end{align*}
This time two branches of eigenvalues have nonzero real parts at the level of $b^2$. Similar phenomenon can be observed for $d=6$, while for $d=4$ only one of these branches has this feature. Hence, for $d=4,5,6$ one can deduce that the second excited states with sufficiently small $b$ are unstable. In other dimensions the respective results are purely imaginary giving us no resolution. In Fig.\ \ref{fig:LGP2nd} we present the numerical values of the eigenvalues of $\mathcal{L}$ for the second excited state in five-dimensional GP system. Once again we can see the presence of the window of stability.

\begin{figure}
\centering
\captionsetup{font=small, width=.78\linewidth}
\includegraphics[width=0.45\textwidth]{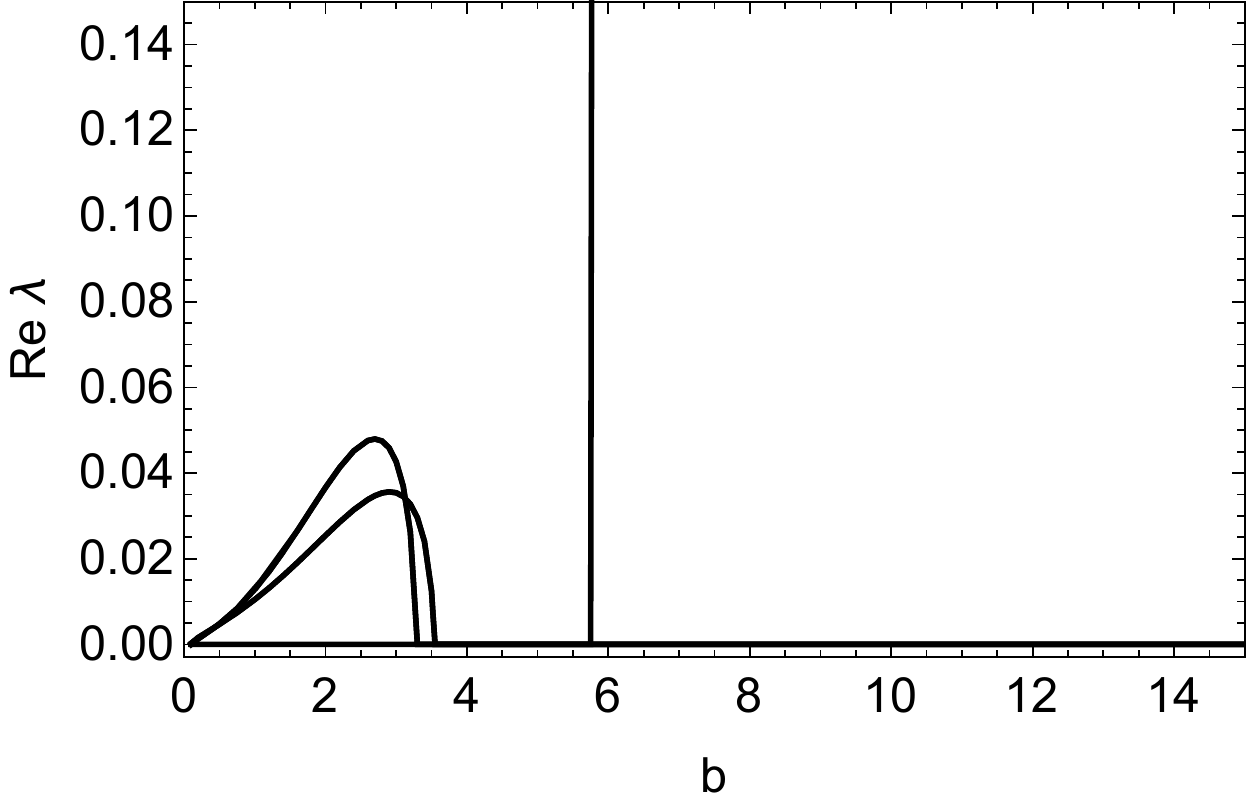}
\includegraphics[width=0.45\textwidth]{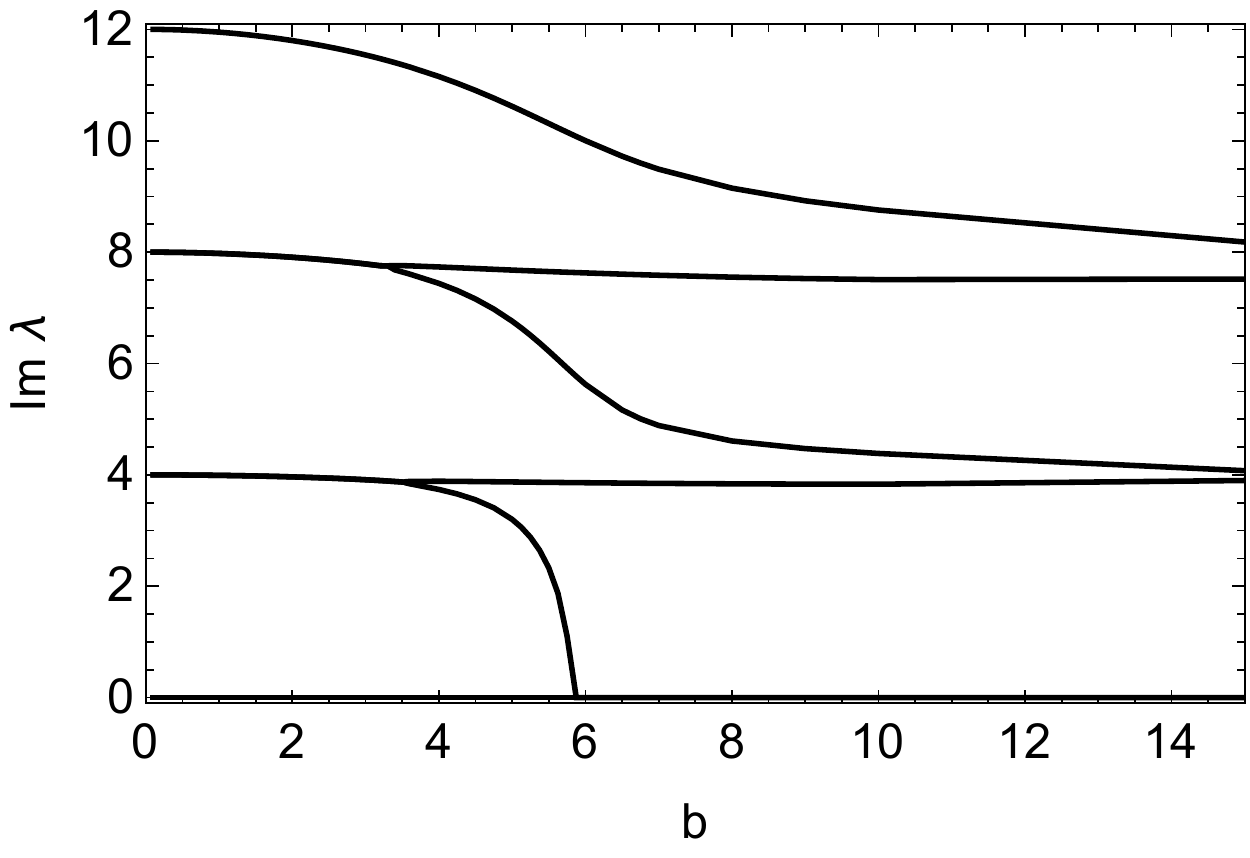}
\caption{Plots of real and imaginary parts of eigenvalues of $\mathcal{L}$ in $d=5$ for the second excited state of GP system.}
\label{fig:LGP2nd}
\end{figure}

\end{adjustwidth}\vspace{\breakFF}

\section{Evolution}\label{sec:evolution}
After the study of the stationary solutions of SNH system and their stability, the next natural step is to investigate the dynamics of this system. As before, we restrict ourselves to spherically symmetric solutions. Then Eq.\ (\ref{eqn:introSNH}) can be written as (we also rescale $\psi$, so there is no factor $(d-2)^{-1}$ in the nonlinearity)
\begin{align}\label{eqn:5SNHt}
i\,\partial_t \psi = -\partial_r^2\psi-\frac{d-1}{r}\partial_r \psi+r^2 \psi-\left(\int_0^\infty \frac{|\psi(t,s)|^2\, s^{d-1}}{\max\{r^{d-2},s^{d-2}\} }\, ds\right) \psi.
\end{align}
There are many possible ways to solve this equation numerically. Probably the most obvious one is to cut some large enough compact subset of the spatial domain $[0,R]$, and evolve this equation inside it with some iterative scheme such as the classical Runge–Kutta method. The main obstacle in such approach is the nonlocality of the nonlinear term. It can be handled directly by using the explicit numerical scheme and calculating the value of the nonlinearity at each node point, or in some other way by substituting the integral with the potential $v$ and solving at each step an auxiliary equation $\Delta v=|\psi|^2$ with a suitable method. Either way, the nature of this nonlinear term greatly increases the resources needed to solve SNH equation numerically, in comparison with local nonlinearities. 

\vspace{\breakFF}
\begin{adjustwidth}{\marwidFF}{\marwidFF}
\small\qquad
Even for the local NLS equations, such as GP equation, the described approach does not seem to be feasible. One can expect that if the investigated system shares some of the interesting behaviours described in Section \ref{sec:turbulence}, they occur at large time scales. The techniques we evoked are not very well suited for performing such long-time simulations, due to the effects present at the boundary of the considered domain.
\end{adjustwidth}\vspace{\breakFF}

\begin{figure}
\centering
\includegraphics[width=0.65\textwidth]{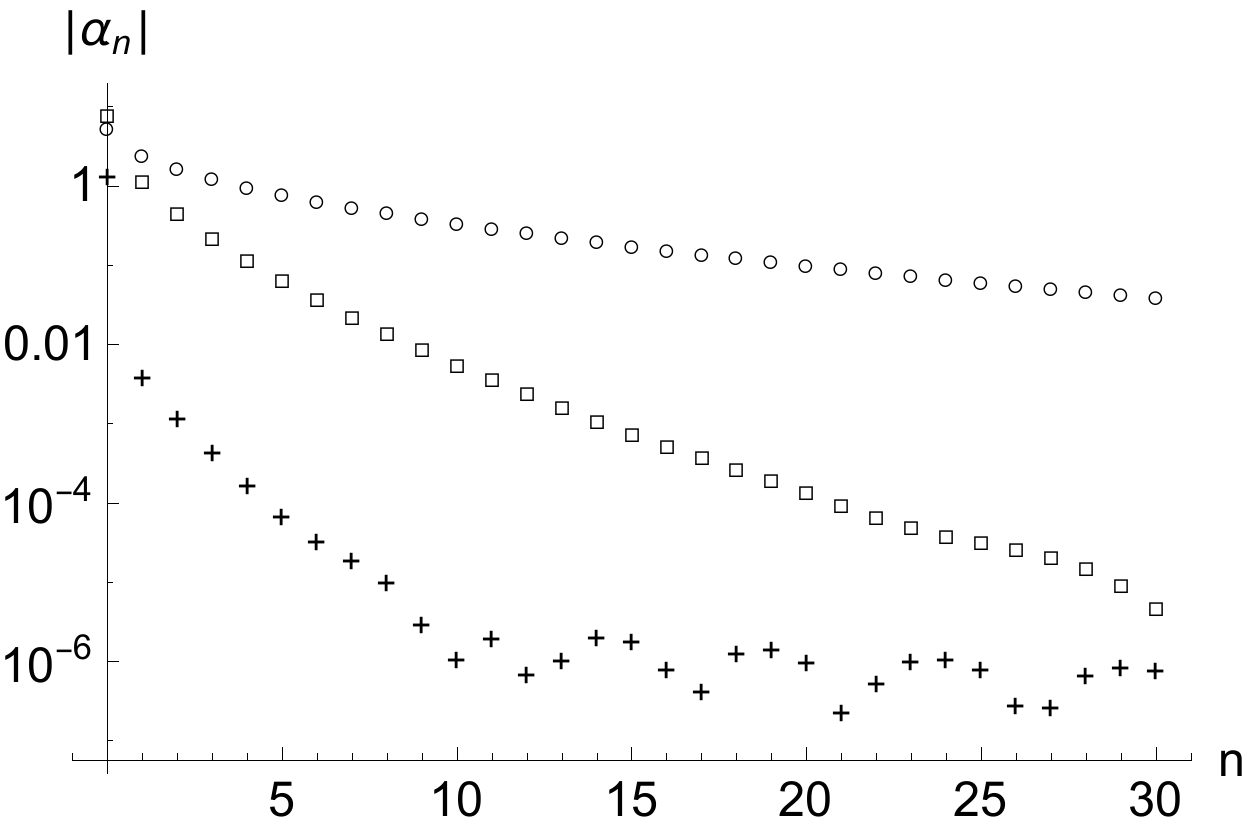}
\caption{Decomposition of the ground state of SNH in $d=7$ with $b=1$ (crosses), $b=10$ (squares), and $b=100$ (circles) into the modes $e_n$.}
\label{fig:specSNH}
\end{figure}

An alternative idea may be to use the spectral methods. Let us expand solution $\psi$ into the series
\begin{equation}\label{eqn:modes}
    \psi(t,r)=\sum_{k=0}^\infty \alpha_k(t)\, e^{-i\Omega_k t} e_k(r).
\end{equation}
Then the linear parts of Eq.\ (\ref{eqn:5SNHt}) can be written as
\begin{align*}
    i\,\partial_t \psi=\sum_{k=0}^\infty \left( i\,\dot{\alpha}_k+\Omega_k \, \alpha_k \right) e^{-i\Omega_k t} e_k,\qquad (-\Delta+V) \psi=\sum_{k=0}^\infty \Omega_k\,\alpha_k e^{-i\Omega_k t} e_k,
\end{align*}
while the nonlinear term changes into the threefold sum
\begin{multline*}
    \left(\int_0^\infty \frac{|\psi(t,s)|^2\, s^{d-1}}{\max\{r^{d-2},s^{d-2}\} }\, ds\right) \psi\\
    =\sum_{j=0}^\infty \sum_{k=0}^\infty \sum_{l=0}^\infty \bar{\alpha}_j \alpha_k \alpha_l e^{i(\Omega_j-\Omega_k-\Omega_l)t}  \left(\int_{\mathbb{R}^d} \frac{e_j(s)\,e_k(s)\, s^{d-1}}{\max\{r^{d-2},s^{d-2}\} }\, ds\right) e_l(r).
\end{multline*}
We can combine these expressions and contract them with $e_n$ to get the equation for the evolution of a single mode (using the orthogonality relation $\langle e_n, e_k \rangle= \delta_{nk}$). We have already introduced the interaction coefficients $S_{klmn}$. However, in this part we want to keep the compatibility with the notation from \cite{Biz18}. Hence, let us define (the only difference is the order of the indices)
\begin{equation}\label{eqn:SSNH}
	S_{njkl}=\int_0^\infty \int_0^\infty \frac{e_n(r)e_j(s)e_k(s)e_l(r)}{\max\{r,s\}^{d-2}} r^{d-1} s^{d-1}\,dr \, ds.
\end{equation}
In the end, we get an infinite system of ordinary differential equations equivalent to Eq.\ $(\ref{eqn:5SNHt})$ that can be written as:
\begin{align}\label{eqn:5spec}
    i\,\partial_t \alpha_n=\sum_{j=0}^\infty \sum_{k=0}^\infty \sum_{l=0}^\infty S_{njkl}\bar{\alpha}_j \alpha_k \alpha_l e^{i(\Omega_n+\Omega_j-\Omega_k-\Omega_l)t}.
\end{align}
This result gives us another approach to the numerical evolution of SNH system, we can decompose the initial data into the modes $e_n$ as in Eq.\ (\ref{eqn:modes}), truncate this sum at some suitable $N$, and then evolve the coefficients $\alpha_0, ..., \alpha_N$ using the truncated Eq.\ (\ref{eqn:5spec}). This method seems to be better suited to our problem, provided that the coefficients $\alpha_n$ decay with $n$ sufficiently fast to justify truncation at some reasonable $N$. This is important, since as $N$ grows, the amount of calculations needed to perform at each step of the numerical scheme (regardless of its specific choice) grows as $N^4$ (there is a threefold sum for each $\alpha_n$). Unfortunately, it is not necessarily the case. As can be seen in Fig.\ \ref{fig:specSNH}, for the ground states the distributions of $|\alpha_n|$ have relatively heavy tails. Additionally, the situation quickly worsens with the increase of the value of $b$ (one could predict, based on the bifurcation analysis, that for small $b$ a single mode dominates). This means, for instance, that the loss of stability of the ground state in $d=7$, that is expected to happen for $b\approx 100$, cannot be reasonably investigated with this method.

\vspace{\breakFF}
\begin{adjustwidth}{\marwidFF}{\marwidFF}
\small\qquad
The analogous derivation can be repeated also for some other NLS equations. In case of GP system it also gives Eq.\ (\ref{eqn:5spec}) with interaction coefficients defined by Eq.\ (\ref{eqn:SGP}). For other cubic systems their evolution is given by the same equation with yet another interaction coefficients $S_{njkl}$. If the NLS equation contains nonlinearity of other order than cubic, then Eq.\ (\ref{eqn:5spec}) also changes its form. For example, if $F(\psi)=|\psi|^4\psi$, it includes a fivefold sum instead.
\end{adjustwidth}\vspace{\breakFF}

We may conclude that simple numerical methods tend to fail when applied to SNH equation, so one shall rather resort to other approaches, possibly modifying some of the methods used for SN system \cite{Bao11, Guz04, Har03, Lub08, Par20}. However, this matter still needs some further work. Instead, we consider here an approximation to Eq.\ (\ref{eqn:5SNHt}), called the resonant approximation, that lets us to drop one sum in Eq.\ (\ref{fig:specSNH}). This not only allows us to perform numerical experiments with reasonable accuracy, but also unveils additional interesting features of SNH system present in $d=4$.

\subsection{Resonant approximation}\label{sec:resonant}
Let us assume that our solution $\psi$ has an amplitude of order $\varepsilon$, so it can be expanded into
\begin{equation*}
    \psi(t,r)=\varepsilon \sum_{k=0}^\infty \alpha_k(t)\, e^{-i\Omega_k t} e_k(r).
\end{equation*}
Performing calculations as before we again arrive at the infinite system of ODEs
\begin{align*}
    i\,\partial_t \alpha_n=\varepsilon^2 \sum_{j=0}^\infty \sum_{k=0}^\infty \sum_{l=0}^\infty S_{njkl}\bar{\alpha}_j \alpha_k \alpha_l e^{i(\Omega_n+\Omega_j-\Omega_k-\Omega_l)t}.
\end{align*}
One can now rescale time by $\varepsilon^2$, then the overall factor $\varepsilon^2$ vanishes, but the exponent under the sum becomes $-i(\Omega_n+\Omega_j-\Omega_k-\Omega_l)t/\varepsilon^2$. The resonant terms in the sum, i.e.\ elements where $\Omega_n+\Omega_j-\Omega_k-\Omega_l=0$, see no change. However, the remaining terms start to oscillate very rapidly as we go with $\varepsilon$ to zero. One can assume that for sufficiently small $\varepsilon$ these oscillations get much smaller than the timescale of $\alpha_n$, hence these terms can be averaged out to zero. By discarding these terms we get an approximate equation that shall model the long time behaviour of our system \cite{Mur, Kuk, Ger16, Bia19res}. For equations with harmonic potential the resonance condition corresponds to $n+j=k+l$, so the resonant system is given by
\begin{align*}
    i\,\dot{\alpha}_n=\sum_{j=0}^\infty \sum_{k=0}^{n+j} S_{njk,n+j-k}\bar{\alpha}_{j} \alpha_k \alpha_l,
\end{align*}
where dots denote derivative over the rescaled time (which will also be called $t$). We also redefine the interaction coefficients 
\begin{align*}
    C_{njkl}= \frac{1}{2} \left(S_{njkl}+S_{njlk}\right).
\end{align*}
One can easily check that this way we get an equivalent equation
\begin{align}\label{eqn:5resonantFin}
    i\,\dot{\alpha}_n=\sum_{j=0}^\infty \sum_{k=0}^{n+j} C_{njk,n+j-k}\bar{\alpha}_{j} \alpha_k \alpha_l,
\end{align}
while the interaction coefficients satisfy $C_{njkl}=C_{njlk}=C_{jnkl}=C_{klnj}$. This infinite system of autonomous ODEs will be called the resonant system.

\vspace{\breakFF}
\begin{adjustwidth}{\marwidFF}{\marwidFF}
\small\qquad
Similar resonant systems can be derived also from some other NLS equations such as GP. There, one also obtains Eq.\ (\ref{eqn:5resonantFin}) but with coefficients given by Eq.\ (\ref{eqn:SGP}) (they are already fully symmetric, so one has $C_{njkl}=S_{njkl}$). The fact that SNH and GP resonant systems differ only by the interaction coefficients comes from both having cubic nonlinearities and both sharing the harmonic potential. The latter means that they have the same $\Omega_n$ and in effect the same resonance condition. For NLS equations with other potentials $V$ the resonance condition may differ.
\end{adjustwidth}\vspace{\breakFF}

The resonant system has a conserved energy given by
\begin{align}\label{eqn:hamiltonianresonant}
   H=\frac{1}{2}\sum_{n=0}^\infty \sum_{j=0}^\infty \sum_{k=0}^{n+j} C_{njk,n+j-k} \bar{\alpha}_n \bar{\alpha}_j \alpha_k \alpha_{n+k-l}.
\end{align}
It also takes a role of the hamiltonian of this system, since $i\partial_t \alpha_n=\partial H/\partial \bar{\alpha}_n$. The other two conserved quantities come from the global ($\alpha_n\to e^{i\phi} \alpha_n$) and local ($\alpha_n\to e^{i n \phi} \alpha_n$) phase symmetries, we call them mass and linear energy, respectively
\begin{align*}
N=\sum_{n=0}^\infty |\alpha_n|^2,\qquad
   J=\sum_{n=0}^\infty n\, |\alpha_n|^2.
\end{align*}
Let us point out that while $N$ and $H$ are conserved also in the full system (\ref{eqn:5spec}), $J$ is an integral of motion only for the resonant system. When $d=4$, the new conserved quantity appears, given by
\begin{align*}
Z=\sum_{n=0}^\infty \sqrt{(n+1)(n+2)} \bar{\alpha}_{n+1} \alpha_n.
\end{align*}
It can either be shown explicitly, or be deduced using the framework introduced in \cite{Bia19res}. Let us also point out that in contrary to the other mentioned integrals of motion, $Z$ must not have a real value.

Proposition 3.2\ in \cite{Bia19res} gives us a set of conditions for a cubic resonant system, i.e.\ system described by Eq.\ (\ref{eqn:5resonantFin}), sufficient for $Z$ to be conserved. One needs to assume that the interaction coefficients $C_{njkl}$ are symmetric under some permutations of indices: $C_{njkl}=C_{njlk}=C_{jnkl}=C_{klnj}$, and additionally when $n+j=k+l+1$ the specific expression $\mathcal{D}_{njkl}$ is equal to zero. This expression is given by
\begin{align}\label{eqn:Dklmn}
    \mathcal{D}_{njkl}=(n+1)\tilde{C}_{n-1,jkl}+(j+1)\tilde{C}_{n,j-1,kl}-(k+1)\tilde{C}_{nj,k+1,l}-(l+1)\tilde{C}_{njk,l+1},
\end{align}
where $\tilde{C}_{njkl}=\sqrt{(n+1)(j+1)(k+1)(l+1)}C_{njkl}$. As rather lengthy calculations presented in Appendix \ref{sec:appDklmn} show, in case of resonant SNH system these conditions hold if $d=4$. Then, we not only get the conservation of $Z$, but also another interesting result formulated in \cite{Bia19res} as Proposition 3.1.: the existence of the finite-dimensional invariant manifold. By this we mean that there exists a finite-dimensional (in this case three-) manifold embedded in the infinite-dimensional space of $\alpha_n$, such that if the initial conditions are posed on this manifold, the whole evolution is confined within it. It is given explicitly by
\begin{align}\label{eqn:5ansatz}
    \alpha_n(t)=\sqrt{n+1}\left(b(t)+\frac{a(t)}{p(t)}\, n\right)p(t)^n,
\end{align}
where $a$, $b$, and $p$ are some complex-valued functions. The further calculations presented in \cite{Bia19res} and \cite{Biz18} let us also extract explicit differential equations describing the evolution of these functions:
\begin{subequations}\label{eqn:5eq3}
\begin{align}
i \dot{p}=&\frac{1}{16} (1+y)^2 (2 |a|^2 p (1+y)+a \bar{b}),\label{eqn:5eq3p}\\
i\dot{a}=&\frac{1}{16} a (1+y)^3 \left(10 |a|^2 (1+3 y)+20\Re(a\bar{b}\bar{p})+4 \bar{a} b p\right)\nonumber\\
&+\frac{7}{16} a (1+y)^2 |b|^2,\\
i \dot{b} =&\frac{3}{8}a \bar{p}(1+y)^4\left(2(1+2y)|a|^2+a\bar{b}\bar{p}\right)\nonumber\\
&+b(1+y)^2\left((1+y)(1+3y)|a|^2+\frac{1}{2}|b|^2+2(1+y)\,\Re(\bar{a}bp) \right),
\end{align}
\end{subequations}
where $y=|p|^2/(1-|p|^2)$.
The conserved quantities can also be written explicitly using $a$, $b$, and $p$. Then the mass, linear energy, and the complex integral of motion are given by
\begin{align*} 
N=&(1+y)^2 \left(2(1+y)(1+3y)|a|^2 +|b|^2+4 (1+y)\,\Re(\bar{a}bp)\right),\\
J=&(1+y)^2 \left(2 (1+y)(1+9y+12y^2)|a|^2+2y|b|^2+4(1+y)(1+3y)\,\Re(\bar{a}bp)\right),\\
Z=&2(1+y)^3\left(6(1+y)(1+2y)|a|^2+|b|^2+6(1+y)\,\Re (\bar{a}bp) \right) \bar{p}+2(1+y)^3 \bar{a}b. 
\end{align*}
Regarding energy $H$, it is more convenient to instead introduce $S$ such that $H=N^2-6S^2$. Then $S=|a|^2(1+y)^4/2$. One can partially invert these expressions and get
\begin{subequations}\label{eqn:5conservedinverse}
\begin{align}
|a|^2=&\frac{2 S}{(1+y)^4},\label{eqn:5conservedinversea}\\
|b|^2=&\frac{(1+3y)N+12yS-J}{(1+y)^3},\\
\Re(\bar{a}bp)=&-\frac{2yN+4(1+6y)S-J}{4 (1+y)^4}.
\end{align}
\end{subequations}
It leads to
\begin{align*}
Z=(2N+J+12S)\bar{p}+2(1+y)^3\, \bar{a}b
\end{align*}
and together with Eq.\ (\ref{eqn:5conservedinversea}) lets us get rid of all functions independent of $p$ in Eq.\ (\ref{eqn:5eq3p}):
\begin{align*}
i \dot{p}=&\frac{1}{32(1+y)}\left(\bar{Z}-(2N+J+4S)p\right).
\end{align*}
Then one can combine the expressions above to get the following line of equalities
\begin{align*}
    \dot{y}^2=&\frac{\left(\dot{p}\bar{p}+p\dot{\bar{p}}\right)^2}{\left(1-|p|^2\right)^4}= -\frac{(1+y)^2}{1024}\left(Zp-\bar{Z}\bar{p}\right)^2\\
    =& -\frac{(1+y)^8}{256}\left(\bar{a}bp-a\bar{b}\bar{p}\right)^2= \frac{(1+y)^8}{64}\left[|a|^2 |b|^2 |p|^2-(\Re(\bar{a}bp))^2\right]^2 \\
    =&-\frac{1}{256}\left[\left(N^2+48S^2\right)y^2-\left(NJ+4JS+4NS-48S^2\right)y+\frac{1}{4}(4S-J)^2\right]\\
    =&-\frac{N^2+48S^2}{256}\left[y+\frac{1}{2}\left(1-\frac{(N+J)(N+4S)}{N^2+48S^2}\right)\right]^2\\
    &+\frac{S(4S-N)\left(48S^2-2NJ-J^2\right)}{128\left(N^2+48S^2\right)}.
\end{align*}
Hence, the function $y$ satisfies equation of the form $\dot{y}^2+\omega^2 (y-y_0)^2=\omega^2 A^2$, with the frequency
\begin{align*}
\omega=\frac{\sqrt{N^2+48S^2}}{16},
\end{align*}
amplitude
\begin{align*} A=\frac{\sqrt{2S(4S-N)\left(48S^2-2NJ-J^2\right)}}{N^2+48S^2},
\end{align*}
and the center of motion
\begin{align*}
y_0=-\frac{1}{2}\left(1-\frac{(N+J)(N+4S)}{N^2+48S^2}\right).
\end{align*}
As a result, the evolution of $y$ is described by the motion of a harmonic oscillator and has a solution of the form $y(t)=A\,\cos(\omega t+\phi)+y_0$. This means that $|p|^2$ is also periodic in time. Moreover, from Eqs.\ (\ref{eqn:5conservedinverse}) one deduces periodicity of $|a|^2$, $|b|^2$, and $\Re(\bar{a}bp)$. Finally, it can be easily checked that this result then also applies to $|\alpha_n|^2$. Hence, we expect that for initial conditions satisfying the ansatz (\ref{eqn:5ansatz}), we observe an evolution where energy periodically returns to the distribution given by the initial condition. In particular, any two-mode initial data with $\alpha_n(0)=0$ for $n\geq 2$ belongs to this class (with $p(0)=0$), so we expect that in such case the energy initially disperses to higher modes, but eventually again concentrates in the two lowest modes.

\vspace{\breakFF}
\begin{adjustwidth}{\marwidFF}{\marwidFF}
\small\qquad
Interestingly, GP system also acknowledges a similar symmetry enhancement, but in this case it takes place in $d=2$. Then the additional conserved quantity is \begin{align*}
Z=\sum_{n=0}^\infty (n+1) \bar{\alpha}_{n+1} \alpha_n,
\end{align*}
while the ansatz giving the three-dimensional invariant manifold takes form of
\begin{align*}
    \alpha_n(t)=\left(b(t)+\frac{a(t)}{p(t)}\, n\right)p(t)^n.
\end{align*}
Solutions laying on this manifold also are periodic. This system belongs to the larger family of resonant systems that can be obtained from the Gross-Pitaevskii equation, as described in \cite{Bia18}. 

Analogous results hold for many other systems, see \cite{Bia19res} for more examples such as lowest Landau level or maximally rotating scalar fields in $(d+1)$-dimensional Anti-de Sitter spacetime.
\end{adjustwidth}\vspace{\breakFF}

Now we want to perform some numerical calculations letting us verify the claims above. To do so, we truncate the first sum in Eq.\ (\ref{eqn:5resonantFin}) at some $N$ and simulate the evolution of the obtained system of $N+1$ equations using some suitable iterative method. However, to do so the exact values of the interaction coefficients $S_{klmn}$ are needed. They are given by a rather complicated integral expression (\ref{eqn:SSNH}) and even though it is possible to evaluate them analytically using a computer algebra system (CAS), the calculation gets progressively more involved as the indices $k$, $l$, $m$, and $l$ increase. The possible remedy is to reformulate the integrals as sums of expressions including hypergeometric functions, as shown in Appendix A of \cite{Biz18}. This allows for further simplification since in even dimensions the relevant hypergeometric functions can be represented by sums of simple coefficients. Using this approach, CAS are usually able to evaluate individual interaction coefficients faster, however, it turns out that if we are interested in the whole table of $S_{klmn}$ with $k,l,m,n<N$ it is better to use the suitable recursive scheme. As its derivation and description is rather technical, we present it in Appendix \ref{sec:appcoeff}. 

In Fig.\ (\ref{fig:SNHres}) we show the plots of $|\alpha_n(t)|$ for the lowest modes in $d=3$ and $d=4$. The initial data for these evolutions are $\alpha_0(0)=1$ and $\alpha_1(0)=1$, so in case of $d=4$ they lay on the invariant manifold. As one can see, then the evolution is indeed periodic, while for $d=3$ it seems chaotic. 
\begin{figure}
\centering
\includegraphics[width=0.45\textwidth]{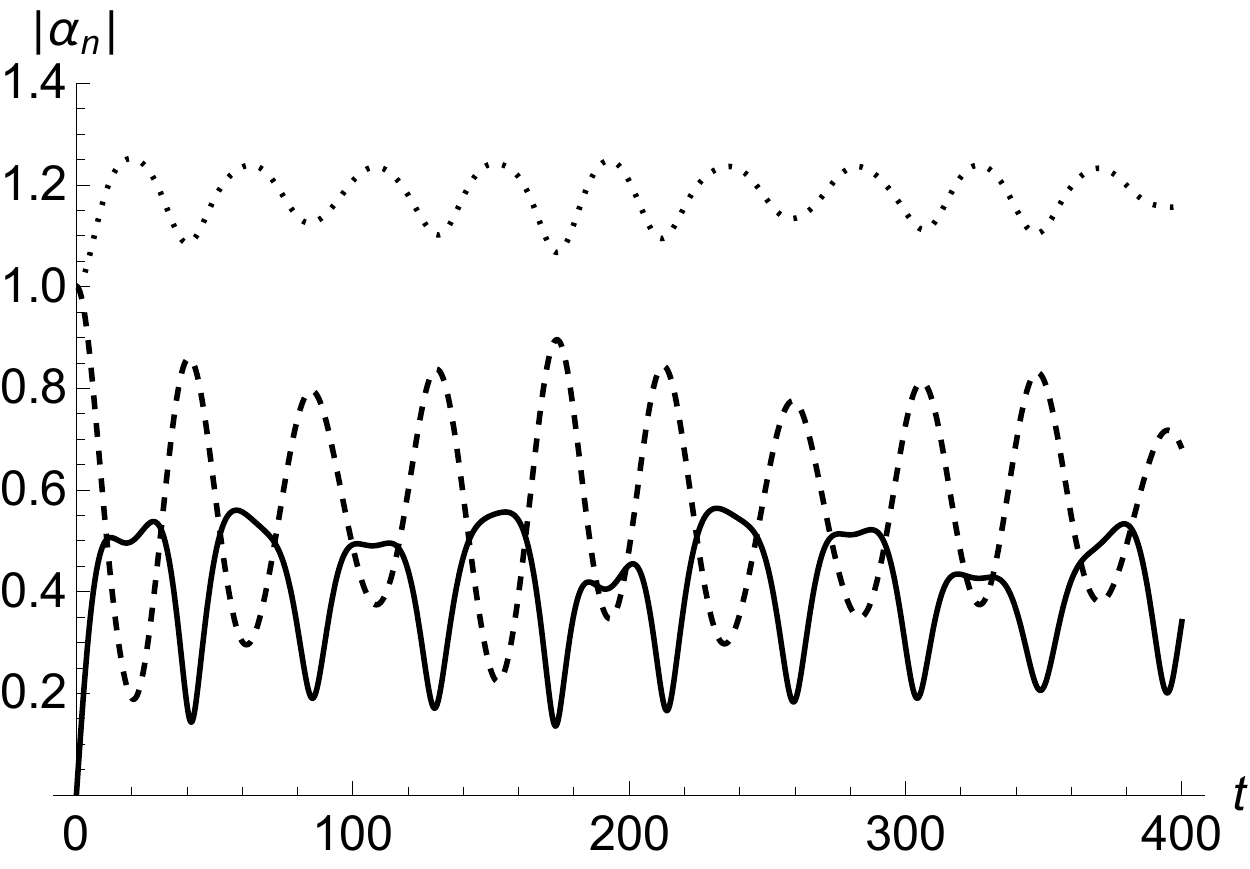}\quad
\includegraphics[width=0.45\textwidth]{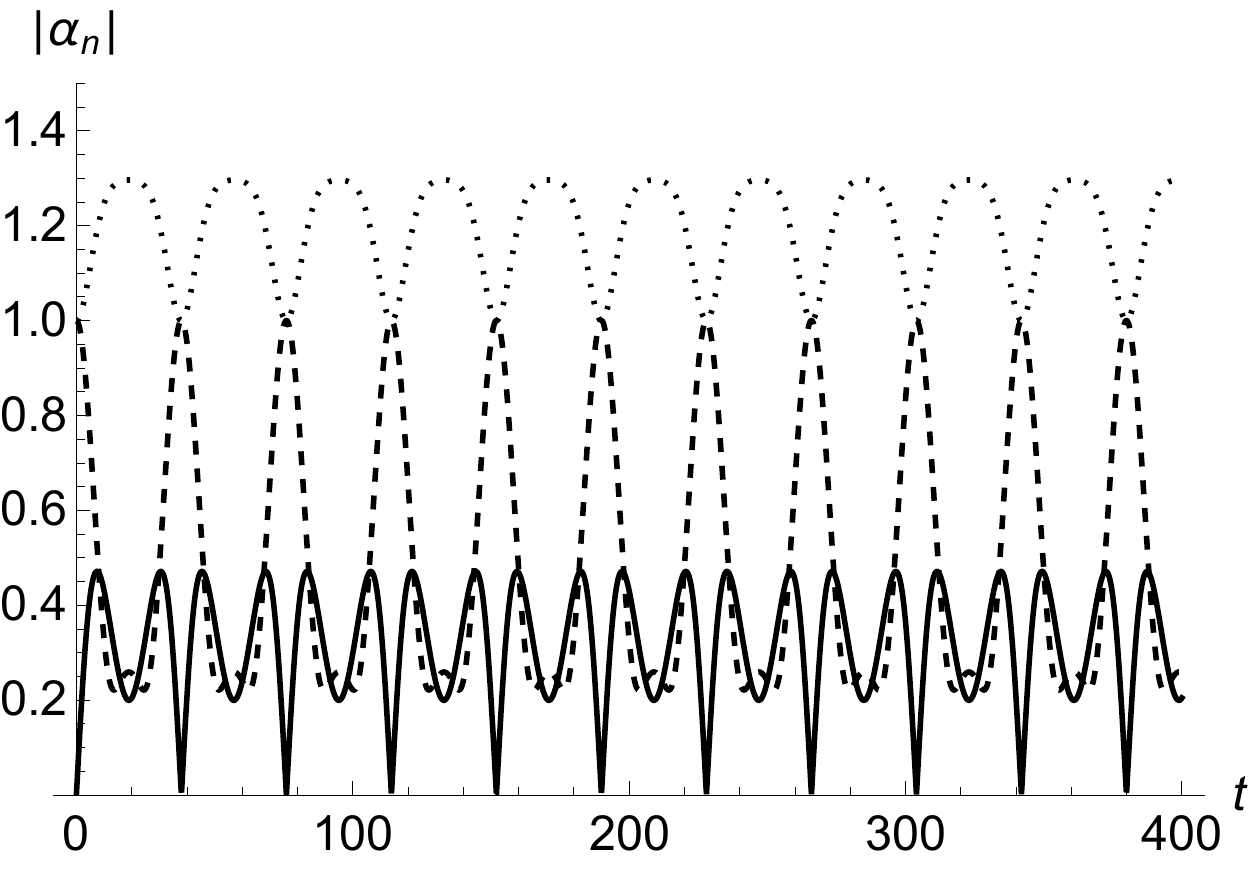}
\caption{Plots of $|\alpha_0|$ (dotted line), $|\alpha_1|$ (dashed line), and $|\alpha_2|$ (solid line) for resonant SNH system in $d=3$ (left plot) and $d=4$ (right plot). The initial data is concentrated in two lowest modes with $\alpha_0(0)=\alpha_1(0)=1$.}
\label{fig:SNHres}
\end{figure}

Yet another indication of the special character of $d=4$ for SNH system can be seen by investigating the quantization of this resonant system. One can consider hamiltonian (\ref{eqn:hamiltonianresonant}) with $\alpha_n$ and $\bar{\alpha}_n$ replaced by the annihilation and creation operators, $\hat{\alpha}_n$ and $\hat{\alpha}^\dagger_n$, respectively, satisfying the standard commutation relation $[\hat{\alpha}_n,\hat{\alpha}^\dagger_m] =\delta_{nm}$. Then it turns out that the spacings between neighbouring eigenvalues of such hamiltonian are well described by the Poisson distribution in $d=4$, while for other dimensions they rather follow the Wigner surmise. For the discussion of this result and its implications we refer to \cite{Biz18}.

\chapter{Conclusion}\label{sec:conclusion}
We conclude with the summary of the most important results presented here and an outline of the possible directions for further research.

We began this thesis by giving some motivation for studying NLS equations in higher dimensions. This included the derivation of the equations describing the nonrelativistic limit of weak scalar perturbations of the anti-de Sitter spacetime, that turned out to be SNH equations. This derivation was partially based on the work by Giulini and Grossardt \cite{Giu12}, however we did not need to assume the spherical symmetry of the solutions. There are still many things not fully understood regarding this limit, for example, how the critical dimension of the equations changes under it, from $d=3$ for Einstein equations to $d=6$ for SNH system. It would be also interesting to investigate this limit with more strict mathematical tools to see how the solutions behave under it. One can also try to get similar nonrelativistic limits for other fields, such as the ones describing vector or spinor particles.

The main content of this thesis was covered in Chapter \ref{sec:stationary} and it concerned stationary solutions of SNH system and other NLS equations. The most important three results we presented there were the existence and uniqueness of the ground states, the existence of the whole ladder of the excited states, and explanation of the behaviour of frequency function $\omega(b)$ including the change of its monotonicity in higher dimensions. This topic can be further investigated in many interesting directions, probably the most obvious one including the search of other NLS systems that can be, similarly to SNH and GP equations, treated with these methods. After we impose the spherical symmetry on our system, the dimension $d$ can be treated as a parameter interpreted as a friction in a second-order ODE. The continuous change of $d$ alters the global structure of solutions, what can be seen, for example, in changes of the shapes of plots of $\omega(b)$ and $\mathcal{M}(\omega)$. It is compelling to look for more simple systems having this behaviour and try to better understand the role played there by the damping term. Finally, in the presented context one can be interested in pursuing the matter of uniqueness of the excited states, a classical open problem \cite{Has}.

In chapter \ref{sec:dynamics} we just started the analysis of the dynamics of SNH equation, focusing on the stability of the stationary solutions and then moving to the resonant approximation. It leaves us with lots of potential future work to do. Most importantly, we would like to be able to perform numerical simulations of this system for large times. This should give us some insight into long-time behaviour of the system, along the lines of motivation described in Section \ref{sec:turbulence}. The other direction worth focusing on is the better understanding of the stability of excited states, i.e.\ the presence of windows of stability and their dependence on the dimensions $d$. We hope to explore it in the future publication including and expanding results presented in Section \ref{sec:stability}. Chapter \ref{sec:dynamics} focused mostly on SNH system with a few mentions of GP system, however, it might be interesting to study the higher-dimensional dynamics (including stability of stationary solutions) of various NLS equations in a more systematic way, similarly as we did for stationary solutions.

\begin{appendices}
\chapter{Solutions to singular ODEs}\label{sec:appODE}
In the main text we often consider equations of the form
\begin{equation}\label{eqn:appODE1}
    u''+\frac{d-1}{r}u'+f(r,u)=0
\end{equation}
with initial conditions posed at $r=0$. By $u$ we understand here either a single real function of $r$ defined on $[0,\infty)$ or a vector consisting of $n$ such functions. In the second case $f$ is understood also as a vector, since each component of $u$ may be a solution to a different elliptic equation. Either way, $f$ are $\mathcal{C}^1$ functions of independent variable $r\in[0,\infty)$ and functions $u$.

To properly apply the shooting method in Section \ref{sec:stationary} we need to know that the solutions to such singular equations exist locally near zero, are unique, and depend continuously on initial values and parameters. In this Appendix we show the relevant results. We begin with the proof of existence and uniqueness.

\begin{thm}\label{thm:A1}
Let $u$ be a solution to the Cauchy problem
\begin{align*}
    (r^{d-1} u')'+r^{d-1}\, f(r,u)=0,\\
    u(0)=u_0,\qquad u'(0)=0,
\end{align*}
where $d\geq 2$. Let $f$ be continuous in $r$ and continuously differentiable in $u$ for $r$ and $u$ such that $0\leq r< r_0$ and $|u-u_0|<b$, where $r_0$ and $b$ are some positive numbers. Then there exists $r_1>0$ such that in the interval $[0,r_1)$ this problem has a unique solution of a class $\mathcal{C}^2$.
\end{thm}
For brevity, we present the proof in the case of $u$ being a single function satisfying equation. If instead $u$ is a vector of functions, the reasoning is analogous.
\begin{proof}
We employ the classical successive approximation scheme. To do so, let us begin with an introduction of integral equations
\begin{align*}
    u(r)&=u_0+\int_0^r v(\rho)d\rho,\\
    v(r)&=-\frac{1}{r^{d-1}}\int_0^r f(\rho,u(\rho)) \rho^{d-1}\, d\rho.
\end{align*}
It is clear that if $u$ and $v$ satisfy these equations, they are continuous for $r>0$. At zero, we get $v(0)=0$ with the L'Hospital's rule, while $u(0)=u_0$. Calculation of the second derivatives yield $u'(r)=v(r)$, and as a result we have 
\begin{align*}\label{eqn:init.existence.integral}
    u''(r)=&v'(r)=-f(r,u(r))+\frac{d-1}{r^{d}}\int_0^r f(\rho,u(\rho)) \rho^{d-1}\, d\rho\\
    =&-f(r,u(r))-\frac{d-1}{r}v(r)=-\frac{d-1}{r}u'(r)-f(r,u(r)).
\end{align*}
Hence, function $u$ satisfying the integral equations is a solution of our problem, that is also twice differentiable for $r>0$.

We define the $n$-th successive approximations as
\begin{subequations}\label{eqn:appODE2}
\begin{align}
    v_{n+1}(r)&=-\frac{1}{r^{d-1}}\int_0^r f(\rho,u_n(\rho)) \rho^{d-1}\, d\rho,\label{eqn:appODE2a}\\
    u_{n+1}(r)&=u_0+\int_0^r v_{n+1}(\rho)d\rho,\label{eqn:appODE2b}
\end{align}
\end{subequations}
with $u_0(r)\equiv u_0$ and $v_0(r)\equiv 0$. Let us emphasize the appearance of $v_{n+1}$ instead of $v_n$ in the definition of $u_{n+1}$. Such choice, even though not mandatory, lets us to avoid the redundancy of $u_1\equiv u_0$.

Since $f$ is continuous in $\{(r,u): 0\leq r <r_0 \mbox{ and } |u-u_0|<b \}$, there exists such $M>0$ that $|f(r,u)|<M$ in this set. As it is also $\mathcal{C}^1$, it is locally Lipschitz, so additionally there exists $K>0$ such that $|f(r,u_1)-f(r,u_2)|<K|u_1-u_2|$ for $r$, $u_1$ and $u_2$ in this set. Let us define $r_1:=\min\{r_0,\sqrt{bd/M},\sqrt{d/2K}\}$. Then one may show inductively that for all approximations of $u$ it holds $|u_n(r)-u_0|<b$ for $r\in[0,r_1)$. It is obvious for $n=0$, then assuming that this inequality holds for some $n\in\mathbb{N}$, we indeed have
\begin{align*}
    |v_{n+1}(r)| \leq \frac{1}{r^{d-1}}\int_0^r |f(\rho,u_n(\rho))| \rho^{d-1}\, d\rho < \frac{M}{r^{d-1}}\int_0^r \rho^{d-1}\, d\rho=\frac{M r}{d}<\frac{M r_1}{d},
\end{align*}
and
\begin{align*}
    |u_{n+1}(r)-u_0|\leq \int_0^r |v_{n+1}(\rho)|d\rho< \int_0^r \frac{M r_1}{d}d\rho =\frac{M r_1 r}{d}<\frac{M r_1^2}{d}<b.
\end{align*}

Thanks to $f$ being locally Lipschitz, for any $n\in\mathbb{N}_+$ and $r\in[0,r_1)$ it also holds
\begin{align*}
    |v_{n+1}(r)-v_{n}(r)|&\leq \frac{1}{r^{d-1}}\int_0^r |f(\rho,u_n(\rho))-f(\rho,u_{n-1}(\rho))| \rho^{d-1}\, d\rho\\
    &\leq\frac{K}{r^{d-1}}\int_0^r |u_n(\rho)-u_{n-1}(\rho)| \rho^{d-1}\, d\rho \\
    &\leq\frac{K}{r^{d-1}}\max_{\rho\in[0,r_1)}\{|u_n(\rho)-u_{n-1}(\rho)|\}\int_0^r\rho^{d-1}\, d\rho\\
    &\leq\frac{K r_1}{d}\max_{\rho\in[0,r_1)}\{|u_n(\rho)-u_{n-1}(\rho)|\}.
\end{align*}
It implies
\begin{align*}
    |u_{n+1}(r)-u_{n}(r)|&\leq\int_0^r |v_{n+1}(\rho)-v_n(\rho)|\, d\rho\\
    &\leq \frac{K r_1^2}{d}\max_{\rho\in[0,r_1)}\{|u_n(\rho)-u_{n-1}(\rho)|\}.
\end{align*}
Our choice of $r_1$ gives us $K r_1^2/d\leq 1/2$, hence we have obtained
\begin{align}\label{eqn:appODE3}
    |u_{n+1}(r)-u_{n}(r)|&\leq \frac{1}{2} \max_{\rho\in[0,r_1)}\{|u_n(\rho)-u_{n-1}(\rho)|\}\nonumber\\
    &\leq\frac{1}{2^n} \max_{\rho\in[0,r_1)}\{|u_1(\rho)-u_0|\}<\frac{b}{2^n}.
\end{align}
For any $n\in\mathbb{N}_+$, function $u_n$ can be represented as
\begin{align*}
    u_n(r)=u_0(r)+\sum_{k=0}^{n-1}[u_{k+1}(r)-u_{k}(r)].
\end{align*}
Then Eq.\ (\ref{eqn:appODE3}) gives the uniform convergence of $u_n$ on $[0,r_1)$. Analogous result can be also obtained for $v$. Hence, we can define $u$ and $v$ as the limits of $u_n$ and $v_n$, respectively, as $n$ goes to infinity. The continuity of $f$ lets us perform the limit in Eqs.\ (\ref{eqn:appODE2}) showing that $u$ and $v$ are indeed desired solutions.

The only thing left is the uniqueness of the constructed solutions. Let us assume that there exist some other solutions $\mu$ and $\nu$ (respectively, to $u$ and $v$). Then there also must hold $|\mu(r)-u_0|<b$ for $r\in[0,r_1)$. To show it, let us assume that inside $[0,r_1)$ there is some $r$ at which $|\mu(r)-u_0|=b$ for the first time, then we have
\begin{align*}
     b=|\mu(r)-u_0|\leq\int_0^r |\nu(\rho)|d\rho& \leq \int_0^r \frac{1}{\rho^{d-1}}\int_0^\rho \sigma^{d-1} |f(\sigma,\mu(\sigma))|d\sigma\, d\rho\nonumber\\
     &\leq M \int_0^r \frac{1}{\rho^{d-1}}\int_0^\rho \sigma^{d-1}d\sigma\, d\rho=\frac{Mr^2}{2d}<b
\end{align*}
giving us a contradiction. As a result, for any $r\in[0,r_1)$ it holds:
\begin{align*}
    |v(r)-\nu(r)|&\leq \frac{1}{r^{d-1}}\int_0^r |f(\rho,u(\rho))-f(\rho,\mu(\rho))| \rho^{d-1}\, d\rho\\
    &\leq \frac{K}{r^{d-1}}\int_0^r |u(\rho)-\mu(\rho)| \rho^{d-1}\, d\rho\\
    &\leq\frac{K}{r^{d-1}}\max_{\rho\in[0,r_1)}\{|u(\rho)-\mu(\rho)|\}\int_0^r \rho^{d-1}\, d\rho\\
    &\leq \frac{K r_1}{d}\max_{\rho\in[0,r_1)}\{|u(\rho)-\mu(\rho)|\}.
\end{align*}
This result gives the following bound
\begin{align*}
    |u(r)-\mu(r)|&\leq \int_0^r |v(\rho)-\nu(\rho)|\, d\rho\leq \frac{K r_1^2}{d}\max_{\rho\in[0,r_1)}\{|u(\rho)-\mu(\rho)|\}\\
    &\leq \frac{1}{2}\max_{\rho\in[0,r_1)}\{|u(\rho)-\mu(\rho)|\}.
\end{align*}
Hence there must be $u(r)=\mu(r)$ for every $r\in[0,r_1)$ giving us uniqueness.
\end{proof}
Thus, we have shown a local existence and uniqueness of the solution near zero. As there is no other singular point, for the remaining part of the half-line one can use the standard results including also the extension of a solution to its maximal domain \cite{Cod, Har}.

In the main text we also heavily use the continuous dependence of the solutions on initial conditions and parameters. Let us assume that the function $f$ additionally depends continuously on some parameter $\alpha$ for each $r$ and $u$. Then, as we have already established the existence and uniqueness of solutions near zero, such dependence is a natural consequence of the Arzel\`{a}–Ascoli theorem (c.f.\ Lemma 3.2 of \cite{Har}). Hence, we have the following result
\begin{thm}\label{thm:A2}
Let $u_{\alpha,\beta}$ be a solution to the Cauchy problem
\begin{align*}
    (r^{d-1} u')'+r^{d-1}\, f(r,u,\alpha)=0,\\
    u(0)=\beta,\qquad u'(0)=0,
\end{align*}
where $d\geq 2$. Let $f$ satisfy the same conditions as in Theorem \ref{thm:A1} and also be continuous in $\alpha$. If the solution exists on some interval $[0,r_0]$ for each value of $(\alpha,\beta)$ from some open set $\Omega\subset\mathbb{R}^2$, then $u_{\alpha,\beta}$ is uniformly continuous on $[0,r_0]$ in $(\alpha,\beta)$ from $\Omega$. 
\end{thm}

\chapter{Asymptotic behaviour of singular SNH near zero}\label{sec:appsingular}
\chaptermark{Singular SNH near zero}
The goal of this appendix is to prove the following key result that was used in the shooting method in Section \ref{sec:exsingular}.
\begin{lem}\label{thm:singas}
Let $\widetilde{u}$ and $\widetilde{h}$ be the solutions of the system
\begin{subequations}\label{eqn:B1}
\begin{align}[left ={ \empheqlbrace}]
\widetilde{u}'' + \frac{d-5}{r}\widetilde{u}' +\frac{2(d-4)}{r^2}(\widetilde{u}\widetilde{h}-\widetilde{u})-r^2 \widetilde{u}&= 0,\label{eqn:B1a}\\
\widetilde{h}''+\frac{d-5}{r} \widetilde{h}'+ \frac{2(d-4)}{r^2}(\widetilde{u}^2-\widetilde{h})&= 0.\label{eqn:B1b}
\end{align}
\end{subequations}
in a supercritical dimension ($d>6$). If these solutions satisfy $\lim_{r\to 0} \widetilde{u}(r)=\lim_{r\to 0} \widetilde{h}(r)=1$, then near zero they have an asymptotic behaviour given by
\begin{align}\label{eqn:B2}
\widetilde{u}(r) = 1 - c\,r^\lambda + \mathcal{O}(r^4), \qquad \widetilde{h}(r) = 1 + 2c\,r^\lambda + \mathcal{O}(r^4),
\end{align}
where $c$ is some real constant shared by both solutions and
\begin{align*}
    \lambda=\frac{-d+6+ \sqrt{d^2 + 4d - 28}}{2}.
\end{align*}
\end{lem}
Of course, in this formulation Eqs.\ (\ref{eqn:B1}) and (\ref{eqn:B2}) are just Eqs.\ (\ref{eqn:utilde}) and (\ref{eqn:asymptilde}) from the main text. The proof we present here was published in \cite{Fic21} and is based on the proof of Lemma 3.1.\ in \cite{Mer91}.

\begin{proof}
In the beginning, it is convenient to slightly reformulate the problem by appropriate changes of variables. Using functions $\eta = \widetilde{u}-1$ and $\xi=\widetilde{h}-1$ simplifies the analysis near zero, as these functions are small in its neighborhood (they converge to zero there). We also introduce $t=\ln r$ as an independent variable. This choice removes $1/r$ from the first-derivative terms at the cost of unfolding the domain into the whole line and moving the focus to the behaviour of the solutions as $t\to-\infty$. In these variables Eqs.\ (\ref{eqn:B1}) become
\begin{subequations}\label{eqn:Betaxi}
\begin{align}[left ={ \empheqlbrace}]
\ddot{\eta} + (d-6)\dot{\eta} +2(d-4)\xi &= e^{4t}(1+\eta) -2(d-4)\eta\xi,\label{eqn:Betaxia}\\
\ddot{\xi} + (d-6)\dot{\xi} +2(d-4)(2\eta-\xi) &= -2(d-4)\eta^2,\label{eqn:Betaxib}
\end{align}
\end{subequations}
where dots denote derivatives over $t$. 

The left hand-side of Eqs.\ (\ref{eqn:Betaxi}) constitutes the system of linear equations with four eigenvalues equal to
\begin{align*}
\frac{1}{2}(-d+6\pm \sqrt{d^2-20d+68}),\qquad \frac{1}{2}(-d+6\pm\sqrt{d^2+4d-28}).
\end{align*}
The first pair in supercritical dimensions has a negative real part. Its imaginary part becomes zero for $d\geq2(5+2\sqrt{2})\approx 15.66$. This change of character of these eigenvalues when coming from $d=15$ to $d=16$ has interesting consequences discussed in Section \ref{sec:omegab}. Regarding the second pair, in supercritical dimensions it is real and is composed of a negative and positive eigenvalue. The latter was dubbed by us as $\lambda$ in the formulation of this lemma. Clearly, for $d\geq 7$ it holds $3\leq\lambda<4$. Hence, the linear part of the considered system is hyperbolic and possesses a one-dimensional unstable subspace.

Now we use this analysis of the linear part of Eqs.\ (\ref{eqn:Betaxi}) to construct the solutions of their homogenous parts. Three eigenvalues of this system have negative real parts so the solutions connected to them are unbounded as $t\to -\infty$. As we are interested in solutions that decay in $-\infty$, the general solution of the left hand side of Eqs.\ (\ref{eqn:Betaxi}) is $\eta_0(t)=-c e^{\lambda t}$, $\xi_0(t)=2c e^{\lambda t}$, where $c$ is some constant. This partial solution can be used to construct the full solution, although in the implicit form, with the method of variation of parameters. Let us introduce functions representing right hand sides of this system:
\begin{align*}
x(t)=e^{4t}\left(1+\eta(t)\right)-2(d-4)\eta(t)\xi(t),\qquad y(t)=-2(d-4)\eta(t)^2.
\end{align*}
It is also convenient to define $\alpha_1=\frac12 \sqrt{d^2-20d+68}$, $\alpha_2= \frac12 \sqrt{d^2+4d-28}$, and $\beta=-\frac{d}{2}+3$. In this notation, we have $\lambda=\beta+\alpha_2$. Then the method of variation of parameters yields
\begin{subequations}\label{eqn:etaxisol1}
\begin{align}
\eta(t)=&-c e^{\lambda t} +\frac{1}{3\alpha_1} \int_{-\infty}^t e^{\beta(t-s)} \sinh\alpha_1(t-s)\cdot [2x(s)+y(s)]ds\nonumber\\
&+\frac{1}{3\alpha_2} \int_{-\infty}^t e^{\beta(t-s)} \sinh\alpha_2(t-s)\cdot [x(s)-y(s)]ds,\label{eqn:etaxisol1a}\\
\xi(t)=&2c e^{\lambda t} +\frac{1}{3\alpha_1} \int_{-\infty}^t e^{\beta(t-s)} \sinh\alpha_1(t-s)\cdot [2x(s)+y(s)]ds\nonumber\\
&-\frac{2}{3\alpha_2} \int_{-\infty}^t e^{\beta(t-s)} \sinh\alpha_2(t-s)\cdot [x(s)-y(s)]ds.\label{eqn:etaxisol1b}
\end{align}
\end{subequations}
Since both $\eta$ and $\xi$ go to zero as $t\to - \infty$, for every $\varepsilon>0$ one may find such $T$ that for all $t<T$ the following bounds are satisfied:
\begin{subequations}\label{eqn:xybounds}
\begin{align}
|2x(t)+y(t)|=&\left|2e^{4t}(1+\eta(t))-2(d-4)\eta(t)[\eta(t)+2\xi(t)]\right|\nonumber\\
\leq& 4 e^{4t}+\varepsilon |\eta(t)|,\label{eqn:xybounds1}\\
|x(t)-y(t)|=&\left|e^{4t}(1+\eta(t))+2(d-4)\eta(t)[\eta(t)-\xi(t)]\right|\nonumber\\
\leq& 2 e^{4t}+\varepsilon |\eta(t)|.\label{eqn:xybounds2}
\end{align}
\end{subequations}
These limits let us to produce constraints on $|\eta(t)|$ by plugging them into Eq.\ (\ref{eqn:etaxisol1}). However, we need to consider separately the cases when $7\leq d\leq 15$ and $d\geq 16$.

For $d\geq 16$ we have $\alpha_1>0$ so for $s\leq t$ it holds $0\leq 2\sinh\alpha_1 (t-s)\leq e^{\alpha_1 (t-s)}$ (the analogous inequality works also for $\alpha_2$). Then we also use the facts that $\beta+\alpha_1<0$ (so $\int_{-\infty}^t e^{-(\beta+\alpha_1)s} |\eta(s)|\,ds$ is convergent) and $e^{(\beta+\alpha_1)(t-s)}\leq e^{\lambda(t-s)}$ to obtain
\begin{align*}
|\eta(t)|\leq |c| e^{\lambda t} + A_1 e^{4 t}+\varepsilon A_2 e^{\lambda t} \int_{-\infty}^t e^{-\lambda s}|\eta(s)|ds.
\end{align*}
Now we may divide both sides of this equation by $e^{\lambda t}$ and use the integral Gr\"{o}nwall's inequality to the function $e^{-\lambda t}|\eta(t)|$ getting for every sufficiently small $t$:
\begin{align}
|\eta(t)| \leq |c| e^{\lambda t} + B_1 e^{4 t},\label{eqn:boundeta}
\end{align}
with $B_1$ denoting some positive constant. To get a similar bound on $|\xi(t)|$ we go back to Eqs.\ (\ref{eqn:xybounds}). Then $|\eta(t)|=\mathcal{O}(e^{\lambda t})$ gives us constraints
\begin{align*}
|2x(t)+y(t)|\leq 4 e^{4t}+\varepsilon |\xi(t)|,\qquad
|x(t)-y(t)|\leq 2 e^{4t}+\varepsilon |\xi(t)|.
\end{align*}
Inserting it into Eqs.\ (\ref{eqn:etaxisol1}) and following calculations done for $|\eta|$ we get
\begin{align}
|\xi(t)| \leq |2c| e^{\lambda t} + B_2 e^{4 t},\label{eqn:boundxi}
\end{align}
where $B_2$ is some positive constant.

Exactly the same bounds on $|\eta|$ and $|\xi|$ can be obtained also for $7\leq d \leq 15$. In this case one can write $\sinh\alpha_1(t-s)$ in Eqs.\ (\ref{eqn:etaxisol1}) as $\sin|\alpha_1|(t-s)$. This function is bounded, so one can once again carefully evaluate and estimate all the needed integrals similarly to the previous case. In the end one gets exactly the same inequalities (\ref{eqn:boundeta}) and (\ref{eqn:boundxi}).

These inequalities give estimates on $x(t)$ and $y(t)$ for sufficiently small $t$:
\begin{align}
x(t)=&e^{4t}+\eta(t)\left[e^{4t}-2(d-4)\xi(t) \right]= e^{4t} + \mathcal{O}(e^{2\lambda t}),\nonumber\\
y(t)=&-2(d-4)\eta(t)^2= \mathcal{O}(e^{2\lambda t}).
\end{align}
The leading term is $e^{4t}$, as for $d\geq 7$ it holds $\lambda<4<2\lambda$. Inserting these bounds into Eqs.\ (\ref{eqn:etaxisol1}) and performing the integrals give
\begin{align*}
\eta(t)=-c e^{\lambda t} + \mathcal{O} (e^{4t}),\qquad \xi(t)=2ce^{\lambda t} + \mathcal{O} (e^{4 t}).
\end{align*}
Going back to the original variables we get
\begin{align*}
\widetilde{u}(r) = 1 - c\,r^\lambda + \mathcal{O}(r^4), \qquad \widetilde{h}(r) = 1 + 2c\,r^\lambda + \mathcal{O}(r^4).
\end{align*}
\end{proof}

\chapter{Relation satisfied by interaction coefficients of SNH in  \texorpdfstring{$d=4$}{d=4}}\label{sec:appDklmn}
\chaptermark{Relation for interaction coefficients}
In this Appendix we prove that in $d=4$ the resonant approximation of SNH system (\ref{eqn:5resonantFin}) satisfies requirements of the framework introduced in \cite{Bia19res}. Obviously it is a cubic resonant system and the desired symmetries in interaction coefficients indices hold by definition, hence the only non-trivial thing to check is whether $\mathcal{D}_{njkl}$ defined in Eq.\ (\ref{eqn:Dklmn}) is equal to zero if $n+j=k+l+1$.

Instead of proving this identity for coefficients $\tilde{C}_{njkl}$ we will do it for their components. Let $\tilde{S}_{njkl}=\sqrt{(n+1)(j+1)(k+1)(l+1)}S_{njkl}$, then $\tilde{C}_{njkl}=(\tilde{S}_{njkl}+\tilde{S}_{njlk})/2$. We will show the identity in question for
\begin{align}\label{eqn:Dprime}
    \mathcal{D}'_{njkl}=(n+1)\tilde{S}_{n-1,jkl}+(j+1)\tilde{S}_{n,j-1,kl}-(k+1)\tilde{S}_{nj,k+1,l}-(l+1)\tilde{S}_{njk,l+1},
\end{align}
so the fact that it follows also for the remaining part of $\mathcal{D}_{njkl}$ is straightforward from the index symmetries.

In four dimensions we have
\begin{align*}
    e_n(r)=\sqrt{\frac{2}{n+1}}L_n^{(1)}\left(r^2\right)\, e^{-r^2/2},
\end{align*}
so the interaction coefficients defined by Eq.\ (\ref{eqn:apSSNH}) can be written as
\begin{align*}
\tilde{S}_{njkl}=4\int_0^\infty \int_0^\infty \frac{L_n^{(1)}\left(r^2\right) L_j^{(1)}\left(s^2\right) L_k^{(1)}\left(s^2\right) L_l^{(1)}\left(r^2\right)}{\max\{r,s\}^2}e^{-r^2} e^{-s^2}\, r^3\, s^3\, dr\, ds.
\end{align*}
This expression can be simplified using $r^3 s^3/ \max\{r,s\}^2=rs\, \min\{r,s\}^2$ and introducing new variables $\rho=r^2$, $\sigma = s^2$. Then
\begin{align*}
\tilde{S}_{njkl}=4\int_0^\infty \int_0^\infty L_n^{(1)}\left(\rho\right)  L_j^{(1)}\left(\sigma\right) L_k^{(1)}\left(\sigma\right) L_l^{(1)}\left(\rho\right) \,\min\{\rho,\sigma\}\,e^{-\rho-\sigma}\, d\rho\, d\sigma.
\end{align*}
Now we can plug it into $\mathcal{D}'_{njkl}$ obtaining
\begin{align*}
\mathcal{D}'_{njkl}=&4 \int_0^\infty \int_0^\infty \min\{\rho,\sigma\}\,e^{-\rho-\sigma} \left[(n+1)L_{n-1}^{(1)}(\rho)L_{j}^{(1)}(\sigma)L_{k}^{(1)}(\sigma)L_{l}^{(1)}(\rho)\right.\\
&+(j+1)L_{n}^{(1)}(\rho)L_{j-1}^{(1)}(\sigma)L_{k}^{(1)}(\sigma)L_{l}^{(1)}(\rho)\\
&-(k+1)L_{n}^{(1)}(\rho)L_{j}^{(1)}(\sigma)L_{k+1}^{(1)}(\sigma)L_{l}^{(1)}(\rho)\\
&\left.-(l+1)L_{n}^{(1)}(\rho)L_{j}^{(1)}(\sigma)L_{k}^{(1)}(\sigma)L_{l+1}^{(1)}(\rho)\right]\,d\rho\, d\sigma.
\end{align*}
Now we use the identity $(n+1)L^{(1)}_{n-1}(\rho)=n L_n^{(1)}(\rho)+\rho L^{(2)}_{n-1}(\rho)$ to get rid of $L^{(1)}_{n-1}$ and $L^{(1)}_{j-1}$. It can be also written as $(k+1) L_{k+1}^{(1)}(\rho)=(k+2)L^{(1)}_{k}(\rho)-\rho L^{(2)}_{k}(\rho)$, this version can be used to substitute for $L_{k+1}^{(1)}$ and $L_{l+1}^{(1)}$. As a result, we get
\begin{align*}
\mathcal{D}'_{njkl}=&4 \int_0^\infty \int_0^\infty \min\{\rho,\sigma\}\,e^{-\rho-\sigma}\\
&\times\left[(n+j-k-l-4)L_n^{(1)}(\rho)L_j^{(1)}(\sigma)L_k^{(1)}(\sigma)L_l^{(1)}(\rho)\right.\\
&\left.+\rho\, L_{j}^{(1)}(\sigma)L_{k}^{(1)}(\sigma) \left(L_{n-1}^{(2)}(\rho)L_{l}^{(1)}(\rho)+L_{n}^{(1)}(\rho)L_{l}^{(2)}(\rho)\right)\right.\\
&\left.+\sigma\, L_{n}^{(1)}(\rho)L_{l}^{(1)}(\rho) \left(L_{j-1}^{(2)}(\sigma)L_{k}^{(1)}(\sigma)+L_{j}^{(1)}(\sigma)L_{k}^{(2)}(\sigma)\right)\right]\,d\rho\, d\sigma.
\end{align*}
In the next step, we remove $L^{(2)}_l$ and $L^{(2)}_k$ with $L^{(2)}_l(\rho)=L^{(1)}_l(\rho)+L^{(2)}_{l-1}(\rho)$. Then all the terms such as $L^{(2)}_{n-1}$ can be transformed into a derivative since $L^{(2)}_{n-1}(\rho)= - \partial_\rho L^{(1)}_n(\rho)$. It yields
\begin{align*}
\mathcal{D}'_{njkl}=&
4 \int_0^\infty \int_0^\infty \min\{\rho,\sigma\}\,e^{-\rho-\sigma}\\
&\times\left[(n+j-k-l-4+\mathcal{T})L_n^{(1)}(\rho)L_j^{(1)}(\sigma)L_k^{(1)}(\sigma)L_l^{(1)}(\rho)\right] \,d\rho\, d\sigma,
\end{align*}
where $\mathcal{T}$ denotes the linear operator equal to $\rho+\sigma-\rho\partial_\rho-\sigma\partial_\sigma$
The further simplification of this expression can be achieved with the assumption $n+j=k+l+1$, then the expression inside the parentheses becomes $(-3+\mathcal{T})$. Now we split the obtained formula into two by unwrapping $\min\{\rho,\sigma\}$:
\begin{align*}
\mathcal{D}'_{njkl}=&\,
4 \int_0^\infty \int_0^\rho \sigma \,e^{-\rho-\sigma} \left[(-3+\mathcal{T})L_n^{(1)}(\rho)L_j^{(1)}(\sigma)L_k^{(1)}(\sigma)L_l^{(1)}(\rho)\right]\,d\rho\, d\sigma\\
&+4 \int_0^\infty \int_\rho^\infty \rho \,e^{-\rho-\sigma} \left[(-3+\mathcal{T})L_n^{(1)}(\rho)L_j^{(1)}(\sigma)L_k^{(1)}(\sigma)L_l^{(1)}(\rho)\right]\,d\rho\, d\sigma.
\end{align*}
Then one can fold these expressions under the derivatives getting:
\begin{align*}
\mathcal{D}'_{njkl}=&
-4 \int_0^\infty \partial_\rho \left(\rho\,e^{-\rho} L_n^{(1)}(\rho) L_l^{(1)}(\rho) \right) \, \left(\int_0^\rho \sigma \,e^{-\sigma} L_j^{(1)}(\sigma)L_k^{(1)}(\sigma)\, d\sigma\right) \, d\rho\\
&-4 \int_0^\infty \,e^{-\rho} L_n^{(1)}(\rho) L_l^{(1)}(\rho) \left(\int_0^\rho\, \partial_\sigma\left(\sigma^2 \,e^{-\sigma} L_j^{(1)}(\sigma)L_k^{(1)}(\sigma) \right)\, d\sigma\right) \,d\rho \\
&-4 \int_0^\infty \partial_\rho \left(\rho^2\,e^{-\rho} L_n^{(1)}(\rho) L_l^{(1)}(\rho) \right) \, \left(\int_\rho^\infty \,e^{-\sigma} L_j^{(1)}(\sigma)L_k^{(1)}(\sigma)\, d\sigma\right) \, d\rho\\
&-4 \int_0^\infty \,\rho\, e^{-\rho} L_n^{(1)}(\rho) L_l^{(1)}(\rho) \left(\int_0^\rho\, \partial_\sigma\left(\sigma \,e^{-\sigma} L_j^{(1)}(\sigma)L_k^{(1)}(\sigma) \right)\, d\sigma\right) \,d\rho.
\end{align*}
The first and third terms can be simplified with integration by parts giving 
\begin{align*}
\int_0^\infty \partial_\rho \left(\rho\,e^{-\rho} L_n^{(1)}(\rho) L_l^{(1)}(\rho) \right) \, \left(\int_0^\rho \sigma \,e^{-\sigma} L_j^{(1)}(\sigma)L_k^{(1)}(\sigma)\, d\sigma\right) \, d\rho\\
=- \int_0^\infty \rho^2\,e^{-2\rho} L_n^{(1)}(\rho) L_j^{(1)}(\rho) L_k^{(1)}(\rho) L_l^{(1)}(\rho)\, d\rho.
\end{align*}
The second and fourth terms can be simply evaluated as
\begin{align*}
\int_0^\infty \,e^{-\rho} L_n^{(1)}(\rho) L_l^{(1)}(\rho) \left(\int_0^\rho\, \partial_\sigma\left(\sigma^2 \,e^{-\sigma} L_j^{(1)}(\sigma)L_k^{(1)}(\sigma) \right)\, d\sigma\right) \,d\rho \\
=\int_0^\infty \rho^2\,e^{-2\rho} L_n^{(1)}(\rho) L_j^{(1)}(\rho) L_k^{(1)}(\rho) L_l^{(1)}(\rho)\, d\rho.
\end{align*}
Hence, the first and second terms cancel each other, similarly as the third and fourth. As a result, we get $\mathcal{D}'_{njkl}=0$ when $n+j=k+l+1$, as desired. It means that also $\mathcal{D}_{njkl}=0$ and SNH resonant system in four dimensions satisfies all the necessary conditions needed for the framework from \cite{Bia19res} to work.

\chapter{Recursive scheme for calculation of interaction coefficients}\label{sec:appcoeff}
\chaptermark{Recursive scheme}
Here we describe a recursive scheme that can be used for efficient calculating large tables of interaction coefficients for resonant SNH system in any dimension $d$. The method we use is similar to the one presented in Ref.\ \cite{Cra14}. We begin with a definition of two functions
\begin{equation*} 
\mu(r)=r^{d-1},  \qquad
\nu(r)=\frac{1}{r^{d-2}}.
\end{equation*}
Then coefficients $S_{ijkl}$ defined in Eq.\ (\ref{eqn:SSNH}) may be rewritten to
\begin{align}\label{eqn:Sapp}
S_{ijkl}=\int_0^\infty \biggl[&e_i(r)e_l(r)\mu(r)\nu(r)\int_0^r e_j(s) e_k(s) \mu(s) ds\nonumber\\
&+e_i(r)e_l(r)\mu(r)\int_r^\infty e_j(s) e_k(s) \mu(s) \nu(s) ds\biggr]\, dr.
\end{align}
The functions $e_i(r)$ satisfy here $\hat{L}e_n=E_n e_n$,
where
\begin{align*}
\hat{L}f(r)=-\frac{1}{2}\frac{1}{\mu(r)}\left(\mu(r) f'(r) \right)'+\frac12 r^2 f(r), \qquad 
E_n=2n+\frac{d}{2}
\end{align*}
are the linear part of SNH equation and its eigenvalue.
The second part of Eq.\ (\ref{eqn:Sapp}) can be reformulated to
\begin{align*}
\int_0^\infty e_i(r)e_l(r)\mu(r)\int_r^\infty e_j(s) e_k(s) \mu(s) \nu(s) ds dr\\
=\int_0^\infty e_j(s)e_k(s)\mu(s)\nu(s) \int_0^s e_i(r) e_l(r) \mu(r) dr ds.
\end{align*}
If we introduce
\begin{align*}
U_{ijkl}=\int_0^\infty e_i(r)e_l(r)\mu(r)\nu(r)\int_0^r e_j(s) e_k(s) \mu(s) ds dr,
\end{align*}
then 
\begin{align}\label{eqn:R6}
S_{ijkl}=U_{ijkl}+U_{jilk}.
\end{align}
The remaining part of this appendix introduces a recursive scheme for calculation of these coefficients. We heavily use the relations
\begin{align*}
e_{n+1}(r)=&\frac{2n+\frac{d}{2}-r^2}{\sqrt{(n+\frac{d}{2})(n+1)}}e_n(r)-\sqrt{\frac{n(n+\frac{d}{2}-1)}{(n+\frac{d}{2})(n+1)}}e_{n-1}(r),\\
e'_{n}(r)=&-r^2 e_n(r)+\frac{2}{r}\sqrt{\frac{n}{n+\frac{d}{2}-1}}\Biggl[(n-1-r^2)e_{n-1}(r)\\
&-\sqrt{\frac{n-1}{(n+\frac{d}{2}-2)}}\left(n-2+\frac{d}{2}\right)e_{n-2}(r)\Biggr],
\end{align*}
that come from the recurrence relation and derivative formula for generalized Laguerre polynomials. It is very handy to write
\begin{align*}
c_n=\sqrt{n\left(n+\frac{d}{2}-1\right)}.
\end{align*}
Then the above relations for $e_{n+1}$ and $e'_{n}$ can be rewritten with the use of functions $\mu$ and $\nu$ as
\begin{subequations}\label{eqn:rec12}
\begin{align}
\mu^2 \nu^2 e_{n}=&-c_{n+1}\, e_{n+1}+\left(2n+\frac{d}{2}\right)e_n-c_n\, e_{n-1}, \label{eqn:rec1}\\
\mu \nu e'_n=& c_{n+1}\,e_{n+1}-\frac{d}{2}e_n-c_n\, e_{n-1}.\label{eqn:rec2}
\end{align}
\end{subequations}

Let us define
\begin{align*}
\chi_{ijkl}=\int_0^\infty e_i(r)e_j(r)e_k(r)e_l(r) \mu(r) dr,
\end{align*}
and
\begin{align*}
X_{ijkl}=\int_0^\infty e_i'(r)e_j(r)e_k(r)e_l(r) \mu^2(r)\nu(r) dr.
\end{align*}
For a coefficient $\chi_{ijkl}$ we introduce $L=i+j+k+l$ as a level of this coefficient. Our goal is to find a formula for a coefficient at the level $L+1$ utilising the coefficients at levels $L$ and lower.
One can use Eq.\ (\ref{eqn:rec2}) to expand $e'_i\mu\nu$ and obtain 
\begin{align}\label{eqn:R2}
X_{ijkl}=c_{i+1}\, \chi_{i+1,jkl}-\frac{d}{2}\, \chi_{ijkl}-c_i\, \chi_{i-1,jkl},
\end{align}
so the knowledge of $\chi_{ijkl}$ values automatically gives us also values of $X_{ijkl}$. To calculate $\chi_{ijkl}$ we consider an integral
\begin{align*}
\int_0^\infty e_i(r)e_j(r)e_k(r)e_l(r) \mu^2(r)\nu'(r) dr.
\end{align*}
Integration over parts, together with the relations 
\begin{equation*} \label{dmunu}
\mu'(r)\nu(r)=(d-1),  \qquad
\mu(r)\nu'(r)=-(d-2),
\end{equation*}
yields
\begin{align}\label{eqn:dchiX}
d \, \chi_{ijkl}=-X_{ijkl}-X_{jkli}-X_{klij}-X_{lijk}.
\end{align}
Every $X$ in this equation can be represented by a sum of $\chi$ coefficients with appropriate indices by Eq. (\ref{eqn:R2}). Then we have a relation between coefficients $\chi_{ijkl}$ at levels $L+1$, $L$, and $L-1$. Now we need to find a way to express a coefficient such as $\chi_{i,j+1,kl}$ with only $\chi_{i+1,jkl}$ and coefficients at levels $L$ and lower. We can consider $\mu^2\nu^2 e_i e_j$ and expand it with Eq.\ (\ref{eqn:rec1}) applied to either $\mu^2 \nu^2 e_i$ or $\mu^2 \nu^2 e_j$. The result is
\begin{align*}
\omega_i\, e_i\, e_j-c_{i+1}\, e_{i+1}\, e_j-c_i\, e_{i-1}\, e_j
=\omega_j\, e_i\, e_j-c_{j+1}\,e_i\, e_{j+1}-c_j\, e_i\, e_{j-1}.
\end{align*}
After multiplying both sides by $\mu e_k e_l$, integrating them from zero to infinity, and some algebraic operations, we obtain
\begin{align*}
\chi_{i,j+1,kl}=\frac{1}{c_{j+1}}\left[ (E_j-E_i)\chi_{ijkl}+c_{i+1}\,\chi_{i+1,jkl}+c_i\, \chi_{i-1,jkl} -c_j\,\chi_{i,j-1,kl}\right].
\end{align*}
We can use this relation with Eqs. (\ref{eqn:R2}) and (\ref{eqn:dchiX}) to obtain a recurrence formula for $\chi_{i+1,jkl}$:
\begin{align}\label{eqn:R1}
\chi_{i+1,jkl}=\frac{1}{2c_{i+1}}[&-2c_i\,\chi_{i-1,jkl}+2c_j\,\chi_{i,j-1,kl}+2c_k\,\chi_{ij,k-1,l}\nonumber\\
&+2c_l\,\chi_{ijk,l-1}+\left(d+6i-2(j+k+l)\right)\chi_{ijkl}].
\end{align}
One can also easily calculate that
\begin{align*}
\chi_{0000}=\frac{1}{2^{\frac{d}{2}-1}\Gamma\left(\frac{d}{2}\right)}.
\end{align*}
This value together with Eq.\ (\ref{eqn:R1}) and the total symmetry in indices of $\chi_{ijkl}$ gives us a recursive scheme for calculations of these coefficients. Moreover, Eq.\ (\ref{eqn:R2}) lets us now to easily compute $X_{ijkl}$ coefficients.

\vspace{\breakFF}
\begin{adjustwidth}{\marwidFF}{\marwidFF}
\small\qquad
One can notice that coefficients $\chi_{ijkl}$ introduced as an intermediate step in the calculations of interaction coefficients SNH, are in fact the interaction coefficients for GP equation. It means that Eq.\ (\ref{eqn:R1}) gives us a feasible recursive scheme for calculation in that case.
\end{adjustwidth}\vspace{\breakFF}

We focus now on $U_{ijkl}$ coefficients. Equation $\hat{L}e_j=E_j e_j$ and integration by parts lead us to
\begin{align*}
&E_j U_{ijkl}=-\frac{1}{2}X_{jikl}\\
&+\frac{1}{2}\int_0^\infty e_i(r)e_l(r)\mu(r)\nu(r)\int_0^r (e_j'(s) e_k'(s)+e_j(s)e_k(s) \mu^2(s) \nu^2(s)) \mu(s) ds dr,
\end{align*}
so
\begin{align}\label{eqn:R3}
(E_j-E_k)U_{ijkl}=\frac{1}{2}(X_{kijl}-X_{jikl}).
\end{align}
This formula gives us explicitly values of $U_{ijkl}$ coefficients in terms of $\chi_{ijkl}$, that can be calculated recursively, as long as $j\neq k$. In case $j=k$ we need to perform some additional computations. Let us consider an integral
\begin{align*}
\int_0^\infty e_i(r)e_l(r)\mu(r)\nu(r)\int_0^r e_j(s)e_k(s) \mu^3(s) \nu^2(s) ds dr.
\end{align*}
Using Eq.\ (\ref{eqn:rec1}) to expand $\mu^2\nu^2 e_j$ or $\mu^2\nu^2 e_k$ and putting $k=j+1$ one obtains
\begin{align*}
&E_j U_{ij,j+1,l}-c_{j+1}\, U_{i,j+1,j+1,l}-c_j\, U_{i,j-1,j+1,l}\nonumber\\
=&E_{j+1} U_{ij,j+1,l}-c_{j+2}\, U_{ij,j+2,l}-c_{j+1}\, U_{i,j-1,j,l}.
\end{align*}
In this formula, all coefficients $U$, except for $U_{ijjl}$ and $U_{i,j+1,j+1,l}$, have different second and third indices, so we can substitute them with coefficients $X$. It results in a recurrence relation for $U_{i,j+1,j+1,l}$:
\begin{align}\label{eqn:R5}
U_{i,j+1,j+1,l}=U_{ijjl}+ \frac{1}{2c_{j+1}}\cdot \Biggl[& X_{j+1,jil}-X_{j,j+1,il}\nonumber\\
&+\frac{1}{4}c_j\,(X_{j+1,j-1,il}-X_{j-1,j+1,il})\nonumber\\
&+\frac{1}{4}c_{j+2}\,(X_{j,j+2,il}-X_{j+2,jil})\Biggr],
\end{align}
so once we know $U_{i00l}$ we can use it to calculate coefficients of the form $U_{ijjl}$ for any $j$. The value of $U_{i00l}$ can be computed with an integral
\begin{align*}
\int_0^\infty e_i(r)e_l(r)\mu^2(r)\nu(r)\nu'(r)\int_0^r e_j(s)e_k(s) \mu(s) ds dr.
\end{align*}
One can either use $\mu(r)\nu'(r)=-(d-2)$ or perform the integration by parts and use $\mu'(r)\nu(r)$ to rewrite this integral. Comparing the effects of these two procedures and using the recurrence relations (\ref{eqn:rec1}) and (\ref{eqn:rec2}) yields
\begin{align}\label{eqn:2U}
2U_{ijkl}=&-c_{i+1}\, _{i+1,jkl}+ d\, U_{ijkl}+c_i\, U_{i-1,jkl}-c_{l+1}\,U_{ijk,l+1} +c_l\, U_{ijk,l-1}\nonumber\\
&-E_i \chi_{ijkl}+c_{i+1}\,\chi_{i+1,jkl}+c_i\,\chi_{i-1,jkl}.
\end{align}
To get rid of $U_{ijk,l+1}$ one may consider
\begin{align*}
\int_0^\infty e_i(r)e_l(r)\mu^3(r)\nu^3(r)\int_0^r e_j(s)e_k(s) \mu(s) ds dr.
\end{align*}
Use of Eq.\ (\ref{eqn:rec1}) to expend $\mu^2\nu^2 e_i$ and $\mu^2\nu^2 e_l$ gives
\begin{align*}
E_i U_{ijkl}-c_{i+1}\, U_{i+1,jkl}-c_i\, U_{i-1,jkl} =E_{l} U_{ijkl}-c_{l+1}\, U_{ijk,l+1}-c_l\, U_{ijk,l-1}.
\end{align*}
This equation, together with Eq.\ (\ref{eqn:2U}) finally yields, after the substitution $j=k=0$, the formula that lets us calculate recursively
\begin{align}\label{eqn:R4}
U_{i+1,00l}=\frac{1}{2c_{i+1}}[&(d-2+E_i-E_l)U_{i00l}+2c_l\, U_{i00,l-1}\nonumber\\
&\left.-E_i\chi_{i00l}+c_{i+1}\,\chi_{i+1,00l}+c_i\,\chi_{i-1,00l} \right].
\end{align}
Together with the initial condition that can be calculated explicitly,
\begin{align*}
U_{0000}=\frac{1}{2^{\frac{d}{2}}\Gamma\left(\frac{d}{2}\right)},
\end{align*}
and the recursive scheme for $\chi_{ijkl}$ calculations, it poses the full system of equations needed to calculate $U_{i00l}$ for any $i$ and $j$.

The final algorithm for computing $S_{ijkl}$ for any value of $i$, $j$, $k$, and $l$ begins with a decomposition into coefficients $U$, as in Eq.\ (\ref{eqn:R6}). Then the calculations of their values depend on whether the second and third indices are different. If so, one uses Eq.\ (\ref{eqn:R3}), where $X$ are calculated with the use of Eq.\ (\ref{eqn:R2}) with $\chi$ that can be quickly calculated with the recurrsive formula (\ref{eqn:R6}). If the indices $j$ and $k$ in $U_{ijkl}$ match, one has to calculate $U_{i00l}$ using the recurrence Eq.\ (\ref{eqn:R4}) first, and then raise it to $U_{ijjl}$ with Eq.\ (\ref{eqn:R5}).
\end{appendices}

\newpage\null\thispagestyle{empty}\newpage

\newpage\null\thispagestyle{empty}\newpage

\includepdf{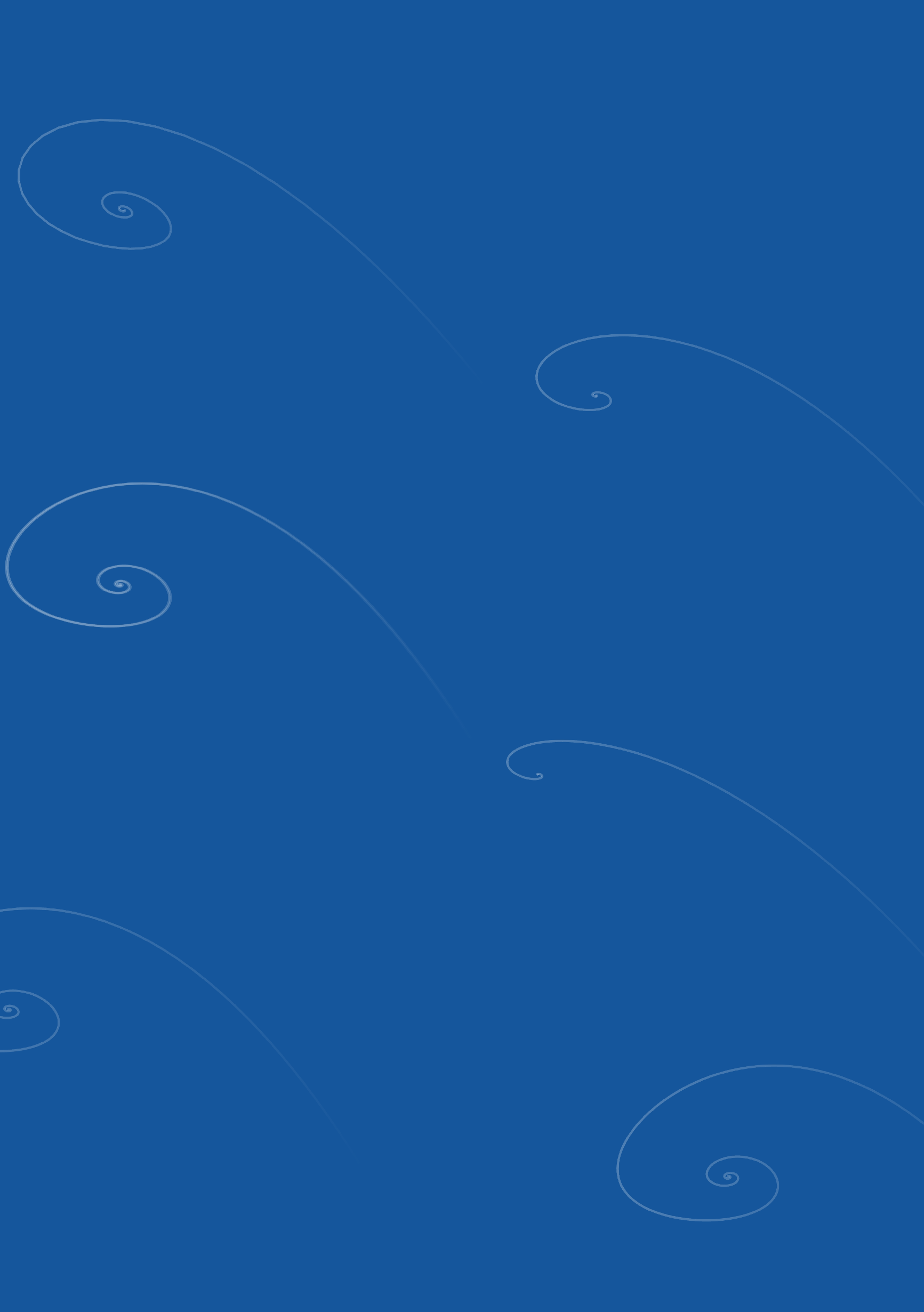}
\end{document}